\documentclass[a4paper,11pt]{report}
\usepackage{rotating}
\usepackage{epsfig}
\usepackage{latexsym}
\usepackage{psfrag}
\usepackage{amssymb}
\usepackage[centertags]{amsmath}
\usepackage{amsfonts}
\usepackage{amssymb}
\usepackage{amsthm}
\usepackage{newlfont}
\usepackage{graphicx}
\usepackage[spanish,american]{babel}
\usepackage{textcomp}
\usepackage{slashed}
\usepackage{dsfont}
\usepackage{caption}
\usepackage{fancyhdr}
\newlength{\defbaselineskip}
\newcommand{\setlinespacing}[1]%
           {\setlength{\baselineskip}{#1 \defbaselineskip}}

\newcommand{\be}{\begin{equation}}
\newcommand{\ee}{\end{equation}}
\newcommand{\bea}{\begin{eqnarray}}
\newcommand{\eea}{\end{eqnarray}}

\newcommand{\cO}{{\cal O}}
\newcommand{\nn}{\nonumber}

\newcommand{\Tr}{\mathrm{Tr}}
\newcommand{\clearemptydoublepage}{\newpage{\thispagestyle{empty}\cleardoublepage}}

\def\baselinestretch{1.1}
\catcode`@=11
\def\marginnote#1{}
\newcount\hour
\newcount\minute
\newtoks\amorpm
\hour=\time\divide\hour
by60
\minute=\time{\multiply\hour by60 \global\advance\minute
by-\hour}
\edef\standardtime{{\ifnum\hour<12 \global\amorpm={am}
\else\global\amorpm={pm}\advance\hour by-12 \fi
 \ifnum\hour=0
\hour=12 \fi
 \number\hour:\ifnum\minute<10
0\fi\number\minute\the\amorpm}}
\edef\militarytime{\number\hour:\ifnum\minute<10
0\fi\number\minute}
\def\draftlabel#1{{\@bsphack\if@filesw
{\let\thepage\relax
 \xdef\@gtempa{\write\@auxout{\string
\newlabel{#1}{{\@currentlabel}{\thepage}}}}}\@gtempa
 \if@nobreak
\ifvmode\nobreak\fi\fi\fi\@esphack}
\gdef\@eqnlabel{#1}}
\def\@eqnlabel{}
\def\@vacuum{}
\def\draftmarginnote#1{\marginpar{\raggedright\scriptsize\tt#1}}
\def\draft{\oddsidemargin
0.0truein
 \def\@oddfoot{\sl preliminary draft \hfil
\rm\thepage\hfil\sl\today\quad\militarytime}
 \let\@evenfoot\@oddfoot
\overfullrule 3pt
 \let\label=\draftlabel
\let\marginnote=\draftmarginnote
\def\@eqnnum{(\theequation)\rlap{\kern\marginparsep\tt\@eqnlabel}
\global\let\@eqnlabel\@vacuum}
}
\catcode`@=12

\begin{document}

\thispagestyle{empty}

\ \vspace{3cm}

\begin{center}

{\LARGE \textbf{Rational Approximations in Quantum Chromodynamics}}

\vspace{1.8cm}

{\Large Pere Masjuan Queralt}

\vspace{0.5cm}
University of Vienna\\
Institut de F\'{\i}sica d'Altes Energies, Universitat Aut\`{o}noma de Barcelona\\
$\cdot$ December 2009 $\cdot$

\vspace{0.2cm}

\end{center}

\vspace{3.8cm}

\begin{center}
\includegraphics[height=3cm,width=5cm]{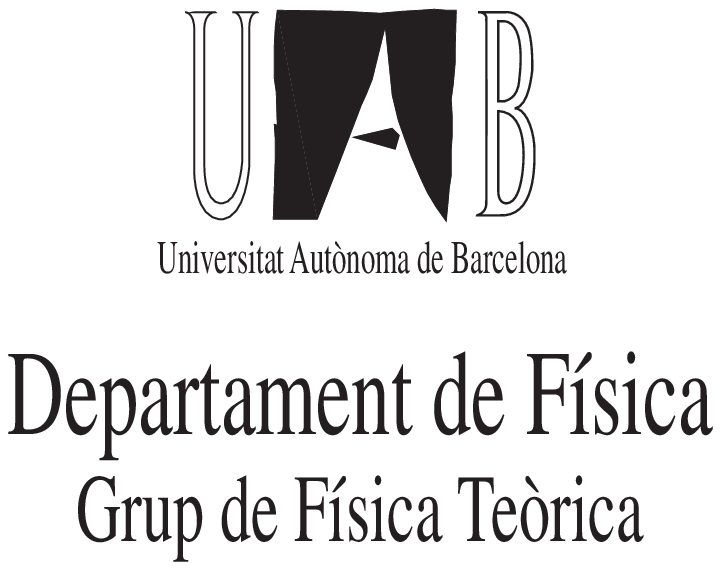}
\includegraphics[height=3cm,width=5cm]{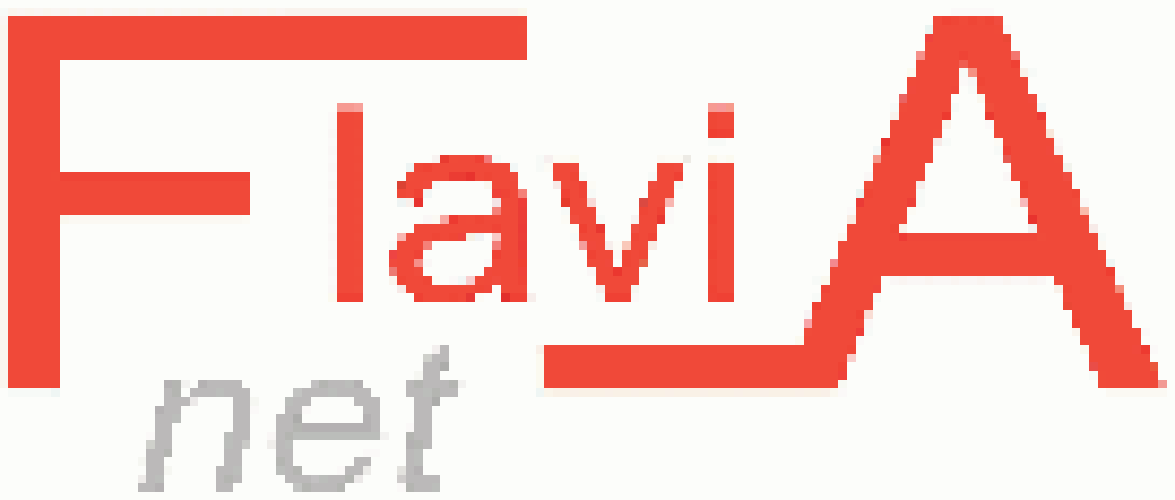}
\end{center}

\vspace{1cm}

\begin{center}
\nn\underline{\ \ \ \ \ \ \ \ \ \ \ \ \ \ \ \ \ \ \ \ \ \ \ \ \ \ \ \ \ \ \ \ \ \ \ \ \ \ \ \ \ \ \ \ \ \ \ \ \ \ \ \ \ \
\ \ \ \ \ \ \ \ \ \ \ \ \ \ \ \ \ \ \ \ \ \ \ \ \ \ \ \ \ \ \ \ \ \ \ \ \ \ \ \ \ \ \ \ \ \ }\\

\vspace{0.3cm}

\parbox{12.5cm}{\footnotesize \emph{Submitted in partial fulfillment of the requirements for the degree of Doctor of Philosophy.}}

\end{center}

\clearemptydoublepage

\thispagestyle{plain}

\pagenumbering{roman}


\chapter*{Foreword}

\quad This PhD thesis has been the work of 3 years at the IFAE and F\'{\i}sica Te\`{o}rica group of the Universitat Aut\`{o}noma de Barcelona. I would like to thank all my colleagues, who have collaborated to create a warm atmosphere of work and friendship over this time. Among them I want to mention Javier Redondo, Germano Nardini and Juan Jos\'{e} Sanz Cillero whom I shared the office for the last three years; and all those colleagues and friends who have joined the group all this time: Rafel Escribano, Javi Virto, Oriol Romero, Antionio Pic\'{o}n, David Diego, Carles Rod\'{o}, Felix Schwab, Alessio Provenza, Jordi Nadal, Lluis Galbany, Javi Serra, Oriol Domenech, Mariona Aspachs, Simone Paganelli, Marc Ramon, Diogo Boito, David Greynat...

Special thanks to my PhD advisor, Dr. Santi Peris, for the patience, understanding and all the time he has spent to teach me. Under his tuition I have learned more than I could imagine and I hope that is just the beginning.

I would also like to acknowledge the warmth and hospitality I have found abroad in my short stay in Wien, Austria. Specially thanks to Prof. Neufeld and Prof. Ecker for the nice discussions and collaborations. This work was supported by EU Contract No.MRTN-CT-2006-035482, "FLAVIAnet".

This PhD Thesis is dedicated to my family, el pare, la mare i la Maria, and with admiration and gratitude a la Gemma.

\tableofcontents

\pagenumbering{arabic}

\chapter{Introduction}

\quad Quantum Chromodynamics, as a gauge theory of strong interaction, is one of the major achievements of particle physics in the last 50 years. However, at sixties of the last century, particle physics was not compelling enough. Quantum Electrodynamics produced increasingly predictions confirmed experimentally. Few years later, with the electroweak gauge theory of the Standard Model, also the weak interactions were included in the well defined framework of Quantum Field Theories. At that time, among all the fundamental interactions, the strong interactions still remained outside that description. With the invention of Bubble and Spark chambers, experimental physics discovered a large number or particles called hadrons. At sixties, a glance at the hadron spectroscopy show that the increasing number of hadrons could be classified successfully by the quark model of Gell-Mann and Zweig. However, the dynamics behind the quark model was still a mystery. The interpretation of strong interactions in the framework of Quantum Field Theories was useful but only at the level of a toy model and Gell-Mann himself, who suggest this approach, declared the quarks to be purely mathematical entities without any physical meaning. Hard work on the experimental side was needed to finally conclude with the acceptance of quarks been elementary particles which, encoded in a $SU(3)_c$ gauge symmetry, end up in the so-called QCD Lagrangian yielding the proper description of the strong interactions.

The QCD Lagrangian is described then in terms of quarks and gluons which, due to confinement, can not be detected in isolation, but only colorless combinations of them, the hadrons. At high energies, however, asymptotic freedom renders the strong coupling to be small enough that perturbation theory becomes the appropriate tool to work with. At energies below 1 GeV, confinement enters in the game binding quarks and gluons together and invalidating the perturbative technics. Models like the Constituent Quark Model, numerical simulations such as lattice QCD or analytical approximations to QCD are among the non-perturbative technics used nowadays to explore the non-perturbative regime.

Also Effective Field theories have successfully contribute to a better understanding of the properties of QCD, specially in its low-energy regime. In particular Chiral Perturbation Theory as a systematical organization of the physics in powers of momenta and quark masses resulted in a good tool for describing the low-energy phenomenology, but requires the knowledge of several low-energy constants to improve on the results.

On the other hand, the large-$N_c$ expansion standed out as a promising analytical approach capable of dealing with the complexities of non-perturbative QCD regime while offering a simple description of the physics. However, a limitation to doing phenomenology was found due to the fact that there is still no solution for large-$N_c$ QCD. On the other hand, the phenomenological approach of resonance saturation works quite well for QCD Green's functions. It was then realized that this successfully result could be encompassed at once as an approximation to large-$N_c$ QCD consisting in keeping only a finite (instead of the original infinite in large-$N_c$) set of resonances in Green's functions, the Minimal Hadronic Approximation.

The present work is devoted in a first part to study how can we understand the Minimal Hadronic Approximation an the limitations that appear when dealing with high-energy matching conditions with a finite number of resonances and how reliable is to extract information on individual mesons from it. We suggest then that resonance saturation in QCD can be understood in the large-$N_c$ limit from the mathematical theory of Pad\'{e} Approximants to meromorphic functions. Due to the success of Pad\'{e} Theory in the framework of large-$N_c$ QCD thanks to the convergence theorems described in the literature, we also apply this technic in a second part to other open questions, such as what is the insight behind the unitarization processes at low energies, how to work out the possible ambiguities appearing in the approximations to the vacumm polarization of a heavy quark and, finally, how to go on improving the Vector Meson Dominance when dealing with experimental data.

The outline of the present work will be as follows: in chapter \ref{capitolintro} we give a brief reminder of the basics of Quantum Chromodynamics together with its low energy chiral realization (Chiral Perturbation Theory) and its high energy expansion through the Operator Product Expansion. We then introduce the $1/N_c$ expansion technic to deal with extensions of the chiral Lagrangian to intermediate energies with the explicit introduction of resonance fields (what is called Resonance Chiral Lagrangian). Since this lagrangian needs an infinite number of resonances and in practice this request can not we easily attended, we also show the approximations that has been used in the past years, mainly the Minimal Hadronic Approximation, which is based in the phenomenological success of resonance saturation.

Chapter \ref{capitol1} introduces the notion of Pad\'{e} Approximant in a brief but precise mathematical way, with several examples to illustrate its usefulness and, more important, all the theorems of convergence that will we needed to tackle all the raised question.

Chapter \ref{capitol2} is focused on the application of Pad\'{e} Theory to meromorphic functions since in the framework of large-$N_c$ QCD it turns out that resonance saturation are rational functions which encompass any saturation with a finite number of resonances. In particular the consequences of the Pommerenke's theorem of convergence applied to rational functions are studied and it is shown that while this rational approximants may reliably describe a Green's functions in the Euclidean, the same is not true for the Minkowski regime. As an example, we estimate the value of low-energy constants for the $\langle VV-AA \rangle$ two-point correlator.

Chapter \ref{capitol4} is devoted to study the reliability of the Pad\'{e} Theory applied to Stieltjes functions when unitarizing low energy amplitudes or when extracting threshold parameters from the vacuum polarization function of a heavy quark.

In chapter \ref{capitol3} we illustrate how Pad\'{e} Approximants can go beyond the vector meson dominance approach with the case of the pion vector form factor in the spacelike region and estimate the constants that appear in its low-energy realization. It is shown how Pad\'{e} Approximants are useful to incorporate information from high energies in a straightforward way.

Finally, we end with the conclusions.

\chapter{Quantum Chromodynamics}\label{capitolintro}

\quad The Standard Model of Particle Physics (SM) is an extremely successful theory of physics at the subnuclear scale\footnote{We will not present a detailed discussion about Standard Model, but just a brief overview of the main points, focused on the strong interaction sector. For comprehensive and general accounts, we refer the reader to \cite{PichSM,NovaesSM}.}. Starting from three generations of quarks and leptons subject to a local $SU(3)_c \times SU(2)_L \times U(1)_Y$ symmetry, it describes all observed phenomena known today (without including gravity effects) within theoretical and experimental errors. The first factor $SU(3)_c$ accounts for the strong interactions, whose discussion will be the main
subject of this chapter. The remaining $SU(2)_L\times U(1)_Y$ factor is responsible for the electroweak interactions.

Our understanding of the electroweak and strong interactions is good and is based on a well defined mathematical theory. Actually, in the electroweak sector (for example Quantum ElectroDynamics, QED) successful mathematical tools are used, i.e., Perturbation Theory (PT) as an expansion of the coupling constant. However, the strong sector (Quantum ChromoDynamics) has a regime where PT can be applied and another regime where it can not. The reason is because the PT breaks down in certain regime called non-perturbative Quantum ChromoDynamics. Therefore, some efforts have been made towards the development of new tools that allow reliable computations in this non-perturbative regime, for example the lattice gauge theories or analytical approximations.
In the present chapter, we will present the main properties of the strong interactions, while focuss in both low-energy and high-energy regimes. We will also present several attempts done in the literature to try to link both regimes in a systematical way. In that sense, we will review beyond Quantum ChromoDynamics, Chiral Perturbation Theory, the Operator Product Expansion technics, the Large-$N_c$ limit in Quantum ChromoDynamics and the Resonance Chiral Theory.

\section{Quantum Chromodynamics: Introduction}

\quad Quantum Chromodynamics (QCD)\footnote{For a pedagogical reviews see for example \cite{PichQCD}, \cite{EckerQCD}} is nowadays the established theory of the strong interactions, a fundamental force describing the interactions of the quarks and the gluons. Mathematically, is a $SU(3)$ Yang-Mills theory of color-charged entities. \textit{Color}, in this framework, is a new quantum number corresponding to a $SU(3)_c$ symmetry, $c$ standing for color. We would like to present some evidence of its existence before writing down the corresponding Lagrangian.

The Particle Data Tables \cite{PDG} reveals the richness and variety of the hadronic spectrum. In the sixties, the rapidly increasing number of hadrons discovered could be classified successfully by the quark model (with the three quarks $u$, $d$ and $s$) \cite{Quarkmodel}, showing the existence of a deeper level of elementary constituents of matter, the quarks. Assuming that mesons are $q\bar{q}$ states (quark-antiquark), while baryons are $qqq$ states, the quark model allow us to classify the entire hadronic spectrum as a Periodic Table of Hadrons.
However, the quark picture faces a basic problem suggesting the presence of a new degree of freedom: a fundamental state of composite system is expected to have angular momentum zero, a baryon ($\Delta^{++}$) with total angular momenta $J=3/2$ corresponds to $uuu$ with the three quark-spins aligned into the same direction and with the relative angular momenta equal to zero. In that case, the wave function is symmetric and our state obeys a wrong Fermi-Dirac statistics. A new degree of freedom, a new quantum number for quarks, \textit{the color}, can solves this problem because our state can be totally antisymmetric in these color indices (at least 3 colors are needed for making an antisymmetric state) respecting the generalized Pauli principle with spatially symmetric wave function.

However, assuming that the number of colors $N_c=3$ for each quark and antiquark, we obtain $9\times9=81$ combination of $q\bar{q}$ only nine of which had been found. In order to avoid the existence of non-observed extra states with non-zero color, one needs to further postulate that all asymptotic state are colorless, singlets under rotations in color space. This postulate, called the \textit{confinement hypothesis}, implies the non-observability of free quarks since they carry color.

The decisive clue to manifest the existence of the color of the quarks came from experiment. Started by the MIT-SLAC collaboration at the end of the
sixties, deep inelastic scattering of leptons on nucleons and nuclei produced unexpected results. Whereas at low energies the cross sections were characterized by baryon resonance production, the behavior at large energies and momentum transfer suggested that the nucleons seemed to consist of
noninteracting partons. Obvious candidates for the partons were the quarks but this idea led to a seeming paradox. How could the quarks be quasi-free at high energies and yet be permanently bound in hadrons, a low-energy manifestation?

That the strength of an interaction could be energy dependent was not really new to theorists. In Quantum ElectroDynamics, the vacuum acts like a polarisable medium leading to the phenomenon of charge screening, but the effective charge increases with energy. In the case of QCD, the deep inelastic experiments seemed to suggest the clue to understand that phenomena.


The deep inelastic scattering measurements actually requires the existence of electrically neutral as well as charged constituents of the proton. Charged partons can be identified with the colored quarks and the neutral partons with gluons. These two identifications are needed to have direct evidence that quantum chromodynamics is the correct physical theory of strong interactions. Its essential properties are:
\begin{itemize}
\item Quarks carry color as well as electric charge; there are three colors: Red, Green and Blue.
\item Color is exchanged by eight bicolored gluons, which are massless and have spin $1$.
\item Color interactions are assumed to be like the electromagnetic once (in a quark-gluon interaction use the rules of $QED$ with the substitution $\sqrt{\alpha}\rightarrow \sqrt{\alpha_s}$).
\item Gluons themselves carry color charge, and so they can interact with other gluons.
\item At short distances (large$-Q^2$), $\alpha_S$ is sufficiently small so that we can compute color interactions using the perturbative techniques.
\end{itemize}




The strong interactions, then, must have the property of asymptotic freedom (Groos and Wilczek \cite{GrossWilczekNobel}, Politzer 1973 \cite{PolitzerNobel}): the interactions in the strong sector should become weaker at short distances, so that quarks behave as free particles for $Q^2\rightarrow \infty$. This also agrees with the empirical observation that the free-quark description of the ratio

\be
\label{eq:R_ee}
R_{e^+e^-} \equiv
{\sigma(e^+e^-\to \mbox{\rm hadrons})\over\sigma(e^+e^-\to\mu^+\mu^-)} \, .
\ee

works better at higher energies.
    Thus, the interaction between a $q\bar q$ pair looks like some kind of
rubber band. If we try to separate the quark form the antiquark
the force joining them increases.
At some point, the energy on the elastic band is larger than $2m_{q'}$, so
that it becomes energetically favorable to create an additional $q'\bar q'$
pair; then the band breaks down into two mesonic systems, $q\bar q'$ and
$q'\bar q$, each one with its corresponding half-band joining the quark pair.
Increasing more and more the energy, we can only produce more and more
mesons, but quarks remain always confined within color--singlet bound states.
Conversely, if one tries to approximate two quark constituents into
a very short-distance region, the elastic band loses the energy and becomes
very soft; quarks behave then as free particles. The ratio in Eq. (\ref{eq:R_ee}) is then given by the sum of the quark electric charges squared:
\begin{equation}
R_{e^+e^-}\approx N_C\, \sum_{f=1}^{N_f} Q_f^2=\,
\left\{
\begin{array}{cc}
\frac{2}{3} N_C = 2\, , \qquad & (N_f=3 \; :\; u,d,s)  \\
\frac{10}{9} N_C = \frac{10}{3}\, , \qquad & (N_f=4 \; :\; u,d,s,c)  \\
\frac{11}{9} N_C = \frac{11}{3} \, ,\qquad & (N_f=5 \; :\; u,d,s,c,b)
\end{array}
\right.
\end{equation}

Notice that strong interactions have not been taken into account, only the confinement hypothesis has been used.

The hadronic decay of the $\tau$ lepton provides additional evidence for $N_C=3$. The branching rations for the different channels are expected to be approximately:

\be
\label{eq:R_tau}
B_{\tau\to lepton} \equiv Br(\tau^- \to \nu_l l^- \bar{\nu}_l)\approx \frac{1}{2+N_c}= \frac{1}{5}
\ee
for $N_c=3$ which should be compared with the experimental average \cite{PDG}, for example. $B_{\tau\to e} = (18.01\pm0.18)\%$.

Finally, we would like to comment also on the anomalies as another compelling reason to adopt $N_c$. An anomaly is a global symmetry which is broken by quantum effects. In particular, the theoretical prediction for the decay $\pi^0 \to \gamma \gamma$, which occurs through a triangular quark loop related by the anomaly, needs $N_c=3$ to predict the experimental value. Also, anomalous triangle diagrams involving electroweak gauge bosons could spoil renormalization
of the Standard Model. This potentially dangerous contributions are canceled due to a subtle combined action of quarks and leptons in each of the three generations, provided that quarks appear with three different colors, $N_c = 3$.

\subsection{QCD Lagrangian}

\quad Strong interactions, then, bind quarks inside the atomic nuclei due to the mediation of gluons,
the gauge bosons of the color group. The interaction is described by a Yang-Mills quantum field
theory under the non-abelian gauge group $SU(3)_C$ of color. Taking the proper normalization for the gluon kinetic term, we finally have
the invariant QCD Lagrangian:

\be
\label{eq:L_QCD}
{\cal L}_{QCD} \,\equiv\, -\frac{1}{4}\, G^{\mu\nu}_a G_{\mu\nu}^a
+ \sum_f\,\bar{q}_f \, \left( i\gamma^{\mu} D_{\mu} - m_f \right)\, q_f \, .
\ee

We required the Lagrangian Eq.~(\ref{eq:L_QCD}) to be also invariant under {\it local} $SU(3)_C$ transformations, $\theta_a = \theta_a(x)$. To
satisfy this requirement, we needed to change the quark derivatives by covariant objects. Since we have now 8 independent gauge parameters,
8 different gauge bosons $G^\mu_a(x)$, also the {\it gluons} are needed:

\be
\label{eq:D_cov}
D^{\mu} q_f \,\equiv\, \left[ \partial^{\mu} - i g_s \frac{\lambda^a}{2}
G^{\mu_a}(x)\right]\, q_f \, \equiv\, \left[ \partial{^\mu} - i g_s G^{\mu}(x)\right] \, q_f \, .
\ee

Notice that we have introduced the compact matrix notation

\be
\label{eq:G_matrix}
  [G^{\mu}(x)]_{\alpha\beta}\,\equiv\,
\left(\frac{\lambda^a}{2}\right)_{\!\alpha\beta}\, G^{\mu_a}(x) \, .
\ee

where $\lambda^a$ ($a=1,2,\ldots,8$) denote the generators of the fundamental representation of the
$SU(3)_C$ algebra (the Gell-Mann matrices $\lambda^a$). The matrices $\lambda^a$ are traceless and satisfy the commutation relations

\be
\label{eq:commutation}
\left[\lambda^a,\lambda^b\right] =
2 i f^{abc} \, \lambda^c \, ,
\ee

with $f^{abc}$ the $SU(3)_C$ structure constants, which are real and totally antisymmetric.

The gauge transformation of the gluon fields is more complicated than the one obtained in QED for the photon. The first part of Eq.~(\ref{eq:L_QCD}) and the non-commutativity of the $SU(3)_C$ matrices gives rise to an additional term involving the gluon fields themselves. The gluon fields belong to the adjoint representation of the color group. Note also that there is a unique $SU(3)_C$ coupling $g_s$, which is called the {\it strong coupling constant}. The existence of self-interactions among the gauge fields is a new feature that was not present in QED; it seems then reasonable to expect that these gauge self-interactions could explain properties like asymptotic freedom and confinement, which do not appear in QED either.

The asymptotic freedom, due to the fact that the strong coupling is small at high energies, let quarks be unbounded and behave as free particles. In that regime, the parton model is explained and perturbation theory techniques are allowed.

Confinement, however, is found at low energies where quarks become more and more tightly bounded due to the increasing of the value of the strong coupling constant, entering in a non-perturbative regime. This is also known as infrared slavery. Confinement makes the quark-gluon picture transform to the hadronic picture we
observe in particle accelerators.

The picture is then quite intricate: with quarks and gluons we are able to write down the QCD Lagrangian Eq.~(\ref{eq:L_QCD}) but we do not know how to evolve it to its dual description in terms of the asymptotic hadronic states.
Several technics were developed to tackle this problem under effective field theories and symmetry arguments. Among them, Chiral Perturbation Theory provides the most general framework to deal with asymptotic states at low energies, and will be the remaining of the next section.

\section{Chiral Perturbation Theory}\label{ChPT}



\quad Chiral Perturbation Theory (ChPT)\footnote{For a pedagogical review see for example \cite{DeRafaelChPT}, \cite{SchererChPT}, \cite{EckerChPT}, \cite{BijnensChPT}, \cite{PichChPT}} provides a systematic framework for investigating strong-interaction processes at \textit{low} energies, as opposed to a perturbative treatment of QCD at high momentum transfer in terms of the $"$running coupling constant$"$. The basis of ChPT is the global $SU(3)_L\otimes SU(3)_R\otimes U(1)_V$ symmetry of the QCD Lagrangian in the limit of massless $u$, $d$, and $s$ quarks. This symmetry is assumed to be spontaneously broken down to $SU(3)_V\otimes U(1)_V$ giving rise to eight massless Goldstone bosons. Actually, below the resonance region ($E<M_{\rho}$), the hadronic spectrum only contains an octet of very light pseudoscalar particles ($\pi, K, \eta$), whose interactions can be easily understood using this symmetry consideration.

The ChPT formalism is then based on two key ingredients: the chiral symmetry properties of QCD and the concept of Effective Field Theory (EFT) \cite{GeorgiEFT,ManoharEFT,PichEFT}.

Effective field theories are the appropriate theoretical tools to take explicitly into account the relevant degrees of freedom for the process under investigation, i.e., those states with $m<<\Lambda$, while the heavier excitations with $M>>\Lambda$ are integrated out from the action. The information on the heavier degrees of freedom is then contained in the couplings of the resulting low-energy Lagrangian, the Low-Energy Constants (LECs). Although effective field theories contain an infinite number of terms at a given order in the energy expansion, the low-energy theory is specified by a finite number of couplings; this allows for an order-by-order renormalization \cite{GLsq}.

On the chiral symmetry side, the other ingredient of Chiral Perturbation Theory, it turns out that in the limit of massless $u,d$ and $s$ quarks the QCD Lagrangian Eq. (\ref{eq:L_QCD}) has a global symmetry spontaneously broken. Actually,

\begin{equation}
{\cal L}_{QCD}^0=-\frac{1}{4}G_{\mu \nu}^a G_a^{\mu \nu}+ i \bar{q}_L\gamma^{\mu}D_{\mu}q_L+ i\bar{q} _R\gamma^{\mu}D_{\mu}q_R
\end{equation}
and is invariant under the global $G\equiv SU(N_f)_L\otimes SU(N_f)_R \otimes U(1)_V\otimes U(1)_A$ transformations of the left and right handed quarks in flavour space:
\begin{equation}
q_L\rightarrow g_L q_L, \qquad q_R\rightarrow g_R q_R, \qquad g_{L,R} \in SU(N_f)_{L,R}\, .
\end{equation}
The two quark chiralities live in separate flavour spaces implying that all previous flavor symmetries get duplicated in two chiral sectors ($SU(N_f)_L$ and $SU(N_f)_R$ spaces). On the other hand, under $U(1)_V$ all quarks have the same change in phase and corresponds to the baryon number while under $U(1)_A$ the right and left-handed quarks have the opposite change in phase. The symmetry is called \textit{chiral} because it acts differently on the left and right-handed quarks. The $U(1)_A$ is only a symmetry of the classical action, not of the full quantum theory of QCD, it is an anomaly and, therefore, the divergence of the associated current does not vanish. It is nonzero by a total divergence but instantons allow for this to have a physical effect. In sum, the final symmetry of QCD in the limit where all quarks are massless, and taken in to account only the meson sector, is thus
\begin{equation}
G_{\chi}=SU(3)_L\otimes SU(3)_R
\end{equation}
The Noether currents associated with the chiral group $G_{\chi}$ are
\begin{equation}
J_X^{a \mu}=\bar{q}_X\gamma^{\mu}\frac{\lambda_a}{2}q_X, \qquad (X=L, R;\quad a=1,..., 8).
\end{equation}

where $\lambda_a$ are the Gell-Mann's matrices with $Tr(\lambda_a\lambda_b)=2\delta_{ab}$. The corresponding Noehter charges $Q_X^a=\int d^3x J_X^{a0}(x)$ satisfy the familiar commutation relations
\begin{equation}
[Q_X^a, Q_Y^b]=i\delta_{XY}f_{abc}Q_X^c
\end{equation}
which were the starting point of the Current Algebra methods of the sixties.
This chiral symmetry, which should be approximately good in the light quark sector $(u, d, s)$, is however not seen in the hadronic spectrum. Although hadrons can be classified in $SU(3)_V$ representations, degenerate multiplets with opposite parity do not exist. Moreover, the octet of pseudscalar mesons happens to be much lighter than all the other hadronic states. To be consistent with this experimental fact, the ground state of the theory (the vacuum) should not be symmetric under the chiral group. The $SU(3)_L\otimes SU(3)_R$ symmetry spontaneously breaks down to $SU(3)_{L+R}$ and, according to Goldstone's theorem, an octet of pseudoscalar massless bosons appears in the theory ($\pi$, K, $\eta$).

\subsection{Effective chiral Lagrangian at lowest order}\label{sec chiral lowest order}

\quad The Goldstone nature of the pseudoscalar mesons implies strong constraints on their interactions, which can be most easily analyzed on the basis of an effective Lagrangian. Since there is a mass gap separating the pseudoscalar octet from the rest of the hadronic spectrum, we can build an effective theory containing only the Goldstone modes. Our basic assumption is the pattern of Spontaneous Chiral Symmetry Breaking ($S\chi SB$) where the quark condensate

\begin{equation}
\langle 0|\bar{u}u|0\rangle = \langle 0|\bar{d}d|0\rangle=\langle 0|\bar{s}s|0\rangle\neq0
\end{equation}

is its natural order parameter. This pattern reads:

\begin{equation}
G_{\chi}=SU(3)_L\otimes SU(3)_R\rightarrow H\equiv SU(3)_V
\end{equation}

Let us denote $\phi^a (a=1,..., 8)$ the coordinates describing the Goldstone fields in the coset space $G_{\chi}/H$, and choose a coset representative $\bar{\xi}(\phi)\equiv(\xi_L\xi(\phi)$,$\xi_R(\phi))\in G_{\chi}$. The change of the Goldstone coordinates under a chiral transformation $g\equiv(g_L, g_R)\in G_{\chi}$ is given by

\begin{equation}
\xi_L(\phi)\rightarrow g_L\xi_L(\phi)h^{\dag}(\phi,g), \qquad \xi_R(\phi)\rightarrow g_R\xi_R(\phi)h^{\dag}(\phi,g)
\end{equation}

where $h(\phi, g) \in H$ is a compensating transformation which is needed to return to the given choice of coset representative $\bar{\xi}$. Since the same transformation $h(\phi, g)$ occurs in the left and right sectors, we can get rid of it by combining the two chiral relations above into the simpler form

\begin{equation}
U(\phi)\rightarrow g_R U(\phi)g_L^{\dag}, \qquad  U(\phi)\equiv \xi _R(\phi)\xi_L^{\dag}(\phi).
\end{equation}

Without lost of generality, we can take a canonical choice of coset representative such that $\xi_R=\xi_L^{\dag}\equiv
u(\phi)$. The $3 \times 3$ unitary matrix

\begin{equation}
U(\phi)=u(\phi)^2=exp\bigg\{\frac{i}{f_0}\sqrt{2}\phi\bigg\}
\end{equation}

gives a very convenient parametrization of the Goldstone fields

\begin{equation}
\phi(x)\equiv \sum_a \frac{\lambda_a}{\sqrt{2}}\phi_a=\left(\begin{array}{ccc}\frac{1}{\sqrt{2}}\pi^0+\frac{1}{\sqrt{6}}\eta_8 & \pi^+ & K^+\\ \pi^- & -\frac{1}{\sqrt{2}}\pi^0+\frac{1}{\sqrt{6}}\eta_8 & K^0 \\ K^- & \bar{K}^0 & \frac{-2}{\sqrt{6}}\eta_8\end{array}\right)
\end{equation}

To get a low-energy effective Lagrangian realization of QCD for the light-quark sector $(u, d, s)$, we should write the most general Lagrangian involving the matrix $U(\phi)$, which is consistent with chiral symmetry. The Lagrangian can be organized in terms of increasing powers of momentum (or number of derivatives):

\begin{equation}
{\cal L}_{eff}=\sum_n {\cal L}_{2n}
\end{equation}

In the low-energy domain we are interested in, the terms with minimum number of derivatives will dominate.
To lowest order, the effective chiral Lagrangian is uniquely given by the term (a proof can be found in \cite{Scherer:2002tk})

\begin{equation}
\label{l2}
{\cal L}_2=\frac{f_0^2}{4}\langle \partial_{\mu}U^{\dag}\partial^{\mu}U \rangle
\end{equation}

(where $\langle A \rangle = tr A$) or expanding $U(\phi)$ in power series in the gold field $\phi$

\begin{equation}
\label{l2exp}
{\cal L}_2=\frac{1}{2}\langle \partial_{\mu}\phi\partial^{\mu}\phi \rangle + \frac{1}{12 f_0^2}\langle (\phi\partial_{\mu}\phi)(\phi\partial_{\mu}\phi)\rangle + O(\phi^6/f_0^4).
\end{equation}

With this expansion one obtains the Goldstone kinetic terms plus a tower of interactions involving and increasing number of pseudoscalars.

This effective field theory technique becomes much more powerful if one introduces couplings to external classical fields. Let us consider an extended QCD Lagrangian, with quark couplings to external Hermitian matrix-valued fields $v_{\mu}$, $a_{\mu}$, $s$, $p$:

\begin{equation}
\label{lqcd}
{\cal L}_{QCD}={\cal L}_{QCD}^0 + \bar{q}\gamma^{\mu}(v_{\mu}+\gamma_5a_{\mu})q -\bar{q}(s-i\gamma_5p)q
\end{equation}

The external fields will allow us to compute the effective realization
of general Green functions of quark currents in a very straightforward
way. Moreover, they can be used to incorporate the
electromagnetic and semileptonic weak interactions, and the
explicit breaking of chiral symmetry through the quark masses:

\begin{equation}
\begin{array}{l}
\label{eq:breaking}
r_\mu \,\equiv\, v_\mu + a_\mu \, = \, e Q A_\mu + \ldots \\
\ell_\mu \, \equiv \, v_\mu - a_\mu \, = \,  e Q A_\mu + {e\over\sqrt{2}\sin{\theta_W}}(W_\mu^\dagger T_+ + {\mbox{\rm h.c.}}) + \ldots \\
s \, = \, {\cal M} + \ldots
\end{array}
\end{equation}

Here, $Q$ and ${\cal M}$ denote the quark-charge and quark-mass matrices, respectively,

\be
Q = {1\over 3}\, \hbox{\mbox{\rm diag}}(2,-1,-1)\, , \qquad\qquad
{\cal M} = \hbox{\mbox{\rm diag}}(m_u,m_d,m_s) \, ,
\label{eq:q_m_matrices}
\ee

and $T_+$ is a $3\times 3$ matrix containing the relevant
Cabibbo--Kobayashi--Maskawa factors

\be
T_+ \, = \, \left(\begin{array}{ccc}
0 & V_{ud} & V_{us} \\ 0 & 0 & 0 \cr 0 & 0 & 0
\end{array}\right).
\label{eq:t_matrix}
\ee

The Lagrangian (\ref{lqcd}) is invariant under the \textit{local} $SU(3)_L\otimes SU(3)_R$ symmetry and in presence of external sources it turns out that the gauge fields $v_{\mu}$ and $a_{\mu}$ can only appear through the covariant derivatives



\begin{equation}
D_{\mu}U=\partial_{\mu}U-ir_{\mu}U+iUl_{\mu},\qquad D_{\mu}U^{\dag}=\partial_{\mu}U^{\dag}+iU^{\dag}r_{\mu}-il_{\mu}U^{\dag},
\end{equation}

and through the field strength tensors

\begin{equation}
F_L^{\mu \nu}=\partial^{\mu}l^{\nu}-\partial^{\nu}l^{\mu}-i[l^{\mu}, l^{\nu}], \qquad
F_R^{\mu \nu}=\partial^{\mu}r^{\nu}-\partial^{\nu}r^{\mu}-i[r^{\mu}, r^{\nu}].
\end{equation}

At lowest order in momenta, the more general effective Lagrangian consistent with Lorentz invariance and (local) chiral symmetry is of the form \cite{GLsq}

\begin{equation}
\label{l2b}
{\cal L}_2=\frac{f_0^2}{4}\langle D_{\mu}U^{\dag}D^{\mu}U+U^{\dag}\chi + \chi^{\dag}U\rangle,
\end{equation}
where
\begin{equation}
\chi=2B_0(s+ip)
\end{equation}
and $B_0$ is a constant which is not fixed by symmetry requirements alone.


At leading order in the chiral expansion we are left with only two low-energy constants, $f_0$ and $B_0$, which can be related to QCD parameters through a matching procedure, which yields

\begin{equation}
f_0 = -i \frac{p_{\mu}}{\sqrt{2}p^2}\langle 0|\frac{\delta {\cal L}_2}{\delta a_{\mu}}|\pi^+(p) \rangle=f_{\pi}\sim 92 MeV\, ,
\end{equation}
\begin{equation}
B_0=\frac{1}{f_0^2}\langle 0|\frac{\delta {\cal L}_2}{\delta s}|0 \rangle = -\frac{1}{f_0^2}\langle 0|\bar{q}q|0\rangle,\qquad \langle 0|\bar{q}q|0\rangle(2GeV)\sim -[(280\pm 30)\mathrm{MeV}]^3\, .
\nonumber
\end{equation}

$f_0$ can therefore be identified with the pion decay constant, and the parameter $B_0$ is proportional to the quark condensate, which takes into account the effect of non-vanishing quark masses. This very first term was already written down by Weinberg \cite{WeinbergChPT}, and to lowest order it reproduces the current algebra results.

Taking $s={\cal M}$ with ${\cal M}=diag(m_u, m_d, m_s)$, and $p=0$ the $\chi$ term in Eq. (\ref{l2b}) gives rise to a quadratic pseudscalar-mass term plus additional interaction proportional to the quark masses. Expanding in powers of $\phi$, one has (dropping irrelevant constants):

\begin{equation}
\frac{f_0^2}{4}2B_0\langle {\cal M}(U+U^{\dag})\rangle=B_0\bigg\{-\langle {\cal M}\phi^2 \rangle +\frac{1}{6 f_0^2} \langle {\cal M}\phi^4 \rangle + O\bigg(\frac{\phi^6}{f_0^4}\bigg)\bigg\}
\end{equation}

The explicit evaluation of the trace in the quadratic mass term provides the relation between the physical meson masses and the quark masses:

\begin{equation}\begin{array}{l}
\label{massesChPT}
M^2_{\pi^{\pm}}=2\hat{m}B_0, \qquad M^2_{\pi^0}=2\hat{m}B_0 - \varepsilon + O(\varepsilon^2),\\
M^2_{K^{\pm}}=(m_u+m_s)B_0, \qquad M^2_{K^0}=(m_d+m_s)B_0,\\
M^2_{\eta_8}=\frac{2}{3}(\hat{m}+2m_s)B_0 + \varepsilon + O(\varepsilon^2),
\end{array}\end{equation}

where

\begin{equation}
\hat{m}=\frac{1}{2}(m_u+m_d), \qquad \varepsilon =\frac{B_0}{4}\frac{(m_u-m_d)^2}{(m_s-\hat{m})}.
\end{equation}

Chiral symmetry relates the magnitude of the meson and quark masses to the size of the quark condensate.
The Eq. (\ref{massesChPT}) imply the old Current Algebra mass ratios (Gell-Mann, Oakes and Renner 1958, Weinberg 1977),

\begin{equation}
\frac{M^2_{\pi^{\pm}}}{2\hat{m}}=\frac{M^2_{K^{\pm}}}{(m_u + m_s)}=\frac{M^2_{K^0}}{(m_d + m_s)}\approx \frac{3M^2_{\eta _8}}{(2\hat{m} + 4 m_s)}
\end{equation}

and (up to $O(m_u-m_d)$ corrections) the Gell-Mann Okubo (1962) mass relation,

\begin{equation}
 3M^2_{\eta _8}=4M_K^2-M_{\pi}^2
\end{equation}

The lowest-order chiral Lagrangian Eq. (\ref{l2b}) encodes
in a very compact way all the Current Algebra results obtained in
the sixties.
The nice feature of the chiral approach is its elegant
simplicity. Moreover, as we will see in the next section,
the effective field theory method
allows us to estimate higher-order corrections in a systematic way.

\subsection{ChPT at $O(p^4)$}\label{ChPTp4}

\quad Eq. (\ref{l2b}) comes from the generating functional Z defined as:

\begin{equation}
e^{iZ[v,a,s,p]}=\langle0|T e^{i\int d^4x\mathcal{L}(x)}|0\rangle.
\end{equation}
This vacuum-to-vacuum transition amplitude generates the Green functions of the vector, axial-vector, scalar and pseudoscalar quark currents built out of the three flavours $u$, $d$ and $s$. $Z[v,a,s,p]$ is associated with the lagrangian of Eq. (\ref{lqcd}) and admits an expansion in powers of the external momenta and of quark masses, $Z=Z_2+Z_4+Z_6...=\sum_n Z_n$. As a consequence of chiral symmetry and its spontaneous breakdown, $Z_2$ coincides in the meson sector at \textit{leading order} in ChPT with the classical action

\begin{equation}
Z_2=\int d^4x \mathcal{L}_2(U,v,a,s,p).
\end{equation}
where $\mathcal{L}_2$ is the same Eq. (\ref{l2b}).


%
\quad At $O(p^4)$, the generating functional consists of a contribution to account for the chiral anomaly, the one-loop functional originating from the lagrangian (\ref{l2b}) and an explicit local action of order $p^4$.
The most general Lagrangian, invariant under parity, charge conjugation, the local chiral transformations, and able to generate a local action of order $p^4$ is given by \cite{GLsq}

\begin{eqnarray}
\label{l4}
\mathcal{L}_{\chi}^{(4)}(U,DU) & = & L_1\langle
D_{\mu}{U}^{\dagger}D^{\mu}U\rangle^2+L_2\langle
D_{\mu}{U}^{\dagger}D_{\nu}U\rangle\langle
D^{\mu}{U}^{\dagger}D^{\nu}U\rangle+\nonumber\\ & + &
L_3\langle D_{\mu}{U}^{\dagger}D^{\mu}U
D_{\nu}{U}^{\dagger}D^{\nu}U\rangle+L_4\langle
D_{\mu}{U}^{\dagger}D^{\mu}U\rangle\langle{U}^{\dagger}\chi+{\chi}^{\dagger}U\rangle+\nonumber\\
& + & L_5\langle
D_{\mu}{U}^{\dagger}D^{\mu}U({U}^{\dagger}\chi+{\chi}^{\dagger}U)\rangle+L_6\langle{U}^{\dagger}\chi+{\chi}^{\dagger}U\rangle^2+\nonumber\\
& + &
L_7\langle{U}^{\dagger}\chi-{\chi}^{\dagger}U\rangle^2+L_8\langle{\chi}^{\dagger}U{\chi}^{\dagger}U+{U}^{\dagger}\chi{U}^{\dagger}\chi\rangle-\nonumber\\
& - & iL_9\langle
F_R^{\mu\nu}D_{\mu}UD_{\nu}{U}^{\dagger}+F_L^{\mu\nu}D_{\mu}{U}^{\dagger}D_{\nu}U\rangle+L_{10}\langle{U}^{\dagger}F_R^{\mu\nu}UF_{L\mu\nu}\rangle+\nonumber\\
  & + & H_1\langle
  F_{R\mu\nu}F_R^{\mu\nu}+F_{L\mu\nu}F_L^{\mu\nu}\rangle+H_2\langle{\chi}^{\dagger}\chi\rangle
\end{eqnarray}

where $F^{\mu\nu}_{R,L}=\partial(v^{\nu}\pm a^{\nu})-\partial(v^{\mu}\pm a^{\mu})-i[v^{\mu}\pm a^{\mu}, v^{\nu}\pm a^{\nu}]$.

The terms proportional to $H_1$ and $H_2$ do not contain the
pseudoscalar fields and are therefore not directly measurable.
Thus, at $O(p^4)$ we need ten additional coupling constants
$L_i$
to determine the low-energy behavior of the Green functions, the low-energy constants (LECs).
These constants  parameterize our
ignorance about the details of the underlying QCD dynamics and are needed in order to make reliable phenomenological predictions. As with any other effective field theory, these LECs play the role of coupling constants and contain the information coming from the integration of the heavy degrees of freedom not explicitly present in the Chiral Lagrangian. We will exemplify this process in subsection \ref{integrateRes} with meson resonances.

However, the increase in the number of operators as one goes to higher orders in the chiral expansion, together with the poorly-known associated low-energy couplings, makes computations difficult already at $\mathcal{O}(p^4)$.
In principle, all the chiral couplings are calculable functions of $\Lambda_{QCD}$ and the heavy-quark masses. At the present time, however, our main source of information about these couplings is low-energy phenomenology. At $\mathcal{O}(p^2)$ there are 2 LECs, at $\mathcal{O}(p^4)$ 10 \cite{GLsq}, and at $\mathcal{O}(p^6)$ the number of constants becomes more than a hundred \cite{p6FearingScherer,p6Bijnens,p6BijnensTalavera}. In the electroweak sector this proliferation of constants appears already at $\mathcal{O}(p^4)$ \cite{ewKambor,ewEspositoFarese}. Although in principle these low-energy constants may be computed on the lattice, in practice this has only recently \cite{lattice} been accomplished in a few cases for the strong Chiral Lagrangian at $\mathcal{O}(p^4)$.


Since ChPT is a quantum field theory, the next step in our description is take into account quantum loops with Goldstone-boson propagators in the internal lines. The chiral loops generate non-polynomial contributions with logarithms and threshold factors, as required by unitarity.

ChPt is an Effective field theory, then have to be provided with a power counting rule. Actually, in our case, at ${\cal O}(p^d)$, the diagrams that contribute are dictated by the relation (\cite{WeinbergChPT}):
\be
d = 2 + \sum_n N_n (n - 2) + 2 N_L\, , \quad  n = 2, 4, 6, . . .
\ee

where $N_n$ is the number of vertices coming from ${\cal O}(p^n)$ operators, and $N_L$ is the number of
loops.

At one loop (in ${\cal L}_2$), the ChPT divergences are $O(p^4)$ and are therefore renormalized by the low-energy couplings in equation (\ref{l4}):

\bea
L_i = L_i^r(\mu) + \Gamma_i \lambda \, , \qquad\qquad
H_i = H_i^r(\mu) + \widetilde\Gamma_i \lambda \, ,
\eea

where

\begin{equation}
\lambda = {\mu^{d-4}\over 16 \pi^2} \left\{
{1\over d-4} -{1\over 2} \left[ \log{(4\pi)} + \Gamma'(1) + 1 \right]
\right\} .
\label{eq:divergence}
\end{equation}

and the evolution under the renormalization group is then given by

\begin{equation}
L_i^r(\mu _2)=L_i^r(\mu _1)+\frac{\Gamma_i}{16 \pi ^2}\log (\frac{\mu_1}{\mu_2})
\end{equation}

where $\Gamma_i$ take the values due to the explicit calculation of the one-loop generating functional $Z_4$
\cite{GLsq} gives:

\bea
\Gamma_1 = {3\over 32}\, , \quad\; \Gamma_2 = \dfrac{3}{16}\, ,
\qquad  \Gamma_3 = 0 \, ,
\qquad \Gamma_4 = {1\over 8} \, , \qquad\;\,
\Gamma_5 = {3\over 8} \, , \qquad \Gamma_6 = \dfrac{11}{144} \, ,\\
\Gamma_7 = 0 \, ,\qquad \Gamma_8 = {5\over 48}\, , \qquad
\Gamma_9 = {1\over 4} \, ,\qquad\! \Gamma_{10} = -\dfrac{1}{4} \, ,
\quad\; \widetilde\Gamma_1 = -{1\over 8} \, , \quad\;
\widetilde\Gamma_2 = {5\over 24}\, .
\label{eq:d_factors}\nonumber
\eea

\subsubsection{Decay constants at $\mathcal{O}(p^4)$}

\quad For illustrative purpose we also show the $O(p^4)$ calculation of the meson-decay constants in the isospin limit ($m_u = m_d = \hat m$) \cite{GLsq}:

\bea\label{decayconst}
f_\pi \, = \, & f \left\{ 1 - 2\mu_\pi - \mu_K +
    {4 M_\pi^2\over f^2} L_5^r(\mu)
    + {8 M_K^2 + 4 M_\pi^2 \over f^2} L_4^r(\mu)
    \right\} ,
\cr
f_K \, = \, & f \left\{ 1 - {3\over 4}\mu_\pi - {3\over 2}\mu_K
    - {3\over 4}\mu_{\eta_8}
    + {4 M_K^2\over f^2} L_5^r(\mu)
    + {8 M_K^2 + 4 M_\pi^2 \over f^2} L_4^r(\mu)
    \right\} ,
\cr
f_{\eta_8} \, = \, & f \left\{ 1 - 3\mu_K +
    {4 M_{\eta_8}^2\over f^2} L_5^r(\mu)
    + {8 M_K^2 + 4 M_\pi^2 \over f^2} L_4^r(\mu)
    \right\} ,
\eea
where
$$
\mu_P \equiv {M_P^2\over 32 \pi^2 f^2} \,
\log{\left( {M_P^2\over\mu^2}\right)} .
\label{eq:mu_p}
$$
masses could be found in \cite{PDG} and $L_i$ in Table \ref{Lis}.

\subsubsection{Electromagnetic Form Factors}

\quad As a second example, we present the main results for the electromagnetic form factors (that will we used in chapter 6). That is a particular case where higher-order local terms in the chiral expansion are important. Ignoring those will lead to a wrong result compared to the experimental value.

At ${\cal O}(p^2)$ the electromagnetic coupling of the Goldstone bosons is just the minimal one, obtained through the covariant derivative. The next-order corrections generate a momentum-dependent form factor:

\begin{equation}
F_V^{\phi^{\pm}}(q^2)=1+\frac{1}{6}\langle r^2 \rangle_V^{\phi^{\pm}} q^2 + ...; \quad F_V^{\phi^0}(q^2)=\frac{1}{6}\langle r^2 \rangle_V^{\phi^{0}} q^2+...
\end{equation}

The meson electromagnetic radius $\langle r^2 \rangle_V^{\phi}$ gets local contributions from the $L_9$ term, plus logarithmic loop corrections (\cite{GLOneLoop}):

\begin{eqnarray}
\langle r^2 \rangle_V^{\pi^{\pm}}=\frac{12 L_9^r(\mu)}{f^2} - \frac{1}{32 \pi^2 f^2}\left(2 \log \left(\frac{M_{\pi}^2}{\mu^2}\right) + \log\left(\frac{M_{K}^2}{\mu^2}\right)\right)\, ,\\ \nonumber
\langle r^2 \rangle_V^{K^0}= - \frac{1}{16 \pi^2 f^2} \log \left(\frac{M_{k}}{M_{\pi}}\right)\, , \\ \nonumber
\langle r^2 \rangle_V^{K^{\pm}}=\langle r^2 \rangle_V^{\pi^{\pm}}+\langle r^2 \rangle_V^{K^{0}}\, .
\end{eqnarray}

\section{QCD Sum Rules, the Operator Product Expansion}\label{secOPEth}

\quad As a consequence of asymptotic freedom the theoretical results obtained from QCD can be compared easily with the experimental situation for the hard processes, which is, at short distances the effective coupling constant $\alpha_s$ becomes small and the interaction can be treated perturbatively. On the other hand, as we have said, any comprehensive theory of the strong interaction must include long distance dynamics as well. Actually, quark interaction within hadrons is strong, since it binds quarks into inseparable groups. However, nowadays we do not have any accurate quantitative framework within QCD for dealing with the strong interaction regime and the evaluation of the hadron spectrum.

An interesting approach was started in 1979 by Shifman, Vainshtein and Zakharov, assuming that confinement exists (\cite{Shifman:1978bx}). The effects of confinement can be described through the use of a few parameters, the so called condensates, and allows to obtain many hadronic properties through an appropriate use of sum rules. One of the main ingredients of this approach is the Operator Product Expansion (OPE), proposed by Wilson in 1969 (\cite{Wilson:1969zs}), who hypothesized that the singular part, as $x\rightarrow y$, of the product $A(x)B(y)$ of two operators is given by a sum over other local operators

\be
\label{OPE}
A(x)B(y)\rightarrow \sum_C F_C^{AB}(x-y)C(y)
\ee

where $F_C^{AB}(x-y)$ are singular c-number functions.
This OPE exists for the free scalar and spinor field theories and for renormalized interacting fields to all orders in perturbation theory. In every case, they are valid for any elementary or composite local fields: $A$ and $B$ can be elementary scalar or spinor fields or local currents or the stress-energy tensor or any local Wick product in a free field theory.
Dimensional analysis suggests that $F_C^{AB}(x-y)$ behaves for $x\rightarrow y$ like power $d_C-d_A-d_B$ of $x-y$, where $d_O$ is the dimensionality of the operator $O$ in powers of mass or momentum. Since $d_O$ increases as we add more fields or derivatives for an operator $O$, the strength of the singularity of $F_C^{AB}(x-y)$ decreases for operators $C$ of increasing complexity. The remarkable thing about the operator product expansion is that it is an \textit{operator} relation. It is the decrease of the singularity in Eq. (\ref{OPE}) with operators $C(y)$ of increasing complexity that makes this expansion useful in drawing conclusions about the behavior of the product $A(x)B(y)$ for $x\rightarrow y$. Renormalization effects modify the power counting: for asymptotic free theories $F_C^{AB}(x-y)$ behaves like the power $d_C-d_A-d_B-\gamma$ of $x-y$ suggested by dimensional analysis where $\gamma$ is a constant. 

The corresponding statement in momentum space is that for $k\rightarrow \infty$,

\be
\int d^4x e^{-ik·x}A(x)B(0)\rightarrow \sum_C V_C^{AB}(k)C(0)
\ee

and correspondingly

\be
\int d^4x e^{-ik·x}T\{A(x)B(0)\}\rightarrow \sum_C U_C^{AB}(k)C(0)
\ee

where $V_C^{AB}(k)$ and $U_C^{AB}(k)$ are functions of $k^{\mu}$ that for large $k$ decrease increasingly rapidly for more and more complicated terms in the series.

Let us show an example, following \cite{PascualTarrach}, to illustrate this process which will be useful to perform our own calculation in section \ref{secOPE}:
Ref. \cite{PascualTarrach} suggests to calculate the $\rho$ meson mass using the QCD sum rules. The first point begins with the isovector part of the electromagnetic current:
\be
\label{Jcurrent}
J^{\mu}_{\rho}=\frac{1}{2}(\bar{u}(x)\gamma^{\mu}u(x)-\bar{d}(x)\gamma^{\mu}d(x))
\ee
These composite operators are the ones which appear in the OPE Eq. (\ref{OPE}). Let us now introduce the two point function:

\be
\Pi_{\rho}^{\mu\nu}(q)=i\int d^4 x\,e^{iq\cdot x}\langle \Omega\mid
T\left(J^{\mu}_{(\rho)}(x)J^{\nu}_{(\rho)}(0)\right)\mid \Omega\rangle \,,
\ee

where $\mid \Omega\rangle$ is the physical vacuum of the theory. Since Eq. (\ref{Jcurrent}) is a conserved current we can write

\be
\Pi_{\rho}^{\mu\nu}(q)\equiv (q^{\mu}q^{\nu}-q^2g^{\mu\nu})\Pi_{\rho}(q^2)
\ee

with

\be
\Pi_{\rho}(q^2)=-\frac{i}{(D-1)q^2}\int d^D x\,e^{iq\cdot x}\langle \Omega\mid
T\left(J^{\mu}_{(\rho)}(x)J_{(\rho)\mu}(0)\right)\mid \Omega\rangle \,,
\ee

where $D$ is the number of space-time dimensions. Notice that the dimensions of the operators appearing here are $d(J_{\rho}^{\mu})=M^{D-1}$ and $d(\Pi_{\rho})=M^{D-4}$. Now, we are in the stage to use the Operator Product Expansion, then we can write:

\be\label{limPi}
\lim_{q\rightarrow\infty}\Pi_{\rho}(q^2)=-\frac{i}{(D-1)q^2}\sum_n \langle \Omega\mid O_n(0)\mid \Omega \rangle \int d^D x\,e^{iq\cdot x} C_n(x)
\ee

where $O_n(0)$ are the local operators that appears in Eq. (\ref{OPE}).
For large values of $q^{\mu}$ the behavior of the integral appearing in the r.h.s. of this equation is:

\be
(q^2)^{[d(O_n)+2-D]/2}
\ee

where it will be necessary to consider in (\ref{limPi}) only operators $O_n(x)$ such that

\be
d(O_n)\leq (D-4)+2N
\ee

where $N$ is used to identify the terms in $\Pi_{\rho}(q^2)$ that decreases faster than $(q^2)^{-N}$ in the limit $q^{\mu}\rightarrow\infty$.

We will be interested in the lowest dimension scalar operators and these are shown in table \ref{tabOPE}.

\begin{table}
\centering
\begin{tabular}{|c|c|}
\hline
$Dimension$ & $Operators$\\
\hline
$0$ & $\mathds{1}$\\
\hline
$4$ & $m_A\bar{q}_{\alpha}^A(x)q_{\alpha}^A(x)$\\
$$ & $F^{\mu\nu}_a(x)F_{\mu\nu}^a(x)$\\
\hline
$$ & $\bar{q}_{\alpha}(x)\Gamma q_{\alpha}(x)\bar{q}_{\beta}(x)\Gamma q_{\beta}(x)$\\
$6$ & $\bar{q}_{\alpha}(x)\Gamma (\lambda ^a)_{\alpha \beta} q_{\beta}(x)\bar{q}_{\gamma}(x)\Gamma
(\lambda ^a)_{\gamma\delta} q_{\delta}(x)$\\
$$ & $m_A\bar{q}^A \lambda^a \sigma _{\mu\nu}(x)q^A(x)F^{\mu\nu}_a(x)$\\
$$ & $f_{abc} F^{\mu\nu}_a(x)F_{\nu\rho}^b(x)F_{\mu}^{c \rho}(x)$\\
\hline
\end{tabular}
\caption{Lowest dimension scalar operators.}\label{tabOPE}
\end{table}

The operators $O_n$ are conveniently classified according to their Lorentz spin and dimension $d$. We will consider only spin-zero operators since only these contribute to the vacuum expectation value. Within the standard perturbation theory only the unit operator would survive in Eq. (\ref{limPi}), but the non-perturbative effects induce non-vanishing vacuum expectation values for other operators as well and they are the so called condensates. Therefore, the non-perturbative effects of QCD introduce power corrections of type $1/(q^2)^N$, $N\geq 1$, to the perturbative calculation.
Since QCD is asymptotically free, the calculation of the coefficients $C_n$ is reliable. The expansion coefficients in Eq. (\ref{limPi}) are calculated as a series in $\alpha_s$ expansion.
The final results concern a perturbative and a non-perturbative contributions. The lowest order perturbative contribution is:





\be\label{pertPi2}
\Pi_{(\rho)}(q^2)=-\frac{1}{8\pi^2}\bigg(1+\frac{\alpha_s}{\pi}\bigg)\ln(-q^2)
\ee

The non-perturbative contributions contains the quark condensate and the gluon condensate:
\be
\label{qcondensPi}
\Pi_{(\rho)}(q^2)=\frac{1}{2}\frac{1}{(-q^2)^2}\big(m_u<\bar{u}u>+\, m_d<\bar{d}d>\big)\, +\, \frac{1}{24(-q^2)^2}\langle\frac{\alpha_s}{\pi}F^{\mu\nu}_a(x)F_{\mu\nu}^a(x) \rangle
\ee



always up to terms of order $\alpha_s$.
From the second order perturbative result due to the contribution of the condensate of four quark fields, we have (using isospin symmetry, $<\bar{u}u>=<\bar{d}d>$):

\be
\label{4qPi}
\Pi_{(\rho)}(q^2)=\frac{112}{81}\pi \alpha_s
<\bar{u}u>^2\frac{1}{q^6}
\ee

Taken into account all these contributions, i.e, Eqs. (\ref{pertPi2}), (\ref{qcondensPi}) and (\ref{4qPi}), we get from QCD the following information (using $-q^2\equiv Q^2$):

\bea
\label{QCDOPE}
\Pi_{(\rho)}(Q^2)=-\frac{1}{8\pi^2}\bigg(1+\frac{\alpha_s}{\pi}\bigg)\ln(Q^2)+\frac{1}{2}\frac{1}{Q^4}(m_u<\bar{u}u>+m_d<\bar{d}d>)\\
+ \frac{1}{24 Q^4}\langle\frac{\alpha_s}{\pi}F^{\mu\nu}_a(x)F_{\mu\nu}^a(x) \rangle - \frac{112}{81}\pi \alpha_s
<\bar{u}u>^2\frac{1}{Q^6}\nonumber
\eea

where we have neglected constants, higher order terms in the $OPE$, radiative corrections to the condensates and $\frac{m^2}{Q^2}$ terms. Except for the first term in Eq. (\ref{QCDOPE}), this is an expansion in terms of $Q^{-2}$ when $Q^2\rightarrow\infty$.

This way to proceed here illustrated will be useful in chapter \ref{capitol2} and in further calculations of the OPE for two-point Green's functions.

\section{The Large-$N_c$ limit}\label{sec:LN}

\quad QCD, as a non-Abelian gauge theory based on the gauge group $SU(3)$ can be understood\cite{largeNtHooft,largeNWitten} by studying a gauge theory based on the gauge group $SU(N_c)$ in the limit $N_c\rightarrow\infty$ \footnote{For a pedagogical review see for example \cite{ManoharLargeN,Lebed}}. One might think that letting $N_c\rightarrow\infty$ would make the analysis more complicated; or $SU(N_c)$ gauge theory could not be related to QCD because the difference between $N_c\to \infty$ and $N_c=3$. We will soon see that $SU(N_c)$ gauge theory simplifies in the $N_c\rightarrow\infty$ limit, that the true expansion parameter is $1/N_c$, not $N_c$. In fact, the $1/N_c$ expansion is equivalent to a semiclassical expansion for an effective theory of color singlet mesons and also for baryons\cite{JenkinsBaryons}. Results for QCD will we obtained from $N_c\rightarrow\infty$ limit by using $1/N_c=1/3$, with good agreement with experiment in both meson and baryon sectors. A way to understand the validity of this approximation is done by 't Hooft and later Witten \cite{largeNtHooft,largeNWitten} and are decomposed in two ways. The first one related to why the perturbation theory is successful in QED. In this theory, the typical expansion parameter is $\frac{e^2}{4\pi}$ where $e=0.3$ is the electric charge. If the typical expansion parameter had turned to be $4\pi e^2$, perturbation theory would not have been very successful for $e$ as large as $0.3$. The other way asks us how small $x$ must be for a series $\sum a_n x^n$ to be dominated by the first few terms. The answer depends entirely on how large are the coefficients $a_n$. If the coefficients are very small, $x=1/3$ can be considered a small number.

Therefore, the large $N_c$ expansion \cite{largeNtHooft,largeNWitten} stands out as a very promising analytic approach capable of
dealing with the complexities of nonperturbative QCD while, at the same time, offering a relatively
simple and manageable description of the physics. This description includes some aspects that are not understood in the QCD context such as:

\begin{enumerate}
\item The suppression in hadronic physics of the $q\bar{q}$ sea; the fact that mesons are approximately pure $q\bar{q}$ states; the absence, or at least suppression, of $q\bar{q}q\bar{q}$ exotic states.
\item The Zweig's rule; the fact that mesons come in nonets of flavor SU(3); the decoupling of glue states.
\item The fact that multiparticle decays of unstable mesons are dominated by resonant two body final states, when these are available.
\item The Regge phenomenology and the success of a phenomenology that describes the strong interactions in terms of tree diagrams with exchange of physical hadrons.
\end{enumerate}

For instance, mesons are $q\overline{q}$ states
with no width, the OZI rule is exact and there is even a proof of spontaneous chiral symmetry
breaking \cite{coleman-witten} (more details will be discussed in sec.\ref{mesonslargeN}). Furthermore, interest in studying SU(N) gauge theories in the large $N_c$ limit has increased due to the discovery of a duality of some highly supersymmetric gauge theories to gravity \cite{Maldacena},
although the real relevance of this connection for QCD still remains to be seen. However, in spite
of all this, the fact that no solution to large-$N_c$ QCD has been found still poses a serious
limitation to doing phenomenology. For instance, in order to reproduce the parton model logarithm
which is present in QCD Green's functions in perturbation theory, an infinity of resonances is
necessary whose masses and decay constants are in principle unknown.

\subsection{QCD in the Large-$N_c$ limit}\label{QCDlargeN}

\quad When describing QCD, we will see that the coupling constant has been chosen to be $g/\sqrt N_c$, rather than $g$,
because this will lead to a theory with a sensible (and non-trivial) large $N_c$
limit (further details will be presented in the next subsection). The field strength is then:
\[
F_{\mu\nu}=\partial_\mu A_\nu - \partial_\nu A_\mu + i {g \over \sqrt N_c}\left[
A_\mu , A_\nu \right],
\]
and the Lagrangian is

\begin{equation}
\label{3.1}
{\cal L} = - \frac{1}{2}\Tr F_{{\mu}{\nu}} F^{{\mu}{\nu}} + \sum_{f=1}^{N_F}
\bar{q}_k \left( i\,\gamma^{\mu} D_{\mu} - m_f \right) q_k.
\end{equation}

The large $N_c$ limit will be taken with the number of flavors $N_F$ fixed. It is
also possible to consider other limits, such as $N_c \rightarrow \infty$ with
$N_F/N_c$ held fixed~\cite{VenezianoLargeN}.

The Lagrangian of the theory stands as in $N_c=3$ QCD (after all, it is still a Yang-Mills Lagrangian) except for the fact that we have changed the gauge group. That means that each quark field, since they sit in the fundamental representation, appears as an $N_c$-plet. Gluons, on the contrary, live in the adjoint representation and enlarge their number to $N_c^2-1$. In practice, it is common to approximate this to $N_c^2$; in other words, we are skipping the tracelessness constraint (we are taking $U(N_c)$ instead of $SU(N_c)$). The missing gluon is numerically unimportant at sufficiently large $N_c$. Besides, it can be shown that the abelian factor is indeed suppressed at large $N_c$. Notice that the theory we are looking at differs from $N_c(=3)$ QCD in that there exist far more gluons than quarks (the former scale with $N_c^2$ while the latter only with $N_c$).

Our aim will be to show the topological ordering of diagrams induced by the large--$N_c$ power counting scheme. For clarity, it is convenient to use {\it{'t Hooft double-line notation}}.

\begin{figure}
    \renewcommand{\captionfont}{\small \it}
    \centering
    \psfrag{A}{$q^i$}
    \psfrag{B}{$\bar{q}_j$}
    \psfrag{C}{$G_{\mu, i}^j$}
    \includegraphics[width=1.5in]{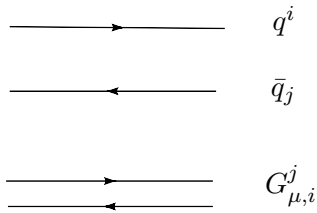}
    \caption{Double line notation for quarks, anti-quarks and gluons.}\label{doubleline}
\end{figure}

\begin{figure}
    \renewcommand{\captionfont}{\small \it}
    \centering
    \includegraphics[width=2.0in]{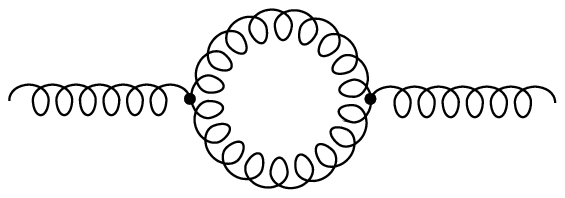}\qquad \qquad
    \includegraphics[width=2.0in]{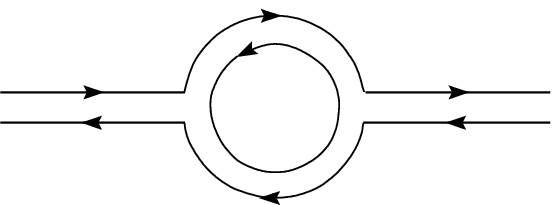}\\
    \includegraphics[width=1.4in]{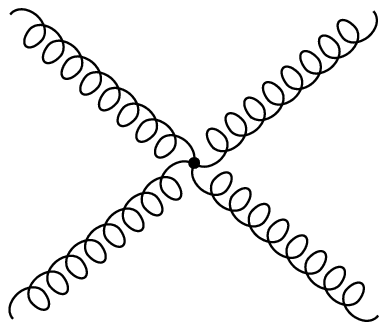}\qquad \qquad \qquad
    \includegraphics[width=1.5in]{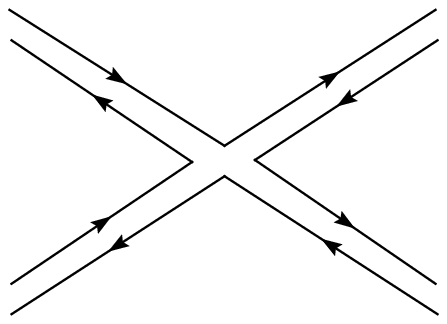}
    \caption{Feynman diagrams with its counterparts in the double line notation introduced by 't Hooft.}\label{planar}
\end{figure}

\subsubsection{The Double-Line Notation and Planarity}

\quad In perturbative QED there is only one coupling constant which shows up to couple fermions and antifermions. That is why Feynman diagrams are so useful to organize calculations in powers of the coupling constant: you only need to count the number of vertices. In QCD, however, the ordinary coupling constant $g_s$ is not really a free parameter, because in view of the renormalization group, it is absorbed into defining the scale of masses. Therefore, we have to change our strategy. In Yang-Mills theory at $N_c\rightarrow\infty$, the mesons and glue states are free, stable, and non-interacting. Meson decay amplitudes are of order $1/\sqrt{N_c}$, and meson-meson elastic scattering amplitudes are of order $1/N_c$. These elastic amplitudes are given, as in Regge phenomenology, by a sum of tree diagrams involving the exchange, not of quarks and gluons, but of physical mesons.


In the $1/N_c$ expansion, we need to keep track not only of the vertices of the theory (we will show later on that the coupling constant at large-$N_c$ is color-dependent) but also of color flow inside the diagram. We would like to have a pictorial approach to be able to determine in an easy way the scaling of physical amplitudes with the color factor. The double-line notation introduced by 't Hooft makes it transparent to extract the color scaling of a given amplitude altering only slightly the Feynman diagrams we are used to. Consider the following pictorial recipe: represent every quark line by a color line, right-faced arrow meaning color flow, left-faced arrow meaning anticolor flow, as is shown in figure \ref{doubleline}.

This does not change the Feynman picture much. When it comes to gluons, however, we have to interpret their color indices as a system of color and anticolor line. Figure \ref{planar} shows some examples of converting Feynman diagrams to double-line diagrams.

From the gluon loop contribution to the gluon propagator, figure \ref{planar}, it is easy to see that even after the color quantum numbers of the initial and final states are specified, there are still N possibilities for the quantum number of the intermediate state gluons. As a result, this diagram receives a combinatoric factor of $N_c$. On the other hand, in the same figure \ref{planar} there is also a factor of coupling at each of the two interaction vertices. If we want the one-loop gluon vacuum polarization to have a smooth limit for large-$N_c$, we must choose the coupling constant to be $g/\sqrt{N_c}$, where $g$ is to be held fixed as $N_c$ becomes large. With this choice, introduced by Witten \cite{largeNtHooft,largeNWitten}, the two vertex factors of $g/\sqrt{N_c}$ in figure \ref{planar} combine with the combinatoric factor of $N_c$ to give a smooth large-$N_c$ behavior: $(g/\sqrt{N_c})^2 \times N_c=g^2$, independent of $N_c$. Thus, the vanishing for large-$N_c$ of the coupling constants cancels the divergence of the combinatoric factor, to produce a smooth large-$N_c$ limit (as expected).

Due to the normalization for the coupling constant we choose, in order to survive as $N_c\rightarrow\infty$, a Feynman diagram must have combinatoric factors large enough to compensate for the vertex factors. Only a certain class of Feynman diagrams, the so-called planar diagrams, have the combinatoric factors large enough to just cancel the vertex factors. The other diagrams vanish for large-$N_c$. The large-$N_c$ limit is therefore given by the sum of the planar diagrams.

As we have mentioned above, figure \ref{planar} has a combinatoric factor of $N_c$. In the right side of this figure the gluon loop is been redrawn in the double line notation. Following the arrows, the color lines at the edge of the diagram are contracted with those of the initial and final states. However, at the center of the figure, there is a closed color line that is contracted only with itself. The color running around this loop is unspecified even when the initial and final states are given, and the sum over the quantum number of this loop gives a factor of $N_c$. This is the combinatoric factor of $N_c$ associated with the upper left diagram of the figure \ref{planar}.

\begin{figure}
    \renewcommand{\captionfont}{\small \it}
    \renewcommand{\captionlabelfont}{\small \it}
    \centering
    \includegraphics[width=4.0in]{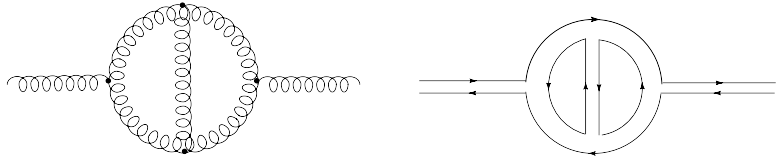}
    \caption{Two-loop contribution to the gluon propagator, in the ordinary notation and in the double line notation, respectively.}\label{2loopN}
\end{figure}

Using the double line notation, it is not difficult to determine whether a given diagram survives in the large-$N_c$ limit. For example, in figure \ref{2loopN} there is drawn, both in the ordinary notation and in the double line notation, a two-loop contribution to the gluon propagator. This diagram has four interaction vertices, each contributing a factor of $1/\sqrt{N_c}$, but it has two closed color loops that are self-contracted, each contributing a factor of $N_c$. Altogether, the diagram is of order $(1/\sqrt{N_c})^4 \times N_c^2=1$ and so survives in the large-$N_c$ limit.

\begin{figure}
    \renewcommand{\captionfont}{\small \it}
    \renewcommand{\captionlabelfont}{\small \it}
    \centering
    \includegraphics[width=2.5in]{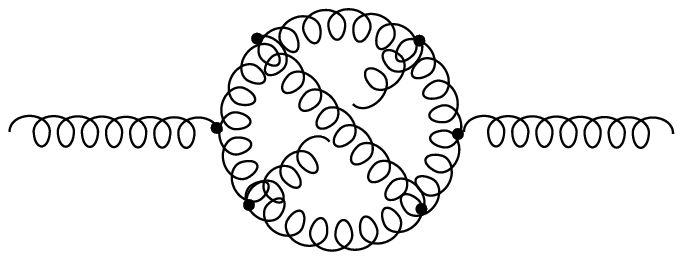}
    \includegraphics[width=2.5in]{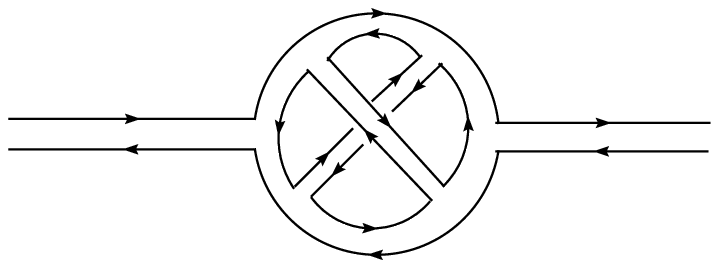}
    \caption{A nonplanar example diagram as a contribution to the gluon propagator in the ordinary and double line notation}\label{supressedN}
\end{figure}

Therefore, exist a relation between the vertices and the loops to indicate when a diagram survives or not. A counterexample of a diagram that do not survives is drawn in figure \ref{supressedN} both in ordinary and double line notation. This diagram has six interacting vertices, but only one large and tangled closed loop. The diagram is, therefore, of order $(1/\sqrt{N_c})^6 \times N_c=1/N_c^2$ and vanishes like $1/N_c^2$ as $N_c$ becomes large. The basic difference between this diagram and the previous one is this diagram is nonplanar, it is impossible to draw this diagram on the plane without line crossing. The diagrams of figures \ref{planar} and \ref{2loopN} are, by contrast, planar, they can be drawn on the plane.

With this example, we arrive to the first "selection rule" in the large-$N_c$ limit, that is nonplanar diagrams are suppressed.

There is also a second "selection rule" that reflects the fact that for large-$N_c$ there are $N_c^2$ gluon states but only $N_c$ quark states, so that diagrams with internal quark lines have fewer possible intermediate states and smaller combinatoric factors.

\begin{figure}
    \renewcommand{\captionfont}{\small \it}
    \renewcommand{\captionlabelfont}{\small \it}
    \centering
    \includegraphics[width=2.0in]{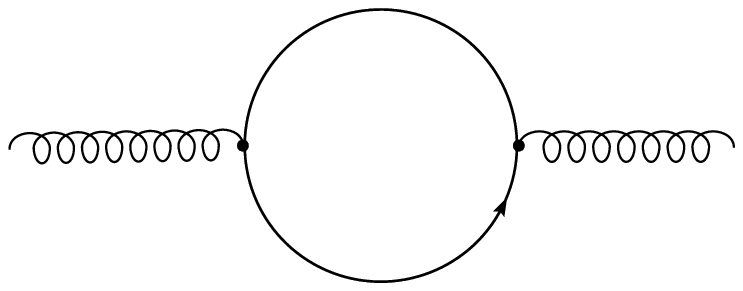}
    \includegraphics[width=2.3in]{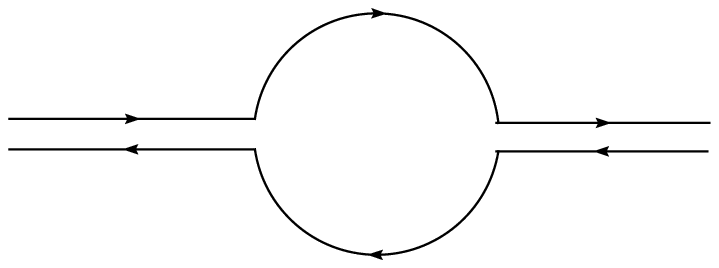}
    \caption{One quark loop contribution to the gluon propagator, in the ordinary and in the double line notations.}\label{quarkgluonN}
\end{figure}

Consider the one quark loop contribution to the gluon propagator. This diagram is drawn in figure \ref{quarkgluonN}, both in the ordinary and the double line notation. Because the quark propagator corresponds to a single color line, not two, the closed color line present in figure \ref{planar} is absent in figure \ref{quarkgluonN}. As a result, figure \ref{quarkgluonN} has no large combinatoric factor, and its only dependence on $N_c$ comes from factors of $1/\sqrt{N_c}$ at each of the two vertices. So figure \ref{quarkgluonN} vanishes like $1/N_c$ for large-$N_c$.

In sum, there are two selection rules for Feynman diagrams in the large-$N_c$ limit;
\begin{itemize}
\item Nonplanar diagrams are suppressed by factors of $1/N_c^2$ .
\item Internal quark loops are suppressed by factors of $1/N_c$.
\end{itemize}

The leading diagrams for large-$N_c$ are the planar diagrams with a minimum number of quark loops.

\subsubsection{Properties of mesons in the Large-$N_c$ limit}\label{mesonslargeN}

\quad Before describing mesons for large-$N_c$, it is necessary to make an important assumption. QCD is a confining theory at $N_c=3$, and we will assume that the confinement persists also at large-$N_c$. Combining the assumption of color confinement and the knowledge from large-$N_c$ and planar diagrams, the large-$N_c$ theory has the following properties:
\begin{itemize}
\item Mesons for large-$N_c$ are free, stable, and noninteracting. Mesons masses have smooth limits for large-$N_c$, and the number of meson states is infinite.
\item Meson decay amplitudes are of order $1/\sqrt{N_c}$; meson-meson elastic scattering amplitudes are of order $1/N_c$, and are given by a sum of tree diagrams involving the exchange of physical mesons.
\item Zweig's rule is exact at large-$N_c$; singlet-octet mixing and mixing of mesons with glue states are suppressed, so that mesons come in nonets; and mesons for large-$N_c$ are pure $q\bar{q}$ states.
\end{itemize}

The first point to establish is that the operator $J(x)$, as a generic quark bilinear, acting on the vacuum, creates, in the large-$N_c$ limit, only one-meson states. It is equivalent to claim that the only singularities of the two-point function of $J$ are one-meson poles. In other words, to lowest order in $1/N_c$,
\begin{equation}\label{2pLN}
\langle J(k)J(-k) \rangle = \sum_n \frac{a_n^2}{k^2-m_n^2}
\end{equation}

\begin{figure}
    \renewcommand{\captionfont}{\small \it}
    \renewcommand{\captionlabelfont}{\small \it}
    \centering
    \includegraphics[width=2.3in]{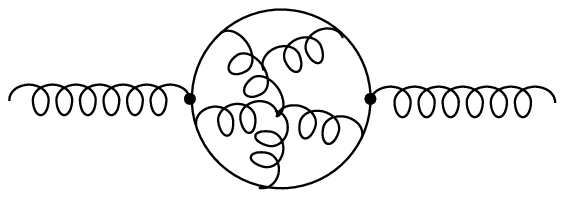}\qquad \quad
    \includegraphics[width=1.7in]{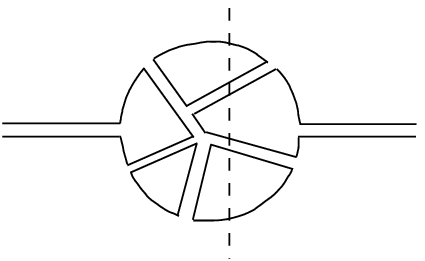}
    \caption{A representative Feynman diagram contributing to the two-point function together with its counterpart in double line notation. The cut illustrates the fact that in planar diagrams only one-particle intermediate states are allowed.}\label{piful}
\end{figure}

Here, $m_n$ is the mass of the $n^{th}$ meson, and $a_n=\langle0|J|n\rangle$. To show this property, one can cut the leading contribution to the two-point function of $J(x)$, an see that the only intermediate states that appears are one-meson intermediate states, as it is shown upper right of figure \ref{piful}, where the intermediate states (the two gluons) go with one quark and one antiquark. In a confining theory, the $q\bar{q}$ pair are always bound together into a meson.

\begin{figure}
    \renewcommand{\captionfont}{\small \it}
    \renewcommand{\captionlabelfont}{\small \it}
    \centering
    \psfrag{A}{$\frac{1}{k^2-m^2}$}
    \psfrag{B}{$a_n$}
    \psfrag{C}{$n$}
    \includegraphics[width=3.5in]{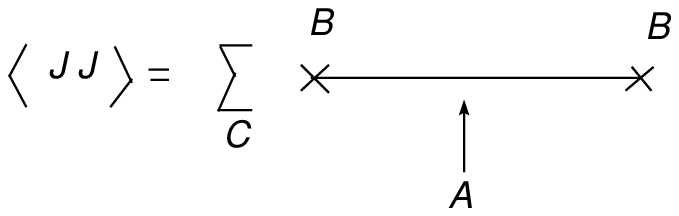}
    \caption{Two-point correlation function in the Large-$N_c$ limit}\label{2pCF}
\end{figure}

\begin{figure}
    \renewcommand{\captionfont}{\small \it}
    \renewcommand{\captionlabelfont}{\small \it}
    \centering
    \psfrag{A}{$\frac{1}{\sqrt{N_c}}$}
    \psfrag{B}{$\sqrt{N_c}$}
    \includegraphics[width=5.0in]{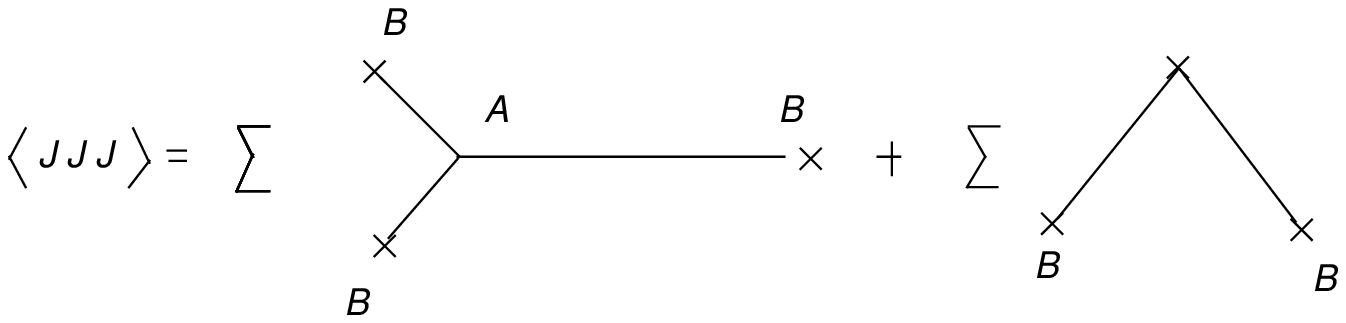}
    \caption{Three-point correlation function in the Large-$N_c$ limit represented as a summation of two possible ways to interact.}\label{3pCF}
\end{figure}

\begin{figure}
    \renewcommand{\captionfont}{\small \it}
    \renewcommand{\captionlabelfont}{\small \it}
    \centering
    \psfrag{N}{$N_c$}
    \includegraphics[width=2.0in]{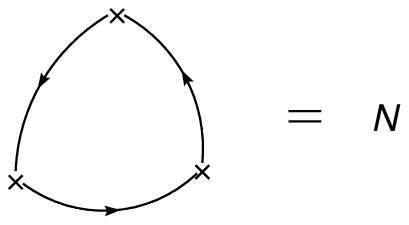}
    \caption{Three-point correlation function loop in the Large-$N_c$ limit as a free field theory.}\label{3pCFloop}
\end{figure}

Following the previous properties, we can write the two-point functions with an one-meson intermediate state as it is shown in figure \ref{2pCF}. Notice that in the case of three-point correlation functions, the diagrams, figure \ref{3pCF}, may be of two types. To determine the amplitude for two-body decays such as $A\rightarrow BC$, we will began with the fact that three point function is of order $N_c$, because in free field theory it is given by the one-loop diagram in figure \ref{3pCFloop}, and we know that the more elaborate planar diagram have the same dependence on $N_c$ that the free field theory has. But this amplitude has a term which is of the form $\langle0|J|m\rangle^3\Gamma_{mmm}$. Since $\langle0|J|m\rangle$ is of order $\sqrt{N_c}$, $\Gamma_{mmm}$ must be of order $1/\sqrt{N_c}$. By similar arguments, we can see that a local vertex with $k$ mesons is of order $1/N_c^{\frac{1}{2}(k-2)}$.

It is also possible to extend this analysis to include glue states. Amplitudes with arbitrary number of mesons and glue states are given, to lowest order in $1/N_c$, by sums of tree diagrams. In these diagrams, the general local vertex with $k$ mesons and $l$ glue states is of order $N_c^{-l-\frac{1}{2}k +1}$.

In conclusion, with the two \textit{selection rules}, it is easy to understand the meson phenomenology explained at the beginning of this section in terms of the Large-$N_c$ limit.

\subsection{Chiral Perturbation Theory in the Large-$N_c$ limit}\label{sec:chiral}

\quad Once we have seen how to understand QCD in the Large-$N_c$ limit, one can proceed to study its low-energy representation. Actually, the $U(N_F)_L \times U(N_F)_R$ chiral symmetry of Large-$N_c$ QCD is spontaneously
broken to a diagonal $U(N_F)_V$ vector symmetry, resulting in (pseudo-) Goldstone bosons\footnote{The axial anomaly is $1/N_c$. See section \ref{resLN}.}. This form for the breaking can be proven in the large-$N_c$ limit~\cite{coleman-witten}. The low-energy interactions of the
pseudo-Goldstone bosons of QCD, the $\pi$, $K$, $\eta$, and now including $\eta'$ (we will se why soon), can be
described, as in section \ref{sec chiral lowest order}, in terms of the effective Lagrangian, the Large-$N_c$ Chiral Lagrangian \cite{LK1,LK2}, which can be computed by evaluating the QCD
functional integral with sources for the pseudo-Goldstone bosons (the source
terms are actually fermion bilinears). In the large-$N_c$ limit we have seen that the
leading order diagrams that contribute to correlation functions of fermion
bilinears are of order $N_c$, and contain a single quark loop. This implies that the leading order terms in the chiral
Lagrangian are of order $N_c$. These leading order terms can be written as a single
flavor trace, since the outgoing quark flavor at one vertex is the incoming
flavor at the next vertex. Similarly, diagrams with two quark loops have two
flavor traces, and are of order unity, and in general, those with $r$ quark
loops have $r$ traces, and are of order $N_c^{1-r}$. This simple rule will help us on the following lagrangian description.

The chiral Lagrangian in the large-$N_c$ limit is written in terms of a unitary matrix
\begin{equation}\label{umatrix}
U = e^{2 i \Pi/f_0},
\end{equation}
where $f_0\approx 92$~MeV is the pion decay constant in the chiral limit, and
\begin{equation}
\Pi = {1\over \sqrt 2}
\left(\begin{array}{ccc}
{\pi^0 \over \sqrt 2} + {\eta \over \sqrt 6} +{\eta' \over \sqrt 3} &
\pi^+ & K^+ \\[0.75em]
\pi^- & -{\pi^0 \over \sqrt 2} + {\eta \over \sqrt 6} +{\eta' \over \sqrt 3} &
K^0 \\[0.75em]
K^- & {\bar K}^0 & -{2\eta \over \sqrt 6} +{\eta' \over \sqrt 3} \\
\end{array}\right),\label{umatrix1}
\end{equation}
is the matrix of pseudo-Goldstone bosons. The $\eta'$ has been included since
it is related to the octet (now nonet) pseudo-Goldstone bosons in the large-$N_c$ limit, by
Zweig's rule (from $SU(3)$ to $U(3)$, the dual group es $N_c^2=9$). The order $p^2$ terms in the chiral Lagrangian (in presence of external fields as in sect. 2.2.1) are

\begin{equation}
\label{3.17}
{\cal L}^{(2)} = \frac{f^2_0}{4} \langle D^{\mu} U D_{\mu} U^{-1}  +
 B_0 ({\cal M}^{\dagger} U + {\cal M} U^{-1}) \rangle,
\end{equation}

where ${\cal M}$ is the quark mass matrix in the QCD Lagrangian. The first term is
order $N_c$, since $f_0 \propto \sqrt N_c$ (as we have said before). The second term in Eq.~(\ref{3.17})
also has a single trace and is of order $N_c$, so $B_0$ is of order unity. The $U$
field can be expanded in powers of $\Pi/f_0$, thus each additional meson
field has a factor of $1/f_0\propto 1/\sqrt N_c$, which gives the required
$1/\sqrt N_c$ suppression for mesons derived earlier. The effective Lagrangian
Eq.~(\ref{3.17}) has then an overall factor of $N_c$. Graphs computed using the chiral Lagrangian have a $1/N_c$ suppression for each loop.

The order ${\cal O}(p^4)$ terms in the chiral Lagrangian were presented in Eq.~(\ref{l4})~\cite{GLsq}. It turns out that one finds the $N_c$-counting rules for the LECs:

\begin{eqnarray*}
\mathcal{O}(N_c): & \qquad & L_1, L_2, L_3, L_5, L_8, L_9, L_{10} \\
\mathcal{O}(1): & \qquad & 2 L_1 - L_2, L_4, L_6, L_7
\label{LislargeN}
\end{eqnarray*}

These are the $N_c$-counting rules given in ref.~\cite{GLsq}, with the exception of
$L_7$, which is taken from ref.~\cite{PerisdeRafaelL7}. In ref.~\cite{GLsq},
$L_7$ was argued to be of order $N_c^2$. The
experimental values for the $L_i$'s in the Large-$N_c$ limit are given in Table~\ref{LsOrderN}. The terms of
order $N_c$ are systematically larger than those of order unity.

\begin{table}
    \centering
    \begin{tabular}{lr@{\hspace{0.2em}}c@{\hspace{0.2em}}lc}
    \hline
    $L_i(M_{\rho})$ & \multispan{3} Value & Order \\
    \hline
    $2 L_1 - L_2 $ & $-0.6$ & $\pm$ & $0.5$ & $1$ \\
    $L_4$ & $-0.3$ & $\pm$ & $0.5$ & $1$ \\
    $L_6$ & $-0.2$ & $\pm$ & $0.3$ & $1$ \\
    $L_7$ & $-0.4$ & $\pm$ & $0.15$ & $1$ \\
    $L_2$ & $1.3$ & $\pm$ & $0.7$ & $N_c$ \\
    $L_3$ & $-4.4$ & $\pm$ & $2.5$ & $N_c$ \\
    $L_5$ & $1.4$ & $\pm$ & $0.5$ & $N_c$ \\
    $L_8$ & $0.9$ & $\pm$ & $0.3$ & $N_c$ \\
    $L_9$ & $6.9$ & $\pm$ & $0.2$ & $N_c$ \\
    $L_{10}$ & $-5.2$ & $\pm$ & $0.3$ & $N_c$ \\
    \hline
    \end{tabular}
    \caption{Experimental values for the coefficients of the order $p^4$ terms in
    the chiral Lagrangian at Large-$N_c$. Values are from ref.~\cite{swissDRV,swissEGPRressonance}}\label{LsOrderN}
\end{table}

Higher derivative terms in the chiral Lagrangian are suppressed by powers of
the chiral symmetry breaking scale $\Lambda_\chi \sim
1$~GeV~\cite{WeinbergChPT,Manohar:1983md}. In the large-$N_c$ limit $\Lambda_\chi$ is of
order unity and so stays at around $1$~GeV. Loop graphs in the chiral
Lagrangian are proportional to $1/(4 \pi f_\pi)^2$ and are of order $1/N_c$. Thus
in the large-$N_c$ limit, the chiral Lagrangian can be used at tree level, and
loop effects are suppressed by powers of $1/N_c$.

\section{Resonances in the Large-$N_c$ limit}\label{resLN}

\quad When we discussed chiral symmetry and its implementation to study the low energy dynamics of the pion octet fields, we estimated the radius of convergence of the chiral expansion to be $\Lambda_{\chi}\sim 1$GeV. This scale lies close to the first resonance multiplet and then a natural extension of the theory would exploit chiral symmetry by incorporating these higher energy excitations states in a Lagrangian. In such a way, this Lagrangian would then allow a prediction of the LECs of the strong interactions in terms of the masses and decay constants of the included resonances.

The building of an effective Lagrangian out of Goldstone bosons and resonance excitation fields in a
chiral-invariant way can be achieved once the resonance fields are embodied with a chiral representation. This is the avenue followed by Resonance Chiral Theory (RChT) which is a description of the Goldstone-resonance interactions in
a chiral invariant framework \cite{swissEGPRressonance,swiss2EGLPR,theworksCiriglianoLEC}. Alternatively to the chiral counting, it uses the $1/N_c$ expansion of QCD in the limit of a large number of colors as a guideline to organize the perturbative expansion. At leading order just tree-level diagrams contribute while loop diagrams yield higher order effects. Integrating out the heavy resonance states leaves at low energies the corresponding chiral invariant effective theory, ChPT. Many works have investigated various aspects of RChT: Green functions (\cite{RuizFemenia,Pablo1,Pablo2,theworksCiriglianoLEC,theworksCirigliano,theworksCiriglianoPich,MoussallamRes,MoussallamRes2,MoussallamRes3,MoussallamRes4,Lipartia,theworksKnechtNyffelerp6}), applications to phenomenology (\cite{RuizFemenia,GuoCillero,JOP,Masjuanetas,Perissubleading,RosellCillerosubleading,RosellCillero2,JuanjoC87}).

In RChT the pseudo-Goldstones enter through the exponential realization $U = exp(i\phi /\sqrt{2}f)$. As the standard effective field theory momentum
expansion is not valid in the presence of heavy resonance states, RChT takes the formal large-$N_c$ limit
expansion as a guiding principle \cite{largeNtHooft,largeNWitten}. The lagrangian can be then organized according to
the number of resonance fields in the interaction terms,
\be\label{LRChT}
{\cal L}_{RChT}={\cal L}_{GB}+{\cal L}_{R_i}+{\cal L}_{R_i R_j}+{\cal L}_{R_iR_jR_k}+...\, ,
\ee
where $R_i$ stands for the resonance multiplets and the first term in the r.h.s. of the equation
contains the operators without resonance fields, Eq.~(\ref{l2}).


Let us consider a chiral-invariant Lagrangian
${\cal L}_R$, describing the couplings of resonance nonet multiplets
$V_i^{\mu\nu}(1^{--})$, $A_i^{\mu\nu}(1^{++})$, $S_i(0^{++})$ and $P_i(0^{-+})$ to
the Goldstone bosons \cite{swiss2EGLPR,BijnensPallante}, and the kinetic terms \cite{swissDRV,swissEGPRressonance},

\begin{eqnarray}\label{eq:L_R}
{\cal L}_R = \sum_i\;\biggl\{ \frac{F_{V_i}}{2\sqrt{2}}\;
\langle V_i^{\mu\nu} f_{+ \, \mu\nu}\rangle\, +\,
\frac{i\, G_{V_i}}{\sqrt{2}} \,\,\langle V_i^{\mu\nu} u_{\mu} u_{\nu}\rangle \, +\, \frac{F_{A_i}}{2\sqrt{2}} \;\langle A_i^{\mu\nu} f_{- \, \mu\nu} \rangle \biggr. \\ \nonumber
\hskip .5cm\biggl.\mbox{}
+\, c_{d_i} \; \langle S_i\, u^{\mu} u_{\mu}\rangle
\, +\, c_{m_i} \; \langle S_i\, \chi_+ \rangle
\, +\, i\, d_{m_i}\;\langle P_i\, \chi_- \rangle
\biggr\}\, ,
\end{eqnarray}

where

$$
u_\mu \equiv i\, u^\dagger D_\mu U u^\dagger, \
f^{\mu\nu}_\pm\equiv u F_L^{\mu\nu} u^\dagger\pm  u^\dagger F_R^{\mu\nu} u
$$

with $F^{\mu\nu}_{L,R}$ the field-strength tensors of the $l^\mu$ and $r^\mu$ flavor fields defined in Eq. (\ref{l4}) and

$$
\chi_\pm\equiv u^\dagger\chi u^\dagger\pm u\chi^\dagger u.
$$

The resonance couplings
$F_{V_i}$, $G_{V_i}$, $F_{A_i}$, $c_{d_i}$, $c_{m_i}$ and $d_{m_i}$
are of \ $O\left(\sqrt{N_c}\,\right)$.

The lightest resonances have an important impact on the
low-energy dynamics of the pseudoscalar bosons.
Below the resonance mass scale, the singularity associated with the
pole of a resonance propagator is replaced by the corresponding
momentum expansion. Therefore, the exchange of virtual resonances generates
derivative Goldstone couplings proportional to powers of $1/M_R^2$.
At lowest order in derivatives, this gives the large--$N_c$ predictions
for the $O(p^4)$ couplings of chiral perturbation theory \cite{swiss2EGLPR,BijnensPallante}.


All these couplings are of $O(N_c)$, in agreement with the
counting indicated in Table~\ref{LsOrderN}, while for the couplings of $O(1)$ we get
$2\, L_1-L_2 = L_4 = L_6 = L_7 = 0$ (as we will see experimentally, except for $L_7$, in table \ref{Lis}).

Owing to the $U(1)_A$ anomaly, the $\eta_1$ field is massive and it is often integrated out from the low-energy chiral theory. In that case, the $SU(3)_L\otimes SU(3)_R$ chiral coupling $L_7$ gets a contribution from $\eta_1$ exchange \cite{GLsq,swiss2EGLPR,BijnensPallante} is:

\be
\label{eq:L7}
L_7 = - \frac{\tilde{d}_m^2}{2\, M^2_{\eta_1}} \, ,
\qquad\qquad\qquad
\tilde{d}_m = -\frac{f}{\sqrt{24}} \, .
\ee

\subsection{Integrating out Resonances}\label{integrateRes}

\quad Once the Lagrangian ${\cal{L}}_{\cal R}$ is chosen, the natural
step is to integrate out the resonances. We are left with the original chiral lagrangian,
the coefficients being functions of hadronic parameters (masses and couplings), to be related
to the $L_i$'s through a matching procedure. The results are as
follows, now with couplings of $O(1)$ and $O(N_c)$ together {\cite{swissDRV,swissEGPRressonance}}:

\begin{equation}
\begin{array}{lcccccccccccccc}
L_1=&\frac{G_V^2}{8M_V^2}&+&
&-&\frac{c_d^2}{6M_S^2}&+&\frac{{\tilde{c}}_d^2}{2M_{S_1}^2}&+& &+& &\\
L_2=&\frac{G_V^2}{4M_V^2}&+& &+& &+& &+& &+& &\\
L_3=&-\frac{3G_V^2}{4M_V^2}&+& &+&\frac{c_d^2}{2M_S^2}&+& &+& &+& &\\
L_4=& &+& &-&\frac{c_dc_m}{3M_S^2}&+&\frac{\tilde{c}_d\tilde{c}_m}{M_{S_1}^2}&+& &+& &\\
L_5=& &+& &+&\frac{c_dc_m}{M_S^2}&+& &+& &+& &\\
L_6=& &+& &-&\frac{c_m^2}{6M_S^2}&+&\frac{\tilde{c}_m^2}{2M_{S_1}^2}&+& &+& &\\
L_7=& &+& &+& &+& &+&\frac{d_m^2}{6M_P^2}&-&\frac{\tilde{d}_m^2}{2M_{{\eta}_1}^2}&\\
L_8=& &+& &+&\frac{c_m^2}{2M_S^2}&+& &-&\frac{d_m^2}{2M_P^2}&+& &\\
L_9=&\frac{F_VG_V}{2M_V^2}&+& &+& &+& &+& &+& &\\
L_{10}=&-\frac{F_V^2}{4M_V^2}&+&\frac{F_A^2}{4M_A^2}&+& &+& &+& &+& &
\end{array}
\end{equation}

\noindent\\

Upon comparison with experiment one gets the values shown in Table \ref{Lis} \footnote{Numerical values
    for masses and couplings are given in Refs.\cite{swissDRV,swissEGPRressonance}.}.

This Table \ref{Lis} shows a remarkable agreement, strongly supporting resonance
saturation. Still, there are two flaws in the argument: first the value $\mu^*=M_{\rho}$ is, though very suggestive, completely arbitrary. The right way to proceed would be to determine $\mu^*$ from
${\cal{L}}_R$ at one loop level. The renormalized couplings $L_i^r(\mu)$ depend on the arbitrary
scale of dimensional regularization $\mu$. This scale dependence is of course canceled by that of the loop amplitude, in any measurable quantity. However, contrary to ${\cal{L}}_{QCD}$ in Eq. (\ref{lqcd}), ${\cal{L}}_{R}$ has no obvious expansion parameter. One way out is rely on the fact that ${\cal{L}}_{R}$ is meant to be an EFT of QCD and then we can resort to Large-$N_c$ methods to tackle the problem.

The second flaw is related to the values used for the masses $M_V$, $M_A$, $M_S$, $M_{S_1}$ and $M_P$ to obtain the estimations shown in Table \ref{Lis}. To describe the Lagrangian Eq.(\ref{LRChT}), one uses instead the Lagrangian ${\cal L_R}$, Eq.(\ref{eq:L_R}), where one has done the assumption that at low energies the main contribution coming from the resonant sector is mainly dominated by the lowest lying states. However, we already know that in the large-$N_c$ limit of QCD, the Green's functions are described by a infinite number of states. The the precedent Lagrangian has also to be described with a infinite number of states. In practice, however, one relies on Eq.(\ref{eq:L_R}) as an approximate Lagrangian which only includes one resonance per channel. We will see in the next chapter than in that situation, i.e., using a finite number of states instead of an infinite one leads to a rational approximation which poles and residues that may have nothing to do with the physical ones. Then, using the values for the lowest meson states from the Particle Data Book to obtain the estimations shown in Table \ref{Lis} may induce a systematical source of error which is not been properly considered up to now.

\begin{table}
\centering
\begin{tabular}{|c|c|c|c|c|c|c|c|c|}
\hline
$L_i(M_{\rho})$ & experimental values ($\times 10^{-3}$) & $V$ & $A$ & $S$ & $S_1$ & $\eta_1$ & total \\
\hline \hline
$L_1$ &     $0.7\pm0.3$ &   $0.6$ & $0$ & $-0.2$ & $0.2$ & $0$ & $0.6$       \\
\hline
$L_2$ &     $1.3\pm0.7$ &   $1.2$ & $0$ & $0$ & $0$ & $0$ & $1.2$      \\
\hline
$L_3$ &            $-4.4\pm2.5$ & $-3.6$ & $0$ & $0.6$ & $0$ & $0$ & $-3.0$         \\
\hline
$L_4$ &             $-0.3\pm0.5$ & $0$ & $0$ & $-0.5$ & $0.5$ & $0$ & $0.0$        \\
\hline
$L_5$ &        $1.4\pm 0.5$ & $0$ & $0$ & $1.4$ & $0$ & $0$ & $1.4$    \\
\hline
$L_6$ &         $-0.2\pm0.3$ & $0$ & $0$ & $-0.3$ & $0.3$ & $0$ & $0.0$       \\
\hline
$L_7$ &          $-0.4\pm0.15$ & $0$ & $0$ & $0$ & $0$ & $-0.3$ & $-0.3$       \\
\hline
$L_8$ &           $0.9\pm0.3$  & $0$ & $0$ & $0.9$ & $0$ & $0$ & $0.9$     \\
\hline
$L_9$ &            $6.9\pm 0.2$ & $6.9$ & $0$ & $0$ & $0$ & $0$ & $6.9$         \\
\hline
$L_{10}$ &          $-5.2\pm 0.3$ & $-10.0$ & $4.0$ & $0$ & $0$ & $0$ &   $-6.0$       \\
\hline
\end{tabular}
\caption{Values of low-energy constants appearing  in Eq. (\ref{l4})}\label{Lis}
\end{table}
\noindent\\

\subsection{Minimal Hadronic Approximation}\label{sectionMHA}

We have seen in sec. \ref{sec:LN} that the operator $J(x)$, as a generic quark bilinear creates, in the Large-$N_c$ limit, only one narrow meson state. In other words, large$N_c$ QCD is able to constrain the analytical form of QCD Green functions to be meromorphic functions, Eq.~(\ref{2pLN}). In order to reproduce the parton model logarithm, the sum of states in Eq.~(\ref{2pLN}) extends to infinity. Analytically, the infinite number of poles and residues of the Green function could be known if the solution to large-$N_c$ QCD is known. However, as we have said, this is still to be determined and then Eq.~(\ref{2pLN}) is not much useful phenomenologically speaking.

The so called \textit{Minimal Hadronic Approximation} (MHA) is, in the context of large-$N_c$ QCD, a much useful phenomenological approximation. Is, in itself, a rational approximation which allows us to make the matching conditions a useful tool where our knowledge of a certain Green's function is centered in its chiral expansion and also the expansion given by its OPE.
In that sense, one writes down Eq.~(\ref{2pLN}) with a finite number of states, i.e.

\begin{equation}\label{MHA}
\Pi(q^2) = \sum_{n=1}^{{\cal N}} \frac{a_n^2}{k^2-m^2}
\end{equation}

The MHA \cite{MHAdeRafaelKnecht,MHAPerisMaching,MHA2,meromorphicPerislargeN,meromorphicdeRafael} to Large-$N_c$ QCD is an interpolating function between these two regimes. These matching conditions imply the fact that we have to use two different languages to describe the same physics: while at short distances we use quark and gluon as a variables, at long distances we use meson fields. These entail that one has to match an expansion in powers of
$\alpha_{s}$ coming from short distances to an expansion in
powers of meson momenta at long distances.

In recent years,
a large amount of work has been dedicated to studying the consequences of these ideas
\cite{theworksPerisRafaelPerrottet,theworksPerisRafaelPerrottetKnecht,theworksPerisGolterman,theworksKnechtPerisRafael,theworksKnechtNyffelerp6,theworksPerisKnechtRafaelg2,theworksGoltermanPerispenguin,theworksHambyePerisRafael,theworksCiriglianoPich,thewrksBijnensGamizPrades,theworksCirigliano,theworksCiriglianoLEC,theworksBijnens}.

It was then realized that all the successful results from a chiral resonant lagrangian
could be encompassed at once as an approximation to large-$N_c$ QCD consisting in keeping only a
\emph{finite} (as opposed to the original infinite) set of resonances in Green's functions.
The main advantage of that approximation is, after ensuring that fulfills all the constrains from low and high energies imposed, one can go improving the result by adding more and more terms in Eq.(\ref{MHA}) expecting a smooth convergence.


\chapter{Pad\'{e} Theory}\label{capitol1}

\section{Introduction to Pad\'{e} Theory}\label{sec:PadeTheory}

\quad In the study of mathematical analysis, the relation between a certain function and the Taylor coefficients of its expansion is still a profound mathematical question. Classically, the answer is that if the Taylor series expansion converges absolutely, then defines uniquely the value of a function which is differentiable an arbitrary number of times. In practice, however, two problems may appear. The first is how to compute the Taylor coefficients of a certain function, and the second what would be the range of applicability of that Taylor expansion.



Moreover, when perturbation methods are used to solve a problem, the answer usually emerges as an infinite series. If the perturbation series converge rapidly, summing the few calculated terms gives a good approximation to the exact solution. However, it is more common for the series to converge slowly, if it converges at all. A regular perturbation series could converge slowly when the modulus of its parameter of expansion $\varepsilon$ is less than the radius of convergence. Singular perturbation series diverge for all values of $\varepsilon\neq 0$, and even if the series is asymptotic, the value of $\varepsilon$ may be too large to obtain much useful information.

While discovering that a perturbation series diverges is really discouraging, it is possible to assign a sensible meaning to the sum of the series and even to use the first few terms to approximate this sum.

Also, a divergent series indicates the presence of singularities and then shows the inability of a polynomial to approximate the function adequately. Therefore, a summation algorithm is needed which requires as input only a finite number of terms of that series.

On top of these kind of methods, Euler summation and Borel summation are examples of summing an infinite series even when diverges and are been used extensively in many fields in physics and biological sciences.






However, both Euler and Borel summation methods have the difficulty to all the terms of the divergent series must be known exactly before the sum be done. These cases are, nevertheless, not always realistic enough since, as we said, sometimes one can only compute few terms of that series.


A well-known summation method having this property is called Pad\'{e} summation.

\subsection{What is a Pad\'{e} Approximant?}

\quad The present work will present this method and its properties, and will be applied to several physical cases where we will learn aspects of resonance saturation, unitarization process, spectral functions, etc. Before going into a detailed mathematical explanation we think that it would be appropriate to show first the mean advantages of the Pad\'{e} Approximants (PAs in the following) method with a nice and simple example. Later on, we will explain in a more rigorous way the main properties of Pad\'{e} Approximants, especially those related with the convergence properties of our approximations.

Lets consider, then, the following function $F(x)$:

\begin{equation}\label{taylorexp}
F(x)=\sqrt{\frac{1+2x}{1+x}}=\sum_{n=0}^{\infty} a_n x^n=1+\frac{x}{2}-\frac{5x^2}{8}+\frac{13x^3}{16}-\frac{144x^4}{128}+{\cal O}(x^5)\,.
\end{equation}
\noindent
As an amusement, lets suppose that the true function $F(x)$ is not known but only its Taylor expansion. However, we would like to compute the value $\lim_{x\to \infty} F(x)$ (for simplicity, called $F(\infty)$). Since the function $F(x)$ is not known (if was, $\lim_{x\to \infty} F(x)=\sqrt{2}$), we will compute it through its Taylor expansion. However, since the radius of convergence of that function is $r=1/2$, for any value of $|x|>1/2$ this expansion fails. In this particular example we can use a special trick to transform the series into one which will let us estimate the value $F(\infty)$. Suppose the change of variables $x=\frac{w}{1-2w}$, then

\begin{equation}\label{taylorw}
F(x(w))=\widetilde{F}(w)=(1-w)^{-1/2}=\sum_{n=0}^{\infty} b_n w^n=1+\frac{w}{2}+\frac{3x^2}{8}+\frac{5w^3}{16}+\frac{35w^4}{128}+{\cal O}(w^5)
\end{equation}
\noindent
For the new variable $w$, the limit $x\rightarrow \infty$ is translated into $w\rightarrow 1/2$ and now the new Taylor series representation, Eq.~(\ref{taylorw}), converges at $w=1/2$. The first few successive approximation to $F(\infty)$, namely $\sum_{n=0}^k b_n w^n$, for $k=0,1,2,3,4,\cdots$ are shown in Table \ref{Tab:taylorvalues} evaluated at $w=1/2$. The sequence of values are certainly converging to $\sqrt{2}\simeq 1.4241$.

\begin{table}
\centering
\begin{tabular}{|c|c|c|c|c|c|}
\hline
$k$ & $0$ & $1$ & $2$ & $3$ & $4$  \\
\hline
$\sum_{n=0}^k b_n w^n$ &     $1$ & $1.25$ &  $1.34375$ & $1.38281$ & $1.39990$\\
\hline
\end{tabular}
\caption{Results for the lowest partial sums of Eq.(\ref{taylorw}).}\label{Tab:taylorvalues}
\end{table}

%
In fact, the partial sums $\sum_{n=0}^k b_n w^n$ can be written in terms of the original variable $x$ following $w=x/(1+2x)$. For the first $k=0,1,2,\cdots$, we find

\be\label{rationalseries}
1,\,\frac{1+(5/2)x}{1+2x}, \,\frac{1+(9/2)x+(43/8)x^2}{(1+2x)^2},\,\cdots
\ee
which are rational fractions in terms of the original variable $x$.

A particular kind of such a rational approximation is the Pad\'{e} Approximant. The mean idea is to match the Taylor series expansion into a rational function so that it would tend to a finite limit as $x$ tends to infinity as in the case of Eq.~(\ref{rationalseries}). The simplest PA to the function $F(x)$ is:

\be
R(x)=\frac{a_0+ a_1 x}{1+ b_1 x}\, .
\ee\label{PAexemple}

The three unknown parameters are fixed in the way that the Taylor expansion of $R(x)$ be the same Taylor expansion of $F(x)$ in Eq.~(\ref{taylorexp}). We find:

\begin{equation}\label{Rx}
R(x)=\frac{1+\frac{7}{4}x}{1+\frac{5}{4}x}=1+\frac{x}{2}-\frac{5x^2}{8}+\frac{25x^3}{32}-\frac{125x^4}{128}+{\cal O}(x^5)\, .
\end{equation}

As shown in Eq.~(\ref{Rx}), if one expands $R(x)$, finds that the first three coefficients are exactly the same that those of Eq.~(\ref{taylorexp}) by construction but the other coefficients are different. Evaluating now $\lim_{x\to \infty} R(x) =1.4$ which is a better determination that any of the approximations of Table~\ref{Tab:taylorvalues}.


The next such approximation, which needs $5$ parameters, is:
\begin{equation}
\frac{1+\frac{13}{4}x+\frac{41}{16}x^2}{1+\frac{11}{4}x+\frac{29}{16}x^2} \stackrel{x\to\infty}{\longrightarrow} \frac{41}{29}=1.41379.
\end{equation}

Further such approximations converge quite well and one sees an improvement of the result while going further in the approximant sequence. In fact, using $7,9$, and $11$ Taylor coefficients we obtain:

\begin{equation}
1.414201183\, , \quad 1.414213198\, \quad \mathrm{and} \quad 1.414213552\, , \nonumber
\end{equation}

respectively, the last one is off by only $10^{-8}$.

We can also form the same type of approximation to the expansion but now in powers of the variable $w$:

\be
1,\,\frac{1-\frac{w}{4}}{1-\frac{3 w}{4}} , \,\frac{1-\frac{3 w}{4}+\frac{w^2}{16}}{1-\frac{5 w}{4}+\frac{5 w^2}{16}},\,\cdots
\ee\label{rationalseriesw}

Evaluating the approximants at $w=1/2$ we get the results $1, 1.4, 41/29,\cdots$, identical values obtained using the series expansion in $x$.

This invariance principal is a general and important property of Pad\'{e} Approximants and is the basis of their ability to sum the $x$ series in our example and give excellent results, even at $x\to \infty$.

The successive approximations we have built increase monotonically. Although this property is not a general one, it can be proven to hold in a wide variety of cases. For these cases, one can prove that certain Pad\'{e} Approximants form converging upper and lower bounds.


\subsubsection{Rigorous definition of a Pad\'{e} Approximant}

Since we already have introduced the notion of a Pad\'{e} Approximant, lets now define them in a more rigorous mathematical way. Let a function $f(z)$ have an expansion around the origin of the complex plane of the form
\begin{eqnarray}\label{fexp}
    f(z)&=& \sum_{n=0}^{\infty} f_n z^n\quad , \quad z\rightarrow 0\ .
   \end{eqnarray}
with a certain radius of convergence $r$. One defines a Pad\'{e} Approximant to $f(z)$, denoted by $P^{M}_{N}(z)$, as a ratio of two polynomials $Q_M(z), R_N(z)$\footnote{Without loss of generality we define, as it is usually done,
$R_N(0)=1$.}, of order $M$ and $N$ (respectively) in the variable $z$, with a \emph{contact} of order
$M+N$ with the expansion of $f(z)$ around $z=0$. This means that, when expanding $P^{M}_{N}(z)$
around $z=0$, one reproduces exactly the first $M+N+1$ coefficients of the expansion for $f(z)$, Eq. (\ref{fexp}):

\begin{equation}\label{Padef}
    P^{M}_{N}(z)=\frac{Q_M(z)}{R_N(z)}
    \approx f_0 + f_1\ z + f_2\ z^2 +...+ f_{M+N}\ z^{M+N}+ \mathcal{O}(z^{N+M+1})\ .
\end{equation}

At finite $z$, the rational function $P^{M}_{N}(z)$ constitutes a resummation of the series
(\ref{fexp}). Is in that sense that sometimes that method is also called a Pad\'{e} summation.

Of special interest for the present work will be the case when $M=N+J$, for a fixed $J$. The corresponding PAs $P^{N+J}_{N}(z)$ belong
to what is called the near-diagonal sequence for $J\neq 0$, with the case $J=0$ being the diagonal
sequence.

As seen with the example, the mean advantages of that method are that the full power series representation of a function need not to be known to construct a Pad\'{e} Approximant -just the first $M+N+1$ terms. We will see that the series can be divergent and still be able to construct a PA sequence that approximates quite well the original function. Moreover, sometimes they work rather well even beyond their proven range of applicability.

\subsection{Examples}

\quad To point out that these advantages are more general that the case we have already explained, we would like to show a few well known examples of convergence for Pad\'{e} Approximants in both convergence and divergence series representation.

\subsubsection{The convergence series case}

\quad The function $e^z$ has a Taylor series $\sum_{n=0}^{\infty} z^n/n!$, and then can be easily transformed into a sequence of Pad\'{e} Approximant. For all complex $z$, both the Taylor series approximant and the Pad\'{e} Approximant approach $e^z$ as $n,M,N\rightarrow\infty$. The rates of convergence are well known in both cases and it can be shown \cite{Bender} that the relative error done by a Pad\'{e} Approximant is $2^n$ times smaller than the error in the Taylor approximant, as $n\rightarrow \infty$.

Another example can be done using the function $1/\Gamma(z)$. In that case, although Pad\'{e} summation enhances the convergence of the Taylor series, the effect is not dramatic. The Pad\'{e} sequence requires about half the number of terms that the Taylor series requires to achieve $1\%$ accuracy. Since $1/\Gamma(z)$ has no singularities, the benefits of Pad\'{e} summation are marginal.

\subsubsection{The divergence series case}

\quad In these two previous examples, one can not see a compelling reason to use Pad\'{e} summation because the Taylor series already converge for all $z$ and the improvement of convergence is not astounding. However, the real power of Pad\'{e} summation arises when applying the method to divergent series.

As a first example of that case, lets examine the function $f(z)=\log(1+z)$, which is a Stieltjes function. A function $f(z)$ is called Stieltjes function if it can be expressed as a Stieltjes integral according to

\begin{equation}\label{stieltjesint}
f(z)=\int_0^{\infty} \frac{d \psi (t)}{1+ z t}, \qquad |\mathrm{Arg}(z)|<\pi.
\end{equation}

Here, $\psi (t)$ is a measure on $0\leq t <\infty$ and has for all $m\geq 0$ finite and positive moments $\mu_m$ defined by

\be\label{momentsstieltjes2}
\mu_m=\int_0^{\infty} t^m d\psi(t).
\ee

The formal series expansion for $f(z)$, which do not need to be convergent,

\be
f(z)=\sum_{m=0}^{\infty} (-1)^m \mu_m z^m\,
\ee

\noindent
is called a Stieltjes series if its coefficients $\mu_m$ are moments of a measure $\psi(t)$ on $0\leq t< \infty$. According to Eq. (\ref{momentsstieltjes2}):

\be
f(z)=\sum_{m=0}^{\infty} (-1)^m z^m \int_0^{\infty} t^m d\psi (t).
\ee

An example of a Stieltjes series is, then, the series for $f(z)=\log(1+z)$:

\be\label{logstieltjes}
\mathrm{Log}(1+z) = \sum_{m=0}^{\infty} \frac{(-1)^m z^{m+1}}{m+1}
\ee

For $|z|<1$, the power series converges absolutely, for $|z|=1$ it converges conditionally, and for $|z|>1$ it diverges. However, as long as the $ |\mathrm{Arg}(z)|<\pi$, the divergent series can at least in principle be summed.

The series shown in Eq. (\ref{logstieltjes}) follows from the integral representation

\be
\mathrm{Log}(1+z) =z\int_0^1\frac{dt}{1+ zt}.
\ee
We only have to set $\psi (t)$ for $0\leq t \leq 1$ and $\psi (t)=1$ for $1< t < \infty$. The moments $\mu_m$ of this positive measure $\psi(t)$ are given by

\be
\mu_m=\int_0^{\infty} t^m d\psi (t) = \int_0^1 t^m dt = \frac{1}{m+1}\, .
\ee

If the positive measure $\psi (t)$ corresponding to the Stieltjes function $f(z)$ could be determined directly from the Stieltjes moments $\mu_m$, the value of $f(z)$ could at least in principle be computed via the integral representation Eq.~(\ref{stieltjesint}), even if the series diverges.

A necessary and sufficient condition which guarantee that a Pad\'{e} Approximant are able to sum a divergent Stieltjes series is that all Hankel determinants $D(m,n)$ formed from the moments $\mu_m$ have to be strictly positive, \cite{Baker1}, i.e.,

\begin{equation}\label{detHankel}
    D(m,n)=\begin{vmatrix}
             \mu_m & \mu_{m+1} & \ldots & \mu_{m+n} \\
             \mu_{m+1} & \mu_{m+2} & \ldots & \mu_{m+n+1} \\
             \vdots & \vdots &  & \vdots \\
             \mu_{m+n} & \mu_{m+n+1} & \ldots & \mu_{m+2n} \\
           \end{vmatrix} >0\quad .
\end{equation}

Unfortunately, the practical application of this condition is by no means simple, in particular if only the numerical values of a finite number of Stieltjes moments $\mu_0, \mu_1, \cdots \mu_n$ are available. However, there is a comparatively simple sufficient condition, the so-called Carleman condition \cite{Baker1}, that says that if the moments $\mu_m$ satisfy
\be
\sum_{m=0}^{\infty} \mu_m^{\frac{-1}{2m}}=\infty \, ,
\ee
then the PA sequence $P_N^{N+J}$ converge for $J>-1$ to the value of the corresponding Stieltjes function as $N\rightarrow \infty$. It was shown in Ref.~\cite{stieltjessums} that the Carleman condition is satisfied if the moments $\mu_m$ do not grow faster that $C^{m+1} (2m)!$ as $N\rightarrow \infty$, with $C$ being a suitable positive constant.
The Carleman condition implies, for example, that the series Eq. (\ref{logstieltjes}) can be approximated by the Pad\'{e} sequence $P_N^{N+J}$ for $J>-1$ for all $z$ in the cut plane $|\mathrm{Arg} (z)|<\pi$, \cite{MasJJVpades}.

The reason why the Pad\'{e} Approximants provide a better description that the Taylor series for the function $f(z)=\ln (1+z)$ is that the poles of $P_N^{N+J}(z)$ have the ability to approximate the effect of its branch-cut singularity. The poles of $P_N^N(z)$ and $P_{N}^{N-1}(z)$ all lie on the real negative axis and, as $N\rightarrow\infty$, the poles become dense and in some sense resemble the branch cut\cite{Bender,MasJJVpades,HeavyquarkPades}.

Among other examples of application of Pad\'{e} summation one can find the well know asymptotic series case, for example the Stirling series, which can be rapidly approximated by a Pad\'{e} summation. Or a simple and extremely efficient computation of a asymptotic series is the case of the parabolic cylinder function, $D_{\nu}(x)$, which for $x\geq1$ is the solution of the parabolic cylinder equation $y''+(\nu + 1/2 +1/4 x^2)y=0$.

\subsection{Convergence theorems}\label{convth}

\quad In despite of these nice examples, since many functions may be asymptotic to the same divergent series, which of these functions, if any, will the Pad\'{e} sequence select as its limit? Moreover, if the limit function is multivalued, which branch of the function will be singled out as that given by the limit of the Pad\'{e} sequence? There is no general theory of Pad\'{e} summation for arbitrary series yet. Nevertheless, the convergence theory of Pad\'{e} Approximants for the special class of Stieltjes series, meromorphic functions\footnote{A function is said to be meromorphic when its singularities are only isolated poles.}, continued fractions, hypergeometric functions, Bessel functions is relatively complete\footnote{In the next section we will briefly schematize few convergence theorems that will we used along these pages.}. For Stieltjes series, one can partially answer these questions.
Since a Stieltjes series is the prototype of an asymptotic series
\be
\label{StieltjesSerie}
\sum_{n=0}^{\infty}(-1)^n n!x^n
\ee
which is a formal solution to a certain differential equation, one can not add up all the terms of the divergent series because the sum does not exist. By "summing" we mean finding a function to which the series is asymptotic. To sum the series (\ref{StieltjesSerie}) we invoke the integral identity $n!=\int_0^{\infty }e^{-t}t^n dt$ and the interchange $\sum_{n=0}^{\infty}(-xt)^n=\frac{1}{1+xt}$, for $|x t|<1$:
\be
\sum_{n=0}^{\infty}(-1)^n n!x^n \rightarrow \sum_{n=0}^{\infty}(-x)^n\int_0^{\infty }e^{-t}t^n dt\rightarrow \int_0^{\infty }dt e^{-t}\sum_{n=0}^{\infty}(-xt)^n
\ee
The last piece can be solved by splitting the integral in two parts:
\be
\int_0^{\infty }dt e^{-t}\sum_{n=0}^{\infty}(-xt)^n = \int_0^{1}dt e^{-t}\sum_{n=0}^{\infty}(-xt)^n+\int_1^{\infty }dt e^{-t}\sum_{n=0}^{\infty}(-xt)^n \, .
\ee
Using $\sum_{n=0}^{\infty}(-xt)^n=\frac{1}{1+xt}$ the first integral becomes $\int_0^{1}\frac{e^{-t}}{1+xt}dt$. The second integral becomes $\int_1^{\infty }\frac{e^{-t}}{1+xt}dt$ using the change of variables $-xt=1/y$ and the sum $\sum_{n=0}^{\infty}y^{-n}=\frac{y}{y-1}$. The resulting integral, summing both parts, $y(x)=\int_0^{\infty }\frac{e^{-t}}{1+xt}dt$ is a Stieltjes integral as Eq. (\ref{stieltjesint}) and satisfies exactly the differential equation $x^2y''+(1+3x)y'+y=0$.

Finally, integrating $y(x)$ by parts iteratively, one recovers the asymptotic series expansion valid as $x\rightarrow 0^+$, Eq.~(\ref{StieltjesSerie}).

A more interesting case is the so called Generalized Stieltjes integral, given by
\be
f(x)=\int_0^{\infty}\frac{\rho(t)}{1+xt}dt
\ee

where the weight function $\rho(t)$ is nonnegative for $t>0$ and approaches zero so rapidly as $t\rightarrow \infty$ that the moment integrals
\be
a_n=\int_0^{\infty }t^n\rho(t)dt
\ee

exist for all positive integers $n$.

The Pad\'{e} sequence of a Stieltjes series representation has some remarkable convergence properties when its weight function is nonnegative. One can show that, for $x>0$:
\begin{itemize}
\item The diagonal Pad\'{e} sequence $P_N^N(x)$ decreases monotonically as $N$ increases.
\item The Pad\'{e} sequence $P^{N-1}_N(x)$ increases monotonically as $N$ increases.
\item The sequence $P_N^N(x)$ has a lower bound, while the sequence $P^{N-1}_N(x)$ has an upper bound.
\end{itemize}
These properties imply that $\lim_{N\rightarrow \infty}P_N^N(x)$ and $\lim_{N\rightarrow \infty}P^{N-1}_N(x)$
exist and also that \\
$\lim_{N\rightarrow \infty}P^{N-1}_N(x) \leq \lim_{N\rightarrow \infty}P_N^N(x)$.

Even more, all Stieltjes functions $F(x)$ with the same finite series representation used in the Pad\'{e} summation satisfy
\be\label{PadesStieltjes}
\lim_{N\rightarrow \infty}\,P^{N-1}_N(x)\, \leq \,F(x)\,\leq\,\lim_{N\rightarrow \infty}P_N^N(x)
\ee
These previous inequalities can be generalized as follows. For any $J\geq -1$ the Pad\'{e} sequence $P^{N+J}_N(x)$, if generated from a Stieltjes function, is monotone increasing when $J$ is odd and monotone decreasing when $J$ is even.

Moreover, since Stieltjes functions are analytic in the cut plane $|\mathrm{Arg}\,(x)|<\pi$ and since the Pad\'{e} approximants $P_{N}^N(x)$ and $P^{N-1}_N(x)$ converge to the original function, all the poles of $P_{N}^N(x)$ and $P^{N-1}_N(x)$ must lie on the negative real axis. Finally, if a series is a Stieltjes series, then the Pad\'{e} sequences $P_{N}^N(x)$ and $P^{N-1}_N(x)$ converge on the cut$-x$ plane and the limit functions are Stieltjes functions.

The convergence properties of the PAs to a given function are much more difficult than those of
normal power series and this is an active field of research in Applied Mathematics. As we have alreadu said,
those which concern continued functions, hyperbolic functions and meromorphic functions are rather well-known and, specially this last case, will be of particular interest
for this work. There are also several well known theorems that will we of our interest, among them Montessus de Ballore and the Nuttal-Pomerenke theorem. The following section presents these theorems without so much detail but accurately, since these theorems will we used along these pages and we will refer it several times.

\textit{Montessus de Ballore}'s theorem applies to a function $f(z)$ which is meromorphic with its $M$ poles within a circle.
In practical terms, provided $M$ is known, Montessus' theorem asserts uniform convergence of the sequence of $P^L_M$  approximants within the circle. The complete statement of Montessus' theorem \cite{Montessus} when the function has simple poles is the following\footnote{An easy proof of the theorem can be found in \cite{Baker1}, chapter 6.2.}:

$\bullet$ Let $f(z)$ be a function which is meromorphic in the disk $|z|\leq R$, with precisely $M$ simple poles at distinct points $z_1, z_2, \cdots, z_M$, where
\be\nn
0<|z_1|\leq|z_2|\leq\cdots\leq|z_M|<R.
\ee
Then
\be\label{montessusdef}
\lim_{L\rightarrow \infty}P^L_M=f(z)
\ee
uniformly on any compact subset of
\be\nn
\mathfrak{D}_M=\{z, |z|\leq R, z\neq z_i, i=1,2,\cdots , M\}\, .
\ee

Later on, R. de Montessus de Ballore extended the theorem for the possibility of multiple poles instead of simple poles:

$\bullet$ Let $f(z)$ be a function which is meromorphic in the disk $|z|\leq R$, with $m$ poles at distinct points $z_1, z_2, \cdots z_m$ with
\be
0<|z_1|\leq|z_2|\leq\cdots\leq|z_m|<R.
\ee
Let the poles at $z_k$ have multiplicity $\mathrm{\nu_k}$, and let the total multiplicity $\sum_{k=1}^m\nu_k=M$ precisely. Then

\be\label{Montessus}
\lim_{L\rightarrow \infty}P^L_M=f(z)
\ee
uniformly on any compact subset of

\be\nn
\mathfrak{D}_M=\{z, |z|\leq R, z\neq z_k, k=1,2,\cdots , m\}.
\ee

The main advantage of that theorem is that guarantees uniform convergence of the approximation within a circle defined by $|z_m|<R$. No spurious poles coming from the approximation will be found inside this disk. Also, the position of the $M$ poles will be correctly determined by taking the limit $\lim_{L\rightarrow \infty}$ in Eq.(\ref{Montessus}), given the number of poles and multiplicities \textit{M} is known. An interesting application of that theorem can be found in Ref.\cite{CilleroPadepoles}.

In despite of these convergence statements, it is difficult to go beyond that since there are no proven rate of convergence for the $P^L_M$ sequence to the original function, or upper and lower boundary constrains as it happens when one constructs the sequences $P^N_N$ and $P^{N-1}_N$ for Stieltjes functions.

When the function is Stieltjes and meromorphic one has to choose which theorem would be better to use. It will depend on the underlying motivation to use Montessus' theorem or the theorem for Stieltjes functions that we have already described. The main difference between both convergence theorems is that the Montessus' theorem guarantees convergence uniformly inside a certain disk while the other one provides a boundary constrains for the approximation.

When the function is meromorphic but is not Stieltjes, the Montessus' theorem says nothing about upper and lower boundary constrains. Then, in practice, one constructs a Pad\'{e} sequence and estimates the accuracy of the method by looking into the difference between consecutive PAs, Refs.\cite{MasjuanPades,MasjuanPerisC87,MasjuanProc,HeavyquarkPades}.

An extension of the Montessus de Ballore theorem is the Nuttall-Pommerenke's Theorem \cite{Pommerenke} (commonly known as Pommerenke's Theorem) which asserts that the sequence of (near) diagonal PA's to a meromorphic function is convergent everywhere in any compact set of the complex plane except, perhaps, in a set of zero area. This set
obviously includes the set of poles where the original function $f(z)$ is clearly ill-defined but
there may be some other extraneous poles as well. For a given compact region in the complex plane,
the Nuttall-Pommerenke's theorem of convergence requires that, either these extraneous poles move very far away
from the origen as the order of the Pad\'{e} Approximant increases, or they pair up with a close-by zero becoming
what is called a \emph{defect} or a \emph{Froissart doublet}\footnote{At thirties, M. Froissart observed that the Pad\'{e} approximants to a power series perturbed by random noise are characterized by some zeros and poles unusually located in the vicinity of the unit circle (which was the range of convergence of the particular series he used). Each such zero was accompanied by a pole at the distance proportional to the scale of the noise, forming the so-called Froissart doublets. Since then, several studies are been done and now this defect phenomena is better understood. Actually, for Stieltjes functions the location $z_d$ of these defects have to obey $|z_d|\geq R$ where $R$ is the convergence radius of its Taylor expansion around $z=0$.} in the mathematical jargon \cite{Baker2}. These are to be considered artifacts of the approximation. Near the location of these extraneous poles the PA approximation
clearly breaks down but, away from these poles, the approximation is safe.

To extend the Montessus' theorem one has to consider that the circle of convergence of a $P_M^L$ to a certain meromorphic function $f(z)$ can grow provided exist $M'>M$, $M'$ been the poles of the extended subset $\mathfrak{D}_{M'}$. Then, the first step (a weaker form of Montessus' theorem) is a theorem applicable when the degree of the denominator is known to be grater than or equal to (instead of precisely equal to) the number of poles within the circle of convergence of the $P_M^L$.

Let $F(z)$ be analytic at the origin and also in a given disk $|z|\leq R$, except for \textit{m} pols, counting multiplicity. Consider a sequence of $P_M^L$ to $f(z)$ with \textit{M} fixed, $M\geq m$, and $L\rightarrow \infty$. Suppose that arbitrarily small, positive $\epsilon$ and $\delta$ are given. Then $L_0$ exists such that $|f(z)-P_M^L|<\epsilon$ for any $L>L_0$ and for all $|z|<R$ except for $z\in \epsilon_L$, where $\epsilon_L$ is a set of points in the z-plane of measure less than $\delta$.

An interesting corollary is deduced at this point: with the hypothesis of the theorem, the more general PA sequence $P_{M_k}^{L_k}$ satisfies $|f(z)-P_{M_k}^{L_k}|<\epsilon$ for any $k>k_0$ and for all $|z|\leq R$ excepting $z \in \epsilon_L$, $\epsilon_L$ a measure less than $\delta$, provided
\begin{itemize}
\item $L_k/M_k \rightarrow \infty$ as $k\rightarrow \infty$ ($M_k\neq 0$), and
\item $M_k\geq M$ for all $k>k_0$.
\end{itemize}

An interesting case consequence of that corollary is when $M_k=L_k$, called the diagonal Pad\'{e} sequence.

This last comment is a weak form of both what is known to be true and what is expected to be true about convergence in measure of the Pad\'{e} sequence $P_M^L$. Nonetheless, it provides a bass for further development.

First, the diagonal sequence may be replaced by the sequence $P_{M_k}^{L_k}$, $k=1,2,...$ provided that, for any $\lambda$ in the range $0<\lambda<1$ however small, $\lambda < \frac{L_k}{M_k} < \lambda^{-1}$. Provided that inequality holds and $L_k+M_K\rightarrow\infty$, this weaker constraint is sufficient to allow convergence in measure.

Second, $f(z)$ need not be meromorphic, but may also have a countable number of isolated essential singularities. This means that for example $f(z)=\exp(-(1-z)^{-1})$ and $g(z)=\exp(z \Gamma(z))$ are allowable functions, but not function whose singularities have a limit point in the finite z-plane.

All these steps leads us to the Pommerenke's theorem, which reads:\\

$\bullet$ Let $f(z)$ be a function which is analytic at the origin and analytic in the entire $z-$plane except for a countable number of isolated poles and essential singularities\footnote{The inclusion of essential singularities was a notable step forward in the extension of the Nuttall's theorem \cite{Nuttall} and in the global convergence theory of Pad\'{e} Approximants.}. Then the sequence of approximants $P^L_M$ with $L/M=\lambda (\lambda \neq 0, \lambda \neq \infty )$ satisfies:

\vspace{-0.5cm}
\begin{center}
\be\label{Pommerenke}
\lim_{M\rightarrow \infty}P^{\lambda M}_M=f(z)
\ee
\end{center}

on any compact set of the $z-$plane except for a set of points of zero measure.

An interesting corollary of this theorem \cite{Baker1} says that it can be generalized for Pad\'{e} sequences of the type $P_N^{N+k}$, for a fixed $k$, and then also convergence for these sequences is found when $N\to \infty$.

The main advantage of the application of this theorem in front of Montessus' theorem is now one does not need to know in advance the number of poles of the original function inside a certain disk. The disadvantage, however, is when you want to construct a sequence of $P^{N+k}_N$, to increase one order on the sequence, one needs, at least, two different inputs (one for the numerator and one for the denominator of the PA) while following Montessus' theorem, increase one order in the sequence only demands one new input (since the denominator of the PA is fixed to a certain degree $M$).




As $N\rightarrow \infty$, the Pommerenke's theorem ensures convergence of the sequence of
PAs to the original meromorphic function, in any compact set in the complex $z$ plane except at a finite
number of poles. Of course, where there is convergence, the PA may be considered an approximate resummation
of the Taylor series around the origin. On the other hand, the set of points where there is no convergence
certainly includes the position of the poles since not even the original function is defined there, but, as we said,
there may appear other artificial poles which have no counterpart in the original function (when, for example, $k<0$). One would naively think that the presence of these artificial poles would cause a major distortion and completely
spoil the rational approximation. However, one can actually show, as the theorem states, that as the order of the
Pad\'{e} increases, i.e. as $N$ grows, these artificial poles either move to infinity in the complex plane and
decouple or they get "almost-canceled" by the appearance of nearby zeros,\textit{ \`{a} la} Froissart. Although, in general, this cancelation is not complete, it is efficient enough to make the region of distortion of the artificial pole
only of zero measure. This is why and how the Pad\'{e} Approximation works. For an explicit example where all
these properties come to play in the context of a Regge-inspired model, we refer to Ref. \cite{MasjuanPades}.






\subsection{Generalization and Extensions of Pad\'{e} Approximants}\label{sec:NpointPade}

\subsubsection{Pad\'{e} Types and Partial Pad\'{e} Approximants}

\quad The last theorems implies somehow certain knowledge of the position of the poles of the original function. Certainly, as $L\rightarrow\infty$, the $M$ poles of Eqs. (\ref{montessusdef}), (\ref{Montessus}) and (\ref{Pommerenke}) will tend to reproduce the original ones. When the position of the poles in the original function is known (for example, for a Green's Function, at least the lowest
lying states), it is interesting to devise a rational approximation which has this information
already built in. The corresponding approximants are called Partial Pad\'{e} Approximants (PPAs) in
the mathematical literature \cite{Canaris} and are given by a rational function
$\mathbb{P}^{M}_{N,K}(z)$:
\begin{equation}\label{three}
    \mathbb{P}^{M}_{N,K}(z)=\frac{Q_M(z)}{R_{N}(z)\ T_{K}(z)}\ ,
\end{equation}
where $Q_{M}(z),R_{N}(z)$ and $T_{K}(z)$ are polynomials of order $M, N$ and $K$
(respectively) in the variable $z$. The polynomial $T_{K}(z)$ is defined by having $K$ zeros
precisely at the location of the lowest lying poles of the original function\footnote{For
simplicity, we will assume that all the poles are simple.} i.e.
\begin{equation}\label{threeprime}
    T_{K}(z)= (z+ z_1)\ (z+ z_2)\ ...\ (z+ z_K) \ .
\end{equation}
As before the polynomial $R_{N}(z)$ is chosen so that $R_{N}(0)=1$ and, together with $Q_M(z)$,
they are defined so that the expansion of $\mathbb{P}^{M}_{N,K}(z)$ matches exactly the first $M+N+1$
terms in the expansion of the original function around $z=0$, i.e. :
\begin{equation}\label{four}
    \mathbb{P}^{M}_{N,K}(z)\approx f_0 + f_1\ z + f_2\ z^2
    +...+ f_{M+N}\ z^{M+N}+ \mathcal{O}(z^{N+M+1})\ .
\end{equation}
At infinity,  $\mathbb{P}^{M}_{N,K}(z)$ obviously falls off like $1/z^{N+K-M}$. Exactly as it
happens in the case of PAs, also  the PPAs will have defects for a general meromorphic
function, the Froissart doblets. 

Finally, another rational approximant defined in mathematics is the so-called Pad\'{e} Type Approximant
(PTA) \cite{Canaris} $\mathbb{T}^M_N(z)$ :
\begin{equation}\label{five}
    \mathbb{T}^M_N(z)=\frac{Q_M(z)}{T_{N}(z)}\ ,
\end{equation}
where $T_{N}(z)$ is also given by the polynomial (\ref{threeprime}), now with $N$ preassigned
zeros at the corresponding position of the poles of the original function, $f(z)$. The
polynomial $Q_M(z)$ is defined so that the expansion of the PTA around $z=0$ agrees with that
of the original function up to and including terms of order $M+1$, i.e.
\begin{equation}\label{six}
    \mathbb{T}^M_N(z)\approx f_0 + f_1\ z + f_2\ z^2
    +...+ f_{M}\ z^{M}+ \mathcal{O}(z^{M+1})\ .
\end{equation}
At large values of $z$, one has that $\mathbb{T}^M_N(z)$ falls off like $1/z^{N-M}$. Clearly
the PTAs are a particular case of the PPAs, i.e. $\mathbb{T}^M_N(z)=\mathbb{P}^M_{0, N}(z)$ and
coincide with what has been called the Hadronic Approximation to large-$N_c$ QCD in the literature
\cite{MHAdeRafaelKnecht,MHAPerisMaching,meromorphicdeRafael,meromorphicPerislargeN,MHA2}.

Since in that last case all the poles are preassigned precisely in the location of the lowest lying poles of the original function, Froissart doblets will not we found.

We may not confuse Pad\'{e} Types with the Montessus' theorem statements. The Pad\'{e} Type sequence converges to the original function when both $M,N \to \infty$. In the Montessus' theorem instead, the denominator was fixed to a certain value and only the degree $L$ of the numerator was enlarge to reach convergence (Eq.~(\ref{Montessus})).


\subsubsection{N-point Pad\'{e} Approximants}

\quad The Pad\'{e} method that we have already introduced could be called \textit{One-point} Pad\'{e} method because the approximants are constructed by matching them with a power series about a particular point $z_0$. However, the function in question may have been investigated in the vicinity of two or more points. One may wish to incorporate information from all these expansions in a single sequence of Pad\'{e} approximants. Suppose $f(z)$ has the asymptotic expansions
\be\label{fz0}
f(z)=\sum_{n=0}^{\infty} a_n(z-z_0)^n, \quad z\rightarrow z_0,
\ee
\be\label{fz1}
f(z)=\sum_{n=0}^{\infty} b_n(z-z_1)^n, \quad z\rightarrow z_1,
\ee
in the neighborhoods of the distinct points $z_0$ and $z_1$, respectively. A Two-point Pad\'{e} approximant\footnote{We are basically interested in Two-Point Pad\'{e}s since they are the closest version of that method commonly used in the physical literature and called Minimal Hadronic Approximation.} to $f(z)$ is a rational function $F(z)=R_N(z)/S_M(z)$ where $S_M(0)=1$. $R_N(z)$ and $S_M(z)$ are polynomials of degree $N$ and $M$, respectively, whose $(N+M+1)$ arbitrary coefficients are chosen to make the first $J$ terms $(0\leq J\leq N+M+1)$ of the Taylor series expansion of $F(z)$ about $z_0$ agree with Eq.~(\ref{fz0}) and the first $K$ terms of the Taylor series expansion of $F(z)$ about $z_1$ agree with Eq.~(\ref{fz1}), where $K+J=N+M+1$.
This idea can be extended to a $N-$point Pad\'{e} when the rational function fits various points, not necessary distinct.
Here we illustrate the method by one Two-point Pad\'{e} example \cite{Bender}. Define the function $f(z)$ as

\be
f(z)=\frac{1}{2\sqrt{z}}e^{-z}\int_0^z\frac{e^t}{\sqrt{t}}dt.
\ee

This function is the unique solution to the differential equation $2zf'(z)\,=\,-(1+2z)f(z)+1$. This solution has power series expansions about both $z=0$ and $z\to \infty$. The series expansion are:

\begin{eqnarray}
f(z)=\sum_{n=0}^{\infty}a_nz^n, \quad a_n=\frac{(-4)^nn!}{(2n+1)!}, \quad |z|\rightarrow 0.\\
f(z)\sim \sum_{n=0}^{\infty}b_nz^{-n}, \quad b_n=2(2n-2)!4^{-n}/(n-1)!,\quad b_0=0, \quad |z|\rightarrow \infty.
\end{eqnarray}
For definiteness we discuss only the diagonal Pad\'{e} sequence with $N=M$ and use as input $J=N+1$ and $K=N$. For this particular function, since $b_0=0$, all the PAs constructed with the expansion $z\to \infty$ will have the same behavior as $1/z$ for $z\rightarrow\infty$ as the function $f(z)$ has.

Two-point Pad\'{e} Approximants are always more constraint than One-point PAs. One can compare the diagonal sequence One-point Pad\'{e} approximants obtained from the expansions at $z=0$ and $z\rightarrow\infty$ with the Two-point Pad\'{e} from both expansions. In particular, the lowest One-point PA constructed with the expansion $z=0$ reads
\begin{equation}\label{PAat0}
\frac{1-\frac{4 z}{15}}{1+\frac{2 z}{5}} \, ,
\end{equation}
while the lowest One-point PA constructed with the expansion $z\to \infty$ reads:
\begin{equation}\label{PAatinf}
\frac{-1}{1-2 z} \, .
\end{equation}

Both PAs needs three inputs. To construct the lowest Two-point PA we will also use three inputs in such a way that the PA will much\footnote{All the PAs we are constructing are diagonal and then we need an odd number of inputs. In that situation, when dealing with a Two-point PA one can always choose $K<J$ or $K>J$ but never $K=J$.} one order from the expansion $z\to \infty$ and two orders from $z=0$. That PA reads:
\begin{equation}\label{2pPA0}
\frac{1}{1+\frac{2 z}{3}} \, .
\end{equation}

The Two-point approximant gives a more uniform approximation to $f(z)$ while in the neighborhood of each point the corresponding One-point Pad\'{e} is slightly more accurate as can be see numerically in the table \ref{Tab:2pPA} (we show the values for the ratio PA$/f(z)$).
\begin{table}
\centering
\begin{tabular}{|c||c|c|c|c|}
\hline
PA & $z=0.1$ & $z=1$ & $z=10$ & $z=100$  \\
\hline \hline
PA$_{z=0}$ Eq.(\ref{PAat0})&     $1.00$ & $0.97$ &  $-6.30$ & $-124$ \\
\hline
PA$_{z\to \infty}$ Eq.(\ref{PAatinf}) &     $-1.34$ & $1.86$ & $0.99$ & $1.00$\\
\hline
2point-PA Eq.(\ref{2pPA0})&    $1.00$  & $1.12$ & $2.46$ & $2.94$ \\
\hline
\end{tabular}
\caption{Comparison between a One-point PA constructed using the Taylor expansion (first raw), One-point PA using the expansion at $z\to \infty$ (second raw), and a Two-point PA (third raw). All the numbers are normalized to the function $f(z)$.}\label{Tab:2pPA}
\end{table}

We can easily improve these results by going to the next PA in the corresponding sequence. In the three cases we will need 5 inputs and the One-point PA constructed using expansion $z=0$ will be:

\begin{equation}\label{PA2at0}
\frac{1-\frac{2 z}{9}+\frac{32 z^2}{945}}{1+\frac{4 z}{9}+\frac{4 z^2}{63}} \, ,
\end{equation}
while the One-point PA constructed using the expansion around $z\to \infty$ will be:
\begin{equation}\label{PA2atinf}
\frac{\frac{-5}{3}+\frac{2 z}{3}}{1-4 z+\frac{4z^2}{3}} \, .
\end{equation}

In its turn, the Two-point PA will be:

\begin{equation}\label{2pPA2}
\frac{1+\frac{2 z}{15}}{1+\frac{4 z}{5}+\frac{4 z^2}{15}} \, .
\end{equation}

In that second case, the numerical results are shown in table \ref{Tab:2pPA2}.

\begin{table}
\centering
\begin{tabular}{|c||c|c|c|c|}
\hline
PA & $z=0.1$ & $z=1$ & $z=10$ & $z=100$  \\
\hline \hline
PA$_{z=0}$ Eq.(\ref{PA2at0}) &     $1.00$ & $1.00$ &  $3.46$ & $92.8$ \\
\hline
PA$_{z\to \infty}$ Eq.(\ref{PA2atinf}) &     $-2.78$ & $1.12$  & $1.00$ & $1.00$\\
\hline
2point-PA Eq.(\ref{2pPA2}) &    $1.00$  & $1.02$ & $1.23$ & $1.04$ \\
\hline
\end{tabular}
\caption{Comparison between a One-point PA constructed using the Taylor expansion (first raw), One-point PA using the expansion at $z\to \infty$ (second raw), and a Two-point PA (third raw). All the numbers are normalized to the function $f(z)$.}\label{Tab:2pPA2}
\end{table}

For small $z$ the Two-point Pad\'{e} is significantly more accurate that the ordinary One-point Pad\'{e} about $z\rightarrow\infty$. For large $z$, the Two-point Pad\'{e} is significantly more accurate that the One-point Pad\'{e} from $z=0$. One has to consider that in those cases a Two-point Pad\'{e} of degree $N$ is comparable in accuracy than the One-point Pad\'{e} of degree $N/2$.

\chapter{Pad\'{e} Theory and Meromorphic functions}\label{capitol2}

\def\baselinestretch{1.66}

\quad In the second chapter we have discussed the main properties of the Pad\'{e} Theory. We have summarized the main convergence theorems we will use in these pages and we have shown few examples of the procedure to follow. Now comes the turn of the physical case.

\section{Rational Approximations: generalities}\label{PAgeneralities}

\quad In the physical case the original function $f(z)$ will be a Green's function $G(Q^2)$ of the
momentum variable $Q^2$. In general, the structure of that kind of function is not know. However, in QCD in the large $N_c$ limit this Green's function is meromorphic with
all its poles located on the negative real axis in the complex $Q^2$ plane. These poles are
identified with the meson masses. On the other hand, the region to be approximated by the Pad\'{e} Approximants will
be that of euclidean values for the momentum, i.e. $Q^2>0$. The expansion of $G(Q^2)$ for $Q^2$
large and positive coincides with the Operator Product Expansion (section \ref{secOPEth}).

In general a meromorphic function does not obey any positivity constraints and, as we will see,
this has as a consequence that some of the poles and residues of the PAs may become
\emph{complex}\footnote{On the contrary, remember the case of a Stieljes functions which obeys positivity constrains. In that case the poles and residues of the diagonal and paradiagonal PA sequences are purely real and with the same sign
as those of the original function \cite{PerisPade,MasJJVpades}.}. This clearly precludes any possibility that
these poles and residues may have anything to do with the physical meson masses and decay
constants. However, and this is very important to realize, this does \emph{not} spoil the validity
of the rational approximation provided the poles, complex or not, are not in the region of $Q^2$
one is interested in. It is to be considered rather as the price to pay for using a rational
function, which has only a finite number of poles, as an approximation to a meromorphic function
with an infinite set of poles.

Given the meromorphic function $G(Q^2)=Q^2 \Pi_{V-A}(-Q^2)$ in the $Q^2$ complex plane with an analytic expansion
around the origin, as in Eq.(\ref{largeNtaylor}), it is possible to construct a Pad\'{e} Approximant,
$P^{M}_{N}(Q^2)$, such that its expansion in powers of $Q^2$ matches that of the original function up to, and including, the term of $\mathcal{O}(Q^{2(M+N)})$. Since the function falls off at large $Q^2$ as $Q^{-4}$ up to logarithms (see
Eq.(\ref{eq:OPE})), we choose\footnote{This choice is not arbitrary as we will see in section \ref{bestPA}.}  $N=M+2$ in order to optimize the matching of the rational approximant at
large $Q^2$ to this behavior\footnote{Due to the presence of logarithms in Eq.(\ref{eq:OPE}), however, this
matching cannot be perfect.}. We emphasize, however, that this choice does not affect the properties of
convergence of Pad\'{e} Approximants, as described in section \ref{sec:NpointPade}. Later on we will show that this procedure is the most efficient way to deal with what we are interested in. However, this is not a mathematical theorem and it depends on the analytical properties of the function and also the motivation of doing PAs. To study Taylor coefficients, poles and residues this strategy is certainly the best option to follow however you have to treat with the Froissart doublets of the approximation. If one is interested in take under control the analytic properties of the PAs under a certain disk on the $Q^2$ variable, one should proceed for example following \textit{Montessus}' theorem, where the dominator of the approximant is fixed in a certain way (sec.~\ref{convth}) and then keep it under control. On the contrary, since we are interested in Taylor and OPE expansions for $G(Q^2)=Q^2 \Pi_{V-A}(-Q^2)$, we will use \textit{Pommerenke}'s theorem (sec.~\ref{convth}) in this chapter.


Pommerenke's theorem (Eq.(\ref{Pommerenke})) is important in this case  because it teaches us useful information about the qualitative behavior of how Pad\'{e}s approximate meromorphic functions\footnote{Remember that Pommerenke's theorem can be applied to any meromorphic function which need not to be Stieltjes.}. Regretfully, when asking more quantitative
questions such as the rate of convergence, which is the first step towards an estimate of the error, such a
theorem is only of limited practical importance. In practice, one can take a more useful approach towards an
estimate of the error by studying the behavior of a set of successive rational approximants, as we will show later on.

\section{Testing rational approximations: a model}\label{sec:model}

\subsection{The model}\label{secOPE}

\quad A particular model should help on the understanding of all the PA properties presented in the previous chapter. To reassemble a physical case we shall consider a correlation function of a vector and axial-vector currents, which is particularly sensitive to properties of Spontaneous Chiral Symmetry Breaking, sec. \ref{ChPT}, namely the two--point function

\be \label{correlator} \Pi^{V,A}_{\mu\nu}(q)=\ i\,\int d^4x\,e^{iqx}\langle J^{V,A}_{\mu}(x)
J^{\dag\ V,A}_{\nu}(0)\rangle\, ,
\ee
with $J_{V}^\mu(x) = {\overline d}(x)\gamma^\mu u(x)$ and $J_A^\mu(x) = {\overline
d}(x)\gamma^\mu \gamma^5 u(x)$.

%
%

In the chiral limit where the light quark masses are set to zero\footnote{See section \ref{ChPT} for details.}, this
two--point function only depends on one invariant function ($Q^2=-q^{2}\ge 0$
for $q^2$ spacelike) which is

\be \label{correlator2}
\Pi^{V,A}_{\mu\nu}(q)= \left(q_{\mu} q_{\nu} - g_{\mu\nu} q^2 \right)\Pi_{V,A}(q^2) \ .
\ee

As it is known, the difference $\Pi_{V-A}(q^2)$ defined as follows, satisfies the
unsubtracted dispersion relation given by\footnote{The upper cutoff which is needed to render the dispersive
integrals mathematically well defined can be sent to infinity provided it respects chiral symmetry
\cite{GP02}.}

\be \label{dispersion2}
\Pi_{V-A}(q^2)\equiv \frac{1}{2}(\Pi_{V}(q^2)-\Pi_{A}(q^2))=
\lim_{\Lambda\rightarrow\infty}\int_0^{\Lambda^2} \frac{dt}{t-q^2-i\epsilon}\ \frac{1}{\pi}\ {\mbox{\rm Im}}\,\Pi_{V-A}(t)\ .
\ee

and is usually called $\Pi_{LR}(Q^2)$, the self--energy function, and its properties are rather well known in the limit of vanishing light quark masses. First, $\Pi_{LR}(Q^2)$ vanishes order by order in perturbation theory and is an order parameter of the spontaneous breakdown of chiral symmetry for all values of the momentum transfer.
It also governs the electromagnetic $\pi^{+}-\pi^{0}$ mass difference~\cite{Das:1967it}

\be\label{eq:piem}
m_{\pi^+}^{2}\vert_{{\mbox{\rm EM}}}=\frac{\alpha}{\pi}\,\frac{-3}{8f_{\pi}^2}\,
\int_0^\infty
dQ^2\,Q^2\Pi_{LR}(Q^2)\ .
\ee

This integral converges in the ultraviolet region as shown by Shifman, Vainshtein and Zakharov~\cite{Shifman:1978bx} using Wilson's \cite{Wilson:1969zs} Operator Product Expansion\footnote{To obtain that expression one can follow the  procedure shown in section \ref{secOPEth} and using the Weinberg Sum Rules (WSR), App. \ref{WSR}.}
\be\label{eq:OPE}
\lim_{Q^2\rightarrow\infty}-Q^2\Pi_{LR}(Q^2)=\frac{1}{Q^4}
\left[8\pi^2\left(\frac{\alpha_s}{\pi}+{\cal O}(\alpha_s^2\log (Q^2))\right)
\langle\bar{\psi}\psi\rangle^2
\right]+{\cal O}\left(\frac{1}{Q^6}\right)\, ,
\ee
\noindent
where $\alpha_s$ is the QCD coupling constant and $\langle \bar{\psi} \psi \rangle$ is the quark condensate.

Unlike the expansion around the origin of energies, the existence of nonvanishing anomalous dimensions, even in the large$-N_c$ limit, gives rise to the $\log Q^2$ terms and, unlike (\ref{largeNtaylor}), renders the expansion around infinity in (\ref{eq:OPE}) \textit{not} analytic.

Witten~\cite{W83, CLT95} has furthermore shown that
\be\label{eq:witten}
-Q^2\Pi_{LR}(Q^2)\ge 0 \qquad for  \quad 0\leq Q^2\leq \infty\,,
\ee
which in particular ensures the positivity of the integral in
Eq.~(\ref{eq:piem}) and thus the stability of the QCD vacuum with respect
to small perturbations induced by electromagnetic interactions. This positivity is also seen experimentally \cite{PDG}.

The low $Q^2$ behavior of this self--energy function~\cite{BR91,ABT}, is
governed by chiral perturbation theory
\be\label{largeNtaylor}
-Q^2\Pi_{LR}(Q^2)=f_{\pi}^2+4L_{10}Q^2+8C_{87}Q^4+{\cal O}(Q^6)\,,
\ee
where $L_{10}$ is the only LEC ~\cite{GLsq}, sec.~\ref{ChPT}, with an axial content in the
${\cal O}(p^4)$ low energy effective chiral Lagrangian, Eq. (\ref{l4}), and $C_{87}$ is the corresponding LEC at ${\cal O}(p^6)$ \cite{p6FearingScherer,p6Bijnens}.

In the large--$N_c$ limit, the spectral function associated with
$\Pi_{LR}(Q^2)$ consists of the difference of
an infinite number of narrow vector states and an infinite number of narrow
axial--vector states, together with the Goldstone pole of the pion \cite{Shifman,duality}:

\bea
\label{spectrum} \frac{1}{\pi}\ {\mbox{\rm Im}}\,\Pi_V(t)&=& 2 F_{\rho}^2
\delta(t-M_{\rho}^2) + 2 \sum_{n=0}^{\infty} F^2_V(n)\delta(t-M^2_V(n))\ ,\nonumber\\
\frac{1}{\pi}\  {\mbox{\rm Im}}\,\Pi_A(t)&=& 2 F_{0}^2 \delta(t) + 2 \sum_{n=0}^{\infty}
F^2_A(n)\delta(t-M^2_A(n))\ .
\eea

Here $F_\rho,M_\rho $ are  the electromagnetic decay constant
and mass of the $\rho$ meson and $F_{V,A}(n)$ are the electromagnetic decay constants of the $n^{th}$
resonance in the vector (resp. axial) channels, while $M_{V,A}(n)$ are the corresponding masses.
$F_0$ is the pion decay constant in the chiral limit. The dependence on the resonance excitation
number $n$ is the following:
\begin{equation}\label{twoprime}
    F^2_{V,A}(n)=F^2= \mathrm{constant}\  ,
    \quad\quad M_{V,A}^2(n) = m_{V,A}^2 + n \ \Lambda^2\ ,
\end{equation}
in accord with known properties of the large-$N_c$ limit of QCD \cite{largeNtHooft,largeNWitten} as well as alleged
properties of the associated Regge theory \cite{ReggeHooft,ReggeCallan,ReggeEinhorn}.

The combination (\ref{dispersion2}) thus reads:

\begin{equation}\label{oneprime}
    \Pi_{LR}(q^2)=\frac{F^2_0}{q^2}+\frac{F_{\rho}^2}{-q^2+M_{\rho}^2}+
    \sum_{n=0}^{\infty} \left\{\frac{F^2}{-q^2+M^2_V(n)}-
    \frac{F^2}{-q^2+M^2_A(n)}\right\}\ .
\end{equation}

or

\be\label{eq:LRN1}
-Q^2\Pi_{LR}(Q^2)=F_0^2-f_{\rho}^2 M_{\rho}^2\frac{Q^2}{M_{\rho}^2+Q^2}+\sum_{A}f_{A}^2 M_{A}^2\frac{Q^2}{M_{A}^2+Q^2}
-\sum_{V}f_{V}^2 M_{V}^2\frac{Q^2}{M_{V}^2+Q^2}\,.
\ee
where $F_i^2=f_i^2 M_i^2$ and $-q^2=Q^2$. This last expression will we useful when studying OPE and Taylor expansions, Eq.~(\ref{eq:OPE}) and Eq.~(\ref{largeNtaylor}) respectively.

This two-point function can be expressed in terms of the Digamma function
$\psi(z)=\frac{d}{dz}\log{\Gamma(z)}$ as \cite{duality}
\begin{equation}\label{onecompact}
    \Pi_{LR}(q^2)=\frac{F^2_{0}}{q^2}+\frac{F_{\rho}^2}{-q^2+M_{\rho}^2}+
    \frac{F^2}{\Lambda^2}\left\{\psi\left(\frac{-q^2+m_A^2}{\Lambda^2}\right)-
    \psi\left(\frac{-q^2+m_V^2}{\Lambda^2}\right)\right\}\ .
\end{equation}

%

To resemble the case of QCD, we will demand that the usual parton-model logarithm is reproduced in
both vector and axial-vector channels and that the difference (\ref{dispersion2}) has an OPE which starts at dimension six (\ref{eq:OPE}) (or at dimension four considering $Q^2 \Pi_{LR}(Q^2)$). A set of parameters satisfying these conditions is
given by\footnote{These numbers have been rounded off for the purpose of presentation. Some of the
exercises which will follow  require much more precision than the one shown here.}
\begin{eqnarray}\label{nature}
\hspace{-2. cm} F_{0}= 85.8 \,\,{\mathrm{MeV}}\ , \quad F_{\rho}= 133.884 \,\,{\mathrm{MeV}}\ ,
\quad F= 143.758 \,\,{\mathrm{MeV}}\, ,\qquad \qquad \\
M_{\rho}= 0.767 \,\,{\mathrm{GeV}}, \quad m_A= 1.182 \,\,{\mathrm{GeV}}, \quad m_V= 1.494
\,\,{\mathrm{GeV}}\, , \quad \Lambda= 1.277 \,\, {\mathrm{GeV}}\ , \nonumber
\end{eqnarray}
This set of parameters has been chosen to resemble
those of the real world, while keeping the model at a manageable level. For instance, the values of
$F_{\rho}$ and $M_{\rho}$ in Eq. (\ref{nature}) are chosen so that the function $\Pi_{LR}$ in
(\ref{onecompact}) has vanishing $1/Q^2$ and $1/Q^4$ in the OPE at large $Q^2>0$, the Weinberg Sum Rules, as in real QCD (see App. \ref{WSR} for more details).
Also, the model admits the introduction of finite widths (which is a $1/N_c$ effect) in the
manner described in Ref. \cite{Shifman}, after which the spectral function looks reasonably similar
to the experimental spectral function. This comparison can be found in Fig. 5 of Ref.
 \cite{duality}. But this model is also interesting for a very different reason. In Ref.
 \cite{duality} several attempts were made at determining the coefficients of the OPE by using the
methods which have become common practice in the literature. Among those we may list: Finite Energy
Sum Rules \cite{FESRRojo,FESRBijnens,FESRPeris}, with pinched weights \cite{pwCirigliano,pwDominguez}, Laplace sum rules \cite{Narison} and MHA \cite{MHAdeRafaelKnecht,MHAPerisMaching,MHA2}. As it turned out, when these methods were tested on the
model, none of them was able to produce very accurate results. We think that this makes the model
very interesting as a way to assess systematic errors \cite{almasy}.

Defining the expansion of the Green's function (\ref{eq:LRN1}) in $Q^2=-q^2$ around
$Q^2=0,Q^2\rightarrow\infty$ as
\begin{equation}\label{exp}
    Q^2\ \Pi_{LR}(-Q^2)\approx \sum_{k} C_{2k}\ Q^{2k}\quad ,\quad \mathrm{with}\quad
    k=0, \pm 1, \pm 2, \pm 3, \ldots
\end{equation}
one obtains that the coefficients accompanying inverse powers of momentum, akin to the OPE at large $Q^2>0$, are given by ($ p=1,2,3,...$ with $k=1-p$):
\begin{eqnarray}\label{expOPE}
     C_{2k} &=& -F_{0}^2 \ \delta_{p,1}+\nonumber \\
     &&\!\!\!\!\!\!\!\!\!\!\!\!
     (-1)^{p+1} \left[ F_{\rho}^2M_{\rho}^{2p-2}  -\frac{1}{p}F^2 \Lambda^{2p-2}
  \left\{B_{p}\left(\frac{m_V^2}{\Lambda^2}\right) -
  B_{p}\left(\frac{m_A^2}{\Lambda^2}\right)\right\}\right]\ ,
\end{eqnarray}
where $B_p(x)$ are the Bernoulli polynomials \cite{GP06}. As stated above, $F_{\rho}$ and
$M_{\rho}$ are defined by the condition that the above expression Eq. (\ref{expOPE}) vanishes for $k=0,-1$
(in agreement with Eq. (\ref{eq:OPE}) above), enforcing that $ Q^2\ \Pi_{LR}(-Q^2)\sim Q^{-4}$ at large
momentum as in QCD, Eq. (\ref{eq:OPE}). We emphasize that the above coefficients of the OPE (\ref{expOPE}) can not be calculated by a naive expansion at large $Q^2$ of the Green's function (\ref{oneprime}). In other words, physical masses and decay constants do not satisfy the naive WSR \cite{GP02}.

On the other hand, for the coefficients accompanying nonnegative powers of momentum, akin to the
chiral expansion at small $Q^2$, one has ($k=1,2,3,...$):
\begin{equation}\label{expChi}
    C_{0}=-F_0^2\quad , \quad C_{2k}= (-1)^{k+1} \frac{F^2_{\rho}}{M^{2k}_{\rho}}- \frac{1}{(k-1)!}
    \frac{F^2}{\Lambda^{2k}}\left\{\psi^{(k-1)}\left(\frac{m_V^2}{\Lambda^2}\right)
    - \psi^{(k-1)}\left(\frac{m_A^2}{\Lambda^2}\right) \right\} \ ,
\end{equation}
where $\psi^{(k-1)}(z)= d^{k-1}\psi(z)/dz^{k-1}$. In Table \ref{table1} we collect the values for
the first few of these coefficients $C_{2k}$ for both expansions Eq.~(\ref{expOPE}) and Eq.~(\ref{expChi}).

\begin{table}
\centering
\begin{tabular}{|c|c|c|c||c|c|c|}
  \hline
  $C_0$ & $C_2$ & $C_{4}$ & $C_6$ & $C_{-4}$ & $C_{-6}$ & $C_{-8}$ \\
  \hline
  $-7.362$ & $21.01$ & $-43.92$ & $81.81$ & $-2.592$ & $1.674$ & $-0.577$ \\
  \hline
\end{tabular}
\caption{Values of the coefficients $C_{2k}$ from the high- and low-$Q^2$ expansions of $Q^2\
\Pi_{LR}(-Q^2)$ in Eq.~(\ref{exp}) in units of $10^{-3}\ GeV^{2-2k}$. Notice that $C_{-2}=0$ and
$C_0=-F_0^2$ (the pion decay constant in the chiral limit), see text for details.}\label{table1}
\end{table}

Following the definitions in section \ref{sec:PadeTheory}, let us start with the construction of the rational
approximants to the function $Q^2\ \Pi_{LR}(-Q^2)$ using as input the coefficients of its chiral
expansion (\ref{expChi}). Since our original function Eq.~(\ref{onecompact}) falls off at large $Q^2$
as $Q^{-4}$, this is a constraint we will impose on all our rational approximants by selecting the
appropriate difference in the order of the polynomials in the numerator and denominator.

\subsection{Construction of Pad\'{e} Approximants}


\quad The simplest PA satisfying the right falloff at large momentum is $P^{0}_{2}(Q^2)$, so we will
begin with this case. In order to simplify the results, and unless explicitly stated otherwise, we
will assume that dimensionful quantities are expressed in units of $GeV$ to the appropriate power.
Fixing the three unknowns with the first three coefficients of the chiral expansion, Eq.~(\ref{expChi}), for the function defined in Eq.~(\ref{onecompact}) (i.e. $C_{0,2,4}$), one gets the following rational function:
\begin{equation}\label{P02}
    P^{0}_{2}(Q^2)=\frac{-\ r_R^2}{(Q^2+z_R) (Q^2+z_R^*)}\ , \quad r_R^2=3.379\times
    10^{-3}\ , \quad z_R=0.6550+i\ 0.1732\ .
\end{equation}
We can hardly overemphasize the striking appearance of a pair of complex-conjugate poles on the
Minkowski side of the complex $Q^2$ plane. Obviously, this means that these poles cannot be
interpreted in any way as the meson states appearing in the physical spectrum
(\ref{spectrum}, \ref{oneprime}). In spite of this, if one expands (\ref{P02}) for large values of
$Q^2>0$, one finds $C_{-4}=-r_R^2=-3.379\times 10^{-3}$ which is not such a bad approximation for
this coefficient of the OPE, see Table \ref{table1}. Even better is the prediction of the fourth
term in the chiral expansion, which is $C_6=79.58\times 10^{-3}$.

\begin{figure}
\renewcommand{\captionfont}{\small \it}
\renewcommand{\captionlabelfont}{\small \it}
\centering
\includegraphics[width=4.5in]{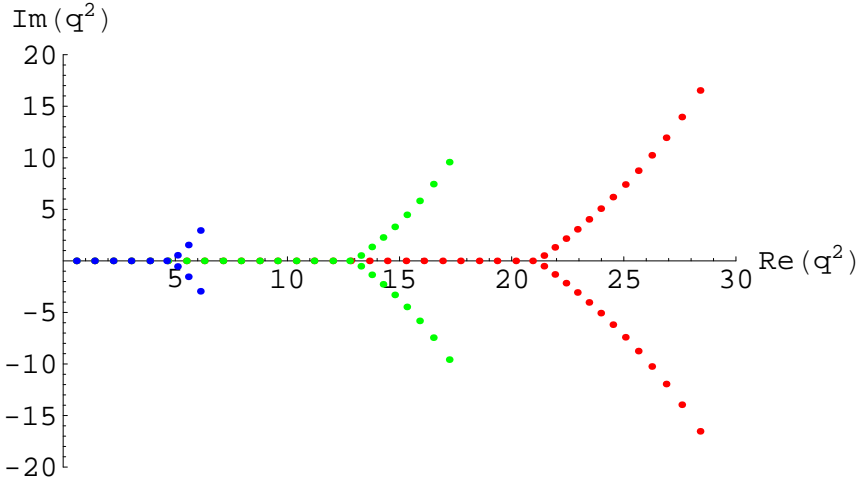}\\
\caption{Location of the poles of the Pad\'{e} Approximants $P^{10}_{12}(-q^2)$(blue), $P^{30}_{32}(-q^2)$(green) and
$P^{50}_{52}(-q^2)$(red) in the complex $q^2$ plane. When the order of the PA is increased, the overall shape of the figure does not change but the two branches of complex poles move toward the right, away from the origin.}\label{poles2}
\end{figure}

This agreement is not a numerical coincidence and the approximation can be systematically improved
if more terms of the chiral expansion are known. In order to exemplify this, we have amused
ourselves by constructing the high-orders PAs $P^{10}_{12}(Q^2)$, $P^{30}_{32}(Q^2)$ and $P^{50}_{52}(Q^2)$ (see Fig~\ref{poles2}), always keeping in mind the appropriate falloff. The particular case of $P^{50}_{52}(Q^2)$ correctly
determines the values for $C_{-4,-6,-8}$ with $52$, $48$ and $45$ decimal figures (respectively). In the
case of $C_{206}$, which is the first predictable term from the chiral expansion for this Pad\'{e} Approximant, the
accuracy reaches some staggering 192 decimal figures. This is all in agreement with Pommerenke's
theorem \cite{Pommerenke}, sec. \ref{convth}.

\begin{figure}
\renewcommand{\captionfont}{\small \it}
\renewcommand{\captionlabelfont}{\small \it}
\centering
\includegraphics[width=4.5in]{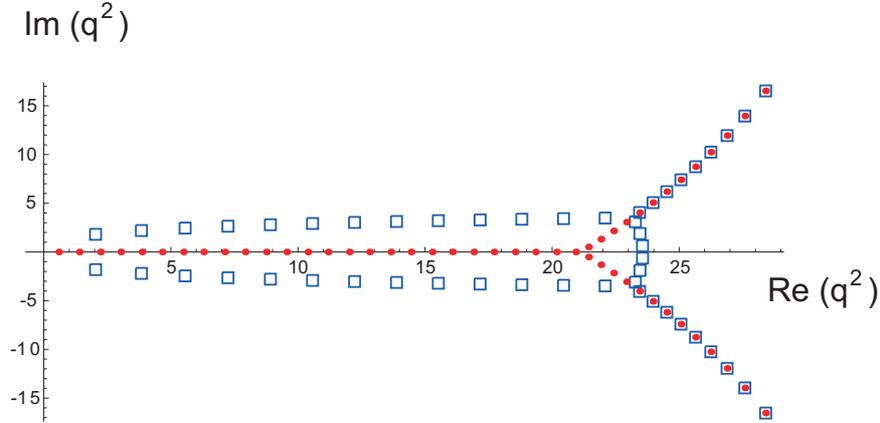}
\caption{Location of the poles (dots) and zeros (squares) of the Pad\'{e} Approximant
$P^{50}_{52}(-q^2)$ in the complex $q^2$ plane.  We recall that $Q^2=-q^2$. Notice how zeros and
poles approximately coincide in the region which is farthest away from the origin.}\label{poles}
\end{figure}

As it happens for the PA (\ref{P02}), also higher-order PAs may develop some artificial poles. In particular, Figure \ref{poles2} shows the locations of the $12$-(blue), $32$-(green) and $52$-(red) poles of the PAs $P^{10}_{12}(Q^2)$, $P^{30}_{32}(Q^2)$ and $P^{50}_{52}(Q^2)$ in the complex $q^2$ plane, respectively. As can be seen in that figure, $P^{10}_{12}(Q^2)$ has 6 poles which are real and close to the lowest part of the spectrum of the model, (\ref{oneprime}-\ref{nature}), and the other 6 poles are complex conjugate pares. At the same time, $P^{30}_{32}(Q^2)$ has 16 real poles and $P^{50}_{52}(Q^2)$ has 26. Notice that the lowest real poles of the three PAs shown in Fig.\ref{poles2} are overlapped and that the complex conjugate pares move away (to infinity) when increasing the order of the Pad\'{e} approximant.

Figure \ref{poles} also shows the locations of the $52$ poles of $P^{50}_{52}(Q^2)$ but now including the location of its zeros (blue squares). In that higher case, the first 26 poles are purely real and the rest are complex-conjugate
pairs. A detailed numerical analysis reveals that the poles and residues reproduce very well the
value of the meson masses and decay constants for the lowest part of the physical spectrum of the
model given in Eqs. (\ref{oneprime}-\ref{nature}), but the agreement deteriorates very quickly as one
gets farther away from the origin, eventually becoming the complex numbers seen in Fig.
\ref{poles}. It is by creating these analytic defects that rational functions can effectively mimic
with a finite number of poles the infinite tower of poles present in the original function
(\ref{onecompact}).

For instance the values of the first pole and residue in $P^{50}_{52}(Q^2)$ reproduce those of the
$\rho$ in Eq.~(\ref{nature}) within 193 astonishing decimal places. However, in the case of
the 26th pole, which is the last one still purely real, its location agrees with the physical mass
only with 3 decimal figures. This is not to be considered as a success, however, because after the
previous accuracy, this is quite a dramatic drop. In fact, the residue associated with this $26^{th}$
pole comes out to be 29 times the true value. The lesson we would like to draw from this exercise
should be clear: the determination of decay constants and masses extracted as the residues and
poles of a PA deteriorate very quickly as one moves away from the origin. There is no reason why
the last poles and residues in the PA are to be anywhere near their physical counterparts and their
identification with the particle's mass and decay constant should be considered unreliable.
Clearly, this particularly affects low-order PAs, for example that of Eq.~(\ref{P02}).

A very good accuracy can also be obtained in the determination of global euclidean observables such
as integrals of the Green's function over the interval $0\leq Q^2<\infty$. Notice that the region
where one approximates the true function is far away from the artificial poles in the PA. For
instance, one may consider the value for the integral
\begin{equation}\label{pi}
    I_{\pi}= \ (-1) \int_0^{\infty} dQ^2\ Q^2 \Pi_{LR}(Q^2)= 4.78719\times 10^{-3},
\end{equation}
which, up to a constant, would yield the electromagetic pion mass difference in the chiral limit
\cite{KPdeR} in the model (\ref{onecompact}). The PA $P^{50}_{52}(Q^2)$ reproduces the value for
this integral with more than 42 decimal figures. This suggests that one may use the integral
(\ref{pi}) as a further input to construct a PA.

For example if we fix the three unknowns in the PA $P^0_2(Q^2)$ by matching the first two terms
from the chiral expansion  but now we complete it with the pion mass difference (\ref{pi}) instead
of a third term ($C_4$) from the chiral expansion as we did in (\ref{P02}), the
approximant results to be
\begin{equation}\label{P02prime}
    \widetilde{P}^0_2(Q^2)=\frac{-\ r_R^2}{(Q^2+z_R) (Q^2+z_R^*)}\ ,\  \mathrm{with}
    \quad r_R^2=2.898\times 10^{-3}\ , \quad z_R=0.5618+i\ 0.2795\ .
\end{equation}
This determines $C_{-4}=-2.898\times 10^{-3}$ and $C_4=-41.26\times 10^{-3}$, which shows that
using the pion mass difference is not a bad idea (see table \ref{Tab:p02} for comparison). Notice how the position of the artificial pole
has changed with respect to (\ref{P02}). We remark that this last procedure, although reasonable from the phenomenological point of view, strictly speaking lies outside the standard mathematical theory of rational approximants \cite{Baker1,Canaris}, and while we will still call $\widetilde{P}^0_2(Q^2)$ a Pad\'{e} Approximant, this is not a strict PA anymore. In section \ref{sec:qcd} we will see how crucial is go beyond the classical Pad\'{e} procedure and we able to control the impact of the modification we are introducing here.

\begin{table}
\centering
\begin{tabular}{|c|c|c|c|c|c|}
  \hline
   $ P^N_M(Q^2)$& $z_R$ & $C_4$$(\times 10^{-3})$ & $C_6$$(\times 10^{-3})$ & $C_{-4}$$(\times 10^{-3})$ & $I_{\pi}$$(\times 10^{-3})$ \\
   \hline
  $P^0_2(Q^2)$ & $0.66+ i0.17$  & input & $79.58$ $(3\%)$ & $-3.379$ $(30\%)$ & $5.04$ $(5.3\%)$ \\
  $\widetilde{P}^0_2(Q^2)$ & $0.56+i0.28$ & $-41.26$ $(6\%)$ & $64.40$ $(21\%)$ & $-2.898$ $(12\%)$ & input \\
  \hline
  Model Eq.(\ref{eq:LRN1}) & $M_{\rho},m_A$ & $-43.92$ & $81.81$ & $-2.593$ & $4.78$ \\
  \hline
\end{tabular}
\caption{Summary of predictions for $P^0_2(Q^2)$ and $\widetilde{P}^0_2(Q^2)$ compared with the real values written in Eqs. (\ref{eq:LRN1}-\ref{nature}). In parenthesis we show the relative error defined as $\frac{\mathrm{predicted value}-\mathrm{real value}}{\mathrm{real value}}$ of each prediction.}\label{Tab:p02}
\end{table}

\subsection{Pad\'{e} Types and Partial Pad\'{e}s: a model example}

\quad In real life, the number of available terms from the chiral expansion for the construction of a PA
is very limited.  Since the masses and decay constants of the first few vector and axial-vector
resonances are known, one may envisage the construction of a rational approximant having some of
its poles at the prescribed values given by the known masses of these resonances. If all the poles
in the approximant are prescribed this way (as in the MHA), we have a PTA. On the contrary, when
some of the poles are prescribed but some are also left free, then we have a PPA (see section \ref{sec:NpointPade}).

\subsubsection{Pad\'{e} Type Approximants (PTAs)}\label{PTs}

\quad Assuming that the first masses are known, let us proceed to constructing the PTAs, Eq. (\ref{five}). The
lowest such PTA is $\mathbb{T}^{0}_{2}(Q^2)$, which contains two poles at the physical masses of
the $\rho$ and the first $A$ in the tower, the lowest one in each tower, Eq.~(\ref{nature}). Fixing the residue through the chiral expansion to be
$C_0=-F_0^2$, one obtains
\begin{equation}\label{PT02}
    \mathbb{T}^{0}_{2}(Q^2)=
\ \frac{-\ F_0^2 M^2_{\rho} M^2_{A} }{(Q^2+M^2_{\rho}) (Q^2+M^2_{A})}\ .
\end{equation}
Even though it has the same number of inputs ($C_0$ and the two masses), this rational approximant
does not do such a good job as the PAs Eq. (\ref{P02}) or Eq. (\ref{P02prime}). For instance, $C_{-4}$ is
2.3 times larger than the true value in Table \ref{table1}. As we have already stated, one way to
intuitively understand this result is the following. The OPE is an expansion at $Q^2\to \infty$ and
therefore knows about the whole spectrum because no resonance is heavy enough with respect to $Q^2$
to become negligible in the expansion, i.e., the infinite tower of resonances does not decouple in
the OPE. Chopping an infinite set of poles down to a finite set may be a good approximation, but
only at the expense of some changes. These changes amount to the appearance of poles and residues
in the PA which the original function does not have. This is how the PA in Eq. (\ref{P02}) manages to
approximate the true function  Eq. (\ref{onecompact}). However, by construction, the PTA Eq. (\ref{PT02})
does not allow the presence of any artificial pole because, unlike in a PA, all its poles are fixed
at the physical values. Consequently, it only has its residues as a means to compensate for the
infinite tower of poles present in the true function and, hence, does a poorer job than the PA Eq.
(\ref{P02}), particularly in determining a large-$Q^2$ observable like $C_{-4}$. Indeed, the role
played by the residues in the approximation can be appreciated by comparing the true values of the
decay constants to those extracted from Eq. (\ref{PT02}). Although the one of the $\rho$ is within 30\%
of the true value, that of the $A$ is off by 100\%.

A different matter is the prediction of low-energy observables such as, e.g., the chiral
coefficients, $C_{2k}$ for $k>0$. In this case heavy resonances make a small contribution and this means that the
infinite tower of resonances does decouple.\footnote{This is because the residues $F^2$ in the
Green's function Eq. (\ref{onecompact}) stay constant as the masses grow. This behavior does not hold
in the case of the scalar and pseudoscalar two-point functions \cite{GP06}.} Truncating the
infinite tower down to a finite set of poles is not such a severe simplification in this case,
which helps understand why a PTA may do a good job predicting unknown chiral coefficients. Indeed,
Eq. (\ref{PT02}) reproduces the value of $C_2$ within an accuracy of 15\%, growing to  22\% in the case
of $C_4$. A global observable like $I_{\pi}$ averages the low and the high $Q^2$ behaviors and ends
up differing from the true value Eq.(\ref{pi}) by 35\%. This gives some confidence that observables
which are integrals over Euclidean momentum may be reasonably estimated with MHA as, e.g.,  in the
$B_K$ calculation of Ref. \cite{PerisBk,PerisCataBk04,PerisCataBk03}.

Improving on the PTA of Eq. (\ref{PT02}) by adding the first resonance mass from the vector tower
produces the following approximant
\begin{equation}\label{PT13}
    \mathbb{T}^1_3(Q^2)=\frac{a+b\ Q^2}{(Q^2+M^2_{\rho}) (Q^2+M^2_{A})
(Q^2+M^2_{V})}\ ,\  \mathrm{with}\ \left\{
                     \begin{array}{ll}
                       a =& -13.5\times 10^{-3},\cr
   b=& +1.33\times 10^{-4} \quad ,  \hbox{}
                     \end{array}
                   \right.
\end{equation}
where the values of the chiral coefficients $C_0$ and $C_2$ have been used to determine the
parameters $a$ and $ b$. The prediction for $C_4$ is much better now (only $2\%$ off), in agreement
with our previous comments. The prediction for $C_{-4}$ is still very bad, becoming now 19 times
smaller than the exact value. Table \ref{Tab:PTs} shows, for the lowest PTAs, the predictions for the Taylor coefficients $C_2,C_4,C_6$, the first OPE coefficient $C_{-4}$, the prediction for the integral of Eq. (\ref{pi}) and also the prediction for the lowest residue, i.e., $F_{\rho}$, Eq.~(\ref{nature}). Nevertheless, it eventually gets much better if PTAs of very high order are constructed. For example, as shown in the Table \ref{Tab:highPTs}, not only the OPE coefficients are improved hierarchically by increasing the order of the PTAs, but also the Taylor coefficients and the integral of Eq.~(\ref{pi}) improve. For instance, we have found $C_{-4}=-2.58\times 10^{-3}$ for the approximant $\mathbb{T}^{7}_{9}$ with 9 poles.

However, another matter is the prediction of the residues. For instance, the prediction for the
decay constant of the state with mass $M_V$ in Eq. (\ref{PT13}) is smaller than the exact value in the
model in Eq. (\ref{nature}) by a factor of $2$. In general, we have seen that the residues of the poles
always deteriorate very quickly so that the residue corresponding to the pole which is at the
greatest distance from the origin is nowhere near the exact value. We again explicitly checked this
up to the approximant $\mathbb{T}^{7}_{9}$, in which case the decay constant for this pole is
almost 5 times smaller than the exact value. The conclusion, therefore, is that PTAs are able to
approximate the exact function only at the expense of changing the residues of the poles from their
physical values (this fact can be check in the last column of Table \ref{Tab:highPTs}). Identifying residues with physical decay constants may be completely wrong in a
PTA for the poles which are farthest away from the origin.

\begin{table}
\centering
\begin{tabular}{|c|c|c|c|c|c|c|}
  \hline
  $\mathbb{T}^N_M(Q^2)$ & $C_2\times 10^{3}$ & $C_4\times 10^{3}$ & $C_6\times 10^{3}$ & $C_{-4}\times 10^{3}$ & $I_{\pi}\times 10^{3}$ & $F_{\rho}\times 10^{3}$ \\
  \hline
  $\mathbb{T}^0_2(Q^2)$ & $17.8$ $(15\%)$  & $-34.0$ $(22\%)$ & $60.5$ $(26\%)$ & $-6.0$ $(133\%)$ & $6.5$ $(35\%)$ & $112.8$ $(16\%)$\\
  $\mathbb{T}^1_3(Q^2)$ & input  & $-43.2$ $(2\%)$ & $79.5$ $(3\%)$ & $-0.1$ $(95\%)$ & $4.2$ $(12\%)$ & $131.0$ $(2\%)$\\
  \hline
  Eq. (\ref{eq:LRN1}) & $21.01$ & $-43.92$ & $81.81$ & $-2.593$ & $4.78$ & $133.8$ \\
  \hline
\end{tabular}
\caption{Summary of predictions for $\mathbb{T}^0_2(Q^2)$ and $\mathbb{T}^1_3(Q^2)$ compared with the real values of Eqs. (\ref{eq:LRN1}-\ref{nature}). In parenthesis we show the relative error of each prediction.}\label{Tab:PTs}
\end{table}

\begin{table}
\centering
\begin{tabular}{|c|c|c|c|c|c|}
  \hline
   $\mathbb{T}^N_M(Q^2)$& $C_{-4}\times 10^{3}$ & $I_{\pi}\times 10^{3}$ & $F_{\rho}\times 10^{3}$& $F_{A}\times 10^{3}$& $F_{last}\times 10^{3}$\\
   \hline
  $\mathbb{T}^3_5(Q^2)$ & $-1.96$ $(24\%)$ & $4.73$ $(1.2\%)$ & $133.87$ $(0.01\%)$& $142.6$ $(0.8\%)$& $51.6$ $(64\%)$\\
  $\mathbb{T}^5_7(Q^2)$ & $-2.48$ $(4.2\%)$ & $4.782$ $(0.12\%)$ & $133.884$ $(??\%)$& $143.7$ $(0.01\%)$& $39.7$ $(72\%)$\\
  $\mathbb{T}^7_9(Q^2)$  & $-2.58$ $(0.6\%)$ & $4.787$ $(0.02\%)$ & $133.884$ $(??\%)$& $143.76$ $(??\%)$& $30.1$ $(80\%)$\\
  \hline
  Eq.(\ref{eq:LRN1}) & $-2.593$ & $4.787$ & $133.884$ & $143.758$ & $143.758$\\
  \hline
\end{tabular}
\caption{Summary of predictions for $\mathbb{T}^3_5(Q^2)$, $\mathbb{T}^5_7(Q^2)$ and $\mathbb{T}^7_9(Q^2)$ compared with the real values of Eqs. (\ref{eq:LRN1}-\ref{nature}). We have included here the prediction of the second residue which corresponds to $m_A$. The last column is the prediction of the last residue of each PT which corresponds to the fifth, seventh and ninth masses of the spectrum for the respective PTs. In parenthesis we show the relative error of each prediction. $(??\%)$ means too close agreement that can not be shown in the Table.}\label{Tab:highPTs}
\end{table}

\subsubsection{Partial Pad\'{e} Approximants (PPAs)}\label{PPAs}

\quad As an intermediate approach between PAs and PTAs, there are the PPAs, Eq. (\ref{three}), where some poles
are fixed at their physical values while some others are left free. From a theoretical point of view one should expect PPAs to have both advantages from PAs and PTs. The simplest of such rational
approximants is $\mathbb{P}^0_{1,1}(Q^2)$ (see section \ref{sec:NpointPade} for notation). Fixing its 3
unknowns with $M^2_{\rho}, C_0$ and $C_2$, one obtains
\begin{equation}\label{PP02}
    \mathbb{P}^0_{1,1}(Q^2)=\frac{-\ r_R^2}{(Q^2+M_{\rho}^2) (Q^2+z_R)}\ ,\ \mathrm{with} \quad
r_R^2= 3.75\times 10^{-3}   \ , \quad z_R=0.8665   \ .
\end{equation}
As can be seen, the mass (squared)  of the first $A$ resonance is predicted to be at $z_R$ which is
sensibly smaller than the true value in Eq. (\ref{nature}).

At that point a comment is in order because, intriguingly enough, this is also what happens in the real case of QCD \cite{swissDRV,swissEGPRressonance,Friot}. The Partial Pad\'{e} $ \mathbb{P}^0_{1,1}(Q^2)$ predicts the location of the $m_A$ at $0.931$GeV when its value in the model is $1.182$GeV, Eq.(\ref{nature}). The same kind of discrepancy was, actually, found in \cite{swissEGPRressonance}. In that reference, RChT (see sec.\ref{resLN}) is used to predict the value of the $m_A$ when saturating the spectrum of the theory to one resonance per channel (only $M_{\rho}$ and $m_A$) and fixing the value of $M_{\rho}$ to the physical value. Technically, this is a Partial Pad\'{e} since $m_A$ is left free and low-energy properties are used to fix the unknowns. Then, to fix its value, the WSR were used and a particular relation between $M_{\rho}$ and $m_A$ was found, i.e, $m_A=M_{\rho}\sqrt{1-f^2_{\pi}/F_V^2}=0.968$GeV. At that time they used $f_{\pi}=93.3$MeV and $F_V=154$MeV. Nowadays, with the values \cite{PDG} we will find $m_A=0.962$GeV which has to be compared with the physical $m_A=1.230\pm0.040$GeV. The authors agreed that the value predicted was a good determination of the physical $m_A$.

A similar example is found in Ref.\cite{Friot} where the MHA is used and the value of $M_{\rho}$ is again fixed to the physical value \cite{PDG}. After imposing several OPE and low-energy constrains, the value of $m_A$ is obtained\footnote{In this reference the authors also studied what would happen if one includes one extra vector state in the approximation, a MHA+V'. They found that the impact of including higher states in the determination of $m_A$ is irrelevant. We will also discuss this result in sec. \ref{complexpolescomment}.}, $m_A=(938.7\pm1.4)$MeV. Since the set of equations used by these authors is different that the set used in Ref.\cite{swissEGPRressonance}, the final value for $m_A$ is also different, but still smaller than the physical one.

Another similar case can we found in Ref.\cite{MateuPortoles}. They used also a Partial Pad\'{e} where the $M_{\rho}$ was fixed to the physical value but to predict $m_A$ they used $F_A$ coming from the radiative decay $a_1\to \pi \gamma$ Eq.(\ref{decay}). After imposing several OPE constraints, they found $m_A=998\pm49$MeV which is, again, smaller than the physical value.

Since all three determinations are done in the framework of large-$N_c$ QCD, one could interpret the results as a large-$N_c$ $m_A$ instead of a real $m_A$ \cite{MateuPortoles}. From the Pad\'{e} Theory point of view, however, one could understand this discrepancy in another way without needing to say that this prediction is a large-$N_c$ effect. As can we seen in Table \ref{Tab:ppas}, the Euclidean quantities like $C_{4}$ or $C_6$ are rather insensible to the value of $m_A$ used, even if it is a complex number. Then, in the other way, imposing $C_{4}$ or $C_6$ could lead to a different values of $m_A$ depending on the set of equations used. We can not exclude the conclusion that these references addressed but it may also be that the location of the second pole $z_R$ in Eq.(\ref{PP02}) has nothing to do with the physical value of $m_A$.

The rational function Eq. (\ref{PP02})
predicts $C_{-4}=-r_R^2=-3.75\times 10^{-3}$ which is a better determination than that of the PTA Eq.
(\ref{PT02}) with the same number of inputs, and $C_{4}=-45.52\times 10^{-3}$ which is not bad
either. Concerning the pion mass difference, one gets $I_{\pi}=5.22\times 10^{-3}$. However, as
compared to the PAs Eq. (\ref{P02}) or Eq. (\ref{P02prime}), the PPA Eq. (\ref{PP02}) does not represent a
clear improvement.

In order to improve on accuracy of the PPA, we can try to go to the Partial Pad\'{e} $\mathbb{P}^1_{2,1}(Q^2)$ or its counterpart $\mathbb{P}^1_{1,2}(Q^2)$. The first one needs $M^2_{\rho}, C_0, C_2$ and $C_4$, and the second one needs $M^2_{\rho},m_A, C_0$ and $C_2$. The results are shown in Table \ref{Tab:ppas}. Since the predictions for the decay constant and for the pion mass difference for both cases are very accurate, one may try to use the mass and decay constant of the first resonance, $M_{\rho}$ and $F_{\rho}$, in addition to $I_{\pi}$ and the chiral
coefficients $C_0, C_2$ and construct\footnote{This is not, strictly speaking, a PPA since we use here $I_{\pi}$  and $F_{\rho}$ instead of Taylor coefficients. We use the \textit{tilde} to identify this approximant. Although we do not lie on a strict mathematical theorem the results are quite good, see Table~\ref{Tab:ppas}.} the $\mathbb{\widetilde{P}}_{2,1}^1(Q^2)$, which can be written as:
\begin{equation}\label{PPA}
    \mathbb{\widetilde{P}}_{2,1}^1(Q^2)=\frac{F_{\rho}^2 M_{\rho}^2}{Q^2+ M_{\rho}^2}+
\frac{a- F_{\rho}^2 M_{\rho}^2\ Q^2}{(Q^2+z_c)\
    (Q^2+z_c^*)}\ ,\ \left\{
                     \begin{array}{ll}
                       a =& 17.43\times 10^{-3},\cr
   z_c=& 1.24+i\ 0.34 \quad .  \hbox{}
                     \end{array}
                   \right.
    \end{equation}
This PPA, upon expansion at large and small $Q^2$, determines $C_{-4}=- 2.47\times 10^{-3} $ and
$C_4= -44.0\times 10^{-3}$ to be compared with the corresponding coefficient in Table
{\ref{table1}. The accuracy obtained is better than that of Eq. (\ref{P02prime}), but this is probably
to be expected since Eq. (\ref{PPA}) has more inputs.

Based on the previous  numerical experiments done on the model in Eqs.
(\ref{onecompact}, \ref{nature}) (and many others), we now summarize the following conclusions.
Although, in principle, the PAs have the advantage of reaching the best precision by carefully
adjusting the polynomial in the denominator to have some effective poles which simulate the
infinite tower present in Eq. (\ref{onecompact}), they have the disadvantage that some of the terms in
the low-$Q^2$ expansion are required precisely to construct this denominator. This hampers the
construction of high-order PAs and consequently limits the possible accuracy.

When the locations of the first poles in the true function are known, there is the possibility to
construct PTAs (with all the poles fixed at the true values) and PPA (with some of the poles fixed
and some left free). As we have seen, although the PTA may approximate low-$Q^2$ properties of the
true function reasonably well, the large-$Q^2$ properties tend to be much worse, at least as long
as they are not of unrealistically high order. The PPAs, on the other hand, interpolate smoothly
between the PAs (only free poles) and the PTAs (no free poles). Depending on the case, one may
choose one or several of these rational approximants. However, common to all the rational
approximants constructed is the fact that the residues and/or poles which are farthest away from
the origin are in general unrelated to their physical counterparts.

\begin{table}
\centering
\begin{tabular}{|c|c|c|c|c|c|c|}
  \hline
  $ $& $m_A$ & $C_4\times 10^{3}$ & $C_6\times 10^{3}$ & $C_{-4}\times 10^{3}$ & $I_{\pi}\times 10^{3}$ & $F_{\rho}\times 10^{3}$\\
  \hline
  $\mathbb{P}^0_{1,1}(Q^2)$ & $0.867$  & $45.5(3.6\%)$ & $88.7(8.4\%)$ & $-3.75(45\%)$ & $5.22(9\%)$ & $151.4(28\%)$ \\
  $\mathbb{P}^1_{2,1}(Q^2)$ & $1.3+i0.3$  & input & input & $-2.32(10\%)$ & $4.74(0.9\%)$ & $133.6(0.2\%)$ \\
  $\mathbb{P}^1_{1,2}(Q^2)$ & input  & input & $81.9(0.2\%)$ & $-2.03(22\%)$ & $4.68(0.5\%)$ & $134.4(0.4\%)$ \\
  $\mathbb{\widetilde{P}}^1_{1,2}(Q^2)$ & $1.2+i0.3$  & $44.0(0.2\%)$ & $82.1(0.3\%)$ & $-2.47(5\%)$ & input & input \\
  \hline
  Eq.(\ref{eq:LRN1}) & $1.182$ & $43.9$ & $81.81$ & $-2.59$ & $4.78$ & $133.88$ \\
  \hline
\end{tabular}
\caption{Summary of predictions for $\mathbb{P}^0_{1,1}(Q^2)$, $\mathbb{P}^1_{2,1}(Q^2)$, $\mathbb{P}^1_{1,2}(Q^2)$ and $\mathbb{\widetilde{P}}^1_{1,2}(Q^2)$ compared with the real values of Eqs. (\ref{eq:LRN1}-\ref{nature}). All the results are shown in absolute value. In parenthesis we show the relative error of each prediction.}\label{Tab:ppas}
\end{table}

\subsection{Comment on Complex poles and their consequences}\label{complexpolescomment}

\quad Artificial poles and analytic defects are transient in nature, i.e. they appear and disappear from
a point in the complex plane when the order of the Pad\'{e} approximant is changed, that is the typical pattern for a \textit{defect} or a Froissart doublet. Sometimes their presence mean certain noise in the Taylor expansion used the built them, sec. \ref{convth}. On the contrary, the typical
sign that a pole in a Pad\'{e} is associated with a truly physical pole is its stability under these
changes in the order of the Pad\'{e}. Of course, when the order in the Pad\'{e} increases there have to be
new poles by definition, and it is natural to expect that some of them will be defects. Pad\'{e}
Approximants place some effective poles and residues in the complex $Q^2$ plane in order to mimic
the behavior of the true Green's function, but it can mimic the function only away from the poles,
e.g. in the Euclidean region. Obviously, PAs cannot converge at the poles,  in agreement with
Pommerenke's theorem \cite{Pommerenke}, since not even the true function is well defined there. The
point is  that what may look like a small correction in the Euclidean region may turn out to be a
large number in the Minknowski region.

To exemplify this in simple terms, let us consider a very
small parameter $\epsilon$ and imagine that a given Pad\'{e} $P(Q^2)$ produces the rational approximant
to the true Green's function $G(Q^2)$ given by
\begin{equation}\label{example}
    G(Q^2)\approx P(Q^2) \equiv R(Q^2) +\frac{\epsilon}{Q^2+M^2}\ ,
\end{equation}
where $R(Q^2)$ is the part of the Pad\'{e} which is independent of $\epsilon$. Although for $Q^2>0$
there is a sense in which the last term is a small correction precisely because of the smallness of
$\epsilon$, for $Q^2<0$ this is no longer true because of the pole at $Q^2=-M^2$. This pole is in
general a defect and may not represent any physical mass. In fact, associated with this pole, there
is a very close-by zero of the Pad\'{e} $P(Q^2)$ at $Q^2=-M^2-\epsilon\ {R(-M^2)}^{-1}$, as can be
immediately checked in (\ref{example}). This is another way of saying that a defect is
characterized by having an abnormally small residue and is the origin of the pairs of zeros and
poles in the y-shaped branches of Fig. \ref{poles}. Therefore, not only are defects unavoidable but
one could say they are even necessary for a Pad\'{e} Approximant to approximate a meromorphic function
with an infinite set of poles.

Similarly to masses, also decay constants may be unreliable. To see this, imagine now that our Pad\'{e}
is given by
\begin{equation}\label{example2}
    P(Q^2)= \frac{F}{Q^2+M^2}+ \frac{\epsilon}{(Q^2+M^2)\ (Q^2+M^2+\epsilon^2)}\,
\end{equation}
again for a very small $\epsilon$. As before, the term proportional to $\epsilon$ may be considered
a small correction for $Q^2>0$. However, at the pole $Q^2=-M^2$ the decay constant becomes $F+
\epsilon^{-1}$ which, for $\epsilon$ small, may represent a huge correction. When the poles are
preassigned at the physical masses, like in the case of PTAs, it is the value of the residues that
compensates for the fact that the rational approximant lacks the infinite tower of resonances. As
we saw before, the residues of the poles in the Pad\'{e} which lie farthest away from the origin are
the ones which get the largest distortion relative to their physical counterparts.

These comments could be related with the results found in Ref.\cite{Friot}. In that reference the properties of the function $\Pi_{LR}(Q^2)$ in the Large-$N_c$ limit are also studied assuming the MHA. MHA has a spectrum consisting of a pion state, a vector state and an axial vector state. Part of the motivation of that study was to understand a discrepancies between various phenomenological analyses Refs. \cite{L10,FESRPeris,Ioffe,FESRBijnens,pwCirigliano,FESRRojo,Zyablyuk} using the same experimental data on hadronic $\tau$-decays published by the ALEPH \cite{Alephdata} and OPAL \cite{Opaldata} collaborations at LEP. Table 1 of Ref.\cite{Friot} shows the conflicting results for the chiral condensates Eq. (\ref{eq:OPE}) which modulate the asymptotic behavior of the $\Pi_{LR}(Q^2)$ function at large $Q^2$ values\footnote{With our analysis we can also provide some insights in this discussion but we left it to the end of sec. \ref{beyondPAs}.}.
To perform their analysis, the authors of Ref.\cite{Friot} study two successive approximations to $\Pi_{LR}(Q^2)$ in the large-$N_c$ limit. After imposing some good high-energy behavior in several Green's functions and form factors throug a fit procedure, Ref.\cite{Friot} obtains, keeping only one vector state $V$ and one axial-vector state $A$, the
precise values $M_V=775.9\pm 0.5\ \mathrm{MeV}$ and $M_A=938.7\pm 1.4\ \mathrm{MeV}$. Actually, they performed a Partial Pad\'{e} $\mathds{P}^0_{1,1}$ since they fixed the $M_V$ to the physical value and led $M_A$ free. Going one step forward, they improved the approximation adding an extra higher vector state, a MHA$+V'$. They again performed a Partial Pad\'{e}, but now a $\mathds{P}^1_{1,2}$ since they fixed the $M_V$ to the physical value and led $M_A$ and $M_{V'}$ to be free. However, in this last case, imposing $M_V = (775.9 \pm 0.5)$ MeV, they found $M_A = (939.4 \pm 1.1)$ MeV and $M_V'= (1258.2 \pm 2.5)$ MeV. The value of $M_A$ is rather stable. More interesting, they found a zero in the numerator of the approximation at $1262.9\pm 2.5$ MeV. Pole and residue are so close that they are almost canceled. Therefore, when they expand at large values of $Q^2$, for both MHA and MHA$+V'$ they found very similar results showing that the inclusion of a higher vector state does not improve the approximation. They argued that this situation \textit{is consistent with the fact that the MHA approximation seems to have already the bulk of the full large-Nc information. In other words, adding an extra V'-pole appears to be compensated, at a very good approximation, by the position of the nearby zero.}

From our point of view, we suspect that this compensation could be understood as a Froissart doublet instead of a physical pole, and then the value $M_V'= (1258.2 \pm 2.5)$ would not correspond to the physical one. We have also explored a Partial Pad\'{e} sequence, Table \ref{Tab:ppas}, and we have seen that letting the poles be complex helps improving on the predictions for the OPE and Taylor coefficients.


\section{Selecting the best approximant sequence}\label{bestPA}

\quad At the beginning of this chapter, at sec. \ref{PAgeneralities}, we argued that we would use PA of the form $P^M_{M+2}$ because the fall off the function $Q^2 \Pi_{LR}(Q^2)$ has a fall off of $Q^{-4}$. We also said that this procedure is not a theorem and could be a source of a systematic error on our predictions. Now we are in position to discuss the quality of this choice with the help of several examples.


In our case we have the advantage of certain knowledge of the behavior at $Q^2\to \infty$ that the function has. However, from a more realistic point of view, we may know just few coefficients of its Taylor expansion and nothing more. Neither poles, nor residues, and even nor OPE coefficients may be known. In that situation, a priori, one can not decide what is the best kind of PA sequence to approximate the function. In our model Eq.~(\ref{eq:LRN1}), since we know a priori certain global information about the function, we can impose extra constraints to our PAs. But now, we would like to test whether assuming $P_{M+2}^M$ is correct or not. For that reason, we will perform an exercise. Lets suppose that only the first Taylor coefficients of our model Eq.~(\ref{eq:LRN1}) are known, i.e. only $C_0 + C_2 Q^2 +\cdots +C_{20} (Q^2)^{10}$ are known from Eq.~(\ref{expChi}). With all of these coefficients, we can construct several PAs, such as $P^9_1, P^8_2, P^7_3, P^6_4, P^5_5, P^4_6, P^3_7, P^2_8, P^1_9$. Then, with all of them we can try to predict the position of the first pole and residue of the original function and also the first not used Taylor coefficient, namely the $C_{22}$. Defining the relative error $\varepsilon$ as $\varepsilon=\frac{predicted\,value-real\,value}{real\,value}$, the results for the prediction of the first pole position, the first residue and the $C_{22}$ are shown in Fig. \ref{detpolrescoef}.

\begin{figure}
\centering
  \includegraphics[width=3in]{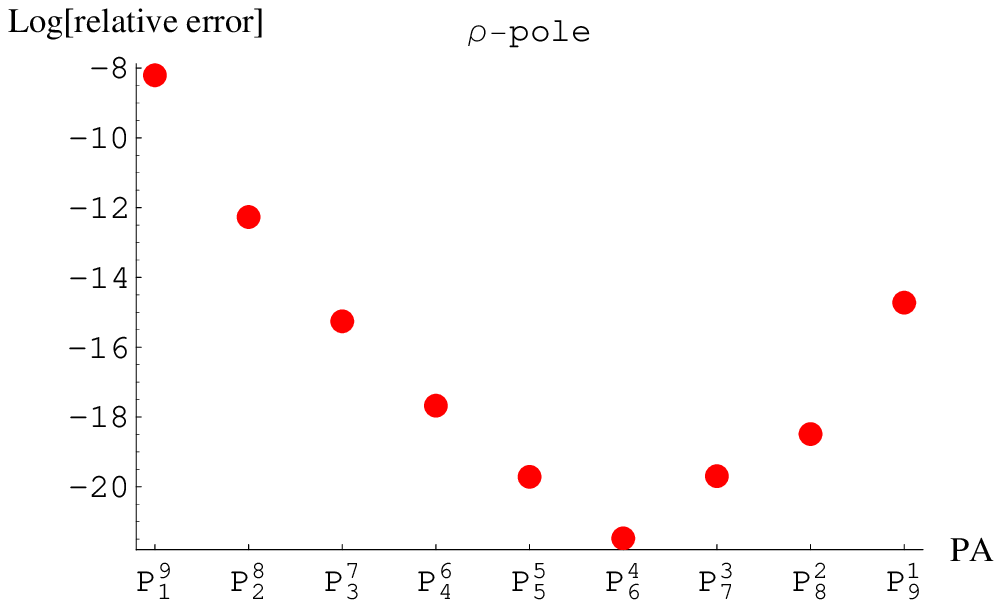}
  \includegraphics[width=3in]{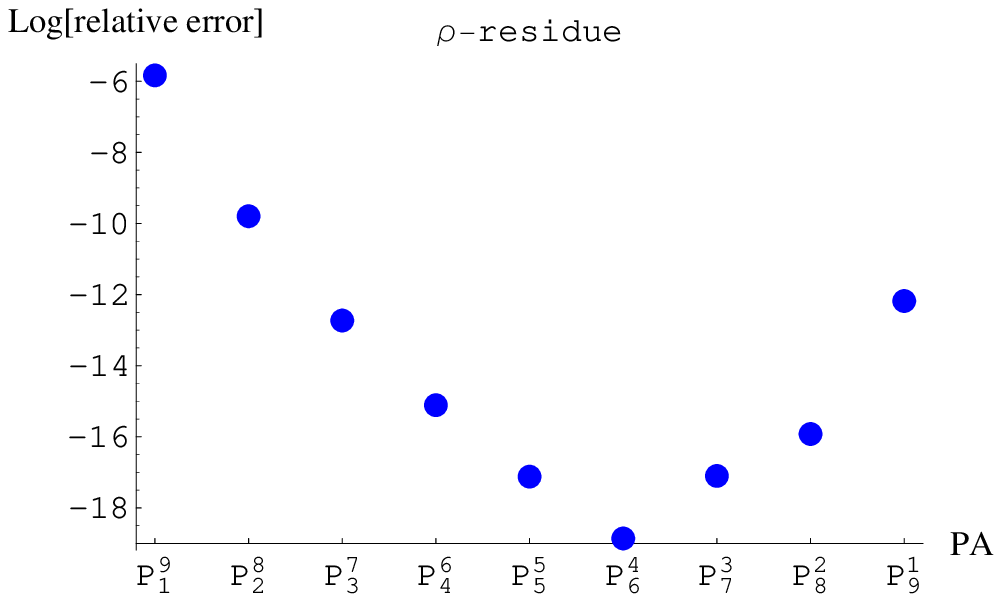}
  \includegraphics[width=3.5in]{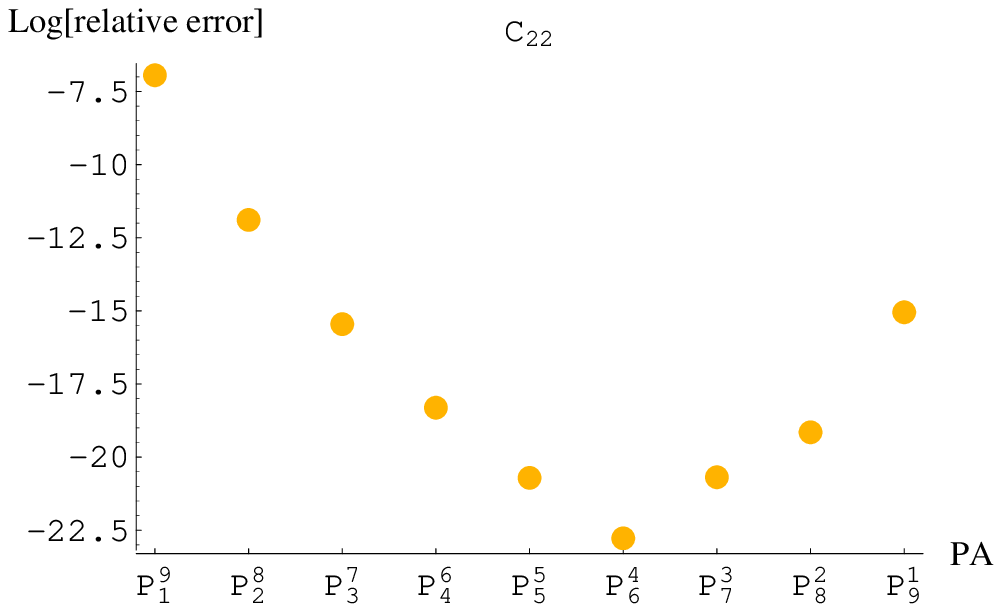}
  \caption{Determination of the position of the first pole (top), the firs residue (center) and the Taylor coefficient $C_{22}$ (bottom) from several PAs. See the text for details.}\label{detpolrescoef}
\end{figure}

In all these pole, residue and $C_{22}$ predictions the best estimation comes from the same $P^4_6$. This PA has an extra particularity, is the only PA from that group with the same fall off behavior as $Q^{-4}$ as our model Eq.~(\ref{eq:LRN1}) has, and then, we can also estimate the value of the first nonvanishing OPE coefficient. In that case, we obtain a prediction for $C_{-4}$ with a relative error $\varepsilon=4\cdot 10^{-4}$.

\begin{figure}
\centering
  \includegraphics[width=4in]{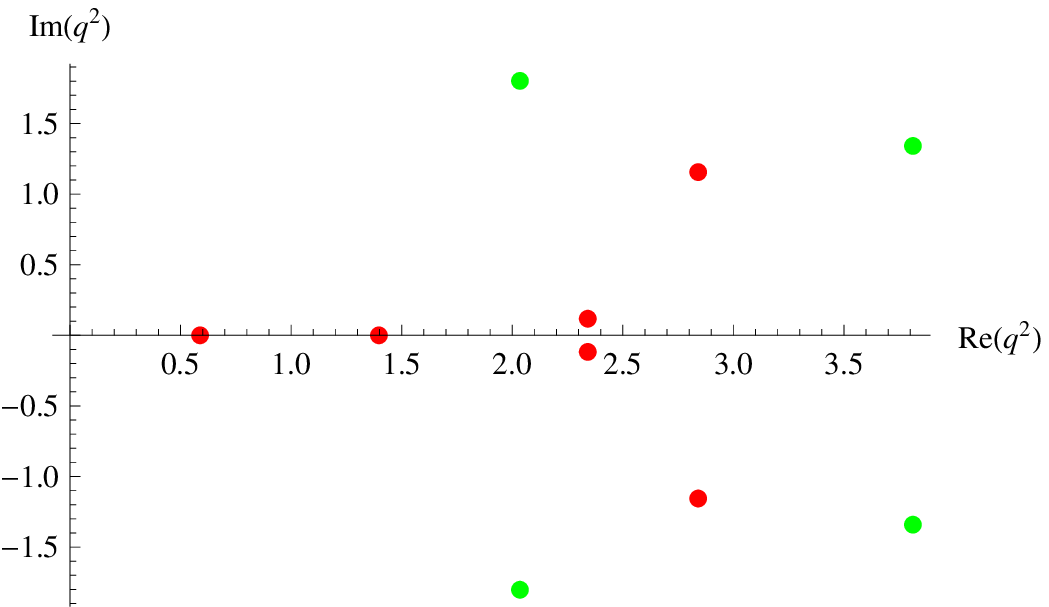}\\
  \caption{Location of the poles (red) and residues (green) in the complex $q^2$-plane for the $P^4_6$.}\label{P46}
\end{figure}

Also, as already shown in section \ref{sec:model}, the analytical structure of the $P^4_6$ follows the same pattern as its counterparts shown in Fig. \ref{poles2}. As can be seen in Fig.\ref{P46} the lowest poles are close to the original once while when one goes away from the origin, its location worsens becoming eventually complex.

PAs are not only useful to predict Taylor coefficients and the lowest masses and residues but also tell us global properties of the original function without the need to impose any extra constrain. In such a way, we obtain not only a prediction for the first OPE coefficient but we predict the right fall off behavior of our model.

To show that this characteristic behavior of the PAs is a general fact and not model dependent, we have also studied a couple of different models with different properties. However, in all of them we find the same conclusion. The first of theses models is close to which we have already studied, that means, is also a meromorphic function with a particular falloff when $Q^2\rightarrow \infty$. Lets define the function $G(Q^2)$ as:

\begin{equation}\label{modelOPE4}
G(Q^2)\,=-\frac{F_{0}^2}{Q^2}+\frac{F_{\rho}^2}{Q^2+M_{\rho}^2}-\frac{F_{a_0}^2}{Q^2+M_{a_0}^2}+
    \sum_{n=1}^{\infty} \left\{\frac{F_V^2}{Q^2+M^2_V(n)}-
    \frac{F_A^2}{Q^2+M^2_A(n)}\right\}\ .
\end{equation}

Again, we impose the dependence on the resonance excitation number $n$ as in Eq. (\ref{twoprime}):

\begin{equation}\label{FaFvOPE4}
    F^2_{V,A}(n)=F^2= \mathrm{constant}\  ,
    \quad\quad M_{V,A}^2(n) = m_{V,A}^2 + n \ \Lambda^2\ .
\end{equation}

Thanks to Eq. (\ref{FaFvOPE4}), the infinite sum in Eq. (\ref{modelOPE4}) is summable. Also, since we are not now concern about the physical properties of the model, just mathematics, we do not need to reassemble QCD. We choose a set of parameters satisfying this time the fall off constraint that the OPE starts at dimension $10$:

\begin{eqnarray}\label{natureOPE4}
\hspace{-2. cm} F_{0}= 85.8 \,\,{\mathrm{MeV}}\ , \quad F_{\rho}= 97.7 \,\,{\mathrm{MeV}}\ ,\quad F_{a_0}= 110.8 \,\,{\mathrm{MeV}}\ ,
\quad F= 143.758 \,\,{\mathrm{MeV}}\, , \\
M_{\rho}= 0.385 \,\,{\mathrm{GeV}}, \quad M_{a_0}= 1.268 \,\,{\mathrm{GeV}}, \quad m_V= 1.494
\,\,{\mathrm{GeV}}\, , \quad \Lambda= 1.277 \,\, {\mathrm{GeV}}\ . \nonumber
\end{eqnarray}

For instance, $F_{\rho}$, $M_{\rho}$, $F_{a_0}$ and $M_{a_0}$ are chosen so that the function $G(Q^2)$ in Eq.~(\ref{modelOPE4}) has vanishing $1/Q^2$, $1/Q^4$, $1/Q^6$, $1/Q^8$ in the OPE at large $Q^2$. Defining the expansion of our $G(Q^2)$ function around $Q^2=0, Q^2\rightarrow\infty$ as
\begin{equation}\label{expG}
    Q^2\ G(-Q^2)\approx \sum_{l} g_{2l}\ Q^{2k}\quad ,\quad \mathrm{with}\quad
    l=0, \pm 1, \pm 2, \pm 3, \ldots
\end{equation}

one obtains the values for the coefficients of both Taylor expansion ($l>0$) and OPE expansion ($l<0$). In Table~\ref{tableOP4} we collect few of these coefficients for the function $G(Q^2)$.

\begin{table}
\centering
\begin{tabular}{|c|c|c|c||c|c|c|}
  \hline
  $g_0$ & $g_2$ & $g_{4}$ & $g_6$ & $g_{-8}$ & $g_{-10}$ & $g_{-12}$ \\
  \hline
  $-7.362$ & $61.18$ & $-429.6$ & $2901.7$ & $-4.95$ & $22.46$ & $-61.87$ \\
  \hline
\end{tabular}
\caption{Values of the coefficients $g_{2l}$ from the high- and low-$Q^2$ expansions of $Q^2\
G(-Q^2)$ in Eq.~(\ref{modelOPE4}) in units of $10^{-3}\ GeV^{2-2l}$. Notice that $g_{-2}=g_{-4}=g_{-6}=0$ and
$g_0=-F_0^2$, see the text for details.}\label{tableOP4}
\end{table}

Following the definitions presented in sec.~\ref{PAgeneralities}, we can proceed with the construction of rational approximants to the function $G(Q^2)$. Again we will play the same game than before, i.e., assuming that we only know the lowest Taylor coefficient of $G(Q^2)$ and nothing more, i.e, $g_0 + g_2 Q^2 +\cdots +g_{20} (Q^2)^{10}$, Eq.~(\ref{expG}). With all these coefficients, we can construct the PAs $P^9_1, P^8_2, P^7_3, P^6_4, P^5_5, P^4_6, P^3_7, P^2_8, P^1_9$. Once all these PAs are know, by reexpanding them we can predict $g_{22}$, the first Taylor coefficient not used in their construction, and also we can try to predict the first pole and residue of $G(Q^2)$ defined in Eq.~(\ref{natureOPE4}). Fig.~\ref{detpolrescoefOPE4} shows the results for the case of the model Eq. (\ref{onecompact}), using also the definition of relative error already presented. The preferred approximant is now the $P^3_7$, which gives the best prediction for the first pole and residue position, and also for $g_{22}$. Also, is the only PA among them that has the right fall off as $Q^{-8}$ as the function $G(Q^2)$ has, giving an extra prediction for the first nonvanishing OPE coefficient with a relative error $\varepsilon_{g_{-8}}=1\%$.


Again, its poles and residues, Fig. \ref{polsres37}, are located in the complex $Q^2$ plane in a similar way than the cases shown in Figs.~\ref{poles2}, ~\ref{poles} and~\ref{P46}.

\begin{figure}
\centering
  \includegraphics[width=3in]{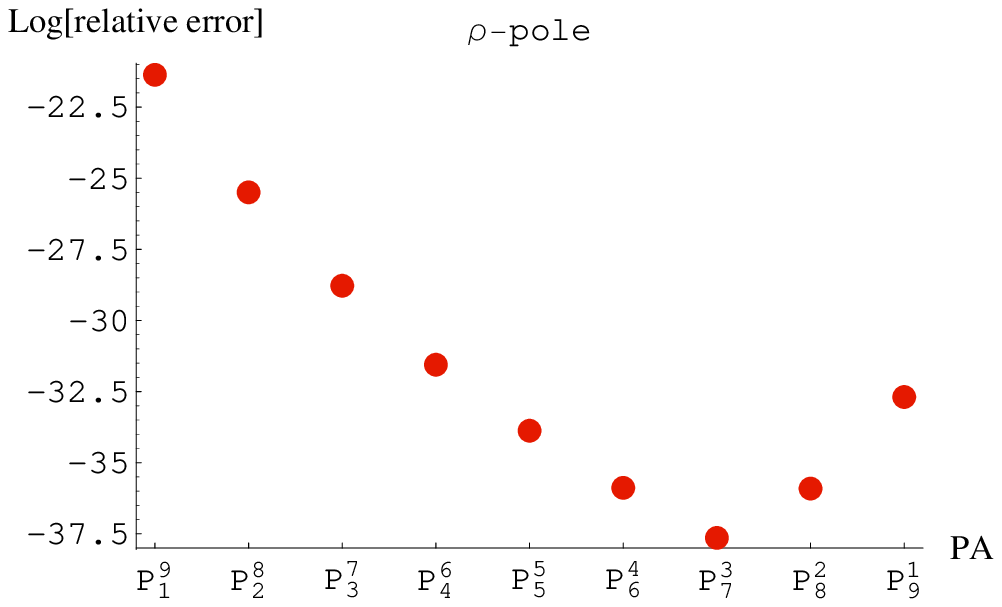}
  \includegraphics[width=3in]{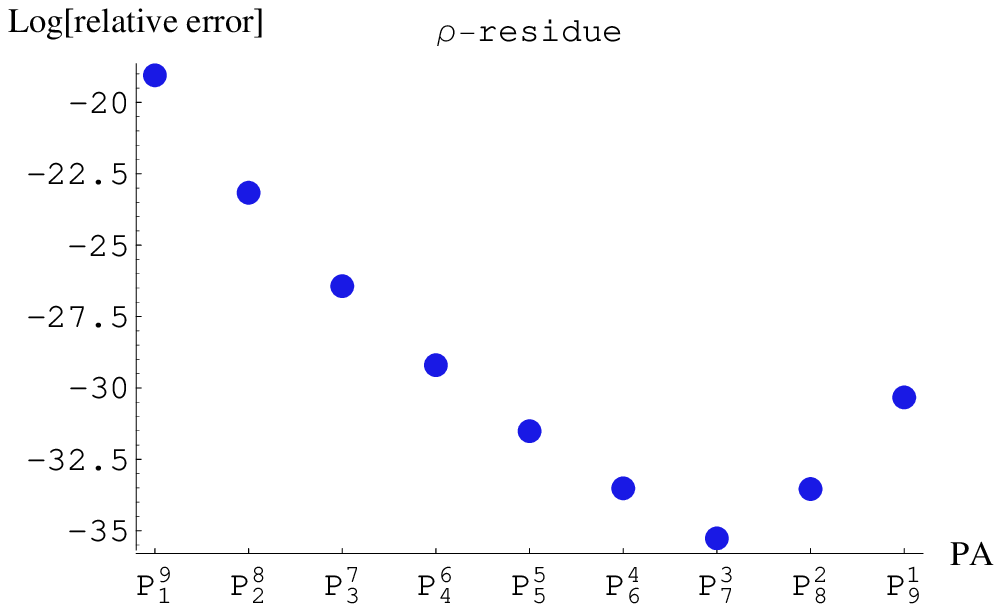}
  \includegraphics[width=3.5in]{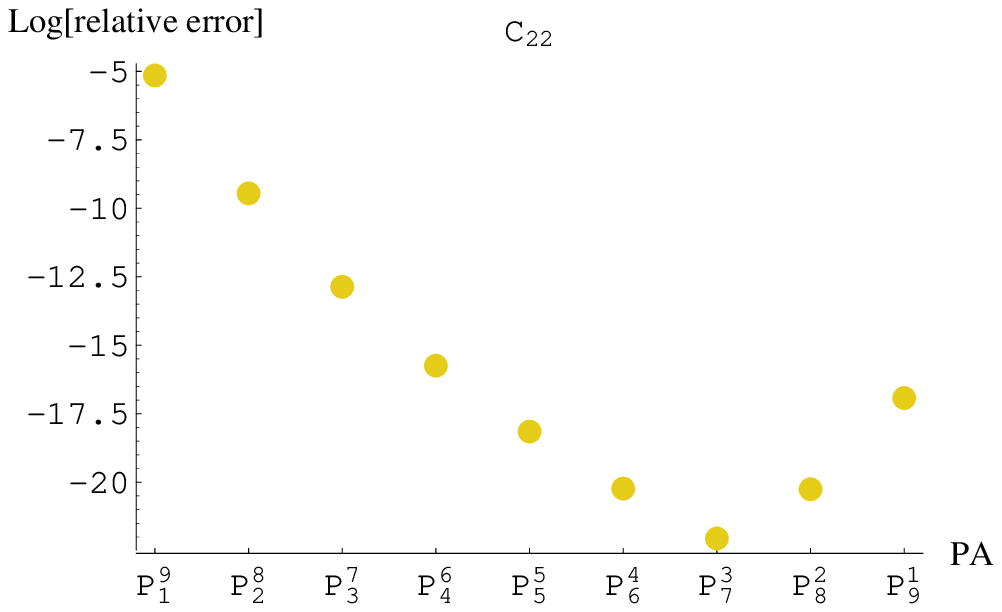}
  \caption{Determination of the position of the first pole (top), the firs residue (center) and the Taylor coefficient $g_{22}$ (bottom) from several PAs, approximating the function $G(Q^2)$. See the text for details.}\label{detpolrescoefOPE4}
\end{figure}

\begin{figure}
\centering
  \includegraphics[width=3.5in]{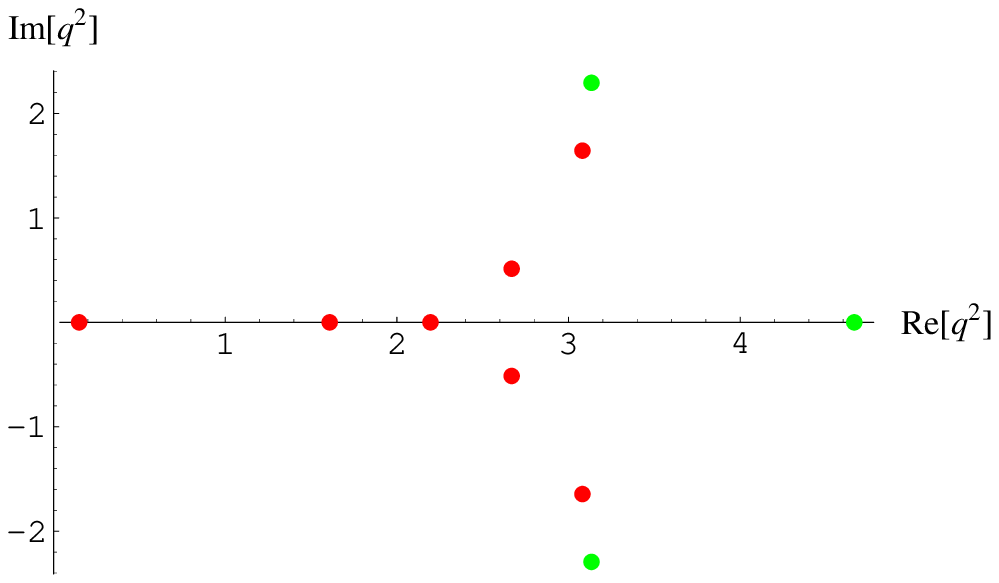}\\
  \caption{Location of the poles (red) and residues (green) in the complex $q^2$-plane for the $P^3_7$.}\label{polsres37}
\end{figure}

At that point, one can think that this selection-rule-behavior shown by PAs comes from the fact that the function we are approximating is meromorphic. To this end, we would like to consider also an example of a Stieltjes function and study again which PAs is the best approximant when only few of the Taylor coefficients are known and no more information about the function is available. Let us consider the following function defined by the sum \cite{Shifman,FESRPeris}

\begin{equation}\label{modellog}
    H(Q^2)=\sum_{n=1}^{\infty} \frac{1}{n (Q^2+n)}\ ,
\end{equation}
as a very simple toy model for the once-subtracted vacuum polarization function in the large-$N_c$ limit, \cite{PerisPade}. Obviously its spectrum consists of masses located at $M^2_n=1,2,3,...$, and one may use the mass of the lowest resonance as the unit of energy. An exact result for $H(Q^2)$ can be readily obtained by summing Eq.~(\ref{modellog}),  \cite{PerisPade}:
\begin{equation}\label{exactmodellog}
    H(Q^2)=\frac{1}{Q^2}\Big\{\psi(Q^2+1)+ \gamma\Big\}
\end{equation}
where the function $\psi(Q^2+1)$ is the Digamma function, defined as
\begin{equation}\label{defgamma}
\psi(z)=\frac{d}{dz} \log \Gamma (z)\ ,
\end{equation}
with $\Gamma (z)$ the Euler Gamma function and $\gamma\simeq 0.577216$ being the
Euler-Mascheroni constant. The function $H(Q^2)$ has the low-energy expansion
\begin{equation}\label{chiralmodellog}
    H(Q^2)= \sum_{p=0}^{\infty} h_{2p}\,Q^{2p}\,=\,\sum_{p=0}^{\infty} (-1)^p \zeta(p+2)\ Q^{2p}\ ,
\end{equation}
where $\zeta(p+2)=\sum_{n=1}^{\infty} n^{-p-2}$ is the Riemann's $\zeta$ function.
On the other hand, the large $Q^2$ expansion is given by ($|\mathrm{Arg} (Q^2)|<\pi$)

\begin{equation}\label{OPEmodellog}
     H(Q^2)\approx \frac{\gamma + \log Q^2}{Q^2}+ \frac{1}{2 Q^4}-
     \sum_{p=1}^{\infty} \frac{B_{2p}}{2p\ Q^{4p+2}}\ ,
\end{equation}
where $B_{2p}=(-1)^{p+1} \frac{2 (2p)!}{(2\pi)^{2p}}\ \zeta(2p)$ are the Bernoulli
numbers for $p=1,2,3,...$.

While the chiral expansion has a finite radius of convergence $|Q^2|< 1$,
the large $Q^2$ expansion has zero radius of convergence (i.e. it is asymptotic) due
to the factorial growth of the Bernoulli numbers. Notice also the presence of the
$\log Q^2$ behavior in (\ref{OPEmodellog}). That is the main different between this example and the previous one. We can not think about a certain falloff when $Q^2\rightarrow \infty$ of the original function since $H(Q^2)$ diverges. However we can again do the same game with PAs. We just take the first Taylor coefficients of Eq.~(\ref{chiralmodellog}), $h_0 + h_2 Q^2 +\cdots +h_{20} (Q^2)^{10}$, and construct the lowest PAs $P^9_1, P^8_2, P^7_3, P^6_4, P^5_5, P^4_6, P^3_7, P^2_8, P^1_9$ . We can, then, predict the position of the first pole and residue, and by reexpanding around $Q^2=0$, predict $h_{22}$. The results are shown in Fig.~\ref{detpolrescoeflog2}. The best approximant in this case is the diagonal $P^5_5$ which also manages to be de one with the closest behavior as a \textit{plateau} as the ratio $Q^{-2}\log(Q^{-2})$ has. Of course, we will be never able to predict the very first coefficient of the OPE expansion since this is not a coefficient but a logarithm. However, PAs somehow organizes the information coming from the Taylor expansion to describe properly the function even for a high values of $Q^2$. That can be easily tested numerically since is a direct application of the convergence theorem for Stieltjes functions, i.e., $\lim_{Q^2\rightarrow Q^{*2}}|H(Q^2)-P^M_N(Q^2)|<\varepsilon$, for any large $Q^{*2}$, $|\mathrm{Arg}(Q^{*2})|>\pi$, and small $\varepsilon$.


Trying to better understand the particular case of model Eq. (\ref{OPEmodellog}), we have also studied a different set of PAs, now restricting ourselves to one less Taylor coefficient, $h_{20}$, giving the opportunity to our set to have a subdiagonal PA since the previous exercise had not. We can construct, then, the set $P^8_1, P^7_2, P^6_3, P^5_4, P^4_5, P^3_6, P^2_7, P^1_8$ and again predict the first pole, the first residue and the first predictable Taylor coefficient. The results are shown in Table~\ref{detpolrescoeflog}. In that last case, the best approximant is the $P^4_5$, which predicts the mean piece of the first coefficient in the OPE, Eq.~(\ref{OPEmodellog}). Figs.~\ref{detpolrescoeflog}, \ref{detpolrescoeflog2} tell us that best PA lies half way as a subdiagonal PA, half way a diagonal PA, and if one can construct both PA sequences, one will see that the original function can be constrain with an upper and a lower limit, as pointed out by the convergence theorem for Stieltjes functions, Eq. (\ref{PadesStieltjes}) in sec. \ref{sec:PadeTheory}.

An interesting feature shared by both PAs is that its analytical structure. Since both are approximating a Stieltjes function, their poles and residues must be all reals, according again to the convergence theorems of sec.~\ref{sec:PadeTheory}. That is shown in Fig.~\ref{poles4555}. At the top, the lowest poles (red points) and residues (green points) of $P^4_5$ and at the bottom, the lowest poles (brown points) and residues (sky-blue points) of $P^5_5$. All of their poles are actually purely reals.

The convergence theorem for Stieltjes functions guarantees convergence for all the Pad\'{e} $P^{N+J}_N$ ($J\geq-1$) sequences. As we have said, for $J\geq -1$ all the poles of our PAs have to lie on the same real axes where the brunch cut is located. Now we have the opportunity to see what would happen if we use $J<-1$. If we look to the analytical properties of $P^3_6, P^2_7, P^1_8$ and $P^3_7, P^2_8, P^1_9$ we see that they have several complex conjugate poles. Actually, there is also a way to count the number of complex conjugate poles given a certain value for $J$. If a certain PA has degree $P^N_M$, and $M=N+k$, the PAs will have, at least, $N+1$ real poles and $(k-1)/2$ complex conjugate pares when possible, i.e., for odd values of $k$.


\begin{figure}
\centering
  \includegraphics[width=3in]{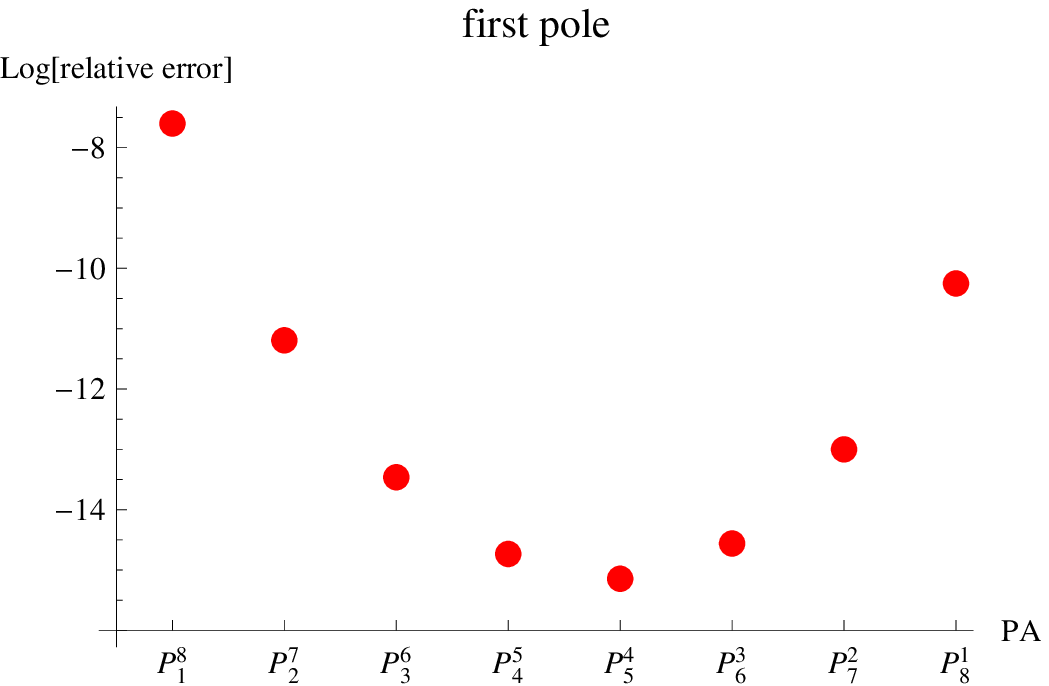}
  \includegraphics[width=3in]{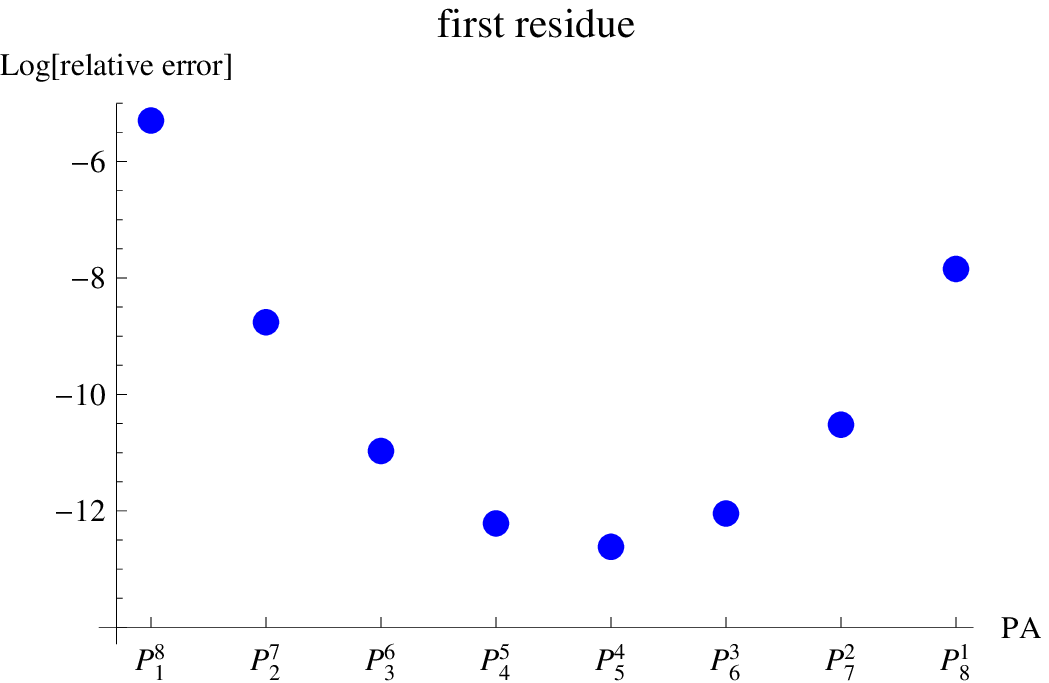}
  \includegraphics[width=3.5in]{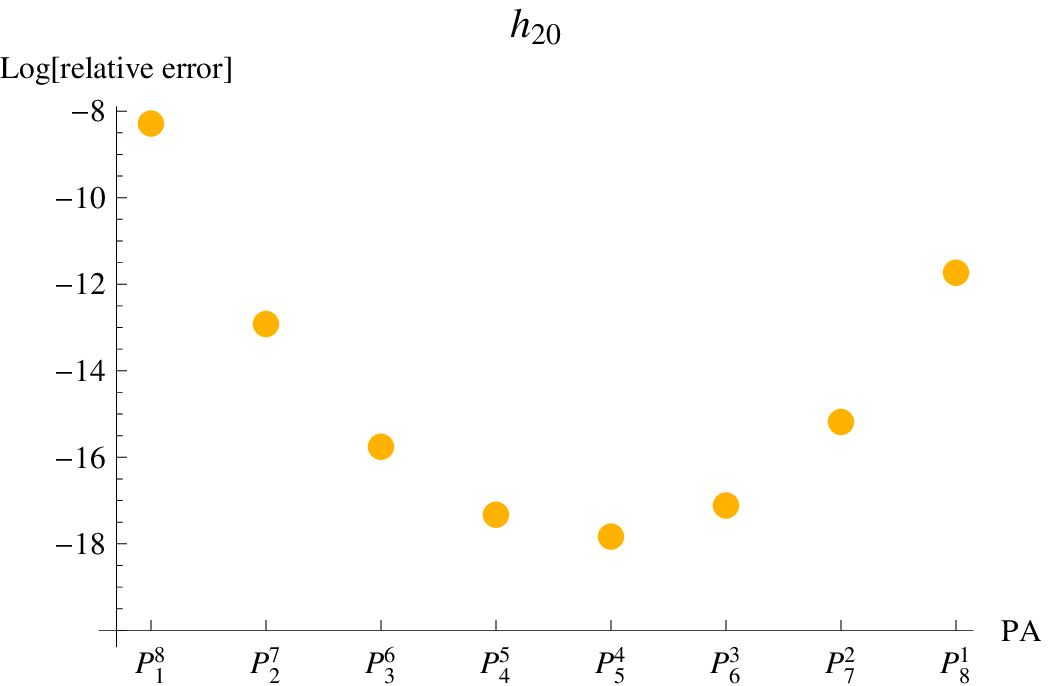}
  \caption{Determination of the position of the first pole (top), the firs residue (center) and the Taylor coefficient $h_{20}$ (bottom) from several PAs, approximating the function $H(Q^2)$. See the text for details.}\label{detpolrescoeflog}
\end{figure}

\begin{figure}
\centering
  \includegraphics[width=3in]{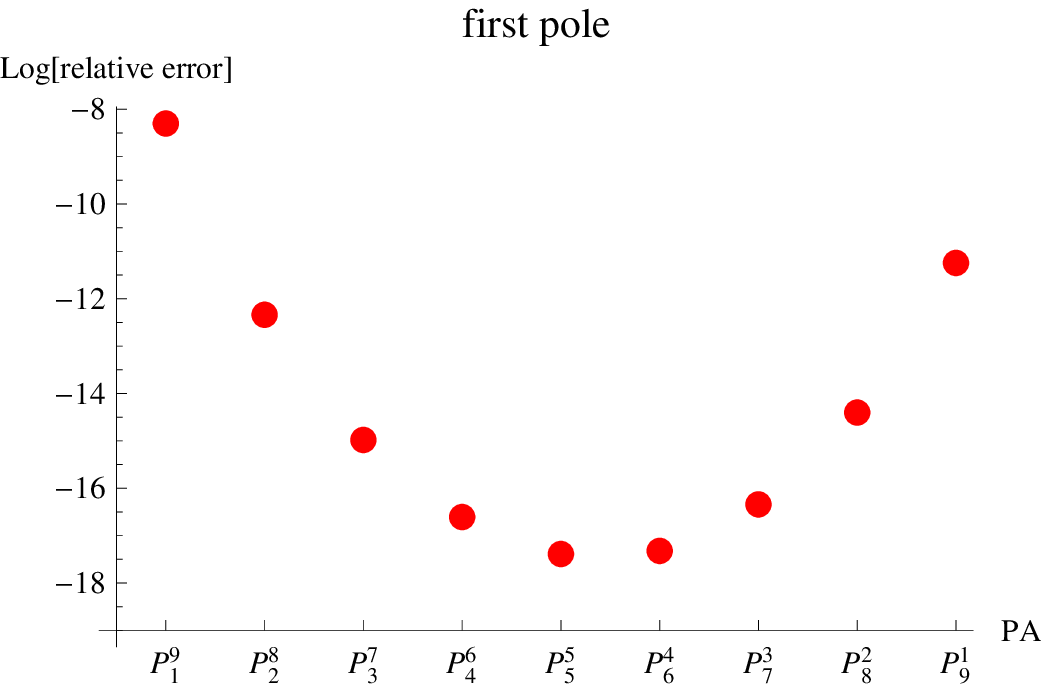}
  \includegraphics[width=3in]{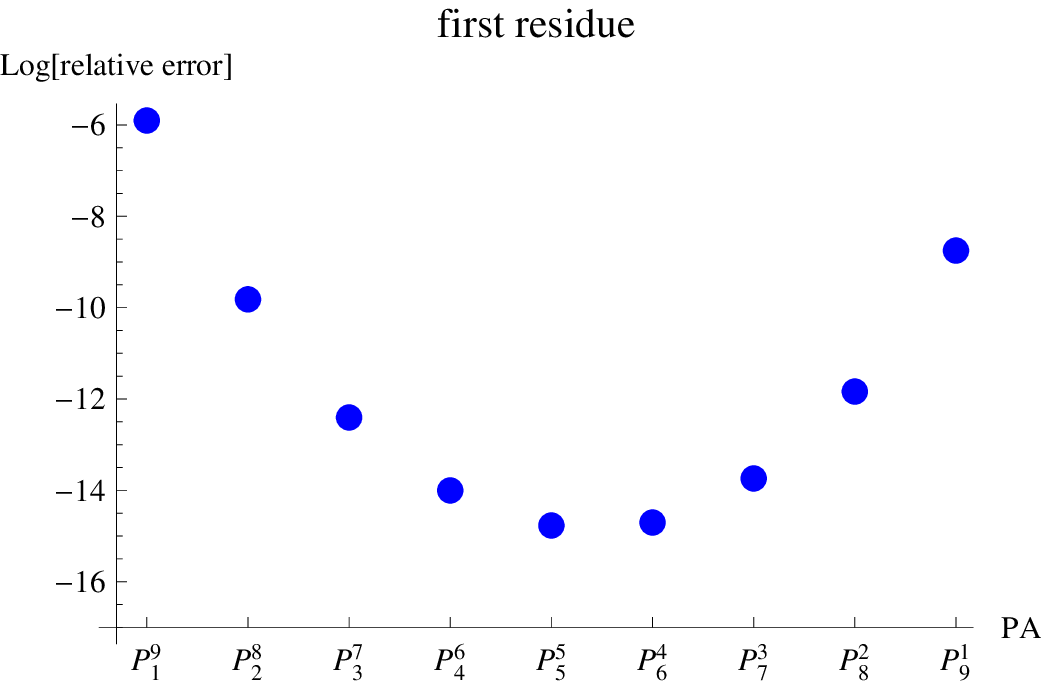}
  \includegraphics[width=3.5in]{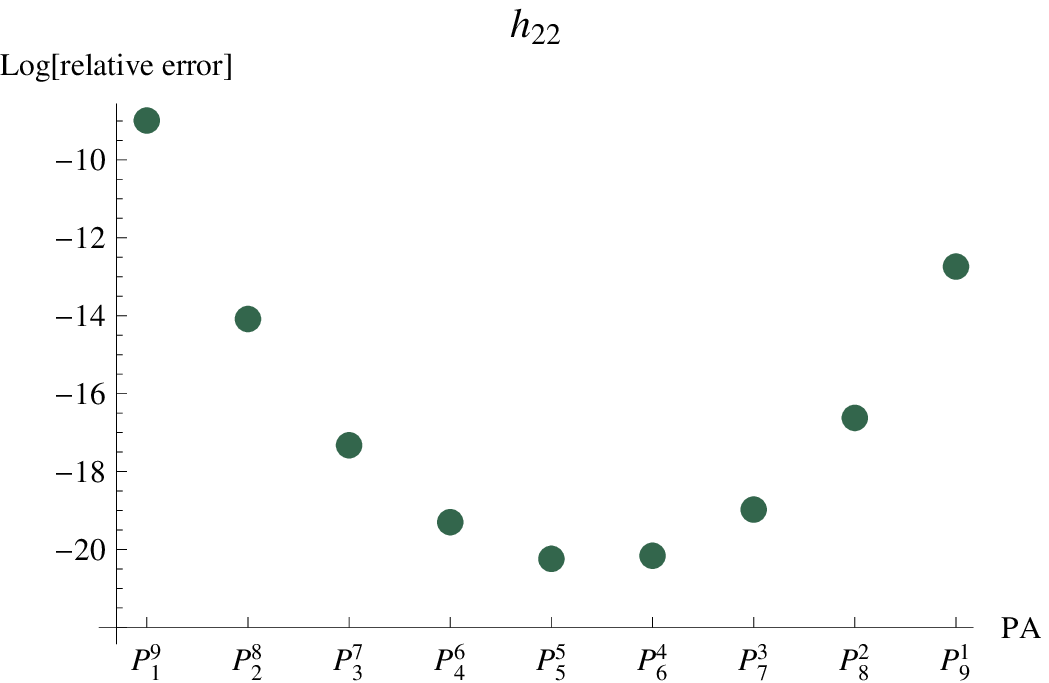}
  \caption{Determination of the position of the first pole (top), the firs residue (center) and the Taylor coefficient $h_{22}$ (bottom) from several PAs, approximating the function $H(Q^2)$. See the text for details.}\label{detpolrescoeflog2}
\end{figure}

\begin{figure}
\centering
  \includegraphics[width=3.5in]{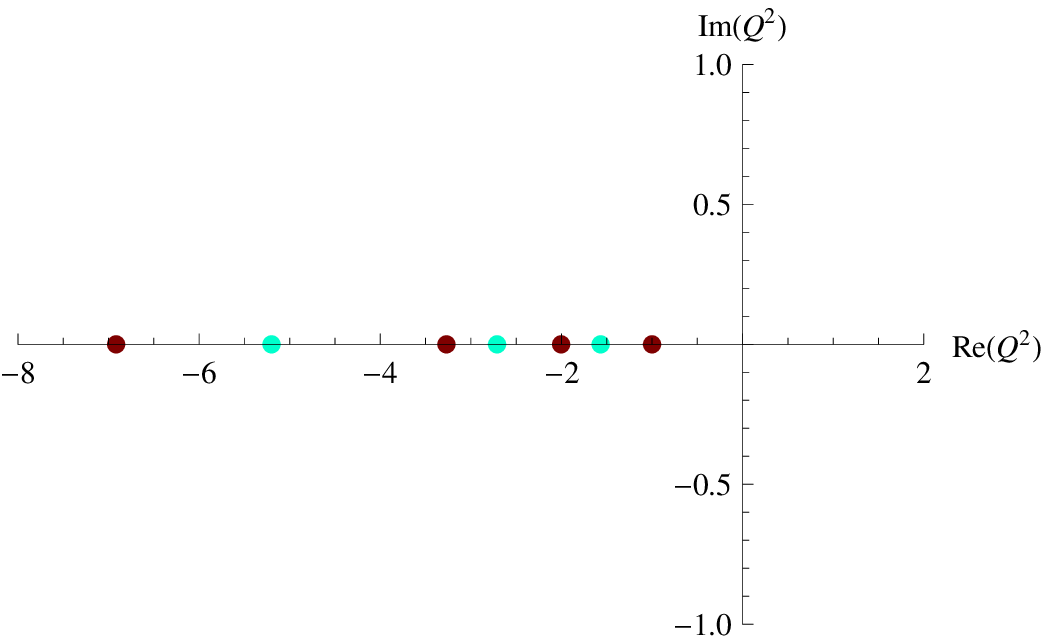}
  \includegraphics[width=3.5in]{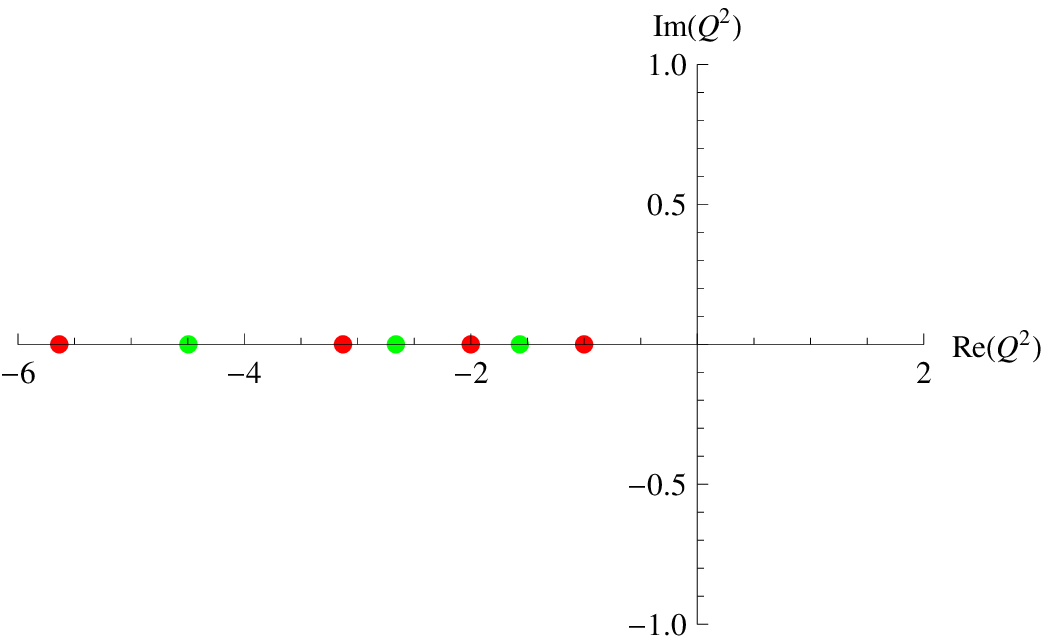}
  \caption{Analytical structure of the $P^4_5$ (top) and $P^5_5$ (bottom), both approaching $H(Q^2)$, Eq.~(\ref{modellog}).}\label{poles4555}
\end{figure}

\subsection{Beyond Pad\'{e} Approximants from Taylor Expansion}\label{beyondPAs}

\quad Throughout the last section we have carefully studied Pad\'{e} Approximants constructed from a Taylor expansion. Now, two questions may be addressed in order to compleat our analysis. The first one is what would be the role of other Pad\'{e} Approximants, such as Pad\'{e}-Type Approximants, compared with the Pad\'{e} Approximants? The second is what may happen if instead of the Taylor expansion defined when $Q^2\to 0$ we use the expansion when $Q^2 \to \infty$ to construct Pad\'{e} Approximants? In the present sub-section we will try to answer both question.

\subsubsection{Pad\'{e} Approximants vs Pad\'{e} Types}

\quad The first answer would be better understood if one compares PTs with PAs defined to be of the same degree, i.e, $P^M_N$ and $\mathbb{T}^M_N$. Using again the model of Eq.~(\ref{oneprime}) and using only the first $11$ coefficients of its Taylor expansion, Eq.~(\ref{largeNtaylor}), i.e., $C_0 + C_2 Q^2 +\cdots +C_{20} (Q^2)^{10}$, we construct the following set of PAs: $P^9_1, P^8_2, P^7_3, P^6_4, P^5_5, P^4_6, P^3_7, P^2_8, P^1_9$. Now imagine that we construct a similar set of PTs with the same Taylor expansion. Now, all these PTs have their denominator fixed in advance. This will lead to the following set of PTs: $\mathbb{T}^9_1, \mathbb{T}^8_2, \mathbb{T}^7_3, \mathbb{T}^6_4, \mathbb{T}^5_5, \mathbb{T}^4_6, \mathbb{T}^3_7, \mathbb{T}^2_8, \mathbb{T}^1_9$. We have now exactly the same number of input parameters, 11, but as a mixture within Taylor coefficients and pols of the function. For example, $\mathbb{T}^9_1$ has only one pole fixed in advance, the $M_{\rho}^2$ and 10 Taylor coefficients. On the contrary, $\mathbb{T}^1_9$ has 9 correctly located poles and only one Taylor coefficient. When all the PTs are constructed, by reexpanding we can predict the Taylor coefficient $C_{22}$ and compare with the results found in Fig.~\ref{detpolrescoef}. The comparison is shown in Fig.~\ref{PAvsPT} where the red points are the predictions for the PTs and the orange points are the predictions for the PAs.

\begin{figure}
\centering
  \includegraphics[width=3.5in]{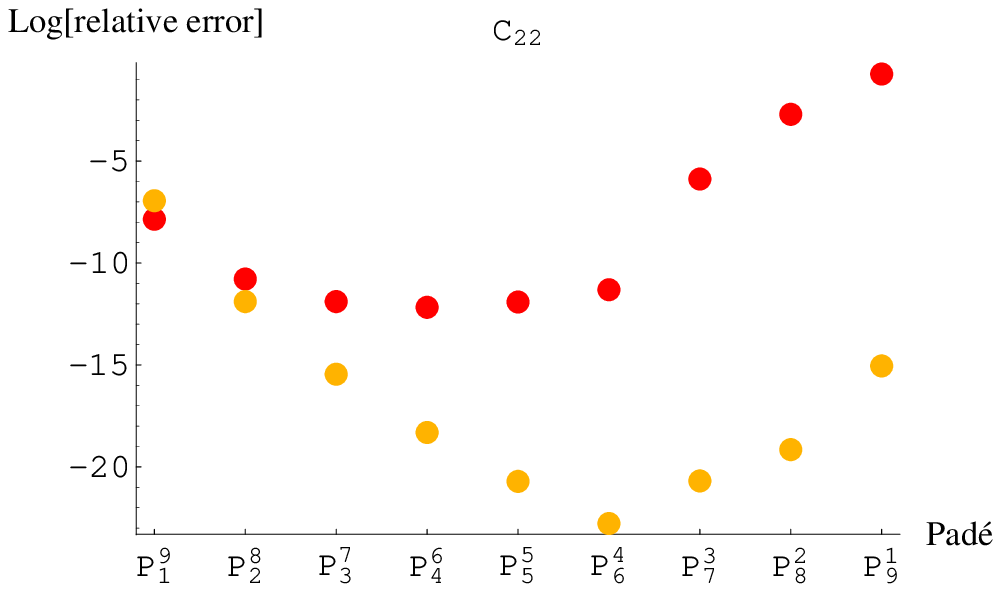}\\
  \caption{Comparison between PAs and PTs when constructed with the same among of input data. For $P^N_M$-Pad\'{e} one must read $\mathbb{T}^N_M$ when is a red point and $P^N_M$ when is an orange point.}\label{PAvsPT}
\end{figure}

From the point of view of PAs the best approximant is the $P^4_6$ which also gives the best prediction for the first pole and residue. It is also the only PA from the selected set able to give a prediction for the first nonvanishing OPE coefficient. However, from the PTs point of view things are different since the best PT is not the $\mathbb{T}^4_6$ as one would expect but $\mathbb{T}^6_4$ which has a divergent behavior for large $Q^2$. This PT also gives the best prediction for the first residue (remember that all the poles are fixed in advance), while is hard to see difference between $\mathbb{T}^6_4$ and $\mathbb{T}^7_3$ or $\mathbb{T}^5_5$. In any case, $\mathbb{T}^4_6$ differs numerically one order of magnitude from $\mathbb{T}^6_4$ in the prediction of $C_{22}$ or for the first residue.

With the same among of input, PAs play a better role on determining Taylor coefficients and residues that their counterparts the PTs. That happens because PAs have an \textit{extra} degree of freedom that PTs do not have. They can have complex conjugate poles to mimic the properties of the model when PTs are constrained to have real poles. The advantage of the first are clear since the difference among them are of several order of magnitudes. Of course, concerning masses PTs are un a different league since masses are inputs in that case.

This comment agrees with what was found in the Ref.\cite{Lipartia}. In that reference, the authors show that many short-distance constraints (which means imposing to the approximant the right fall off when $Q^2\to \infty$) can be easily incorporated in a general model based on resonance saturation. At the same time, however, they pointed out that these approximations cannot reconcile all short-distance constraints due to a general conflict between those constrains coming from Green functions, and those coming from form factors and cross-sections. The authors also show that if one wants to remedied this situation without spoiling the matching into the OPE, one can not proceed using a single or any finite number of resonances per channel type of approximations. In that sense, an earlier example where single resonance does not allow to reproduce all short-distance constraints can be found in Ref. \cite{theworksKnechtNyffelerp6}.


\subsubsection{Using the OPE to construct Pad\'{e} Approximants}

\quad Concerning the second question we addressed ourselves, in 1977, A.A. Migdal \cite{Migdal} suggested PAs as a method to extract the spectrum of large-$N_c$ QCD from the leading term in the OPE of the $\langle VV\rangle$ correlator, i.e., from the parton model logarithm. However, nowadays this proposal should be considered unsatisfactory for a number
of reasons \cite{Catamigdal,Cataregge}, the most simple of them being that different spectra may lead to the same
parton model logarithm \cite{GoltermanPerisspectra,AndrianovEspriuspectra}. In fact, the full OPE series is expected to be only an asymptotic expansion at $Q^2\rightarrow\infty$ (i.e. with zero radius of convergence), and PAs constructed
from this type of expansions cannot in general reproduce the position of the physical poles
 \cite{Baker3}. Migdal's approach has been recently adopted in some
models exploiting the so-called AdS/QCD correspondence \cite{Erlich} and, consequently, the same
criticism also applies to them. And now we can address this criticism based on Pad\'{e} Theory just constructing PAs from the $1/Q^2$ expansion at infinity.

We will show how the PAs constructed from this expansion (akin to the OPE)
do not in general reproduce even the first resonances in the spectrum of the model Eq. (\ref{onecompact}), unlike those constructed from the chiral expansion. Recalling the definition of the OPE for the model Eq. (\ref{onecompact}), given in Eq. (\ref{eq:OPE}), with the corresponding coefficients, Eq. (\ref{expOPE}), it is straightforward to construct a PA in $1/Q^2$ around infinity, i.e. by matching powers of the OPE in $1/Q^2$. Since the function $Q^2 \Pi_{LR} (-Q^2)$ behaves like a constant for $Q^2\rightarrow 0$, we will consider diagonal Pad\'{e} Approximants, i.e., of the form $P^{N}_{N}(1/Q^2)$, in order to
reproduce this behavior. Figure \ref{polesfromOPE} shows the position of the poles and zeros of the PA
$P^{50}_{50}(-1/q^2)$ in the complex $q^2$ plane (recall that $Q^2 = -q^2$). As it is clear from this plot, the positions of
the poles have nothing to do with the physical masses in the model, given by Eqs.
(\ref{twoprime}-\ref{nature}), even for the lightest states. This is to be contrasted with what
happens with the PA constructed from the chiral expansion around $Q^2$, which is shown in Fig.
\ref{poles}. The difference between the two behaviors is due to the fact that, while the chiral
expansion has a finite radius of convergence, the radius of convergence of the OPE vanishes because
this expansion is asymptotic. In spite of this, one can see that the PA $P^{50}_{50}(1/Q^2)$ is an
excellent approximation to the function $Q^2 \Pi_{LR} (-Q^2)$ in the euclidean region.

\begin{figure}
\centering
  \includegraphics[width=3.5in]{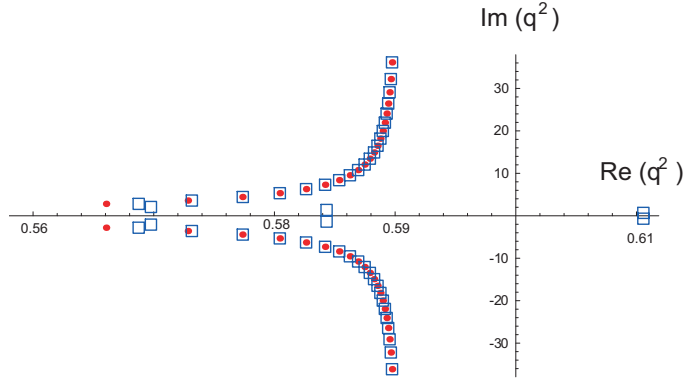}\\
  \caption{Location of the poles (red-dots) and zeros (blue-squares) of the Pad\'{e} Approximant $P^{50}
_{50}(-1/q2)$, constructed from the OPE of $Q^2 \Pi_{LR} (-Q^2)$ defined in Eqs.~(\ref{expChi} and \ref{expOPE}), in the complex $q^2$ plane. We recall that $Q^2 = -q^2$. The poles are all complex-conjugate pairs.}\label{polesfromOPE}
\end{figure}

On interesting feature of that expansion is that one can try to predict chiral coefficients.

The simplest PA with only one resonance per channel constructed using the OPE expansion is the PA:

\be
P_2^2(1/Q^2)=-F_0^2+\frac{F_V^2 Q^2}{Q^2+m_V^2}-\frac{F_A^2 Q^2}{Q^2+m_A^2}
\ee\label{PadeOPEs}

For the four unknown parameters, we can use the first four OPE coefficients, i.e, the following set of equations
\cite{Catapades}:
\begin{eqnarray}\label{setOPE}
F_V^2-F_A^2=F_0^2\\\nonumber
F_V^2m_V^2-F_A^2m_A^2=0\\\nonumber
F_V^2m_V^4-F_A^2m_A^4=e_6\\\nonumber
F_V^2m_V^6-F_A^2m_A^6=e_8\\
\end{eqnarray}

where $e_6$ and $e_8$ are the OPE coefficients of order ${\cal O(Q^{-6})}$ and ${\cal O(Q^{-8})}$ respectively, Eq.(\ref{eq:OPE}). Once the four undetermined parameters of Eq. (\ref{PadeOPEs}) are known, Eq. (\ref{setOPE}), by reexpanding at $Q^2\to 0$ one can estimate the LECs $F_0$ and $L_{10}$, one obtain this nice relation among OPE coefficients and LEC coefficients:

\be
L_{10}=F_0^4\frac{e_8}{4 e_6^2}\, ,
\ee

which determines uniquely the sign of the condensate of order ${\cal O(Q^{-8})}$. One can go improving this result by constructing higher and higher PAs from the OPE. Fig. \ref{ChPTfromOPE} shows the prediction for $F_0$ and $L_{10}$ for all the PAs up to the $P^{50}_{50}(1/Q^2)$. The sign of $L_{10}$ never change, the prediction improves by increasing the order of the PA, and the location of the poles and residues are completely far away from their physical counterparts, Fig. \ref{polesfromOPE}.
In sec. \ref{complexpolescomment} we have commented on Ref. \cite{Friot} and the discrepancy on the values of the condensates of order ${\cal O(Q^{-6})}$ and ${\cal O(Q^{-8})}$ found in the literature. After our analysis, we can give a new insight on that discussion since we found a relation between the $L_{10}$ and the OPE coefficient of order ${\cal O(Q^{-8})}$, $e_8$. Nevertheless, we can not say anything about the relative sign of ${\cal O(Q^{-6})}$ versus ${\cal O(Q^{-8})}$ since in our result $e_6$ always appears squared. In any case, since we have seen that $L_{10}$ has the same sign as $e_8$, that fact may contradict the predictions for the OPE coefficient of order ${\cal O(Q^{-8})}$ obtained in Refs. \cite{FESRBijnens,FESRRojo,pwCirigliano} which found different sign for it.

\begin{figure}
\centering
  \includegraphics[width=2.8in]{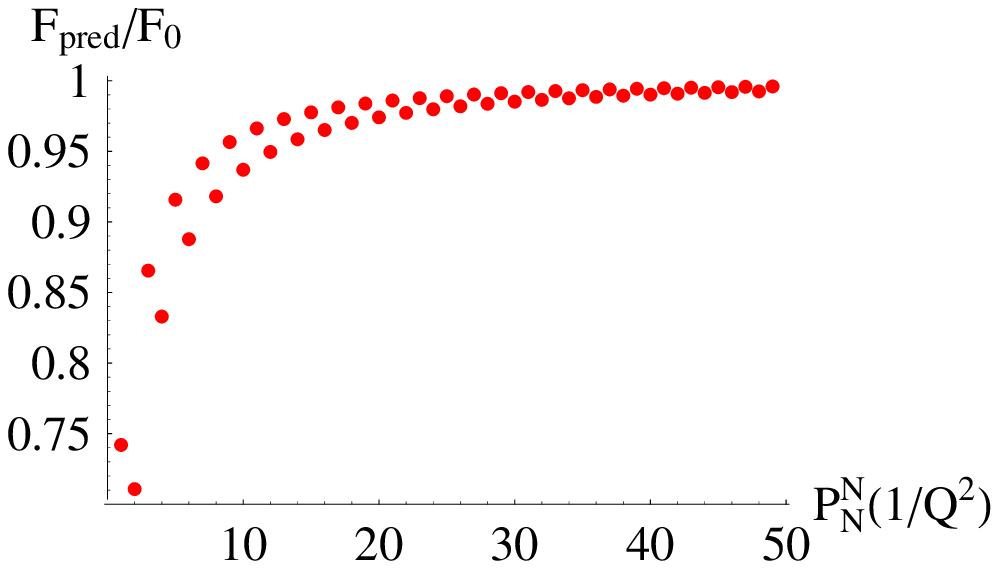}
  \includegraphics[width=2.8in]{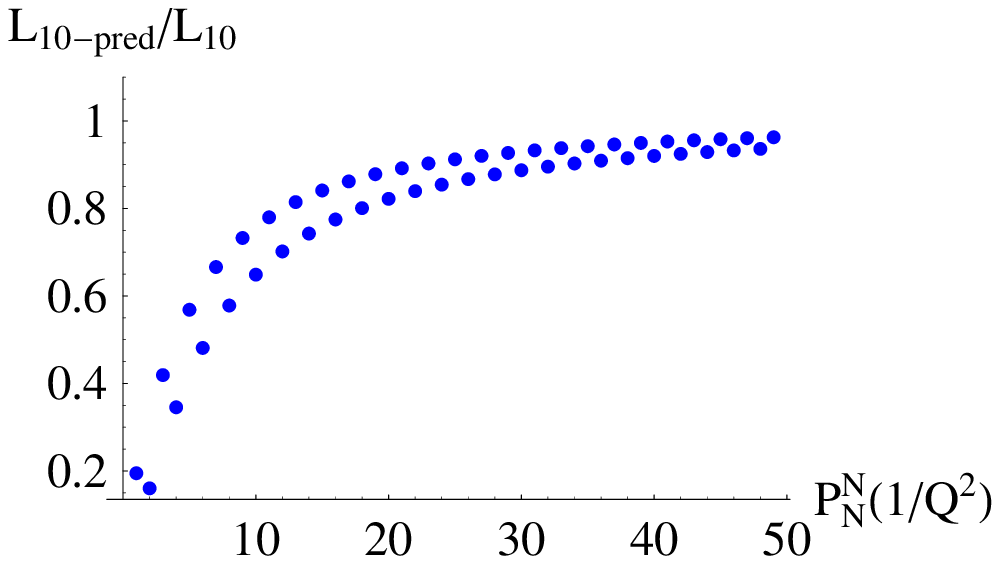}\\
  \caption{Predictions for $F_0$ (left) and $L_{10}$ (right) from PAs constructed using the OPE expansion.}\label{ChPTfromOPE}
\end{figure}

\section{The QCD case}\label{sec:qcd}

\quad After all the mathematical discussion about Pad\'{e} Theory applied to the model Eq. (\ref{onecompact}), let us now discuss the real case of large-$N_c$ QCD in the chiral limit. In contrast to the case of
the previous models, any analysis now is limited by two obvious facts. First, any input
value will have an error (from experiment and because of the chiral and large-$N_c$ limits), and
this error will propagate through the rational approximant. And second, it is not possible to go to
high orders in the construction of rational approximants due to the rather sparse set of input
data. In spite of these difficulties one may feel encouraged by the phenomenological fact that
resonance saturation approximates meson physics rather well and the quite well understanding of Pad\'{e} Theory.

The simplest PA to the function $Q^2 \Pi_{LR}(-Q^2)$ with the right fall-off as $Q^{-4}$ at large
$Q^2$ is  $P^{0}_{2}(Q^2)$:
\begin{equation}\label{onepade}
    P^{0}_{2}(Q^2)=\frac{a}{1+ A \ Q^2 + B\ Q^4}\ .
\end{equation}


By simple re-expansion around $Q^2=0$ it is then
possible to predict an estimate for the term of $\mathcal{O}(Q^4)$ in (\ref{largeNtaylor}). In the QCD case, the values of the three unknowns $a,A$ and $B$ may be fixed by requiring that this PA reproduces the correct values for $F_0, L_{10}$\footnote{Since $m_s$ decouples from $F_{\pi}$
in the large-$N_c$ limit, the value of $F_0$ is estimated in Eq. (\ref{realnature}) by extracting the chiral
corrections from $F_{\pi}$ using $SU(2)\times SU(2)$ chiral perturbation theory, but doubling the error as
compared to Ref. \cite{Colangelolattice}.} and $I_{\pi}$ \footnote{Recall that $I_{\pi}$ is, up to a constant, the electromagnetic pion mass difference $\delta m_{\pi}$ \cite{Friot} and is defined in
terms of $\Pi_{LR}$ in Eq. (\ref{pi}).} given by
\begin{eqnarray}\label{realnature}
F_0&=& 0.086\pm  0.001\ \mathrm{GeV} \quad , \nonumber\\
   \delta m_{\pi}= 4.5936\pm 0.0005\ \mathrm{MeV}\
&\Longrightarrow  &\  I_{\pi}= (5.480 \pm 0.006)\times 10^{-3} \mathrm{GeV}^4\ , \nn \\
L_{10}(0.5\ \mathrm{GeV}) \leq L_{10}\leq L_{10}(1.1\ \mathrm{GeV})& \Longrightarrow &
L_{10}=\left( -5.13\pm 0.6\right) \times 10^{-3} \ .
\end{eqnarray}
The low-energy constant $L_{10}$ is related to the chiral coefficient $C_2$, in the notation of Eq.
(\ref{exp}), by $C_2=-4L_{10}$. Since $L_{10}$ does not run in the large-$N_c$ limit, it is not
clear at what scale to evaluate $L_{10}(\mu)$ \cite{Perissubleading,Rosellsubleading,Portolessubleading,RosellCillerosubleading}. In Eq. (\ref{realnature}) we have
varied $\mu$ in the range $0.5\ \mathrm{GeV}\leq \mu \leq 1.1\ \mathrm{GeV}$ as a way to estimate
$1/N_c$ systematic effects. The central value corresponds to the result for $L_{10}(M_{\rho})$
found in Ref. \cite{L10}. The other results in (\ref{realnature}) are  extracted from Refs \cite{GLsq,PDG}.

Obviously, the PA  (\ref{onepade}) can also be rewritten as
\begin{equation}\label{P02real}
     P^{0}_{2}(Q^2)=\frac{-\ r^2}{(Q^2+z_V) (Q^2+z_A)}\ ,
\end{equation}
in terms of two poles $z_{V,A}$. In order to discuss the nature of these poles, we will define the
dimensionless parameter $\zeta$ by the combination
\begin{equation}\label{zeta}
    \zeta\equiv  -4 L_{10}\ \frac{\ I_{\pi}}{F_0^4}= 2.06\pm 0.25\quad ,
\end{equation}
where the values in (\ref{realnature}) above have been used in the last step.
 Imposing the constraints (\ref{realnature}) on
the PA (\ref{P02real}) one finds two types of solutions depending on the value of
 $\zeta$: for $\zeta> 2$ the two poles $z_{V,A}$ are real, whereas for
$\zeta< 2$ the two poles are complex. At $\zeta=2$, the two solutions coincide. To see this, let us
write the set of equations satisfied by the PA (\ref{P02real}) as:
\begin{eqnarray}\label{set}
  F_0^2 &=& \frac{r^2}{z_V z_A} \nn \\
  -4L_{10} &=& F_0^2\left(\frac{1}{z_V}+\frac{1}{z_A} \right) \nn \\
  I_{\pi} &=& F_0^2\frac{z_V z_A}{z_A-z_V}\log\frac{z_A}{z_V}\ .
\end{eqnarray}
The first of these equations can be used to determine the value of the residue $r^2$ in terms of
$z_Vz_A$. In order to analyze the other two, let us first assume that both poles $z_{V,A}$ are
real. In this case, they also have to be positive  or else the integral $I_{\pi}$ will not exist
because it runs over all positive values of $Q^2$. Let us now make the change of variables
\begin{equation}\label{change}
    z_V=R\ (1- x) \qquad , \qquad  z_A=R\ (1+x)\quad .
\end{equation}
The condition $z_{V,A}>0$ translates into $R>0, |x|< 1$. In terms of these new variables, the
second and third equations in (\ref{set}) can be combined into
\begin{equation}\label{master}
    \zeta= \frac{1}{x}\log \frac{1+x}{1-x}\ ,
\end{equation}
where the definition (\ref{zeta}) for $\zeta$ has been used. With the help of the identity $\log
(1+x/1-x)=2\ \mathrm{th}^{-1}x$ (for $|x|< 1$), one can finally rewrite this expression as
\begin{equation}\label{master2}
    \zeta=\frac{2}{x}\ \mathrm{th}^{-1}x\quad , \quad (x\ \mathrm{real})
\end{equation}
which is an equation with a solution for $x$ only if $\zeta \geq 2$. Once this value of $x$ is
found, the value of $R$ can always be obtained from one of the last two equations (\ref{set}) and
this determines the two real poles $z_{V,A}$ from (\ref{change}).

On the other hand, when $\zeta <2$, Eq. (\ref{master2}) does not have a solution. However,
according to (\ref{zeta}), $\zeta$ can \emph{also} be smaller than 2. In order to study this case,
we may use the identity $\mathrm{th}^{-1}(i\ y)= i\ \tan^{-1} (y)$ to rewrite the above equation
(\ref{master2}) in terms of the variable $x=i\ y$ ($y$ real) as
\begin{equation}\label{master3}
    \zeta=\frac{2}{y}\ \mathrm{tan}^{-1}y\quad , \quad (y\ \mathrm{real}) .
\end{equation}
One now finds that this equation has a solution for $y$ when $\zeta \leq 2$. In this case the poles
of the PA (\ref{onepade}) are complex-conjugate to each other and can be obtained as
$z_{V,A}=R(1\pm i\ y)$. These poles, obviously, cannot be associated with any resonance mass and
this is why this solution has been discarded in all resonance saturation schemes up to now.
However, from the point of view of the rational approximant (\ref{onepade}) there is nothing wrong
with this complex solution, as the approximant is real and well behaved. From the lessons learned
in the previous section with the model, there is no reason to discard this solution since, as we
saw, rational approximants may use complex poles to produce accurate approximations. Therefore, we
propose to use \emph{both} the complex as well as the real solution for the poles $z_{V,A}$, at
least insofar as the value for $\zeta\gtrless 2$. In this case we obtain, using the values given in
Eqs. (\ref{realnature}),
\begin{eqnarray}
 &&\!\!\!\!\!\!\!\!\!\!\!\!\!\!\!\!\!\!\!\!\!\!\!\!\!(\zeta\geq 2)\ ,\quad   r^2 = -(4.1\pm 0.5)\times 10^{-3}\ ,\
   z_V = (0.77\pm 0.10)^2\  ,\ z_A = (0.96\pm 0.21)^2  \label{results1}\\
  &&\!\!\!\!\!\!\!\!\!\!\!\!\!\!\!\!\!\!\!\!\!\!\!\!\!(\zeta \leq 2)\  ,\quad   r^2=-(3.9 \pm 0.1) \times 10^{-3}\ , \
    z_V=z_A^* = (0.81\pm 0.04)^2+ i\ (0.25\pm 0.25)\ \label{results2},
\end{eqnarray}
in units of $\mathrm{GeV}^6$ for $r^2$, and $\mathrm{GeV}^2$ for $z_{V,A}$. The two solutions in
Eqs. (\ref{results1},\ref{results2}) have been separated for illustrative purposes only. It is
clear that they are continuously connected through the boundary at $\zeta=2$, at which value the
two poles coincide and $z_V=z_A \simeq (0.85)^2$. The errors quoted are the result of scanning the
spread of values  in (\ref{realnature}) through the equations (\ref{set}).

With both set of values in Eqs, (\ref{results1}, \ref{results2}), one can get to a prediction for the
chiral and OPE coefficients by expansion in $Q^2$ and $1/Q^2$, respectively. These expansions of
the PA can be done entirely in the Euclidean region $Q^2>0$, away from the position of the poles
$z_{V,A}$, whether real or complex. Recalling the notation in Eq. (\ref{exp}), the above
$P^{0}_{2}(Q^2)$ produces the coefficients for these expansions collected in Table \ref{table2}.
The values for the OPE coefficients $C_{-4,-6,-8}$ in this table are compatible with those of Ref.
 \cite{Friot}, after multiplying by a factor of two in order to agree with the normalization used by
these authors. However, the spectrum  in our case is different because of the complex solution in
(\ref{results2}). As we saw in the previous section with a model, this again shows that Euclidean
properties of a given Green's function, such as the OPE and chiral expansions, or integrals over
$Q^2>0$ are safer to approximate with a rational approximant than Minkowskian quantities, such as
resonance masses and decay constants.

\begin{table}
\centering
\begin{tabular}{|c|c|c|c|c||c|c|c|c|}
  \hline
  $C_0$ & $C_2$ & $C_{4}$ & $C_6$ & $C_8$ & $C_{-4}$ & $C_{-6}$ & $C_{-8}$  \\
  \hline
  $-F_0^2$ & $-4\,L_{10}$ & $-43\pm 13$ & $81\pm 53$ & $-145\pm 120$ & $-4.1\pm 0.5$
& $6\pm 2 $& $-7\pm 6$   \\
  \hline
\end{tabular}
\caption{Values of the coefficients $C_{2k}$ in the high- and low-$Q^2$ expansions of $Q^2\
\Pi_{LR}(-Q^2)$ in Eq. (\ref{exp}) in units of $10^{-3}\ GeV^{2-2k}$. Recall that $C_{-2}=0$.
}\label{table2}
\end{table}

\subsection{Improving $C_{87}$}\label{C87}

\quad Once the rational approximant is known, upon reexpansion around $Q^2=0$, higher order unknown
coefficients of the chiral expansion may be predicted, as shown in the left side of Table \ref{table2}, \cite{MasjuanPades}, \cite{MasjuanProc}, \cite{MasjuanPerisC87}. If the rational approximant is a better description
of the original function than the partial sums of the chiral expansion, one may expect this prediction to be
reliable.
In a real QCD case, although the LEC $L_{10}$ is pretty well known \cite{GLsq,L10}, this is not so for the following coefficient $C_{87}$. It is therefore important to obtain a new determination of this LEC with its associated error.

Using $P^{0}_{2}(Q^2)$ we obtained an estimate of the $\mathcal{O}(Q^4)$ term in the expansion (\ref{largeNtaylor}), or $C_4$ in Table \ref{table2}, which translates into the value $C_{87}\,= \,(5.4 \pm 1.6 )\times 10^{-3}\ \mathrm{GeV^{-2}}$, $C_4=-8\,C_{87}$. In the present section, we would like to reassess this particular value with a more complete analysis. The same can be done for higher order coefficients, e.g. $\mathcal{O}(Q^6)$, but with less precision as higher the coefficient.

Improving the precision of our prediction means going further among the sequence of approximants. In order to be able to construct that sequence of rational approximants it is of course crucial to have enough
number of inputs. Since PAs are constructed from the coefficients of the Taylor expansion
(\ref{largeNtaylor}) one immediately faces an obvious difficulty. Since what one wishes is an estimate of
$C_{87}$, only the two coefficients $F_0$ and $L_{10}$ may be used. With these two coefficients as input,
the only PA vanishing at large $Q^2$ is $P^{0}_{1}$ and gives a prediction for $C_{87}=(7.1\pm0.5)\times 10^{-3}\ \mathrm{GeV^{-2}}$. Actually, another possibility is the $P^{1}_{0}$ but this case is just the Taylor expansion itself and, of course, will give the prediction $C_{87}=0$. Since we know from section~\ref{bestPA} that the best approximant has to falloff as $Q^{-4}$, and $P^{0}_{1}$ only falls off as $Q^{-2}$ which is too slow as
compared to Eq. (\ref{eq:OPE}), is  necessary to consider more general rational approximants
than the standard PAs.

The model confirms that one may estimate the unknown LECs with PTAs and PPAs where, in the first case, the physical masses were chosen in increasing order, i.e. $M_1< M_2<M_3 ...$ For
instance, with the PTA $\mathbb{T}^M_{M+2}(Q^2)$ we could see that one has a good prediction for the term of
$\mathcal{O}(Q^{2(M+1)})$ in the low-$Q^2$ expansion, which is the first one not used as input, with a
precision which improves as the order of the approximant, $M$, increases. Furthermore, the accuracy obtained
for the unknown coefficients of the Taylor expansion is very hierarchical: the accuracy obtained for the
term $\mathcal{O}(Q^{2(M+1)})$ is better than that for the term of $\mathcal{O}(Q^{2(M+2)})$, and that
better than for the term $\mathcal{O}(Q^{2(M+3)})$, with a quick deterioration for higher-order terms. The
case of PAs follows the same pattern. As to the description of the spectrum, we found that PAs also
reproduced the values for the residues and masses in a hierarchical way: while the first masses and residues
are well reproduced, the prediction quickly worsens so that the last pole and residue of the PA has no
resemblance whatsoever with its physical counterpart. The same is true for the residues of a PTA (since the
masses are fixed to be the physical ones by construction).

In working with PTs we have seen with the help of the model in section \ref{PTs} that using the residue of the heaviest resonance as input is not a good strategy (see, for example, Tables~\ref{Tab:PTs} and \ref{Tab:highPTs}). Consequently, in the following PTs we will not use the residue of the last (and even next to last) resonance as input.

Based on the above, one can envisage the following strategy for getting a sequence of estimates for the
$\mathcal{O}(Q^4)$ LEC $C_{87}$. Assuming that the vector and axial-vector meson masses stay approximately
the same in the large-$N_c$ and chiral limits, one can use their values extracted from the PDG
book \cite{PDG} to construct several PTAs. We think that this assumption is reasonable for both limits.
First, for the chiral limit, this is because the up and down quark masses are very small, \cite{PDG}.
Second, for the large-$N_c$ limit, there is a non negligible amount of phenomenological evidence in favor of
the $\rho$ meson being a $q\overline{q}$ state \cite{Jaffe}. Besides, the success in the spectroscopy of
the quenched lattice results for the lightest vector mesons is also suggestive that $1/N_c$ corrections may
not be very large \cite{Sharpe}\footnote{Be that as it may, whether the assumption is correct or not will
ultimately be judged by the final results obtained.}. Therefore, we will use for the masses
\begin{eqnarray}\label{masses}
  m_{\rho}= 0.7759 \pm 0.0005\ ,
  m_{\rho'}=1.459 \pm 0.011 &\!\!\!\!\!\!\!\!\!,&\!\!\!\!\!\!\!\!\!\!\!\  m_{\rho''}=1.720 \pm 0.020,
  m_{\rho'''}= 1.880 \pm 0.030 \nn \\
   m_{a_1}=1.230 \pm 0.040&,&\  m_{a'_1}= 1.647 \pm 0.022 ,
\end{eqnarray}
where all the numbers have been expressed in GeV.

For instance, with only $F_0^2$ and the masses of the $\rho$ and $a_1$, one can construct the PTA
$\mathbb{T}^0_{2}(Q^2)$ and predict the value for $L_{10}=(-4.32 \pm 0.02)\times 10^{-3}$, which is not bad
when compared, e.g., with Eq. (\ref{realnature}). The small error comes from the small error of $F_0^2$ since the impact of the error of the masses is rather negligible. The next term in the expansion gives the following value for
$C_{87}= (4.00 \pm 0.09)\times 10^{-3}\ \mathrm{GeV^{-2}}$ which is similar to that obtained in section \ref{sec:qcd}
with the Pad\'{e} $P^{0}_{2}$. However, since this value for $C_{87}$ comes from the second unknown term in the
expansion of $\mathbb{T}^0_{2}(Q^2)$ rather than the first (we predicted also $L_{10}$), it is quoted here only for illustrative purposes
and will not be included in our final estimate. Adding $L_{10}$
and the $\rho'$ mass to the previous set of inputs one can then construct $\mathbb{T}^1_{3}(Q^2)$, which
produces $C_{87}= (5.13 \pm 0.26 )\times 10^{-3}\ \mathrm{GeV^{-2}} $. The PTA $\mathbb{T}^{2}_{4}$ can be
constructed if one also uses the pion mass difference Eq. (\ref{eq:piem}) and $m_{a'_1}$, yielding in this case
$C_{87}= (5.24 \pm 0.33)\times 10^{-3}\ \mathrm{GeV^{-2}} $. We find the stability of these predictions
quite reassuring.

A comment on the quoted error estimates is in order. These quoted errors are the result of the propagation
of errors from the input via the montecarlo method \cite{Eadie}. As such, they do not reflect the intrinsic
systematic error due to the approximation itself which will be estimated, at the end, as the spread of
values obtained with the sequence of different approximants. On the other hand, the propagation of the error
from the input via the montecarlo method consists in the following. Taking each input in Eq. (\ref{realnature})
and Eq. (\ref{masses}), we have constructed a sample of data with a gaussian probability distribution yielding
as the average and standard deviation precisely the corresponding input value and its quoted error,
respectively. For each member of this sample, the rational approximant is then constructed and, upon
reexpansion, the LEC is obtained. The distribution of the different values for $C_{87}$ so obtained happens
to be also gaussian to a very good approximation. Therefore it will have an average value $X$ and a standard
deviation $Y$ which are then used to quote the result for $C_{87}$ as $X\pm Y$.

To be able to construct further rational approximants one needs an extra assumption. Although, as we have
emphasized above, the residues of the heaviest poles in a rational approximant do not come out anywhere
close to the corresponding physical decay constants, this is not true for the lightest ones. In particular,
section \ref{PTs} shows that the value of the residue for the first pole in a PTA could reproduce
the exact value in the model with very good precision \emph{if} the order of the PTA is high enough and,
more importantly, it is improving as the order of the PTA grows. Consequently, if we are willing to use the
decay constant $F_{\rho}$, and perhaps also the $F_{a_1}$, one can go for the construction of higher PTAs.
These two residues can be gotten from the decays $\rho\rightarrow e^+ e^-$ and $a_1\rightarrow \pi \gamma$,
respectively, and their values are \cite{Friot}
\begin{equation}\label{decay}
    F_{\rho}=0.156 \pm 0.001 \quad  F_{a_1}=0.123 \pm 0.024
\end{equation}
in GeV units.

\begin{table}
\centering
\begin{tabular}{|c||c|c|}
\hline
$\mathds{T}^n_m$ & inputs & $C_{87}$  \\
\hline \hline
$\mathds{T}^0_2$ &     $f_{0}$ ; $m_{\rho}$, $m_a$ & $4.00\pm0.09$ \\
\hline
$\mathds{T}^1_3$ &     $f_{0}$, $L_{10}$ ;  $m_{\rho}$, $m_a$, $m_{\rho^{'}}$  & $5.24\pm0.33$ \\
\hline \hline
$\mathds{T}^{2\, (a)}_4$ &    $f_{0}$, $L_{10}$, $\delta M_{\pi}$ ;  $m_{\rho}$, $m_a$, $m_{\rho^{'}}$, $m_{a^{'}}$  & $5.25\pm0.10$ \\
\hline
$\mathds{T}^{2\, (b)}_4$ &     $f_{0}$, $L_{10}$, $F_{\rho}$ ;  $m_{\rho}$, $m_a$, $m_{\rho^{'}}$, $m_{a^{'}}$ & $6.00\pm0.15$ \\
\hline \hline
$\mathds{T}^{3\, (a)}_5$ &     $f_{0}$, $L_{10}$, $F_{\rho}$, $\delta M_{\pi}$ ; $m_{\rho}$, $m_a$, $m_{\rho^{'}}$, $m_{a^{'}}$, $m_{\rho^{''}}$  &  $5.78\pm0.21$ \\
\hline
$\mathds{T}^{3\, (b)}_5$ &    $f_{0}$, $L_{10}$, $F_{\rho}$, $F_a$ ;  $m_{\rho}$, $m_a$, $m_{\rho^{'}}$, $m_{a^{'}}$, $m_{\rho^{''}}$ & $6.24\pm0.38$ \\
\hline \hline $\mathds{T}^{4}_6$ &     $f_{0}$, $L_{10}$, $F_{\rho}$, $F_a$, $\delta M_{\pi}$ ; $m_{\rho}$,
$m_a$, $m_{\rho^{'}}$,
$m_{a^{'}}$, $m_{\rho^{''}}$, $m_{\rho^{'''}}$   &$6.03\pm0.38$\\
\hline
\end{tabular}
\caption{Set of inputs used for the construction of the different Pad\'{e} Type Approximants in the
text. The last column is the summary of predictions for $C_{87}$ explained in the text.}\label{Tab:PadeTypes}
\end{table}

For instance, using $F_0^2, L_{10}, \delta M^2_{\pi}$ and $F_{\rho}$, as well as the five masses $m_{\rho},
m_{a_1}, m_{\rho'}, m_{a'_1}$ and $m_{\rho''}$, one can construct the PTA $\mathbb{T}^{3}_{5}$. Upon
expanding this approximant, one obtains the value $C_{87}=(5.78\pm 0.21)\times 10^{-3}\ \mathrm{GeV^{-2}} $.
Alternatively, one can also use $F_0^2, L_{10}, F_{\rho}$ and only the first four masses to construct a
$\mathbb{T}^{2}_{4}$ approximant, which is different from the other $\mathbb{T}^{2}_{4}$ considered above.
The value obtained for $C_{87}$, i.e.  $C_{87}=(6.00\pm 0.15)\times 10^{-3}\ \mathrm{GeV^{-2}} $, is
nevertheless very similar, which again brings confidence on the prediction.

In this way we have constructed a variety of rational approximants which we have listed on Table \ref{Tab:PadeTypes}, in increasing order of the degree in the denominator, together with the set of inputs used and the prediction for $C_{87}$. We have gone all the way up until the $\mathds{T}^{4}_6$, with the six masses listed on (\ref{masses}).

Figure \ref{C4graph} shows the prediction for the LEC $C_{87}$ from the corresponding rational approximant shown on the
abscissa, upon expansion around $Q^2=0$. We also included the previous result obtained with the PA $P^0_2$ Ref. \cite{MasjuanPades},\cite{MasjuanProc}, but with the montecarlo method for the treatment of errors. As one can see, the
stability of the result is quite striking. After averaging over all these points, we obtain as our final
result in the large-$N_c$ limit,
\begin{equation}\label{resultC87}
    C_{87}= (5.7 \pm 0.5)\times 10^{-3}\ \mathrm{GeV^{-2}}.
\end{equation}
The error in Eq. (\ref{resultC87}) is mainly dominated by the error on the input for $L_{10}$ in Eq.
(\ref{realnature}) and is rather insensitive to the errors on the other inputs. For instance, one could
increase the error on $F_0$ to $5$ MeV in Eq. (\ref{realnature}), or the error on $m_{\rho}$ to $50$ MeV in
Eq. (\ref{masses}), or the error on $\delta M_{\pi}$ to $0.5$ MeV  in Eq. (\ref{realnature}), without falling out of
the error band given in Eq. (\ref{resultC87}).

For comparison, we also show in Fig. \ref{C4graph} the result of several other estimates for this LEC. Reference
\cite{ABT} (shown as `A') uses the residues in Eq. (\ref{decay}) and the $\rho$ and $a_1$ physical masses to
construct, in effect, what we could call the PTA $\mathbb{T}^{2}_{2}$ to $Q^2 \Pi_{V-A}$. The difference
between this result and ours stems from the fact that this rational approximant falls off like a constant at
large $Q^2$, unlike Eq. (\ref{eq:OPE}). Also, as we have already emphasized, the use of the physical
decay constant $F_{a_{1}}$ (\ref{decay}) in a rational approximant which has the $a_1$ as the heaviest pole
is a potential source of error.

Reference \cite{theworksKnechtNyffelerp6} also obtains an estimate for $C_{87}$ (shown as `B') based on the construction of a
rational approximant which effectively coincides with the PTA $\mathbb{T}^{0}_{2}$ but using the physical
value of $F_{\pi}=92.4$ MeV \cite{PDG} instead of the value of $F_0$ in Eq. (\ref{realnature}). Had they
used $F_0$, the result would have been lower, and would have agreed with the value we mentioned in the
paragraph right after Eq. (\ref{masses}). Therefore, our comments on the Pad\'{e} Type $\mathbb{T}^{0}_{2}$
found in that paragraph also apply to this determination in \cite{theworksKnechtNyffelerp6}.

One can also get still another estimate for $C_{87}$ from the PTA $\mathbb{T}^{0}_{2}$ in \cite{theworksKnechtNyffelerp6}
by assuming that the $a_1$ mass in the large-$N_c$ limit is not approximated by the physical value in Eq.
(\ref{masses}), but by a value which comes from the radiative pion decay saturated with the $\rho$ and the
$a_1$. This value turns out to be $m_{a_{1}}\sim 998$ MeV \cite{MateuPortoles}. This lower number for the $a_1$
mass is the reason for a higher value for $C_{87}$ than that obtained in \cite{theworksKnechtNyffelerp6}, and is shown as `C'
in Fig. \ref{C4graph}. However, there is no compelling reason to associate this different mass of the $a_1$ with the
large-$N_c$ limit as we have discussed in sec.\ref{PPAs}. In fact, our results show how similar values for $C_{87}$ can be obtained with the
physical masses of the mesons used for the poles. Moreover, one of the advantages of our method is that one
can get a rough idea about the systematic error involved by looking at the dispersion of the values
obtained.

Recently, two more estimates for $C_{87}$ appear in the literature. The first, Ref. \cite{JuanjoC87}, is based on Resonance Chiral Theory \cite{swissEGPRressonance,swissDRV} and obtains a estimate $C_{87}=(3.9\pm1.4)\cdot10^{-3}$ by studying the Next-to-leading order term in the large-$N_c$ limit of the function $Q^2\Pi_{LR}(Q^2)$. The second one, Ref.\cite{GonzalezAlonso}, based on $\tau$-decay analysis, gives $C_{87}=(4.89\pm0.19)\cdot10^{-3}$ trough a direct fit to the experimental data using a sum rule approach.

\begin{figure}
  \includegraphics[width=5in]{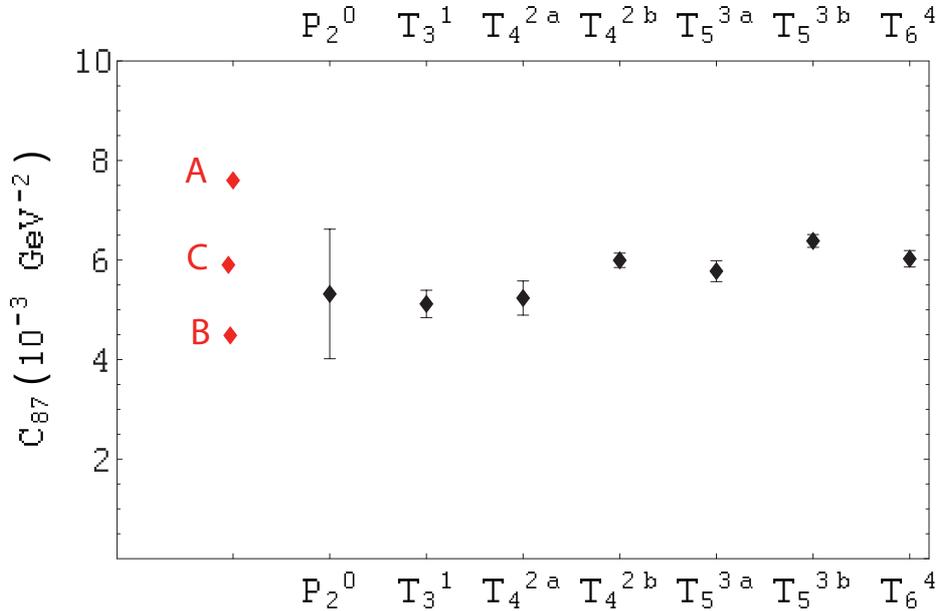}\\
  \caption{Prediction for $C_{87}$ in the large-$N_c$ limit from  the PA $P^0_2$ in Refs. \cite{MasjuanPades}, \cite{MasjuanPerisC87}, and the different PTAs
  discussed in the text and appearing in Table 1.
   For comparison we also show the estimate from Refs. \cite{ABT,theworksKnechtNyffelerp6,MateuPortoles},
   which we label `A', `B' and `C' (resp.).}\label{C4graph}
\end{figure}

\begin{figure}
  \includegraphics[width=5in]{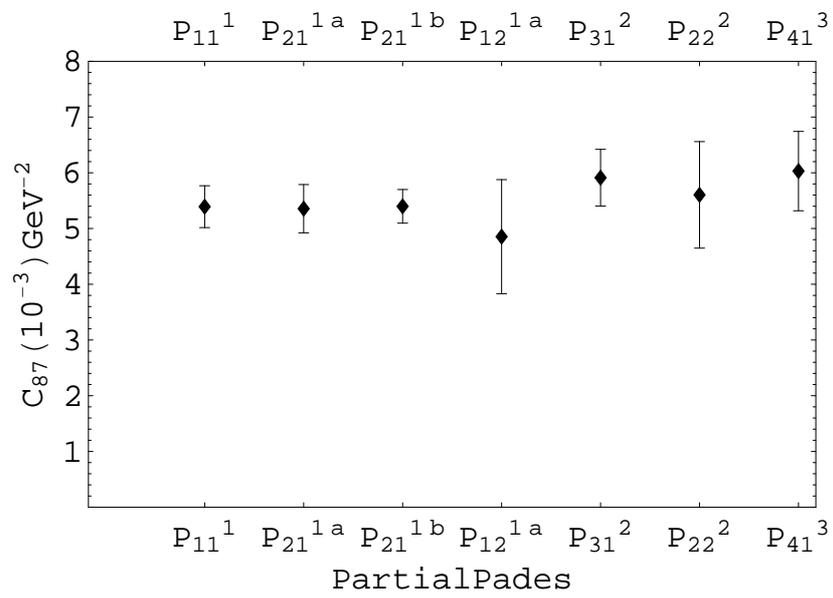}\\
  \caption{Prediction for $C_{87}$ in the large-$N_c$ limit from the different PPAs
  discussed in the text.}\label{C4graphPP}
\end{figure}

Of course, in the large-$N_c$ limit $C_{87}$ does not run with scale whereas in the world at $N_c=3$ it
does. This is an additional source of systematic error in the result (\ref{resultC87}). However,
phenomenological evidence as well as theoretical prejudice \cite{swiss2EGLPR} suggests that a reasonable guess for
this systematic error may be obtained by varying the scale in $C_{87}(\mu)$ between the range $0.5\
\mathrm{GeV}\lesssim \mu \lesssim 1\ \mathrm{GeV}$ (compare with $L_{10}$ in (\ref{realnature})). Using the
running obtained in Ref. \cite{run}, this error turns out to be $\sim 30$ per cent, right in the ballpark
expected for a typical $1/N_c$ effect. This systematic error should be added to our large-$N_c$ result in
Eq. (\ref{resultC87}) in order to obtain an estimate for $C_{87}(\mu \sim 0.7)$ in the real world. In this
case, all the different results in Fig. \ref{C4graph} can be encompassed by this error.

We would like to finish this section by recalling that PAs and PTAs are, in a way, two extreme versions of a rational
approximant. While in the latter all poles are fixed at the physical masses, in the former the poles are
left free, and they are obtained by demanding that the expansion around $Q^2=0$ reproduces that of the
original function to the highest possible order. Besides these two rational approximants, there are also the
Partial Pad\'{e} Approximants\cite{MasjuanPades,Canaris} (introduced in section \ref{PPAs}) which, from a certain point of view, lie half way between PAs and PTAs. As the poles of its denominator are only partially preassigned, there is no reason why, in general, the poles of a Partial Pad\'{e} should come out to be
purely real\footnote{Although, when complex, they always come in complex conjugate pairs. This just means
that, in general, the poles of a rational approximant are not necessarily physical.}, unlike those of a PTA,
which are of course real by construction. We have constructed seven of these Partial Pad\'{e}s, see Figure \ref{C4graphPP}, with a polynomial in the denominator up to fifth order in $Q^2$. In some of the cases the poles were actually complex, as it was also the case of the PA $P^0_2$, Eqs. (\ref{P02},\ref{P02prime}). However, the results obtained for $C_{87}$ are almost identical to those in Fig. \ref{C4graph}, although with errors which are somewhat larger. This feature reinforces the stability of the result shown in Fig. \ref{C4graph}, and gives us reassurance about the reliability of our result. Finally, we would like to mention that predicting $\mathcal{O}(p^8)$ LECs may also be another straightforward application of this method.

\chapter{Pad\'{e} Theory applied to Meromorphic functions of Stieltjes type}\label{capitol4}

\section{Introduction}

\quad In the present chapter we will steer things in a different direction. Up to now, we have focused our attention on meromorphic (but not Stieltjes) functions, but as we have already said in the introduction Pad\'{e} Approximants can be successfully applied also to a Stieltjes functions. In the following chapter, we will present two different examples of that kind of application. The first one, related with the Linear Sigma Model and the prediction of a resonance mass (all the details can be found in \cite{MasJJVpades}, \cite{Virtopades}, \cite{CilleroPadepoles}), and the second one related to the vacuum polarization function of a heavy quark (see \cite{HeavyquarkPades} for further details).

One of the mean advantages of PAs with regard to Taylor expansion is its range of applicability (see the examples of sec. \ref{sec:PadeTheory} and the underlying motivation of chapter $3$). In chapters $2$ we saw that PA can deal with meromorphic functions when trying to predict LECs. What happens, however, when instead of a set of poles, your function has a brunch cut? When the function is Stieltjes Eq. (\ref{stieltjesint}), i.e., its spectral function is positive defined (see de discussion after Eq. (\ref{stieltjesint})), the convergence of the PAs are guaranteed in a region out of the branch cut. When its spectral function is not positive defined (either meromorphic), we do not know of any mathematical result assuring the convergence of the PA sequence. Nevertheless, in few of these cases, PAs are surprisingly useful even when the convergence of PA are not properly guaranteed for any theorem. There is also a particular function that its analyticity is defined by a cut and a finite set of poles, the so-called meromorphic functions of Stlietjes type, for which also convergence theorems are known and can be applied \cite{Canaris,Lagomasino1,Gonchar}. In the first section of this chapter we will go into these kind of functions and in section \ref{heavypades} we will concentrate our efforts on an example of a  Stieltjes function.

\section{The Linear Sigma Model case}

\quad The physical motivation for these studies is based on the need of further ingredients when your Effective Field Theory (EFT) is unable to describe the resonance region. One example of that situation occurs in Chiral Perturbation Theory. As an EFT is a very useful tool for the description of low-energy physics~\cite{GeorgiEFT,ManoharEFT,PichEFT} but its range of validity is just a few hundred MeV, breaking down as one approaches new states not included in the EFT.


Several works, then, have tried to extend the range of validity of the EFT by means of the unitarization of the low-energy amplitude. Unitarization methods have been applied extensively to Quantum Chromodynamics and $\chi$PT~\cite{unitariza}, where the issue of the scalar resonances is particularly interesting. However, their range of applicability is wider; for example they have also been applied to $WW$--scattering~\cite{BSM1,BSM2}. Among the most usual ones, Pad\'e Approximants were already used before the nineties ~\cite{pipi-Pade,remarks-unitariza} and more recently one of its variants, the Inverse Amplitude Method (IAM)~\cite{IAM1,IAM2,IAM3,IAM4,IAM-op6}, although some remarks and criticisms on the reliability of these and other unitarization methods have been raised~\cite{remarks-unitariza,Leutwyler-criticism,IAM-P12-critics}. In the present section we study up to what point one can rely on them to describe the resonant properties of the theory. By means of a couple of models (the Linear Sigma Model and a vector resonance model),  we show that these unitarization procedures may lead to improper determinations of the resonance pole position (masses and
widths)~\cite{IAM-poles1,IAM-poles2}.


The starting point of our analysis is, therefore, a model where the properties of the resonances are known. Then we derive and unitarize the
corresponding low-energy amplitude. The predictions for the masses and widths obtained from the PA sequence $P^{N+J}_N$, ($J\geq-1$)\footnote{While one can use $J\geq-1$ an obtain similar results, we choose the diagonal case to exemplify the procedure.} quickly converge for Stieltjes functions, as $N$ is increased. We show that also for a meromorphic function of Stieltjes type (such as the first line of Eq.~(\ref{eq.LSM-PW}) the convergence is also obtained thanks to a theorem of convergence \cite{Canaris,Lagomasino,Lagomasino1,Gonchar}.

We also show that the same convergence patterns does not occur for any PA sequence, i.e., the restriction $J\geq -1$ should be fulfilled. Actually, when applied to a Stieltjes function it is known that all the sequences $P^{N+J}_N$, ($J\geq -1$), have guaranteed convergence (see sec.\ref{convth}). For the case $J<-1$, your sequence is not anymore protected. In particular, could happen your PAs have undesirable poles. Sometimes these poles will be complex-conjugate poles in the form of defects or will tend to infinity when increasing the order of the PA. Eventually, they could lie on the real axis of the complex plane where the function has no branch cut. One of this cases is when $J=1-N$, the sequence $P^1_N$(referred to as IAM in some works~\cite{remarks-unitariza,IAM1,IAM2,IAM3,IAM4,IAM-op6,IAM-P12-critics}). In the present section we will study a Stieljes function and also a meromorphic function of Stieltjes-type with both $P^N_N$ and $P^1_N$ sequences. The results are compared and found to be quite different from those of the original model. For simplicity the chiral limit will be assumed all along the section, but this will not alter the main results.


\section{One-loop Linear Sigma Model and its unitarization} \label{LSM}

\subsection{Unitarization of the $\chi$PT amplitude} \label{LSM3}

\quad To proceed we will first compute the $P^1_N$ sequence. To show that certainly the Inverse Amplitude Method ($IAM$) is a $P^1_N$ sequence we will first define $IAM$.

\quad The IAM provides an amplitude that is unitary not only at the perturbative
level but exactly. Considering the partial waves projection

\begin{equation}
t^I_J(s) \,\, =\,\, \frac{1}{64\pi} \int_{-1}^1 d\cos{\theta}
\, \, P_J(\cos{\theta})\,\,  T(s,t,u)^{\mbox{\rm I}}\, ,
\label{eq.PW}
\end{equation}

where $T(s,t,u)^I$ are the isospin amplitudes and $\theta$ the scattering angle in the $\pi \pi$ center-of-mass rest frame. $P_J$ are the Legendre polynomial. In the elastic limit one has for $s>0$

\begin{equation}
\mbox{\rm Im}\,  t(s)  \, =\, |t(s)|^2  \, ,
\end{equation}

where the indices $IJ$ are assumed, i.e. $t=t^I_J$.

This relation can be reexpressed as a relation for the inverse amplitude:
\begin{equation}
\mbox{\rm Im} \, t(s)^{-1}  \, =\, - 1  \, .
\end{equation}
Thus, the imaginary part of $t(s)^{-1}$ becomes completely determined
and one only needs to specify the real part Re$\,t^{-1}$. The IAM relies then on
a low-energy matching to $\chi$ PT (since IAM has been applied in the literature to that EFT) with

\begin{equation}
t_{\mbox{\rm $\chi$ PT}}^{-1}= t_{(2)}^{-1}\left[  1 - t_{(4)}/t_{(2)}+...\right]\, ,
\end{equation}

in order to fix the unknown part of the amplitude. For the first partial waves $t^I_J(s)$,
with $IJ=00,11,20$, one finds the following $\cO(p^2)$ amplitudes,

\begin{eqnarray}
t_0^0(s)_{(2)} &=& \frac{s}{16\pi F^2} \, , \nn \\[2mm]
t_1^1(s)_{(2)} &=& \frac{s}{96\pi F^2}\, , \nn \\[2mm]
t_0^2(s)_{(2)} &=& -\frac{s}{32\pi F^2}\, ,
\label{eq.t2}
\end{eqnarray}

and at $\cO(p^4)$

\begin{eqnarray}
t_0^0(s)_{(4)}\, = \, t_0^0(s)_{(2)} \, \, \times\,\, \frac{11 s}{6 M_\sigma^{  2}} \left[  1  - \frac{g}{264\pi^2} \left(
18\ln\frac{-s}{M_\sigma^{  2}} + 7 \ln\frac{s}{M_\sigma^{  2}} +\frac{193}{3}\right) +\cO(g^2) \right] \, , \nonumber\\
t_1^1(s)_{(4)} \, =\, t_1^1(s)_{(2)} \,\, \times\,\, \left(\frac{-s}{M_\sigma^{2}}\right)\,\, \left[1  + \frac{g}{48\pi^2} \left(
\ln\frac{-s}{M_\sigma^{  2}} - \ln\frac{s}{M_\sigma^{ 2}} -\frac{26}{3}\right) +\cO(g^2) \right] \, , \nonumber \\[2mm]
t_0^2(s)_{(4)} = t_0^2(s)_{(2)} \,\,\times\,\, \left( \frac{- 2 s}{3 M_\sigma^{  2} } \right)\,\, \left[  1  - \frac{g}{24\pi^2} \left(
\frac{9}{4} \ln\frac{-s}{M_\sigma^{  2}} + \frac{11}{4} \ln\frac{s}{M_\sigma^{  2}} +\frac{163}{24}\right)+\cO(g^2) \right] \, ,
\label{eq.t4}
\end{eqnarray}
where we have used the relation $2 g F^2 =M_\sigma^2$~\cite{GLOneLoop}.

Thus, at $\cO(p^4)$, one has the unitarized amplitude,
\begin{equation}
t_{{\mbox{\rm IAM}}}\, \, =\, \, \frac{t_{(2)}}{1\, -\, \frac{t_{(4)}}{t_{(2)}}}\, .
\end{equation}
This expression is sometimes also known as a $P_1^1$ Pad\'e
Approximant  of the partial-wave amplitude.
The IAM has been also
extended up to $\cO(p^6)$ by means of what is sometimes named as a
$P^1_2$ approximant~\cite{IAM-op6,IAM-P12-critics}:
$$
t_{{\mbox{\rm IAM}}}\, \, =\, \, \frac{t_{(2)}}{1\, -\, \frac{t_{(4)}}{t_{(2)}}
\, - \,  \frac{t_{(6)}}{t_{(2)}}  + \left(\frac{t_{(4)}}{t_{(2)}}\right)^2   }\, .
$$

and the $P^1_3$ Pad\'e  at $\cO(p^8)$,
$$
t_{{\mbox{\rm IAM}}}\, \, =\, \, \frac{t_{(2)}}{1\, -\, \frac{t_{(4)}}{t_{(2)}}
\, - \,  \frac{t_{(6)}}{t_{(2)}}  + \left(\frac{t_{(4)}}{t_{(2)}}\right)^2
\, - \,  \frac{t_{(8)}}{t_{(2)}}  + \frac{ 2 t_{(6)} t_{(4)}}{t_{(2)}^2}
- \left(\frac{t_{(4)}}{t_{(2)}}\right)^3
}\, .
$$

However, we want to remark that  $t(s)_{{\mbox {\rm IAM}}}$ is not a PA in the variable $s$:
It is not a rational approximant since it also contains the logarithms from the
pion loops, as one can see in the specific expressions of the corresponding amplitudes Eq.(\ref{eq.t4})\footnote{See Ref. \cite{MasJJVpades} for farther details about the calculations of these amplitudes.}.
Thus, strictly speaking no theoretical argument ensures the recovery of
the physical amplitude.
Only in the tree-level limit $t(s)_{{\mbox {\rm IAM}}}$  becomes a PA.
In any case, we will see that both the whole and the tree-level IAM amplitudes
are unable to reproduce the original partial waves in the resonance region.

To proceed given the $\cO(p^2)$ and $\cO(p^4)$ amplitudes, $t_{(2)}$ and $t_{(4)}$, Eqs. (\ref{eq.t2}) and (\ref{eq.t4}), is possible to extract the poles of the corresponding $t(s)_{{\mbox{\rm IAM}}}$ for the LSM, satisfying $1= t(s)_{(4)}/t(s)_{(2)}$ at $s=s_p$. For the firsts partial waves $t^I_J(s)$, one finds (see Ref. \cite{MasJJVpades}):


{\bf IJ=00}
\begin{eqnarray}
s_p&=&   \frac{ 6}{11}  M_\sigma^{  2} \, \left[  1  +  \frac{g}{264\pi^2}
\left( \frac{193}{3} + 25 \ln\frac{6}{11} - 18 i \pi \right)
+\cO(g^2) \right] \, \, ,
\label{eq.LSM00}
\end{eqnarray}
{\bf IJ=11}
\begin{eqnarray}
s_p &=&  - M_\sigma^{  2} \,\,
\left[1  +  \frac{g}{48\pi^2}
\left(   \frac{26}{3} + i\pi \right)
+\cO(g^2) \right] \, ,
\label{eq.LSM11}
\end{eqnarray}
{\bf IJ=20 }
\begin{eqnarray}
s_p &=&    -\frac{3}{2} M_\sigma^{  2} \,\,
\left[  1  + \frac{g}{24\pi^2}
\left( \frac{163}{24}+5 \ln\frac{3}{2}+ \frac{11 i \pi}{4}\right)+\cO(g^2) \right]
\, .
\label{eq.LSM20}
\end{eqnarray}
These are the poles that appear in the unphysical Riemann sheet as one approaches from upper half of the first Riemann sheet. There is also a conjugate pole at $s_p^*$ if one approaches the real $s$--axis from below.

The first thing to be noticed is that poles appear in the $IJ=11$ and $20$
channels even for small values of $g $, contrary to what one expects in
the LSM, where no meson  with these quantum numbers exists. Furthermore,
these ``states'' are not resonances, as they are located on the left-hand
side of the complex $s$--plane, out of the physical Riemann sheet,
and carrying a negative squared mass.

As for the  $IJ=00$ channel, one finds a resonance with pole mass and width,
\begin{eqnarray}
\frac{M_p^2}{M_\sigma^2}&=&  \frac{6}{11} \, \, \left[  1  +  \frac{g}{16\pi^2} \left( \frac{50}{33} \ln\frac{6}{11} +\frac{386}{99}\right)
+\cO(g^2) \right] \, \, , \nn \\[2mm]
\frac{M_p \Gamma_p}{M_\sigma^2}  &=& \frac{24}{121}\,\cdot \,  \frac{ 3 g}{16\pi}\,\, +\, \cO(g^2)  \, .
\end{eqnarray}

These expressions have to be compared with the original ones in the LSM, which are given in Ref. \cite{MasJJVpades}:

\begin{eqnarray}
\frac{M_p^2}{M_\sigma^{  2} } &=& 1 \, +   \frac{3 g}{16\pi^2}\, \left( -\frac{13}{3} +  \pi\sqrt{3}  \right)+ \cO(g^2)\, \nonumber \\[2mm]
\frac{M_p \Gamma_p}{M_\sigma^{  2}} &=&  \frac{3 g }{16\pi} \,+\, \cO(g^2) \, .
\label{eq.LSM-pole}
\end{eqnarray}

The IAM predictions for $M_p^2$ and $M_p \Gamma_p$ result,
respectively, 40\% and 80\% smaller than the original
ones in the LSM, shown in Eq.~(\ref{eq.LSM-pole}). The IAM poles remain badly located even in the weakly interacting limit, so this failure cannot be attributed to non-perturbative effects. In the
limit when $g \to 0$ and $M_\sigma $ is kept fixed one finds that the poles predicted in all the different channels fall down to the real $s$--axis. This points out the low reliability of this particular  method in order to recover
the hadronic properties of the theory from its effective
low-energy description.

We are left with just tree-level amplitudes and the expressions become greatly simplified. Due to the smoothness of this limit, it will be assumed in
the next analysis of higher order Pad\'e Approximants $P^M_N$ and in the study of the vector model in Section \ref{VM}.

\subsection{Higher order Pad\'{e} Approximants for tree-level amplitudes}
\label{padessec}

\quad In this subsection we consider higher order Pad\'{e} Approximants to
the partial wave amplitudes, with the hope that this will provide
some insight on the nature of the unitarization process discussed
above. We will see that the PA sequence associated with the IAM does
not converge properly, and that diagonal and paradiagonal sequences, sec.\ref{sec:PadeTheory} are much more
suitable for this purpose.

\subsubsection{Tree-level PAs in the LSM}

\quad In order to be able to handle the amplitude at higher orders,
we will consider the $\pi\pi$ scattering at tree-level. This is
equivalent to working in the limit $g \ll 1$ and keeping just the first
non-trivial contribution in the $g $ expansion. Thus, the
$\pi\pi$--scattering is determined in the LSM by the function
\begin{equation}
A(s,t,u)\,\, =
\,\, \frac{s}{F^2}\,\frac{M_\sigma^2}{M_\sigma^2\,-\,s}\, ,
\end{equation}
By means of the partial wave projection in Eq.~(\ref{eq.PW}), this provides
\begin{eqnarray}\label{eq.LSM-PW}
t_0^0(s)  &=& \frac{M_\sigma^2}{32 \pi F^2} \, \left[  - 5 + \frac{ 3M_\sigma^2}{M_\sigma^2-s} + \frac{ 2 M_\sigma^2}{s}
\ln\left(1+\frac{s}{M_\sigma^2}\right)\right] \, , \nn\\[2mm]
t_1^1(s) &=&\frac{M_\sigma^4}{32 s \pi F^2 } \, \left[  - 2 + \left(\frac{ 2 M_\sigma^2}{s}+1\right)\, \ln\left(1+\frac{s}{M_\sigma^2}\right)\right]
\, , \nn\\[2mm]
t_0^2(s) &=&- \frac{M_\sigma^2}{16 \pi F^2 } \,
\left[ 1 - \frac{  M_\sigma^2}{s}\, \ln\left(1+\frac{s}{M_\sigma^2}\right)\right] \, .
\end{eqnarray}
The $\ln{\left[1+s/M_\sigma^2\right]}$ logarithms come from the partial-wave projection of the tree-level exchanges of resonances in the crossed
channel. They have absolutely nothing to do with the logarithms of the $\chi$PT amplitudes in Eq.~(\ref{eq.LSM00}), which come from the $\pi\pi$ loops.

From a mathematical point of view both $t_1^1$ and $t_0^2$ are Stieltjes functions (see Eq.\ref{stieltjesint} and below) with a brunch cut in $-M_{\sigma}^2\geq s> -\infty$. In that situation, the convergence of a Pad\'{e} sequence $P_N^{N+J}$, $J\geq-1$ is guaranteed (see sec.\ref{convth}). However, $t_0^0$ is what is called a meromorphic function of Stieltjes type \cite{Canaris,Lagomasino,Lagomasino1,Gonchar}, that is a function that can be decomposed in two parts, a Stieltjes function and a rational function. This last piece must have a finite number of poles all located out of the branch cut of the Stieltjes part. In our case, $t^0_0$ has a simple pole at $s=M_{\sigma}^2$ and a brunch cut $-\infty < s \leq-M_{\sigma}^2$. Besides, Refs. \cite{Canaris,Lagomasino,Lagomasino1,Gonchar} demonstrate a theorem of convergence for Pad\'{e} Approximants for diagonal Pad\'{e} sequences \cite{Gonchar,Lagomasino} and also for $P_N^{N+J}$, $J\geq-1$, \cite{Lagomasino1,Canaris} that will we used in this section.

The relation between Stieltjes functions and its corresponding Pad\'{e} Approximant is already explained in chapter \ref{capitol1}. Further details will be studied in section \ref{heavypades}. The most interesting part here is the study of $t_0^0$ who has different analytical properties than its counterparts Eq.(\ref{eq.LSM-PW}). From now on we will basically focuss on this particular function.


At low energies the amplitude becomes
\begin{equation}
A(s,t,u)\,\, =\,\, \frac{s}{F^2}\,\left[ 1\, +\, \frac{ s}{M_\sigma^2} \, +\, \frac{ s^2}{M_\sigma^4} \, \, +\,\, ...\right]\, ,
\end{equation}
so the partial waves are given by:
\begin{eqnarray}\label{t00}
t_0^0(s)  &=& \frac{s}{16 \pi F^2} \, \left[ 1 + \frac{11 s}{6 M_\sigma^2} +\frac{15 s^2}{12 M_\sigma^4}\,\,+\,\,...\right] \, , \nn\\[2mm]
t_1^1(s) &=&\frac{s}{96\pi F^2} \, \left[ 1- \frac{s}{M_\sigma^2} +\frac{ 9 s^2}{10 M_\sigma^4} \,\,+\,\, ... \right] \, , \nn\\[2mm]
t_0^2(s) &=&- \frac{s}{32 \pi F^2}\, \left[ 1 - \frac{ 2 s}{ 3 M_\sigma^2} + \frac{s^2}{ 2 M_\sigma^4}\,\, +\,\,...\right] \, .
\end{eqnarray}

The comparison between the low-energy expansions and the whole result provides a first insight of the piece of information that is lost in the
unitarization procedure.  At high energies, the partial waves contain poles on the right-hand side of the $s$--plane, related to $s$--channel
resonance exchanges, and a left-hand cut, related to the crossed--channel resonance exchanges.
At low energies, both kinds of exchanges
contribute  equally to the low energy couplings, so the crossed resonance exchanges
shift the IAM poles from their  physical value.
Although $t$ and $u$ channels are not so relevant in the region close to the resonance pole, at low
energies they are  as important as the $s$--channel.

The simplest Pad\'e Approximant,  $P_1^1$, gives the prediction
\begin{eqnarray}
s_0^0 &=& \frac{6}{11} M_\sigma^2\, ,\nn\\[2mm]
s_1^1 &=& - M_\sigma^2\, ,\nn\\[2mm]
s_0^2 &=& -\frac{3}{2} M_\sigma^2 \, ,
\end{eqnarray}
which agrees with the one-loop calculation from Eqs.~(\ref{eq.LSM00})--(\ref{eq.LSM20}) if one remains at leading order in $g$.

\subsection{Higher order PAs in the LSM}

\quad Now we will proceed to study higher order PAs. In particular, we will employ and compare the $P^1_N$ and $P^N_N$ sequences for the study of the $\pi\pi$ partial wave scattering amplitudes. We will also comment
on PAs of the $P^{N+J}_N$, e.g  $P^{N-1}_N$ and $P^{N+1}_N$. In the next lines
we will focus our attention on the $IJ=00$ partial wave, but
analogous results are found for the other channels.

Former works pointed out that the PAs and other unitarizations fail to incorporate
the crossed channel resonance exchanges~\cite{PW,guo06}.
Nonetheless, we will see that as $N$ grows, the
poles of the sequence $P^N_N$ actually tend to mimic not only the
$s$--channel poles but also the left-hand cut contribution from
diagrams with resonances in the $t$ and $u$ channels.

Another advantage of the sequence $P^N_N$ is that, as already explained in sec. \ref{bestPA}, is the best PA sequence one can use when trying to predict global properties of the original function. Fig. \ref{detpolsLSM} shows the determination of the position of the $M_{\sigma}^2$ pole, its residue and the Taylor coefficient $a_{11}$ from several PAs, all constructed using only the first Taylor coefficients from Eq.(\ref{t00}). Exactly following sec. \ref{bestPA} we take the $t_0^0(s)\approx a_1 s^1 + a_2 s^2 + \cdots a_{10} s^{10}$ and we study all the possible PA constructed using only these coefficients. These lets to a set of nine PA: $P^9_1, P^8_2, P^7_3, P^6_4, P^5_5, P^4_6, P^3_7, P^2_8, P^1_9$. The best PA is the diagonal one, who gives the best prediction for the three parameters we are studying. Also, this diagonal PA gives an extra insight about the behavior of $t^0_0(s)$ for $s\to \infty$, Eq.(\ref{opes}). Of course, a rational function will never we able to predict a logarithm function but, the diagonal PA finds the compromise between the $1/s$ piece and the $\log (1/s)$ piece:

\begin{equation}\label{opes}
\frac{-3 -2 \log \left(\frac{1}{s}\right)}{s}-(1/s)^2+{\cal O}((1/s)^3)
\end{equation}


\begin{figure}
\centering
  \includegraphics[width=3in]{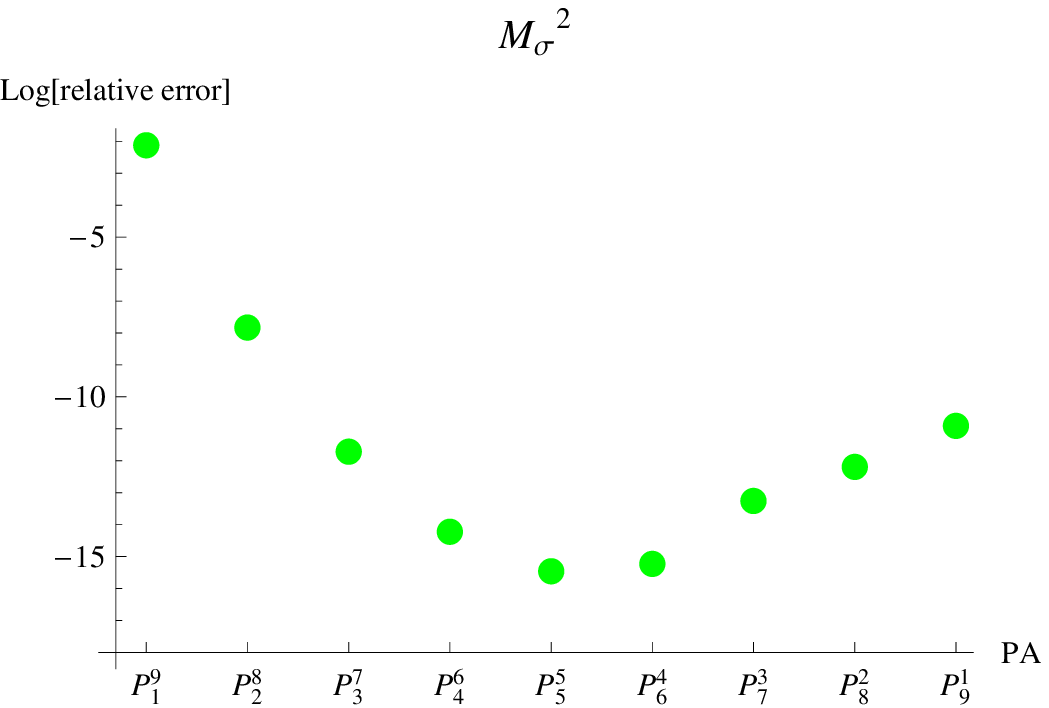}
  \includegraphics[width=3in]{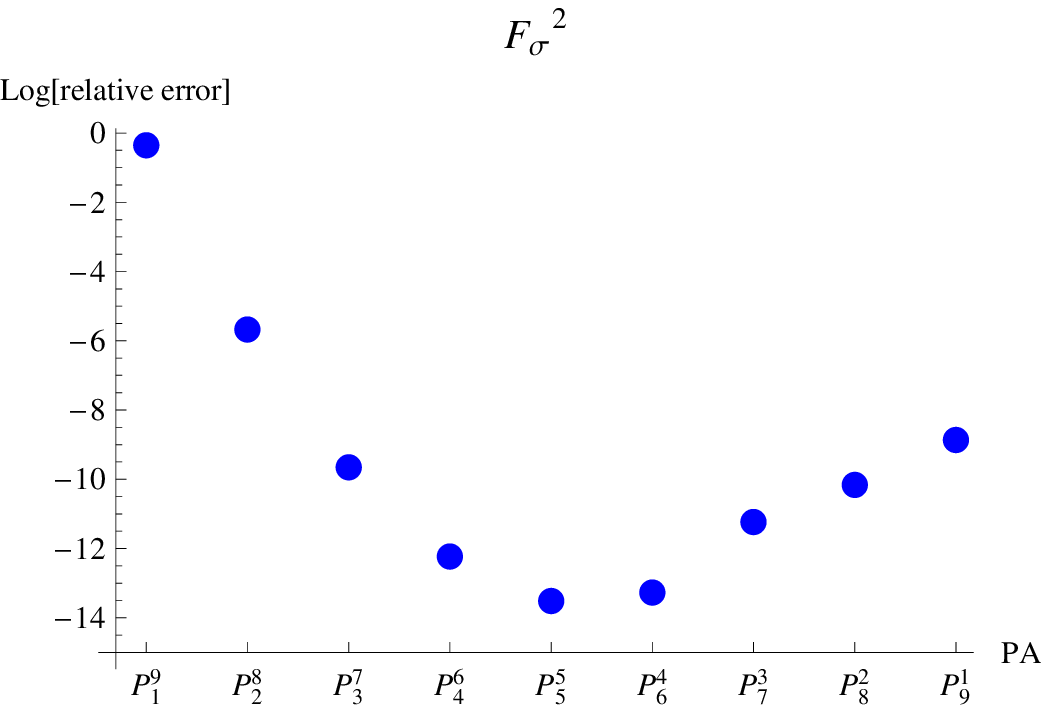}
  \includegraphics[width=3in]{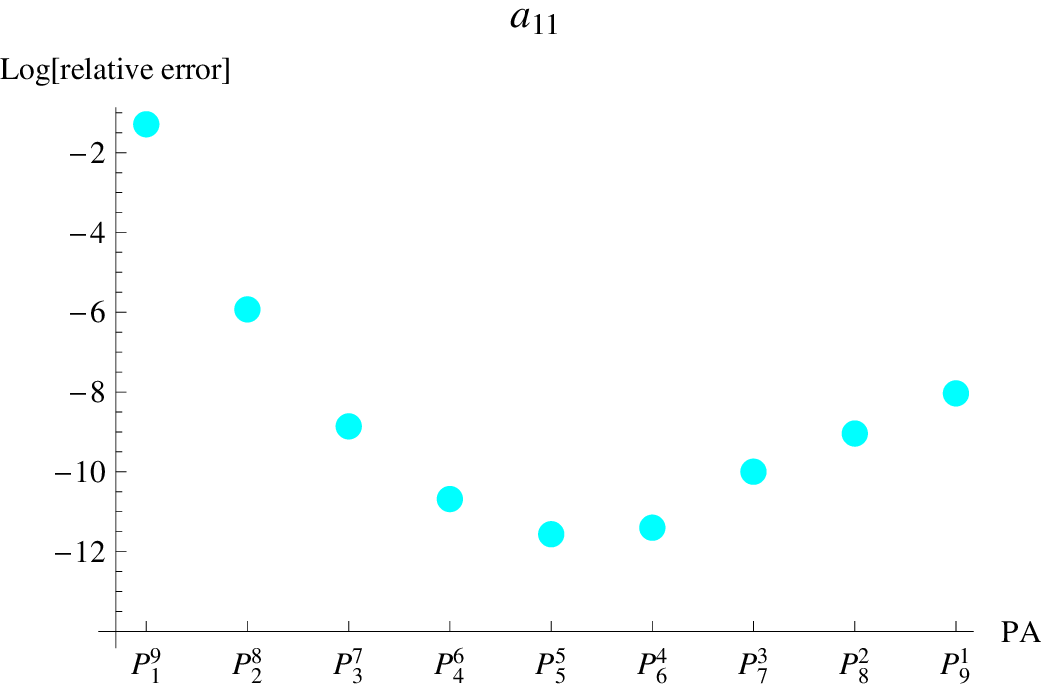}
  \caption{Determination of the position of the $M_{\sigma}^2$ pole (top), the first residue (center) and the Taylor coefficient $s_{22}$ (bottom) from several PAs, approximating the function $t^0_0(s)$, Eq. (\ref{t00}). See the text for details.}\label{detpolsLSM}
\end{figure}

The results shown in these plots reassess the conclusions extracted from sec.\ref{bestPA}.

Our results for the sequence $P^1_N$ are summarized in Fig.~\ref{LSM30}: No convergence is found with this
sequence.
In the case of $N$ odd, Fig.~\ref{LSM30}.a. shows that the $P^1_N$ pole
closest to $M_\sigma^2$ does not approach this value even for very
large $N$, always remaining a 30\% below. The analytical
structure of the original amplitude ($s$--channel sigma pole plus
left-hand cut) is never recovered since the $P^1_N$ PAs always set
the poles in the circular pattern  shown in
Fig.~\ref{LSM30}.b. This  suggests that
the use of further $P^1_N$ approximants to extend the IAM is not the optimal way to
proceed, even if  we had an accurate knowledge of the low-energy  expansion up to
very high orders as is now the case.

\begin{figure}[!t]
\begin{center}
  \includegraphics[width=8cm]{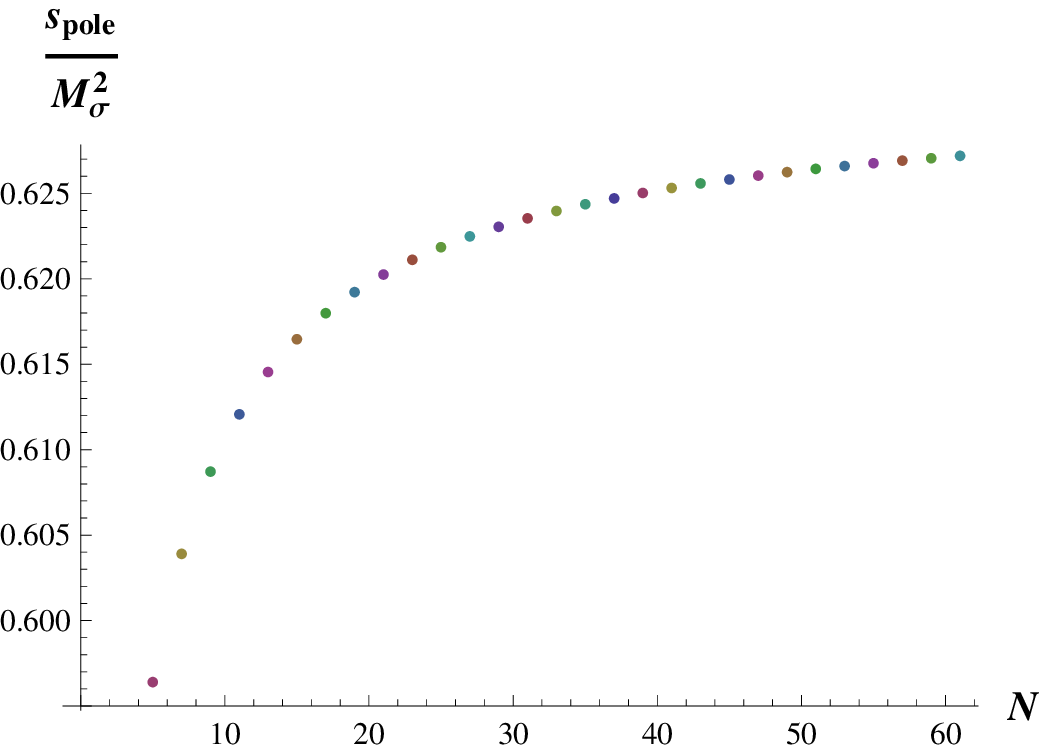}
  \hspace{2cm}
  \includegraphics[width=8cm]{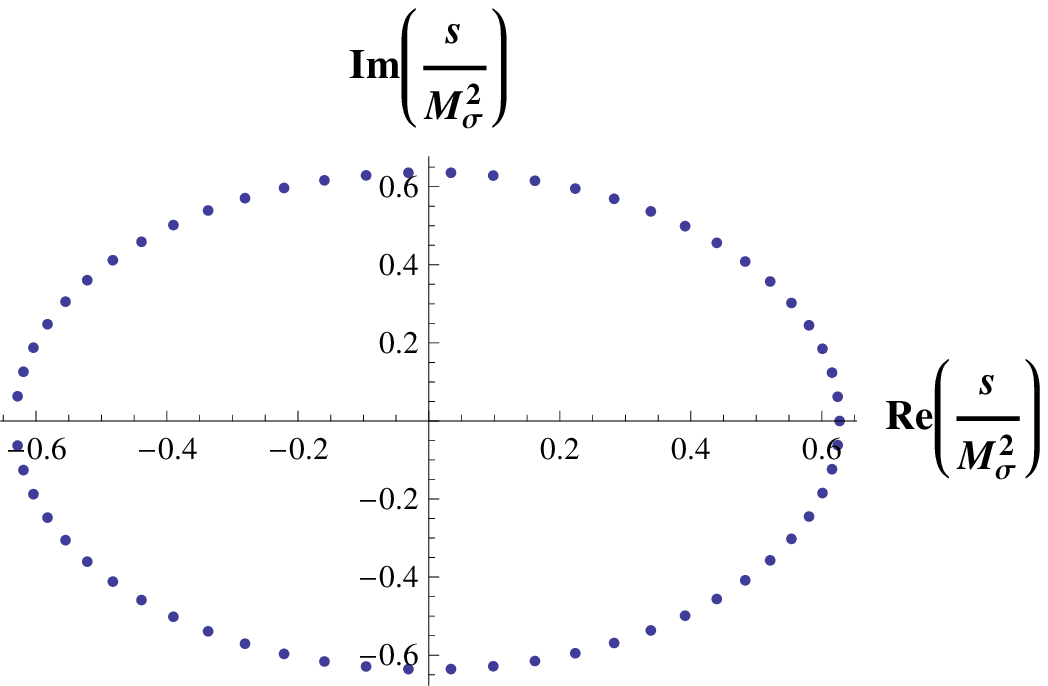}
  \caption{Top: Position of the nearest pole to $M_\sigma^2$
  for the first PAs of the form $P^1_N$ with $N$ odd
  (for even $N$ all the poles are complex).
  Bottom: Poles of the $P^1_{61}$ in the complex plane.}
  \label{LSM30}
\end{center}
\end{figure}

Alternatively,  the use of sequences such as
$P^{N+J}_N$ (e.g. $P^{N-1}_N, \, P^{N}_N,\,  P^{N+1}_N\ldots$)
seems to be a better strategy.
In the following we analyze the sequence $P^N_N$, as it ensures the
appropriate behavior at high energies\footnote{Imposing unitarity to our PA sequence is a constrains that lies on a physical motivation but as far as concern only with PAs, it is not a necessary assumption to be able to establish a converge sequence as explained in sec.\ref{sec:PadeTheory}.}, $|t(s)|<1$.
The $P^N_N$ pole closest to
$M_\sigma^2$ is shown in Fig.~\ref{PadeNN20}.a. One finds a quick
convergence of the sequence: $P_1^1$ reproduces the sigma pole a
$40\%$ off but $P^2_2$ disagrees by less than $1\%$, $P^3_3$ by less
than $0.1\%$, etc. Notice that already  $P_2^2$ provides a much
better description than $P^1_{61}$, although one includes far more
low-energy information in the latter. All this points out the
sizable discrepancy of the first element of the sequence ($P^1_1$)
with respect to the original amplitude. It also indicates that the $P^{1}_N$ PAs do
not produces a serious improvement. On the contrary, the $P^{N}_N$
sequence provide a far more efficient strategy with a quick
convergence.

\begin{figure}[!t]
\begin{center}
  \includegraphics[width=8cm]{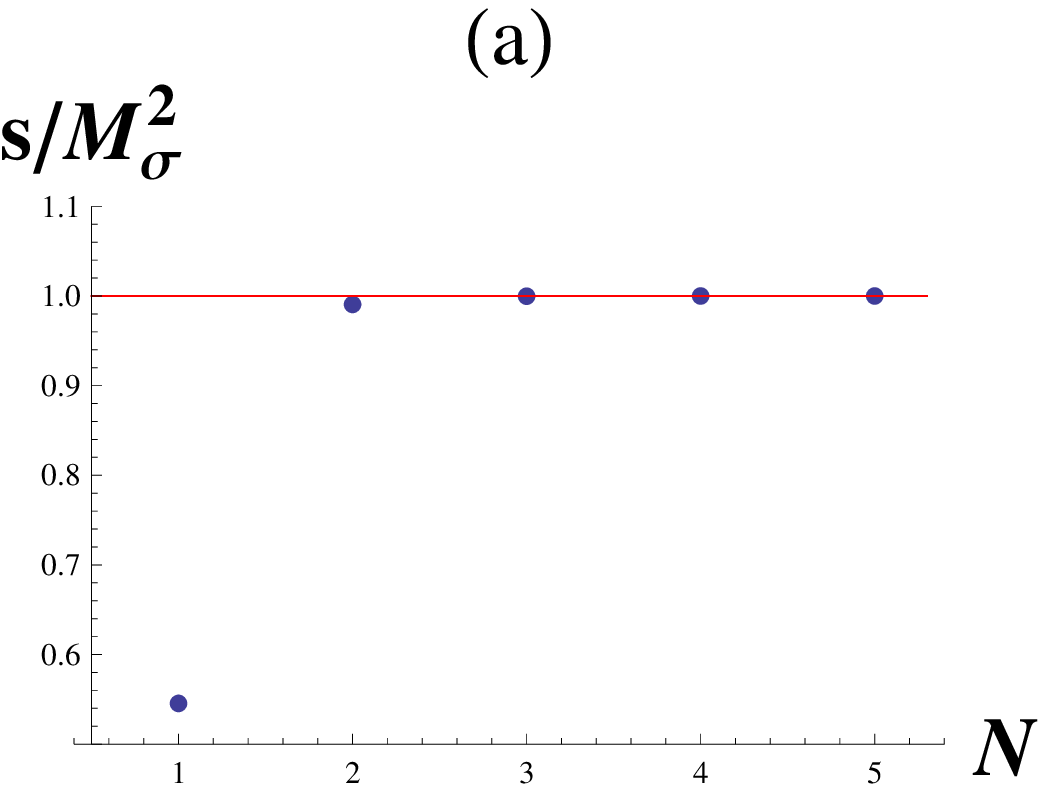}
  \hspace{1.5cm}
  \includegraphics[width=8cm]{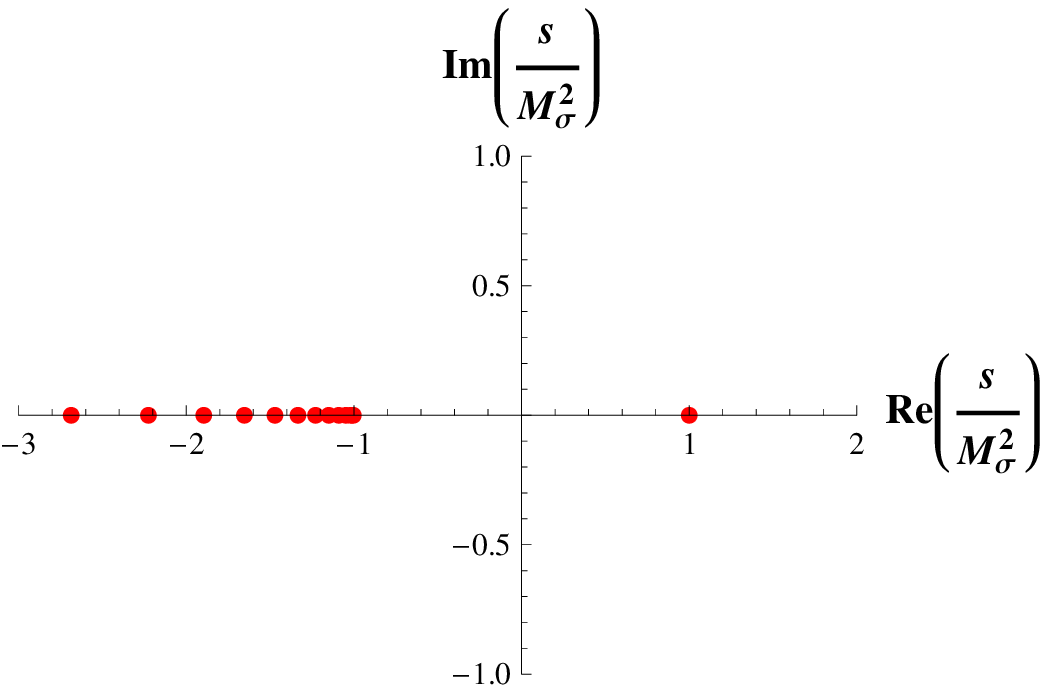}
  \caption{{\small
  Top: Location of the closest pole to $M_\sigma^2$
  for the first $P^N_N$ PAs. The red line showns the exact value $s/M_\sigma^2=1$.
  Bottom: Poles of $P^{20}_{20}$.
  }}
\label{PadeNN20}
\end{center}
\end{figure}

Likewise,  Fig.~\ref{LSM30}.b shows how the $P^{1}_N$ PAs  are unable to recover
the analytical structure of the original amplitude, whereas the $P^{N}_N$ sequence,
besides providing the isolated pole of the sigma, tends to reproduce the left-hand cut as $N$ increases, as expected since $t^0_0$ is a Stieltjes-type function. The poles of $P_{20}^{20}$ are
plotted in Fig.~\ref{PadeNN20}. Although a PA is a rational function without cuts,
these are mimicked by placing poles where the cuts should lie.
The $P_{20}^{20}$ has one isolated pole near $M_{\sigma}^2$ (with an accuracy of $10^{-30}$) and nineteen poles over the real axis at
$s_p<-M_{\sigma}^2$, i.e. on the left-hand cut of the original function. As $N$ is increased, the number of poles lying on the branch cut increases
too.

A remarkable feature found for the first $P_N^N$ approximants ($P^1_1$, $P^2_2$, $P^3_3$)
is that they obey exact unitarity, as it happened with the IAM sequence $P^{1}_N$.

In several situations the PAs set all poles over the left-hand cut position and one isolated pole that
approached $M_\sigma^2$ when $N\to\infty$. In the worst cases, in
addition to this we found extraneous poles that either moved away as
$N$ increased or they tend to be canceled by nearby zeros at
$N\to\infty$.

As an amusement inspired by the \textit{Montessus}'s theorem and the fact that there is only one ressonance $\sigma$, we have also probed the PA sequence $P^N_1$ which
has the same number of inputs as the $P^1_N$ has for a given $N$. Also has the advantage that to increase one order on the sequence we just need one extra parameter while in the $P^N_N$ case, we need two extra parameters.
In this new case we have found convergence in both LSM and the
resonance model presented in the following subsection but slower than
the $P_N^N$. For instance, the prediction for the $M_{\sigma}^2$ for
the first $P_1^N$ are
$\frac{s_p}{M_{\sigma}^2}=0.55,1.47,0.73,1.27,0.81...$ A criticism
that can be done to this sequence is its lack of unitarity, and the impossibility to mimic the brunch cut of the original function, in contrast to the other studied sequences.

\section{Vector Resonance Model}\label{VM}

\quad In order to broaden our analysis, we consider now a model with
just vector mesons~\cite{PW}. It could be derived either from the
gauged chiral model~\cite{swissDRV} for the couplings $3 g_\rho F^2=M_\rho^2$,  or
from resonance chiral theory~\cite{swissEGPRressonance} with only vectors and the relation
$3 G_V^2=F^2$. The $\pi\pi$--scattering is given in this
model by
\begin{equation}
A(s,t,u) \, = \,  \frac{M_\rho^2}{3 F^2}
\left[ \frac{s-u}{M_\rho^2 -t} + \frac{s-t}{M_\rho^2-u}  \right]\, .
\end{equation}

The study of the $IJ=11$ partial wave leads to the same conclusions found for the LSM.
It is at first sight remarkable that, on the contrary to the
previous case, one recovers $s_p=M_\rho^2$ from the first-order approximant $P_1^1$. However, the sequence  $P_N^1$ already worsens at $N=2$, where
the two complex-conjugate poles are located at $s_p=(0.71\pm 0.96 i) \, M_\rho^2$ on the physical Riemann sheet. On the other hand, $P_N^N$ exactly
recovers $s_p=M_\rho^2$ for any odd $N$. For $N$ even, the prediction from $P^2_2$ is a 30\% off, but one has again a quick convergence to
$s_p=M_\rho^2$ as $N$ increases: $P_4^4$ disagrees by less than 0.1\%, $P_6^6$ disagrees by less than $10^{-6}$, etc.

Furthermore, the  $P_N^1$ and $P_N^N$ sequences produce, respectively, the same analytical structure of poles found for the LSM. This is,  $P_N^1$ generates the circular analytical structure of poles of Fig.~\ref{LSM30}.b. and the sequence  $P_N^N$ places one pole at $s_p\simeq M_\rho^2$ and the remaining ones reproducing the left-hand cut in analogy to Fig.~\ref{PadeNN20}.b.

\section{The Vacuum Polarization of a Heavy Quark: Motivation}\label{heavypades}

\quad In this second example, the physical motivations is the study of the vacuum polarization function of two electromagnetic currents, in the framework of heavy quark physics. This study requires high order perturbative calculations which, because of the obvious need to keep a nonzero mass $m$ for the quark, become extremely difficult to perform. This is why, while the $\mathcal{O}(\alpha_s^0)$ and $\mathcal{O}(\alpha_s^1)$ contributions have been known for a long time \cite{Kallen}, state of the art calculations can only produce a result at $\mathcal{O}(\alpha_s^2)$ in the form of an expansion at low energies (i.e. $q^2=0$), at high energies (i.e.  $q^2\rightarrow\infty$) or at threshold (i.e. $q^2=4 m^2$), but not for the complete function, which is still out of calculational reach. In this circumstances, it would of course be very interesting to be able to reconstruct this function by some kind of interpolation between the former three expansions.

After the work in refs. \cite{Broadhurst1,Broadhurst2,Broadhurst3,ChetyrkinMethod,ChetyrkinH01}, it has become customary to attempt this reconstruction of the vacuum polarization function with the help of Pad\'{e} Approximants. Since these
approximants are ratios of two polynomials in the variable $q^2$, they are very suitable for the
matching onto the low-energy expansion. This is so because this low-$q^2$ expansion is truly an
expansion in powers of $q^2$, as a consequence of the finite energy threshold\footnote{In
perturbation theory, this is true so long as purely gluonic intermediate states are not considered.
Beyond perturbation theory, the threshold occurs at $4 m^2_{\pi}$, where $m_{\pi}$ is the pion
mass. } starting at $4m^2$. However, it is clear that they cannot fully recover the nonanalytic
terms which appear, e.g., in the form of logarithms  of $q^2$  in the expansion at high energies
(or at threshold, where there is also a squared root). Therefore, what is really done in
practice is to first subtract all these logarithmic pieces from the full function (impossible to
match exactly with a Pad\'{e} Approximant) with the help of a \emph{guess} function with the appropriate
threshold and high-energy behavior, and then apply Pad\'{e} Approximants to the remaining regular
expression. In fact, this is done after a conformal mapping from $z$ to $\omega$ whereby all the (cut)
complex plane is mapped into a circle of unit radius. In this way, the authors of Ref. \cite{HoangMateu}
were able to compute, e.g.,  the value of the physical constant  $K^{(2)}$ appearing in the $\mathcal{O}(\alpha_s^2)$
expansion of the vacuum polarization  at threshold (Eq.(\ref{threshold})), which has not yet been possible to obtain from a Feynman diagram calculation.

Although this result is very interesting, the construction is not unique.  As recognized in Ref.
\cite{HoangMateu}, some amount of educated guesswork is required in order to resolve the inherent
ambiguity in the procedure. For instance, a certain number of unphysical poles are encountered, and
some additional criteria have to be imposed in order to decide how to discard these poles. Since
the resulting ambiguity leads to a systematic error which needs to be quantified, this error is
then estimated by varying among several of the possible arbitrary choices in the construction.
Although all these choices are made judiciously and in a physically motivated manner, it is very
difficult to be confident of the error made in the result, which obviously has an impact on the
value extracted for the constant $K^{(2)}$.

In this section we would like to point out that, regarding the vacuum polarization function,
one can do away with all the above ambiguities when applying Pad\'{e} Theory.

Since the vacuum polarization function is a Stieltjes function (as we will see in the next subsection), the convergence of all the Pad\'{e} Approximants $P^{N+J}_N$, ($J\geq-1$) is guaranteed. The result of this theorem together with the fantastic amount of information obtained on the Taylor expansion around $q^2=0$, for which 30 terms are known \cite{Maier}, will allow us to predict a value for $K^{(2)}$ Eq.(\ref{threshold}). As it turns out, our result is very close to that of Ref. \cite{HoangMateu}, although slightly smaller.


\subsection{Pad\'{e} Approximants and the Vacuum Polarization function: Definitions}

\quad The vacuum polarization function $\Pi(q^2)$ through the correlator of two
electromagnetic currents $j^\mu(x)=\bar q(x)\gamma^\mu q(x)$, is defined as:
\begin{align}
\label{pidef}
\left(g_{\mu\nu}q^2-q_\mu q_\nu\right)\, \Pi(q^2)
\, = \, \,
- \,i
\int\mathrm{d}x\, e^{iqx}\left\langle \,0\left|T\, j_\mu(x)j_\nu(0)\right|0\,
\right\rangle
\,,
\end{align}
where $q^\mu$ is the external four-momentum. The optical theorem tells us that the  $e^+e^-$ cross section is proportional to the imaginary part of $\Pi(q^2)$. As a result, $\mbox{Im}\,\Pi(q^2)$ is a positive definite function, i.e.
\begin{equation}
\label{positive}
\mbox{Im}\,\Pi(q^2+i \varepsilon)\geq 0 \ ,
\end{equation}
a property which will become crucial  in what follows.

In perturbation theory $\Pi(q^2)$ may be decomposed to ${\mathcal O}(\alpha_s^2)$ as
\begin{align}
\label{Pi}
\Pi(q^{2}) \, = \,&\,
\Pi^{(0)}(q^{2})
\, + \,\left(\frac{\alpha_{s}}{\pi}\right)\,
\Pi^{(1)}(q^{2})+\left(\frac{\alpha_{s}}{\pi}\right)^{2}\,
\Pi^{(2)}(q^{2})
\, + {\mathcal O}(\alpha_s^3)\, .
\end{align}

For definiteness, $\alpha_s$  denotes the strong coupling constant in the $\overline{\mathrm{MS}}$ scheme at the scale $\mu=m_{pole}$, but this is not important for the discussion which follows. Equation (\ref{Pi}) will be understood in the on-shell normalization scheme  where a subtraction at zero momentum has been made in such a way as to guarantee that $\Pi(0)=0$. In that way, the vacuum polarization in Eq. (\ref{Pi}) satisfies a once subtracted dispersion relation (see App.\ref{secdr} for details), i.e.
\begin{equation}\label{disprel}
    \Pi(q^{2}) = q^2 \int_{0}^{\infty} \frac{dt}{t (t-q^2-i \varepsilon)}\ \frac{1}{\pi} \mbox{Im}\,\Pi(t+i \varepsilon)\ .
\end{equation}
Since all diagrams with intermediate gluon states are absent up to $\mathcal{O}(\alpha_s^2)$, the lower limit for the dispersive integral (\ref{disprel}) starts, in fact,  at a finite value given by the threshold for pair production, i.e. $4m^2$. This fact only carries over to higher orders in $\alpha_s$ provided these intermediate gluon states are neglected. From now on, we will restrict ourselves to the vacuum polarization in Eq. (\ref{Pi}) to $\mathcal{O}(\alpha_s^2)$, neglecting higher orders in $\alpha_s$.

In terms of the more convenient variable
\begin{equation}
\label{z}
z \, \equiv \, \frac{q^2}{4 m^2}\ ,
\end{equation}
one can rewrite Eq. (\ref{disprel}), after redefining $u=4 m^2/t$, as\footnote{We are simplifying the notation by replacing $ \Pi(4 m^2 z) \rightarrow  \Pi(z) $.}
\begin{equation}\label{disprelz}
    \Pi(z) = z \int_{0}^{1} \frac{d u}{1- u z-i \varepsilon}\ \frac{1}{\pi} \mbox{Im}\,\Pi\left(4 m^2 u^{-1}+i \varepsilon\right)\ .
\end{equation}
Recalling that a Stieltjes function is defined as \cite{Baker1,Baker2,Brezinski,Canaris,Bender}
\begin{equation}\label{Stieltjes}
    f(z)=\int_{0}^{1/R} \frac{d\phi(u)}{1-u z}
\end{equation}
where $\phi(u)$ is any \emph{nondecreasing} function, one sees that the identification

\begin{equation}\label{id}
    d\phi(u)=\frac{1}{\pi} \mbox{Im}\,\Pi\left(4 m^2 u^{-1}+i \varepsilon\right) \ du
\end{equation}
allows one to recognize that the integral in Eq. (\ref{disprelz}) defines the Stieltjes function $z^{-1}\Pi(z)$.
It is been common in the literature to work with the variable $\omega$ defined as $z=4\omega/(1+\omega)^2$ instead of $z$ \cite{Broadhurst1,Broadhurst2,Broadhurst3,ChetyrkinMethod,ChetyrkinH01, HoangMateu}. This change of variables allows a mapping to the cut $z$ plane into a unit circle in the $\omega$ plane. In particular, the three expansions at low energies (i.e. $z=0$), at high energies (i.e.  $z \to \infty$) and at threshold (i.e. $z=1$) translate into expansions at $\omega = 0$, $\omega =-1$ and $\omega = 1$ respectively. This conformal mapping from $\omega$ to $z$ allows the approximation to be valid also on the boundary, providing a smooth approximation of the absorptive part of the integral. We note, however, that this change of variables invalidates the Stieltjes property and then there is no theorem that we know of ensuring the desired convergence for the Pad\'{e} Approximants in the new conformal variable.

As one can see, the positivity property Eq. (\ref{positive}) is crucial for the identification (\ref{id}) to be possible. The connection between the imaginary part of the vacuum polarization and the cross section $\sigma(e^+e^-\rightarrow \mathrm{had})$ (a clearly positive definite quantity) assures that this  property is guaranteed to all orders of perturbation theory.

The representation of the function $f(z)$ in Eq. (\ref{Stieltjes}) clearly shows a cut in the $z$ complex plane on the positive real axis for $R\leq z <\infty$. For the physical function $\Pi(z)$, this of course corresponds to the physical cut in momentum for $4m^2\leq q^2< \infty$, i.e. the physical case corresponds to $R=1$ in Eqs. (\ref{Stieltjes},\ref{id}). Furthermore, just like the function $f(z)$ in Eq. (\ref{Stieltjes}) has  a power series expansion convergent in the disk $|z|<R$, so does the function $ \Pi(z)$ in Eq. (\ref{disprelz}) have a power series expansion convergent in the disk $|z|<1$.


The position of the poles in the Pad\'{e} Approximant $P_N^{N+J}(z)$ (with $J\geq -1$) accumulate on the positive real axis starting at threshold, $q^2=4 m^2$, mimicking the presence of the  physical cut in the original function (provided that the function is a Stieltjes function). When PAs are applied to the vacuum polarization, this means, in particular, that there can be no spurious poles outside of the positive real axis in the $z$ plane and, consequently, no room for ambiguities. Furthermore, the convergence of the approximation (and the error) can be checked as a function of $N$, as we will see.

In Eq. (\ref{Pi}), the full functions $\Pi^{(0,1)}(q^2)$ are known. They are given by the following expressions \cite{Kallen}:

\begin{align}
\label{eq:Pi01}
\Pi^{(0)}(z) & \, = \,
\frac{3}{16\pi^{2}}\left[\frac{20}{9}+\frac{4}{3z}-\frac{4(1-z)(1+2z)}{3z}G(z)\right],
\nn \\[2mm]
\Pi^{(1)}(z) & \, = \,
\frac{4}{16\pi^{2}}\left[\frac{5}{6}+\frac{13}{6z}-\frac{(1-z)(3+2z)}{z}G(z)+
\frac{(1-z)(1-16z)}{6z}G^{\, 2}(z)\right. \nn  \\
&\qquad \qquad \qquad -\,\left.\frac{(1+2z)}{6z}\left(1+2z(1-z)\frac{d}{dz}\right)\frac{I(z)}{z}\right] \quad ,
\end{align}
where
\begin{align}
\label{eq:Gz}
I(z) & \, = \,
6\Big[\zeta_{3}+4\,\mbox{Li}_{3}(-u)+2\,\mbox{Li}_{3}(u)\Big]-
8\Big[2\,\mbox{Li}_{2}(-u)+\mbox{Li}_{2}(u)\Big]\ln u\nn \\
&\qquad \qquad \qquad \qquad  -2\Big[2\,\ln(1+u)+\ln(1-u)\Big]\ln^{2}u\,, 
\nn \\[2mm]
 G(z) & \, = \, \frac{2\, u\,\ln u}{u^{2}-1}\ ,
\quad
\mbox{with}\quad
 u \, \equiv \, \frac{\sqrt{1-1/z}-1}{\sqrt{1-1/z}+1}
\ ,
\end{align}

and with $\zeta_{n}$, the Riemann zeta function, $\mbox{Li}_{n}(u)$ the Polylogarithm function defined as $\mbox{Li}_{n}(z) = \sum_{k=1}^{\infty}\frac{z^k}{k^n}$.

However, the situation with the function $\Pi^{(2)}(q^2)$ is different. In fact, $\Pi^{(2)}(q^2)$ is
only partially known through its low-energy power series expansion around $q^2=0$, its high-energy
expansion around $q^2\rightarrow\infty$ and its threshold expansion around $q^2=4 m^2$, but the full function has
not yet been computed. Unlike the latter two
expansions, for which only a few terms are known, our knowledge of the expansion of $ \Pi^{(2)}(q^2)$ around $q^2=0$ is
very impressive, after the work of Ref. \cite{Maier} where 30 terms of this expansion were computed.

Although the full vacuum polarization function $\Pi(q^2)$ is Stieltjes, there is no reason why all
the individual contributions $\Pi^{(0,1,2,...)}(q^2)$ should also have this property. Amusingly,
however, this happens to be true both for $\Pi^{(0)}(q^2)$ and $\Pi^{(1)}(q^2)$ \cite{Broadhurst1,Broadhurst2,Broadhurst3}. The case of $\Pi^{(0)}(q^2)$ is trivial as it coincides with the full vacuum polarization $\Pi(q^2)$ for $\alpha_s$=0, and the case of $\Pi^{(1)}(q^2)$ was verified in \cite{Broadhurst1}. As we will now show, this is no longer the case for $\Pi^{(2)}(q^2)$ because its power series expansion around $q^2=0$ does not satisfy the Hankel determinants' condition Eq.~(\ref{detHankel}) which holds for a Stieltjes function. As a reminder of that condition, let $f(z)$ be a Stieltjes function with a power expansion around $z=0$ as:

\begin{equation}\label{expsum}
    f(z)= \sum_{n=0}^{\infty} f_n z^n\ .
\end{equation}
Let $D(m,n)$ be the determinant constructed with the Taylor coefficients $f_n$
\begin{equation}\label{det}
    D(m,n)=\begin{vmatrix}
             f_m & f_{m+1} & \ldots & f_{m+n} \\
             f_{m+1} & f_{m+2} & \ldots & f_{m+n+1} \\
             \vdots & \vdots &  & \vdots \\
             f_{m+n} & f_{m+n+1} & \ldots & f_{m+2n} \\
           \end{vmatrix} \quad .
\end{equation}
 As $f(z)$ is a Stieltjes function by construction, its coefficient must satisfy $D(m,n)>0$, for all $m,n$ \cite{Baker1,Baker2,Brezinski,Canaris,Bender}.

The particular case of $\Pi^{(2)}(q^2)$, however, is different since using the  $f_n$ coefficients given in Ref. \cite{Maier} (in the on-shell scheme, with the number of light flavors $n_\ell=3$):
\begin{equation}\label{taylor}
    z^{-1} \Pi^{(2)}(z)\approx 0.631107 + 0.616294 \ z + 0.56596 \ z^2 + 0.520623 \ z^3+ \ldots \quad ,
\end{equation}
one can immediately see that, e.g.,
\begin{equation}\label{D11}
    D(0,1)=\begin{vmatrix}
                             0.631107  &0.616294 \\
                              0.616294 &  0.56596 \\
                            \end{vmatrix}= -0.0226376  < 0\quad  .
\end{equation}
This proves that the individual function $\Pi^{(2)}(q^2)$ is, all by itself,  not a Stieltjes
function, even though the combination $\Pi(q^2)$ in Eq.~(\ref{Pi}) is.  Therefore, we will now focus on
applying the Theory of Pad\'{e} Approximants to the full combination $\Pi(q^2)$  in Eq. (\ref{Pi}) in
order to extract information on the individual term $\Pi^{(2)}(q^2)$.

 \begin{figure}
\renewcommand{\captionfont}{\small \it}
\renewcommand{\captionlabelfont}{\small \it}
\centering
\includegraphics[width=3in]{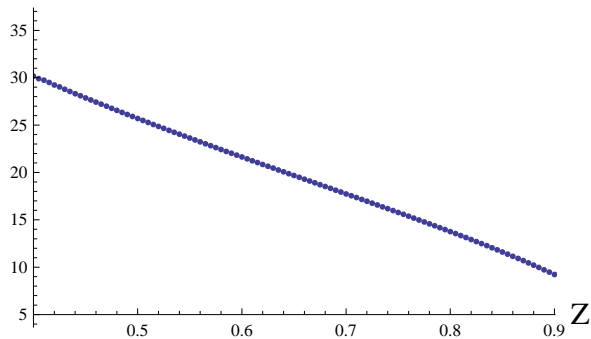}
\caption{Number of decimal places reproduced by the rational approximation in Eq. (\ref{aprox}) to the function $\Pi^{(1)}(z) $  as a function of $z$, in the interval $0.4\leq z\leq 0.9$. In this plot we show, overlaid, the results for $\beta=0.1,0.2,0.3,...,1$. As one can see, the particular value of $\beta$ chosen makes no significant impact on the result.  }\label{precision}
\end{figure}

\subsection{The method}

\quad Since, as it is obvious from Eq. (\ref{Pi}), the function $\Pi(q^2)$ depends
on the value of $\alpha_s$, any Pad\'{e} Approximant to it will also depend on the
value of $\alpha_s$, i.e.  $P_N^{N+J}(z;\alpha_s )$. This means that it is possible
to construct a rational approximation to the three functions $ \Pi^{(0,1,2)}(q^2)$
from three different sequences of Pad\'{e} Approximants to $\Pi(q^2)$ constructed at three
arbitrary values of  $\alpha_s$, let us say $\alpha_s=0, \pm \beta$, with $\beta$
sufficiently small so as to be able to neglect the terms of $\mathcal{O}(\alpha_s^3)$
in Eq. (\ref{Pi}).  In this way one obtains
\begin{eqnarray}
\label{aprox}
  z^{-1}\Pi^{(0)}(z) &\approx & P_N^{N+J}(z;\alpha_s=0 ) \nn \\
  z^{-1}\Pi^{(1)}(z)  &\approx & \frac{\pi}{2 \beta}\left\{P_N^{N+J}(z;\alpha_s=\beta )-P_N^{N+J}(z;\alpha_s=-\beta ) \right\}\nn \\
  z^{-1}\Pi^{(2)}(z)  & \approx & \frac{\pi^2}{2 \beta^2}\left\{P_N^{N+J}(z;\alpha_s=\beta )+P_N^{N+J}(z;\alpha_s=-\beta )- 2  P_N^{N+J}(z;\alpha_s=0 )\right\}\, ,
\end{eqnarray}
  where $J\geq -1$ and $N\rightarrow \infty$. Since the value of $\beta$ chosen is arbitrary, the $N\rightarrow \infty$ limit should produce results which are independent of $\beta$,  due to the convergence of the Pad\'{e} Approximants to $\Pi(q^2)$. Therefore, one should see that the three combinations (\ref{aprox}) are increasingly independent of $\beta$ as $N$ grows.\footnote{This independence of $\beta$ in the case of $ \Pi^{(0)}(q^2)$ is trivially true.} This is indeed what happens.

Furthermore, since we know the exact functions $\Pi^{(0)}(z) $ and $\Pi^{(1)}(z) $, we can compare them to the rational approximation on the right hand side of the first and second Eqs. (\ref{aprox}) in order to test the approximation. Actually, the case of $\Pi^{(0)}(z)$ is rather trivial because is independent on $\alpha_s$ and also because is a direct application of the theorem of convergence. More interesting is the case of $\Pi^{(1)}(z)$ where we can really see how the method works. Figure \ref{precision} shows the number of decimal places reproduced by this rational approximation in the interval $0.4 \leq z\leq 0.9$, when $N=14$ and $J=0$ (i.e. the diagonal Pad\'{e} Approximant $P_{14}^{14}$),  for values of $\beta$ in the interval $0.1 \leq \beta \leq 1$. As one can see, the dependence on $\beta$ cannot be distinguished in the Fig. \ref{precision}, and the accuracy reaches, e.g.,  $\sim 10$ decimal places at $z=0.9$.

We remark that $\beta\sim 0.1$ corresponds to $\beta \sim \alpha_s(M_Z)$, a value which is clearly expected to be perturbative. Of course, the method can be used for smaller values of $\beta$ as well but we use what we think are realistic numbers. For the values of $\beta$ we choose, we have checked that all the determinantal conditions in Eq. (12), which the Taylor coefficients of a Stieltjes function must satisfy, are indeed satisfied  by the 30 Taylor coefficients made available in Ref. \cite{Maier}. As it turns out, there are 392 determinantal conditions for this number of coefficients.

The third Eq. (\ref{aprox}) yields the desired approximation to $\Pi^{(2)}(z) $. To be precise, it gives us a rational approximation to $\Pi^{(2)}(z) $ in any compact set of the $z$ complex plane, away from the cut $1\leq z< \infty$. Since the threshold expansion at $z\approx 1$ can be written as \cite{HoangMateu,Czarnecki}:
\begin{eqnarray}\label{threshold}
    \Pi^{(2)}_{\mathrm{th.}}(z) &= &\frac{1.72257}{\sqrt{1-z}}
    +\left[0.34375-0.0208333\ n_{\ell}\right] \ \ln^{2}(1-z)\nn \\
    &+& \left[0.0116822\ n_{\ell} + 1.64058 \right]\ \ln(1-z) + K^{(2)}\nn \\
&&\!\!\!\!\!\!\!\!\!\!\!\!\!\!\!\!\!\!\!\!\!\!\!\!\!\!\!\!+ \left[-0.721213 - 0.0972614\ n_\ell +  3.05433 \ \ln(1-z)\right] \sqrt{1-z}
\, + \, {\mathcal O}(1-z)\ ,
\end{eqnarray}
in terms of an unknown constant $K^{(2)}$, our Pad\'{e} Approximation (\ref{aprox}) may be used to determine this constant, as we will next discuss. In this threshold expansion we take $ n_{\ell}=3$ as the number of light flavors. Even though the numerical coefficients have been rounded off for simplicity, they may be extracted exactly from the results in Ref. \cite{Czarnecki}.

Since PAs are not convergent on the physical cut, it is impossible to match the rational approximants (\ref{aprox}) to the threshold expansion (\ref{threshold}) as a function of $z$. This fact is obvious from the presence of logarithms and squared roots in Eq. (\ref{threshold}).  However, both approximations (\ref{aprox}) and (\ref{threshold}) are valid for values of $z$ at a finite distance from the cut and, in particular, in a certain window in the interval $0\leq z<1$. Within this window, a numerical matching of (\ref{aprox}) and (\ref{threshold}) is possible and will in fact  allow us to determine the unknown constant $K^{(2)}$.

 \begin{figure}\label{pi2thr}
\renewcommand{\captionfont}{\small \it}
\renewcommand{\captionlabelfont}{\small \it}
\centering
\includegraphics[width=3in]{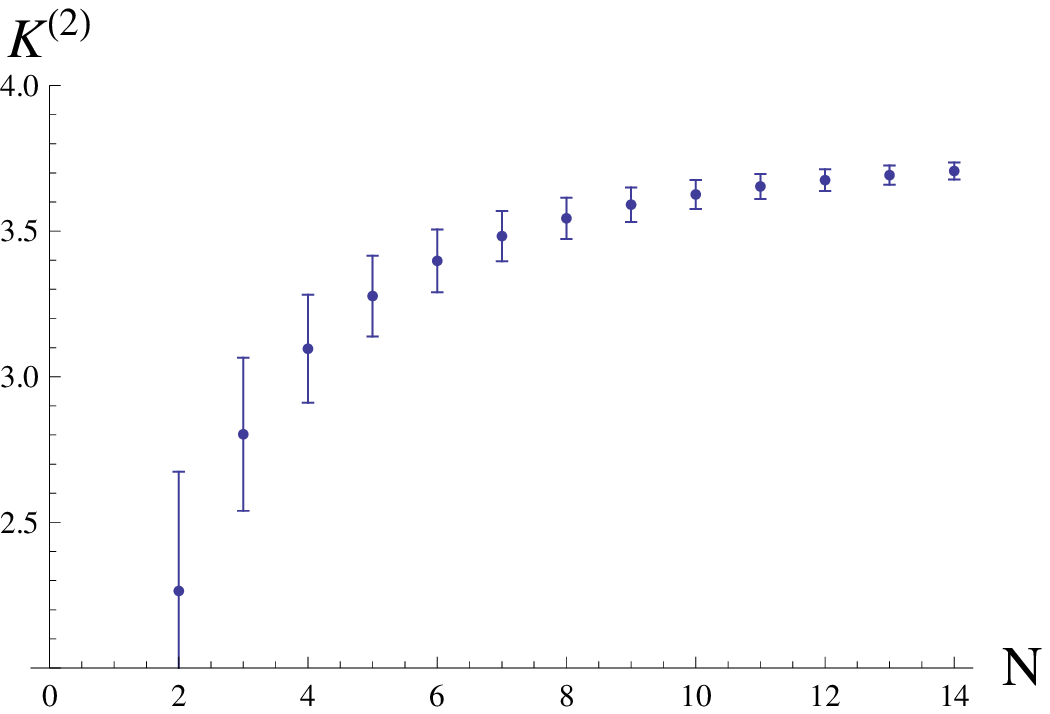}
\includegraphics[width=3in]{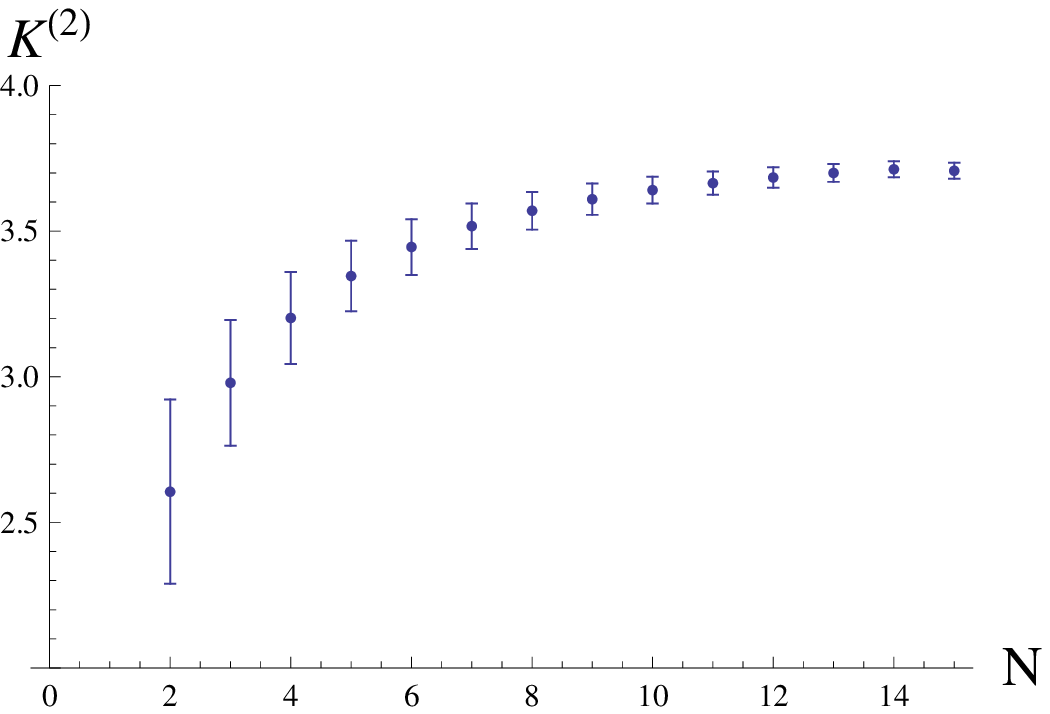}
\caption{Predicted value for $K^{(2)}$ from the sequence of diagonal, $P^{N}_{N}$ (left panel), and first paradiagonal, $P^{N-1}_{N}$ (right panel),  Pad\'{e} Approximants. This figure corresponds to $\beta=0.5$.}\label{convergence}
\end{figure}

In order to determine this window, we make the following observations. First, although the rational approximation (\ref{aprox}) is convergent as $N\rightarrow \infty$ in the interval $0\leq z< 1$, it is more accurate the closer one gets to $z=0$ in this interval, for a given value of $N$. On the other hand, the threshold expansion (\ref{threshold}) is more accurate the closer one gets to the branching point at $z=1$. From these two competing effects it is possible to determine an optimal window in $z$ by minimizing a combined error function. We will call this error function $\mathcal{E}(z)$.

The function $\mathcal{E}(z)$ has to take into account the error from the Pad\'{e} Approximants as well as the error from the threshold expansion. To estimate the error from the PAs, we consider the difference between two consecutive elements in the same sequence, i.e.  $|P_{N}^{N+J}-P_{N-1}^{N-1+J} |$. As to the threshold expansion, we estimate its error as $|1-z|$, since the expression (\ref{threshold}) is accurate up to terms of $\mathcal{O}(1-z)$.  Therefore, in order to avoid possible accidental cancelations between the two errors, we define our combined error function as the following sum:
 \begin{eqnarray}\label{error}
   \!\!\!\! \mathcal{E}(z)&\!\!\!\!\!=&\!\!\!\!\!\left| \frac{\pi^2z}{2 \beta^2}\Big\{P_N^{N+J}(z;\alpha_s=\beta )+P_N^{N+J}(z;\alpha_s=-\beta )- 2  P_N^{N+J}(z;\alpha_s=0 )\Big\}- \Big\{N\!\rightarrow \! N\!-\!1\Big\}\right|\nn\\
    &&\qquad +\quad |1-z|\ .
 \end{eqnarray}
Minimizing $\mathcal{E}(z)$ with respect to $z$ in the interval $0 \leq z< 1$, for every given values of $N$ and $\beta$, we may determine a value of $z$ at the minimum, namely $z^*$. This $z^*$ is then the one used to determine the constant $K^{(2)}$ as
\begin{equation}\label{K}
\!\! K^{(2)}\approx \frac{\pi^2z^*}{2 \beta^2}\left\{P_N^{N+J}(z^*;\alpha_s=\beta )+P_N^{N+J}(z^*;\alpha_s=-\beta )- 2  P_N^{N+J}(z^*;\alpha_s=0 )\right\}- \widehat{\Pi}^{(2)}_{\mathrm{th.}}(z^*) \quad ,
\end{equation}
for the given $N$ and $\beta$. In Eq. (\ref{K}), $ \widehat{\Pi}^{(2)}_{\mathrm{th.}}(z^*) $ stands for the expression in Eq. (\ref{threshold}) without the constant $K^{(2)}$ and, of course, without the term $\mathcal{O}(1-z)$, evaluated at $z=z^*$. The knowledge of 30 terms from the low-energy expansion gives us enough information  to be able to construct up to the Pad\'{e} Approximant $P_{14}^{14}$ from the diagonal sequence, and up to the Pad\'{e} Approximant $P_{15}^{14}$ from the first paradiagonal sequence. This corresponds to $J=0$ and $J=-1$ in Eq. (\ref{K}). The theorem of convergence for Stieltjes functions, sec.\ref{convth}, guarantees convergence for all the sequences $P_{N}^{N+J}$, $J\geq-1$. We focus our effort on the sequences which have more elements, i.e., the diagonal and the paradiagonal once, to obtain the largest possible sequence. $J=-1$ and $J=0$ translate into 14 elements for each sequence, while for bigger values of $J$, the number of elements decrease at the same rate as $J$ increase, with the limit of $J=29$ been the Taylor expansion.

In all cases considered we have varied $\beta$ in the generous range $0\leq \beta \leq 1$, but our results are insensitive to this variation within errors, as expected.

\begin{figure}
\renewcommand{\captionfont}{\small \it}
\renewcommand{\captionlabelfont}{\small \it}
\centering
\includegraphics[width=3in]{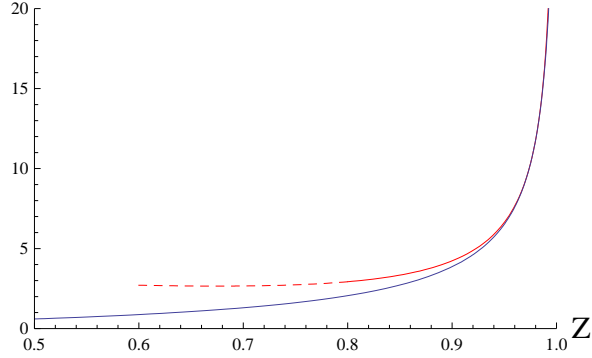}
\caption{Matching of the rational approximant in Eq. (\ref{aprox}) (solid blue line) to the
threshold expansion $\Pi^{(2)}_{\mathrm{th.}}(z)$ in (\ref{threshold}) (solid-dashed red line)  for
$N=14, J=0$ and the value of  $K^{(2)}$ in (\ref{result}). This figure shows the result for different values of $\beta$ in the range $0\leq \beta\leq1$, but the dependence on $\beta$ is so small
that cannot be discerned.}\label{matching}
\end{figure}

\subsection{Result and checks}

\quad Our results for the constant $K^{(2)}$ are shown in Fig. \ref{convergence}. This figure shows
the convergence of the diagonal sequence and the first paradiagonal sequence as a function
of the order in the PA. As one can see, we find a very nice convergence in both cases, with compatible results. Based on
this analysis, we obtain the following  value of  $K^{(2)}$:
\begin{equation}\label{result}
    K^{(2)}=3.71\pm 0.03\ .
\end{equation}
This result is very close to, although slightly smaller than, the value
obtained in Ref. \cite{HoangMateu}, i.e. $K^{(2)}=3.81\pm 0.02$. 

The error bars shown on Fig. \ref{convergence} have been  calculated as $\pm
\mathcal{E}(z^*)$. Looking at the figure we see that the change in the value of  $K^{(2)}$ from
one element of the sequence to the next is of the order of the errors shown, which is a good sign that the estimate for the error we have made is correct. In fact, we have explored what happens if the $|1-z|$ term in the error function in Eq. (\ref{error}) is multiplied by an arbitrary constant $c$ and this constant is varied within a generous range, say, between $c=10$ and $c=1/10$. In this case the particular form shown in Eq. (\ref{error}) would correspond to the (natural) choice $c=1$ (see Eq. (\ref{threshold})). While the central value for $z^*$ and, correspondingly,  $K^{(2)}$ are rather insensitive to these changes in $c$ (within the errors quoted), the error function of course does change. However, one sees that for $c=10$ the error would become much larger than the change in the central value between two subsequent PAs, whereas for $c=1/10$ the error would become much smaller. The two cases would correspond either to an overestimation of errors (the former) or  an underestimation of errors (the latter). We conclude, therefore, that our estimate of errors in Eq. (\ref{result}) is reasonable, and essentially the best we could make.
Eq.~(\ref{K}) seems rather sensitive to the particular value of $z^*$. We have already said that variations of the parameter $c$ do not cause a major impact in the determination of $z^*$ (which moves between the window 0.976-0.987 depending on $c$). What would be, however, the impact on $z^*$ of variations of $0\leq\beta\leq 1$? For the PA $P_{14}^{14}$ that variation on $\beta$ implies a variation on $z^*$ from 0.975 to 0.981, which translates to a variation of $5\times 10^{-3}$ of the value of $K^{(2)}$.

Although we have taken symmetric errors for simplicity, it is also clear from the figure that the approach to the true value is made from below, so that a slightly more accurate determination could be achieved with the use of an asymmetric
error. Apart from that, given the present knowledge of the expansions at low energy (\ref{taylor}) and at
threshold (\ref{threshold}), we find it difficult to believe any error estimate which could significantly go below
our figure in Eq. (\ref{result}). Of course, should more terms in either expansion be known, a
rerun of  our analysis could immediately produce a more precise determination of  $K^{(2)}$.

Figure \ref{matching} shows the matching of the rational approximant (i.e. the right hand side of
the third of the Eqs. (\ref{aprox})) to the threshold expansion given by
$\Pi^{(2)}_{\mathrm{th.}}(z)$ in Eq. (\ref{threshold}), for $N=14$ and $J=0$,
i.e. with the PA $P_{14}^{14}$, and for the value of $K^{(2)}$ we have obtained. As one can see,
this PA is able to reproduce, with high accuracy, the threshold expansion behavior in
a window $0.92 \lesssim z < 1$. At $z=1$ and above, the two lines in Fig. \ref{matching} will again diverge from
each other, just as they do at low $z$. The  value of $z^*$ minimizing the error function $\mathcal{E}(z)$ in (\ref{error}) was found at $z^* \simeq 0.98$ in this particular case.

For illustration, in Fig. \ref{poles14} we show the position of the poles in the PA $P_{14}^{14}$. As one can see, all the poles are sitting on the positive real axis above $z=1$, as it should be. Notice how they
accumulate in the region $z\gtrsim 1$. This is how PAs approximate the physical cut present
in the original function. As ensured from Pad\'{e} Theory, this behavior was found in all the PAs
considered.

\begin{figure}
\renewcommand{\captionfont}{\small \it}
\renewcommand{\captionlabelfont}{\small \it}
\centering
\includegraphics[width=3in]{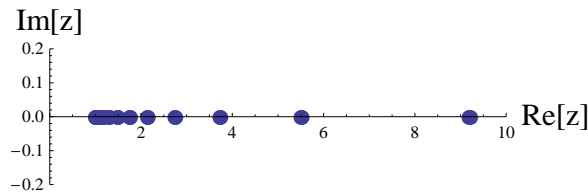}
\caption{Location of the poles in the Pad\'{e} $P_{14}^{14}$ in the complex plane. Notice the
accumulation of poles at $z\gtrsim 1$, simulating the physical cut. }\label{poles14}
\end{figure}

In order to further illustrate the reliability of our method we will include here several other examples.

Let us consider the function  $G(z)$ in Eq. (\ref{eq:Gz}) which, as $z\rightarrow 1$, has the same behavior as the function $\Pi^{(2)}(z)$ we are interested in. One obtains
\be
G(z)|_{thr}=\frac{\pi }{2 \sqrt{1-z}}+K^{(G)}+\frac{1}{4} \pi  \sqrt{1-z} -\frac{2}{3} (1-z) + \mathcal{O}\left((1-z)^{3/2}\right)\ ,
\ee
where the constant  $K^{(G)}$ is of course exactly known, i.e. $K^{(G)}=-1$. We find that the PA $P_{14}^{14}$  constructed from the Taylor expansion of $G(z)$  yields a very accurate prediction for $K^{(G)}$, namely
$ K^{(G)}=-0.999\pm 0.005$. The accuracy increases as the order of the Pad\'{e} gets larger. For instance, the PA $P_{50}^{50}$ yields $ K^{(G)}=-1.0001\pm 0.0003$. These error estimates are obtained on the basis of an error function of the form (\ref{error}), but with $|1-z|^{3/2}$ instead of $|1-z|$ in accord with the threshold expansion of $G(z)$ above. And, of course, the result can be improved if instead of an error $|1-z|^{3/2}$ one uses $|1-z|^{2}$. In this last case, with the PA $P_{14}^{14}$ the prediction for the constant $K^{(G)}$ is $K^{(G)}=-1.001\pm0.001$. For the PA $P_{50}^{50}$, the prediction reads $K^{(G)}=-1.00003\pm0.00003$.


One can repeat the same exercise with the function $\Pi^{(0)}(z)$ itself and get similar results. That function, at threshold, reads:

\be
\Pi^{(0)}(z)|_{thr}=-\frac{3 \sqrt{1-z}}{8 \pi }+\frac{1-z}{\pi ^2}-\frac{5 (1-z)^{3/2}}{16 \pi }+\,K^{(0)}+\frac{(1-z)^2}{\pi ^2}\, ,
\ee
where $K^{(0)}=\frac{2}{3 \pi ^2}=0.0675475$.

In that case, the $P^{14}_{14}(z)$ gives a prediction for $K^{(0)}=0.0674\pm 0.0001$ and the PA $P^{50}_{50}(z)$ gives a prediction for $K^{(0)}=0.06754\pm 0.00002$.

A further example is obtained with the function $\Pi^{(1)}(z)$ which, at threshold, reads

\begin{eqnarray}\label{pi1thr}
\Pi^{(1)}(z)|_{thr}&=&0.477465 \sqrt{1-z}+K^{(1)} -\frac{3}{16} \log (1-z)\nn \\
 \qquad &+& (1-z) \left[\frac{1}{8} \log (1-z)+0.354325\right]+ \mathcal{O}\left((1-z)^{3/2}\right)\, ,
\end{eqnarray}

where again the constant $K^{(1)}$ is known, i.e. $K^{(1)}= -0.314871$. A rerun of our analysis with the PA $P_{14}^{14}$ gives in this case the prediction $K^{(1)}=-0.31493\pm 0.00060$, from the second expression in Eq. (\ref{aprox}). In Fig. one can see the rate of convergence if the PAs $P_N^N$ to the constant $K^{(1)}$. Alternatively, one could also construct the PAs directly from the Taylor expansion of $\Pi^{(1)}(z)$. Within the error quoted, the two numbers agree. These examples confirm that our estimation of errors is reasonable.

\begin{figure}
\renewcommand{\captionfont}{\small \it}
\renewcommand{\captionlabelfont}{\small \it}
\centering
\includegraphics[width=3in]{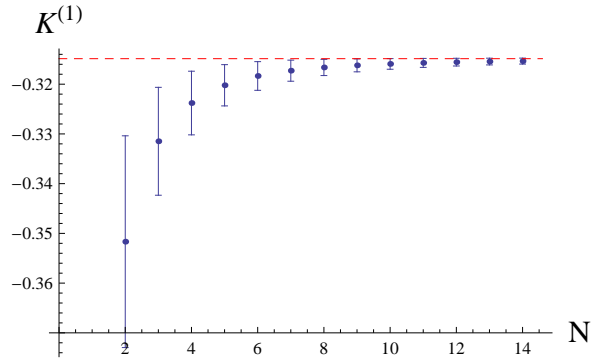}
\caption{Sequence of approximants to the constant $K^{(1)}$ in Eq.\ref{pi1thr}, together with the exact result as a red line.}\label{poles14}
\end{figure}

Furthermore, since 30 Taylor coefficients of $\Pi^{(2)}(z)$ are known (\cite{Maier}), one can entertain oneself by constructing a PA of low enough order that does not require all of the 30 terms in its construction,  in order to then predict the rest of the Taylor coefficients and compare with their exact result. For instance, with the first 21 Taylor coefficients we have constructed a PA  $P^{10}_{10}$ for  $\Pi^{(2)}(z)$ using the last expression in our Eq. (\ref{aprox})  with $N=10$ and $J=0$ which, upon re-expansion about $z=0$, predicts the value for the coefficient of the term $z^n$  of  $\Pi^{(2)}(z)$ for $n=22,23,...,30$. Using the same choice of values for $\beta$ as before, i.e. $0.1 \leq \beta \leq 1$, the order of magnitude in the relative error, $\epsilon \equiv \left|\frac{c^{predicted}_n-c^{exact}_n}{c^{exact}_n}\right|$,  is given in Table \ref{table123}. As one can see, the results turn out to be very accurate.

\begin{table}
\centering
\begin{tabular}{|c|c|c|c|c|c|c|c|c|c|}
  \hline
  n & $22$ & $23$ & $24$ & $25$ & $26$ & $27$ & $28$ & $29$ & $30$ \\
   \hline
  $\epsilon$ & $10^{-12}$ & $10^{-11}$ & $10^{-10}$ & $10^{-9}$ & $10^{-9}$ & $10^{-8}$ & $10^{-8}$ & $10^{-7}$ & $10^{-7}$ \\
  \hline
\end{tabular}
\caption{Order of magnitude in the relative error, defined as  $\epsilon\equiv  \left|\frac{c^{predicted}_n-c^{exact}_n}{c^{exact}_n}\right|$,  of the prediction for the coefficient $c_n$ of the term $c_n z^n$ in the Taylor expansion of $\Pi^{(2)}(z)$ from the PA $P_{10}^{10}$ in Eq. (\ref{aprox}).}\label{table123}
\end{table}

Finally, as a further test of our method, we have calculated the value of the constants $H^{(2)}_0$ and
$H^{(2)}_1$ which appear in the large-$z$ expansion of the function $\Pi^{(2)}(z)$ (we take
$n_f=n_{\ell}+1=4$ in the following expression):

\begin{eqnarray}
\label{Pihigh}
&&\Pi_{High-z}^{(2)}(z) \,= \,  \,
(0.034829 - 0.0021109\ n_f)\ln^{2}(-4z)
+(-0.050299 + 0.0029205\ n_f)\ln(-4z) \nn \\
& &+\quad  H^{(2)}_0
+ (0.18048 - 0.0063326\ n_f)\frac{\ln^{2}(-4z)}{z}
+ (-0.59843 + 0.027441\ n_f)\frac{\ln(-4z)}{z}\nn \\
& &+\quad  \frac{H^{(2)}_1}{z}
+\,{\mathcal O}\Big(z^{-3} \ln^3(-z)\Big)\ .
\end{eqnarray}

Using our method, we find $H_0^{(2)}=-0.582 \pm 0.008$. This result is to be compared to the true value $H^{(2)}_0=-0.5857$ \cite{ChetyrkinH01}. If we now input this exact value of $H_0^{(2)}$, by a rerun of the method, we may then determine the value of  $H^{(2)}_1$. In this way, we find  $H_1^{(2)}=-0.194 \pm 0.033$, which is to be compared to the exact value $H^{(2)}_1=-0.1872$ \cite{ChetyrkinH01}. Again, we find this agreement rather reassuring.

\chapter{Pad\'{e} Theory and Experimental Data}\label{capitol3}

\def\baselinestretch{1.66}

\section{Motivation}

\quad The fourth chapter of the present work has extensively talked about the application of Pad\'{e} Theory to meromorphic functions (but not Stieltjes functions) in the context of two-point Green's functions. We focused on meromorphic functions due to the approach we used, the Large-$N_c$ limit. In that limit, Green's functions are meromorphic functions (sec.~\ref{sec:LN}) and then the Pommerenke's theorem can be applied. While the limit of large-$N_c$ is close to realistic \cite{largeNtHooft}, it is, in itself, an approximation.

The fifth chapter has focused on Stieltjes functions and meromorphic functions of Stieltjes type. In both cases, a theorem of convergence for PA sequences is been applied to the low energy amplitudes in the Linear Sigma Model and for a Vector Meson Model, and also to the vacuum polarization of a heavy quark. And in both cases we obtain nice convergence patterns.

In the present chapter, we would like to apply Pad\'{e} Theory to a more realistic case, using now experimental data and nothing more, neither Large-$N_c$ limit nor chiral limit.

The particular case we would like to study is the pion vector form factor (defined in sec. \ref{ChPTp4}) in the space-like region, which has been known for along time that is very well described by a monopole ansatz of the type given by Vector Meson Dominance (VMD) in terms of the $\rho$-meson ($M_{\rho}^2$). While since its first phenomenological evidence, some effort has been done to extended \cite{BramonGreco}, it has remained unclear whether there is a good reason for this from QCD or it is just a mere coincidence and, consequently, it is not known how to go about improving on this ansatz, \cite{MasjuanPerisSCVFF}, \cite{SanzCilleroVFF}.

\subsection{The Pion Vector Form Factor}

\quad To begin our discussion, let us define the vector form factor (VFF), $ F(Q^2)$, by the matrix element
\begin{equation}\label{def}
    \langle \pi^{+}(p')| \ \frac{2}{3}\ \overline{u}\gamma^{\mu}u-\frac{1}{3}\ \overline{d}\gamma^{\mu} d-
    \frac{1}{3}\ \overline{s}\gamma^{\mu} s\ | \pi^{+}(p)\rangle= (p+p')^{\mu} \ F(Q^2)\ ,
\end{equation}
where $Q^2=-(p'-p)^2$, such that $Q^2>0$ corresponds to space-like data.
Since  the spectral function for the corresponding dispersive integral for $F(Q^2)$
starts at twice the pion mass, the form factor can be approximated by a Taylor expansion
in powers of the momentum for $|Q^2|< (2 m_\pi)^2$.
At low momentum, Chiral Perturbation Theory is the best tool for organizing the pion interaction in a
systematic expansion in powers of momenta and quark masses~(sec. \ref{ChPT}). This means that the coefficients in the Taylor expansion can be expressed in terms of LECs and powers of the quark masses (sec. \ref{ChPTp4}).


In principle, the coefficients in the Taylor expansion
may be obtained by means of a polynomial fit to the experimental data
in the space-like region\footnote{Time-like data is provided
by $\pi\pi$ production experiments and, consequently,
they necessarily correspond to values of the momentum  above the $\pi\pi$ cut, i.e. $ |Q^2|> 4m_\pi^2$ with $Q^2<0$.}
below $|Q^2|= (2 m_\pi)^2$. However, such a polynomial fit
implies a tradeoff. Although, in order to decrease the (systematic) error of the truncated Taylor expansion,
 it is clearly better to go to a low-momentum region, this also downsizes the set
 of data points included in the fit which, in turn, increases the (statistical) error.
In order to achieve a smaller statistical error one would have to include experimental data from higher energies, i.e. from $|Q^2|> (2 m_\pi)^2$. Since the Taylor expansion can not be applied at that high $Q^2$, one can not recover correctly its coefficients through a polynomial fit, then the use of alternative mathematical descriptions
may be a better strategy.

One such description, which includes time-like data as well, is based on the use of the Roy equations and
Omn\'es dispersion relations. This is the avenue followed by \cite{Colangelo1,Colangelo2,ColangeloB}, which has already
produced interesting results on the scalar channel \cite{Caprini}, and which can also be applied to the
vector channel. Other procedures have relied on conformal transformations for the joint analysis of both
time-like and space-like data~\cite{Yndurain1,Yndurain2}, or subtracted Omn\'es relations~\cite{Guerrero,Portoles}. Further analyses may be found in Ref. \cite{Caprini2a,Caprini2b}.

On the other hand, as already mentioned above, one may also consider an ansatz of the type
\begin{equation}\label{vmd}
    F(Q^2)_{_{\mbox{\rm VMD}}}=\left(1+\frac{Q^2}{M^2_{_{\mbox{\rm VMD}}} }\right)^{-1}\ .
\end{equation}
to describe the VFF. Even though the simplicity of the form of Eq.~(\ref{vmd}) is quite astonishing, it reproduces the space-like data rather well, even for a range of momentum of the order of a few GeV, i.e. $Q^2\gg 4 m_\pi^2$.  If this
fact is not merely a fluke, it could certainly be interesting to consider the form (\ref{vmd}) as the first
step in a systematic approximation, which would then allow improvement on this VMD ansatz.

The presence of a resonance such as $M_{VMD}^2$ means that the Taylor expansion of $F(Q^2)_{VMD}$ has a radius of convergence below $|M_{VMD}^2|$. For values of $|Q^2|> |M_{VMD}^2|$ this representation will diverge.
We would like to point out that the pion vector form factor is a good laboratory to apply the Pad\'{e} method in front of the usual Taylor expansion. PA allow the inclusion of low, medium and high energy information in a rather simple way which will let us cover a larger range of experimental data that with a simple polynomial.

Also, the VMD ansatz can be viewed as the first element in a sequence of Pad\'{e} Approximants which can be constructed in a systematic way. By considering higher-order PA in the sequence, one may be able to describe the space-like data with an increasing level of accuracy~\footnote{Obviously, unlike the space-like data, one should not expect to reproduce the time-like data since a Pad\'e Approximant contains only isolated poles and cannot reproduce a time-like cut.}.




In the present chapter we will show that, due to the precision allowed by the experimental data, there are sequences of PAs which improve on the lowest order VMD result in a rather systematic way. This has allowed us to extract
the values of the lowest-order coefficients of the low-energy expansion with an associated error.

However, precisely these lowest-order coefficients of the low-energy expansion that we want to extract is what one needs as an input in doing PAs (see sec.~\ref{PAgeneralities}).
Then, instead of the typical procedure, one can consider to apply a $N-$point Pad\'{e} using each experimental data. This means to build a rational function of degree $L$ and $M$ where $L+M+1=N$, being $N$ the amount of available data. Nevertheless, this leads to a theoretical uncertainty. What is the best PA sequence to apply? The parameter $L$ runs from $L=0$ to $L=N-1$ while the parameter $M$ runs from $M=N-1$ to $M=0$. We then have $N$ different possible configurations and no way to improve the result after choosing a particular one. Also, as one can easily notice, this implies a large computation due to the amount of available data.
Instead of this procedure, our strategy will consist in determining these coefficients by a least-squares fit of a Pad\'e Approximant to the vector form factor data in the space-like region. This will let as improve the approximation and estimate the systematic error.

There are also several types and sequences of PAs that may be considered in a fit procedure. In order to achieve a fast numerical convergence, the
choice of which one to use is largely determined by the analytic properties of the function to be
approximated (see, for example, the discussion in section \ref{PAgeneralities}). In this regard, a glance at the time-like data of the pion form factor makes it obvious that the form factor is clearly dominated by the $\rho$-meson contribution. The effect of higher resonance states,
although present, is much more suppressed. In these circumstances the natural choice is a $P^{L}_{1}$ Pad\'{e}
sequence~\cite{Baker1,Bender}, i.e. the ratio of a polynomial of degree $L$ over a polynomial of degree
one~\footnote{Conventionally, without loss of generality, the polynomial in the denominator is normalized to unity at the origin.}. Notice that, from this perspective, the VMD ansatz in (\ref{vmd}) is nothing but the
$P^0_1$ Pad\'{e} Approximant.

In that point we may say that strictly speaking there are no mathematical theorem that guarantees convergence of the sequence $P^L_1(Q^2)$ to the function $F(Q^2)$ with the fit procedure. However, a hint to understand why our sequence will converge may be as follows. Since $\Gamma_{\rho}/M_{\rho}\ll 1$ one can thing of $F(Q^2)$ to be quasi-meromorphic function with only one simple pole at $M_{\rho}^2$. We have already shown how to deal with meromorphic functions. Had this the case, we would have applied the \textit{Montessus}'s theorem and the convergence would have been guaranteed when $L\rightarrow \infty$. Since this is not the case, this theorem can not be directly applied and only remains as a mathematical inspiration to understand the results.

To test then the aforementioned single-pole dominance, one should check the degree to which the
contribution from resonances other than the $\rho$-meson may be neglected. Consequently, we have also considered the
sequence $P^{L}_{2}$, and the results confirm those found with the PAs $P^{L}_{1}$. Furthermore, for
completeness, we have also considered Pad\'e-Type approximants (PTs)~\cite{Brezinski,MasjuanPerisC87}, with the feel that one can trust the single-pole dominance of the $\rho$-resonance. Finally, we have also considered an intermediate case, the Partial-Pad\'e approximants (PPs)~\cite{Brezinski,MasjuanPerisC87}. We have fitted all these versions of rational approximants to all the available pion VFF space-like data~\cite{Amendolia,JLAB1,JLAB2a,JLAB2b,Bebek1,Bebek2,Bebek3,Bebek4,Brauel,Dally}. The result of
the fit, table~\ref{Tab:VFFpades}, is rather independent of the kind of rational approximant sequence used and all the results show consistency among themselves.

\section{Evaluation of the systematic error}\label{sec:modelVFF}

\quad In the previous section we have presented our strategy of fitting to the experimental data. What is new in that case is the way to estimate the systematic error of the approximation when dealing with real data. In the present section we will illustrate the usefulness of the PAs as fitting functions in the way we propose here, by using, first, a phenomenological model as a theoretical laboratory. Furthermore, the model will also give us an idea about the size of possible systematic uncertainties.

To consider the method as quite realistic, we use a once-subtracted Omn\'es relation to recover the form-factor,
\begin{equation}\label{model2}
    F(Q^2)=\exp\left\{-\frac{Q^2}{\pi} \int_{4 \hat{m}_{\pi}^{2}}^{\infty}\ dt\ \frac{\delta(t)}{t (t+Q^2)}\right\}.
\end{equation}

The $\delta (t)$ function
\begin{equation}
    \delta(t)=\tan^{-1}\left[\frac{\hat{M}_{\rho} \hat{\Gamma}_{\rho}(t)}{\hat{M}_{\rho}^2-t} \right]\ ,
\end{equation}

plays the role of the vector form factor phase-shift~\cite{Guerrero,Portoles,Cillero1,Cillero2} and its $t$-dependence comes from the width ($\sigma(t)=\sqrt{1-4 \hat{m}_{\pi}^{2}/t}$):

\begin{equation}\label{width}
    \hat{\Gamma}_{\rho}(t)= \Gamma_{0}\ \left( \frac{t}{\hat{M}_{\rho}^2} \right)\ \frac{\sigma^3(t)}{\sigma^3(\hat{M}_{\rho}^2)}\ \theta\left( t- 4 \hat{m}_{\pi}^{2} \right).
\end{equation}
To emphasize that the model should be considered realistic, we chose the input parameters close the physical ones:
\begin{equation}\label{param}
 \Gamma_{0} = 0.15\ \mathrm{GeV}\quad ,\quad
 \hat{M_{\rho}}^2= 0.6\ \mathrm{GeV}^2\quad ,\quad
 4 \hat{m}_{\pi}^{2}= 0.1 \ \mathrm{GeV}^2\, .
\end{equation}
Actually, this model has been used in Ref.~\cite{Guerrero,Portoles,Cillero1,Cillero2} to extract the values for the
physical mass and width of the $\rho$ meson through a direct fit to the (timelike) experimental data.

Expanding $F(Q^2)$ in Eq. (\ref{model2}) in powers of $Q^2$ we readily obtain a Taylor expansion for the VFF
\begin{equation}\label{expmodel}
    F(Q^2)\, =\, 1 \, - \,  a_1\ Q^2 \, + \,  a_2\ Q^4 \,  - \ a_3\ Q^6 + ... \,\, ,
\end{equation}
with known values for the coefficients $a_i$\footnote{$a_0=1$ to resemble the vector current conservation condition.}. The goal is to obtain these $a_i$ parameters trough expanding the rational functions after fitting to the experimental data.

The real experimental data are obtained from ~\cite{Amendolia,JLAB1,JLAB2a,JLAB2b,Bebek1,Bebek2,Bebek3,Bebek4,Brauel,Dally} and to work close to the real case, we will simulate the same amount of data as in the physical case\footnote{To estimate the systematic error derived purely from our approximate description of the form factor, these points will be taken with vanishing error bars.}. Then, we generate fifty ``data'' points in the region $0.01\leq Q^2\leq 0.25$, thirty data points in the interval $0.25\leq Q^2 \leq 3$, and seven points for $3\leq Q^2\leq 10$ (all these momenta in units of GeV$^2$)

In Table~\ref{table1VFF} we show the predicted coefficients $a_i$ from the $P^{L}_{1}(Q^2)$ sequence we have fitted. The last PA we have obtained to these data is $P^6_1$. As expected, the pole position of the Pad\'{e}s differs from the true mass of the model Eq.~(\ref{param}), since the \textit{Montessus de Ballore}'s theorem is not fulfilled.

\begin{table}[t]
\centering
\begin{tabular}{|c|c|c|c|c|c|c|c|c|}
  \hline
   & $P^{0}_{1}$ &  $P^{1}_{1}$ &  $P^{2}_{1}$ &  $P^{3}_{1}$&$ P^{4}_{1}$  & $P^{5}_{1}$  &$P^{6}_{1}$ & $F(Q^2)$(exact)\\ \hline
  $a_1$(GeV$^{-2}$) & 1.549 & 1.615 & 1.639 & 1.651 & 1.660&1.665 & 1.670 & 1.685 \\
  $a_2$(GeV$^{-4}$)& 2.399 & 2.679& 2.809 & 2.892 & 2.967&3.020 & 3.074& 3.331\\
  $a_3$(GeV$^{-6}$)& 3.717 & 4.444 & 4.823 & 5.097 & 5.368&5.579 & 5.817& 7.898\\
  \hline
  \hline
  $s_p$(GeV$^{2}$) &$0.646$&$0.603$&$0.582$&$0.567$&$0.552$&$0.540$&$0.526$&$0.6$\\
  \hline
\end{tabular}
\caption{Results of the various fits to the form factor $F(Q^2)$ in the model, Eq. (\ref{model2}).
The exact values for the coefficients $a_i$ in Eq. (\ref{expmodel}) are given on the last column. The last
row shows the predictions for the corresponding pole for each Pad\'{e} ($s_p$), to be compared to the true mass
$\hat{M}_{\rho}^{2}=0.6\ $GeV$^2$ in the model.} \label{table1VFF}
\end{table}

A quick look at Table \ref{table1VFF} shows that the sequence seems to converge to the exact result, although
in a hierarchical way, i.e. much faster for $a_1$ than for $a_2$, and this one much faster than $a_3$,
etc, showing exactly the same convergence behavior than in the case worked in section \ref{C87}. The relative error achieved in determining the coefficients $a_i$ by the last Pad\'e, $P^6_1$, is
respectively $0.9\%$, $8\%$ and $26\%$ for $a_1, a_2$ and $a_3$. Naively, one would expect these results to
improve as the resonance width decreases since the $P^{L}_{1}$ contains only a simple pole, and  this is
indeed what happens. Repeating this exercise with the model, but with a $\Gamma_0=0.015$~GeV ($10$ times
smaller than the previous one), the relative error achieved by $P^6_1$ for the same coefficients as before
is $0.12\%$, $1.1\%$ and $4.7\%$. On the other hand, a model with $\Gamma_0$ five times bigger than the
first one produces, respectively, differences of $2.1\%$, $14.4\%$ and $37.8\%$.

To complete the study, we have also studied the convergence of Pad\'e-Type approximants with the model. Thus, in
this case, we have placed the $T^L_1$ pole at $s_p=\hat{M}^2_\rho$ and found a similar pattern as in
Table~\ref{table1VFF}. For $T^6_1$, the Pad\'e-Type coefficient $a_1$ differs a $2.5\%$ from its exact value,
$a_2$ by $16\%$ and $a_3$ by $40\%$.

This procedure gives us a rough estimate of the systematic uncertainties when fitting to experimental data. Since, as we will see, the best fit to the experimental data comes from the Pad\'{e} $P^4_1$, we will take the error in Table \ref{table1VFF} from this Pad\'{e} as a reasonable estimate and add to the final error an extra systematic uncertainty of $1.5\%$ and $10\%$ for $a_1$ and $a_2$ (respectively).

\section{Fitting the pion Vector Form Factor}\label{sec:FF}

\quad We will use all the available experimental data in the space-like region, which may be found in
Refs.~\cite{Amendolia,JLAB1,JLAB2a,JLAB2b,Bebek1,Bebek2,Bebek3,Bebek4,Brauel,Dally}.
These data range in momentum from $Q^2=0.015$ up to 10~GeV$^2$.

As discussed previously, the prominent role of the $\rho$-meson contribution motivates
that we start with the $P^{L}_{1}$ Pad\'e sequence.

\subsection{Pad\'e Approximants $P^{L}_{1}$}\label{sec:PAL1}

\begin{figure}
  \center
  \includegraphics[width=13cm]{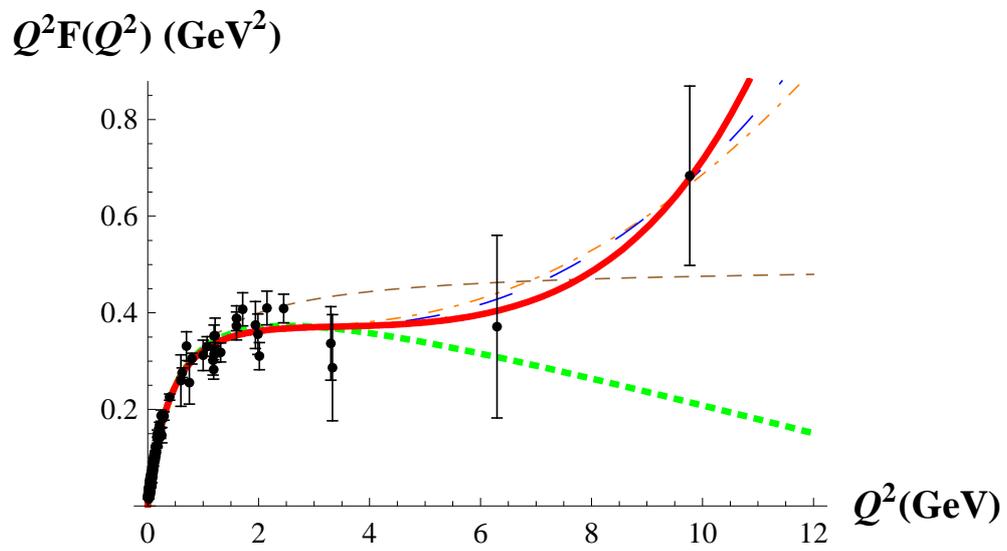}
  \caption{The sequence of $P^L_1$ PAs is compared to the available space-like
  data~\cite{Amendolia,JLAB1,JLAB2a,JLAB2b,Bebek1,Bebek2,Bebek3,Bebek4,Brauel,Dally}:
  $P^0_1$ (green thick-dashed), $P^1_1$ (brown dashed),
  $P^2_1$ (orange dot-dashed), $P^3_1$ (blue long-dashed),
  $P^4_1$ (red solid).}\label{fig:VFF}
\end{figure}

\quad Without any loss of generality, a $P^{L}_{1}$ Pad\'e Approximant is given by
\begin{equation}
P_1^L(Q^2) \,\, \,= \,\,\,
1\, +\,  \sum_{k=1}^{L-1}a_k (-Q^2)^{k} \,\,
+ \, (-Q^2)^{L} \,  \frac{\ a_L}{1+\frac{a_{L+1} }{a_L}  \, Q^2}\  ,
\label{PL1}
\end{equation}
where $P^L_1(0)=1$ has been imposed (in fact, the vector current conservation condition)
and the coefficients $a_{k}$ are the low-energy coefficients of the corresponding Taylor expansion of the VFF (compare with (\ref{expmodel}) for the case of a model).

The fit of $P^L_1$ to the space-like data points in Refs. ~\cite{Amendolia,JLAB1,JLAB2a,JLAB2b,Bebek1,Bebek2,Bebek3,Bebek4,Brauel,Dally}
determines  the coefficients  $a_{k}$ that best interpolate them. According to Ref.~\cite{brodsky-lepage1,brodsky-lepage2,brodsky-lepage3},
the form factor is supposed to fall off as $1/Q^2$ (up to logarithms) at large values of $Q^2$. This means
that, for any value of $L$, one may expect to obtain a good fit only up to a finite value of $Q^2$, but not
for asymptotically large momentum. This is clearly seen in Fig. ~\ref{fig:VFF}, where the Pad\'{e} sequence
$P^L_1$ is compared to the data up to $L=4$.

Fig.~\ref{fig:a1PL1} shows the evolution of the fit results for the Taylor
coefficients $a_1$ and $a_2$ for the $P^L_1$ PA  from $L=0$ up to $L=4$.
As one can see, after a few Pad\'es
these coefficients become stable. Since the experimental data have non zero error
it is only possible to fit a $P^L_1$ PA up to a certain value for $L$. From this order on, the large error bars in the highest coefficient in the numerator polynomial make it compatible with zero and, therefore, it no
 longer makes sense to talk about a new element in the sequence. For the
 data in Refs. ~\cite{Amendolia,JLAB1,JLAB2a,JLAB2b,Bebek1,Bebek2,Bebek3,Bebek4,Brauel,Dally}, this happened
 at $L=4$ and this is why our plots stop at this value. Therefore, from the PA  $P^4_1$
 we obtain our best fit and, upon expansion around $Q^2=0$,  this yields
\begin{equation}
a_1\, =\, 1.92 \pm 0.03\,\,\mbox{GeV}^{-2} \, ,  \qquad\qquad a_2\, =\, 3.49 \pm 0.26\,\,\mbox{GeV}^{-4} \,
;
\end{equation}
with a $\chi^2/\mathrm{dof}=117/90$.

\begin{figure}[!t]
  \center
  \includegraphics[width=7cm]{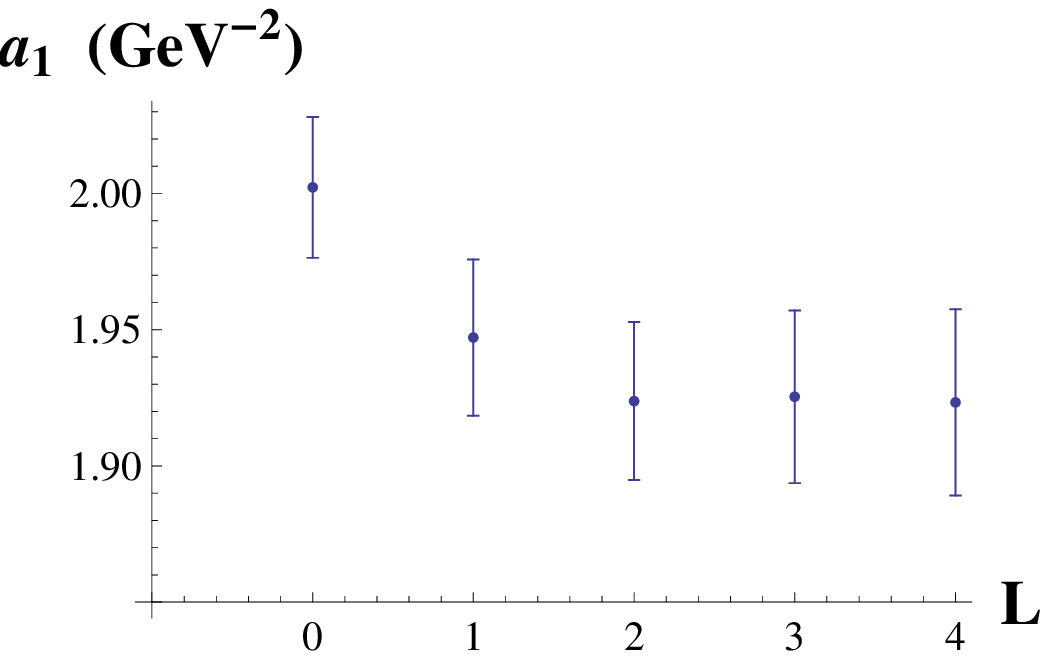}
  \hspace*{1.cm}
  \includegraphics[width=7cm]{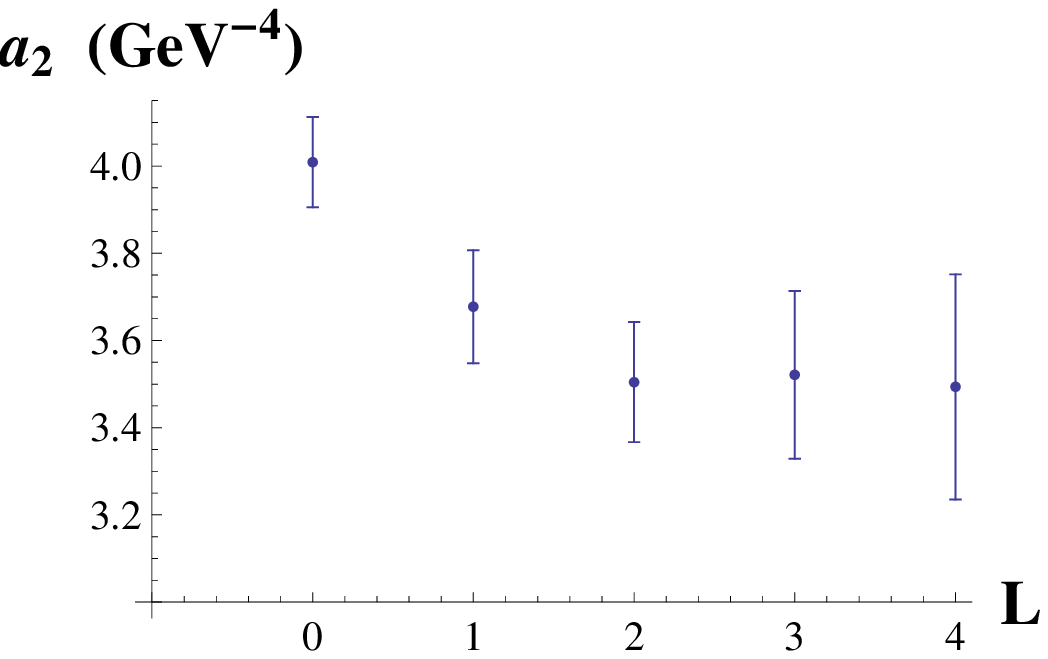}
  \caption{$a_1$ and $a_2$ Taylor coefficients for the
  $P^L_1$ PA sequence.}\label{fig:a1PL1}
\end{figure}

Eq.~(\ref{PL1}) shows that the pole of each $P^L_1$ PA is determined by the ratio $s_p=a_L/a_{L+1}$.  This
ratio is shown in  Fig.~\ref{fig:spPL1}, together with a gray band whose width is given by $\pm
M_\rho\Gamma_\rho$ for comparison. From this figure
 one can see that the position of the pole of the PA is close to the
physical value $M_\rho^2$~\cite{PDG}, although it does not necessarily agree with it, as we already saw in the model of the previous section.

\begin{figure}
  \center
  \includegraphics[width=6.5cm]{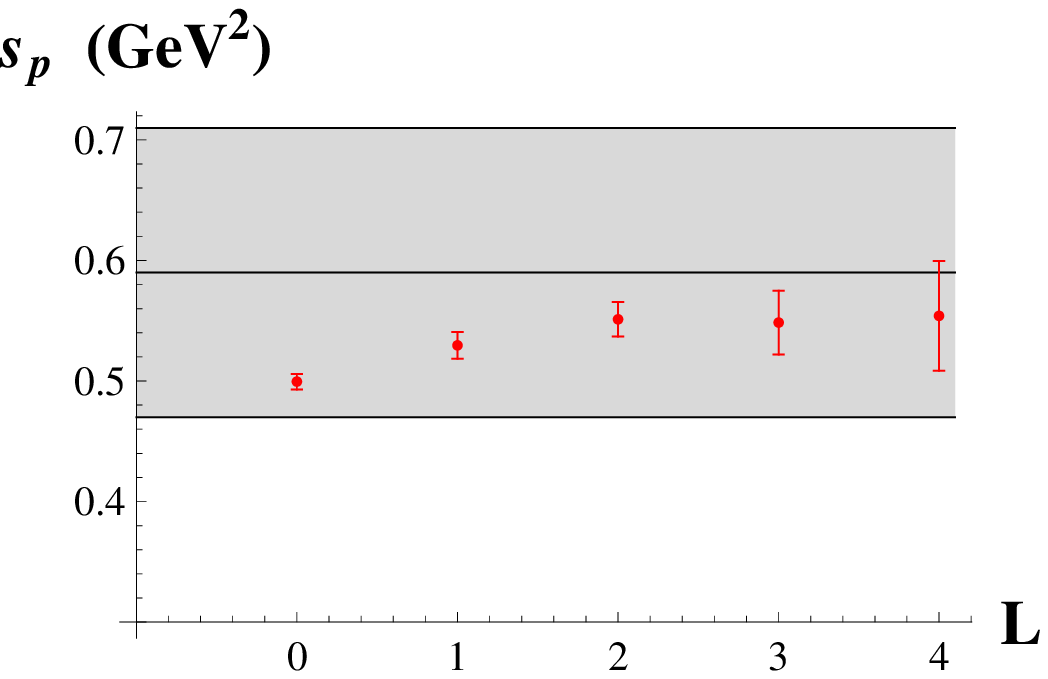}
  \caption{Position $s_p$ of the pole for the different $P^L_1$.
  The range with the physical values
  $M_\rho^2\pm M_\rho\Gamma_\rho$ is shown (gray band)
  for comparison.}\label{fig:spPL1}
\end{figure}

\subsection{Comment on $P^L_2$ Pad\'es}  \label{sec:PAL2}

\quad Although the time-like data of the pion form factor
is clearly dominated by the $\rho(770)$ contribution,
 consideration of two-pole
$P^L_2$ PAs will give us a way to assess any possible systematic bias in our previous analysis, which was limited to only single-pole PAs.

We have found that the results of the fits of $P^L_2$ PAs to the data tend to reproduce
the VMD pattern found for the $P^L_1$ PAs in the previous section. The $P^L_2$ PAs place the first of the two poles around the $\rho$-mass, while the second wanders around the complex momentum plane together with a close-by zero in the numerator, what we have already explained to be a defect \cite{Baker1,Bakerconvergence}, sec. \ref{convth}. As a local perturbation, at any finite distance from it the effect is essentially negligible. This has the net effect that the $P^L_2$ Pad\'e in the Euclidean region looks just like a $P^L_1$ approximant and, therefore, yields essentially the same results. For example, for the $P^2_2$, one gets
\begin{equation}
a_1\, =\, 1.924 \pm 0.029 \,\,\mbox{GeV}^{-2} \, ,  \qquad\qquad
a_2\, =\, 3.50  \pm 0.14 \,\,\mbox{GeV}^{-4} \,  ,
\end{equation}
with  a $\chi^2/\mathrm{dof}= 120/92$.

\subsection{Pad\'{e} Type and Partial Pad\'e Approximants}\label{sec:PTPP}

\begin{figure}[!t]
 \center
 \includegraphics[width=8cm]{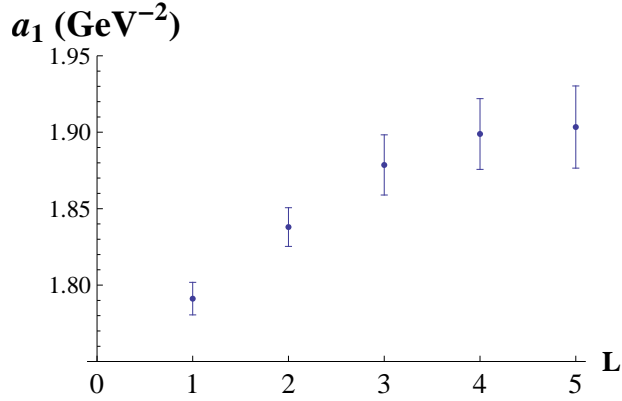}\\
 \caption{Low-energy coefficient $a_1$ from the
 $T^L_1$ Pad\'e-Type sequence}\label{fig:a1PTL1}
\end{figure}

\quad As in the case of the model, one may consider other kinds of rational approximants, such as Pad\'e Type and Partial Pad\'e Approximants. Since the value of the physical $\rho$-mass is known ($M_{\rho}=775.5$~MeV), it is natural to attempt a fit of PTAs to the data with a pole fixed at that mass. Actually, the results found with the model and also whit $P^4_1$ and $P^2_2$ inspire this way to proceed. The corresponding sequence will be called $T^L_1$. This has the obvious advantage that the number of parameters in the fit decreases by one and allows one to go a little further in the sequence. Our best value is then given by the Pad\'e Type Approximant $T^5_1$, whose expansion around $Q^2=0$ yields the following values for the Taylor coefficients:
\begin{equation}
a_1\, =\, 1.90 \pm 0.03\,\,\mbox{GeV}^{-2} \, ,  \qquad\qquad a_2\, =\, 3.28  \pm  0.09 \,\,\mbox{GeV}^{-4}
\, ,
\end{equation}
with a $\chi^2/\mathrm{dof}=118/90$.

The previous analysis of PTAs may be extended by making further use of our knowledge of the vector
spectroscopy~\cite{PDG}. For instance, by taking $M_{\rho}=775.5$~MeV, $M_{\rho'}=1459$~MeV and
$M_{\rho''}=1720$~MeV,\footnote{As will be seen, results do not depend on the precise value chosen for these masses.}  we may construct further Pad\'e-Type sequences of the form $T^L_2$ and $T^L_3$.

In the PTA sequence $T^L_2$ one needs to provide the value of two poles. For the first pole, the natural
choice is $M_{\rho}^2$. For the second pole, we found that choosing either $M_{\rho'}^2$ or $M_{\rho''}^2$
(the second vector excitation) does not make any difference. Both outcomes are compared in
Fig.~(\ref{fig:a1PT2}). Using $M_{\rho'}^2$, we found that the $T^3_2$ PTA yields the best values as
\begin{equation}
a_1\, =\, 1.902 \pm 0.024\,\,\mbox{GeV}^{-2} \, ,  \qquad\qquad a_2\, =\, 3.29  \pm  0.07
\,\,\mbox{GeV}^{-4} \, ,
\end{equation}
with a $\chi^2/\mathrm{dof}=118/92$.

Using $M_{\rho''}^{2}$ as the second pole one also gets the best value from the $\widetilde{T}^3_2$ PTA, with
the following results:
\begin{equation}
a_1\, =\, 1.899 \pm 0.023\,\,\mbox{GeV}^{-2} \, ,  \qquad\qquad a_2\, =\,  3.27 \pm 0.06 \,\,\mbox{GeV}^{-4}
\, ,
\end{equation}
with a $\chi^2/\mathrm{dof}=119/92$. We find the stability of the results for the coefficients $a_{1,2}$ quite
reassuring.

\begin{figure}[!t]
 \center
 \includegraphics[width=7cm]{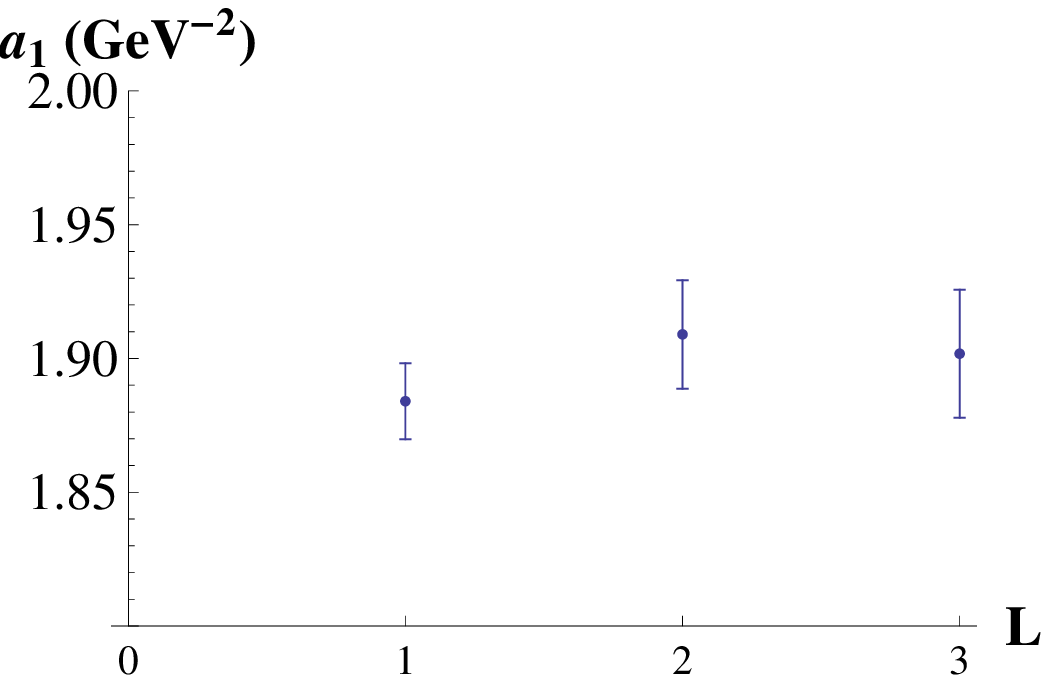}
 \hspace*{.5cm}
 \includegraphics[width=7cm]{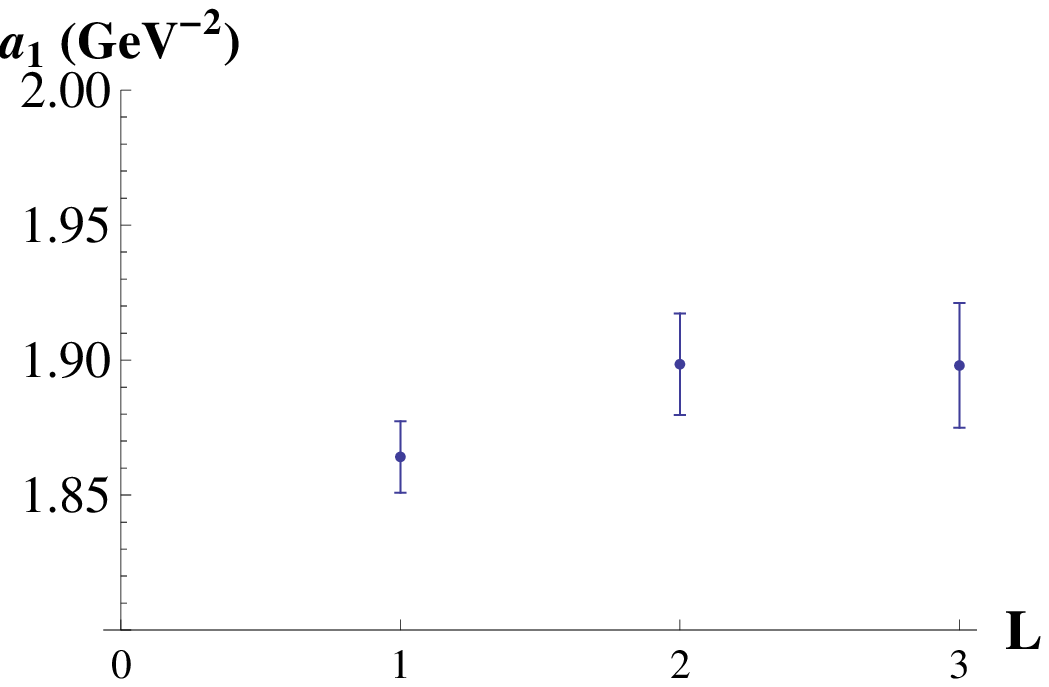}
 \caption{{\small
 Low energy coefficient $a_1$  for the $T^L_2$ Pad\'e-Type
 sequence with $M_{\rho}$ and $M_{\rho'}$ (left),
 and with $M_{\rho}$ and $M_{\rho''}$ (right).
 }}
 \label{fig:a1PT2}
\end{figure}

We have also performed an analysis of the PTA sequence $T^L_3$, with similar conclusions. From the $T^3_3$
we obtain the following values for the coefficients:
\begin{equation}
a_1\, =\, 1.904 \pm 0.023\,\,\mbox{GeV}^{-2} \, ,  \qquad\qquad a_2\, =\, 3.29  \pm 0.09 \,\,\mbox{GeV}^{-4}
\, ,
\end{equation}
with a $\chi^2/\mathrm{dof}=119/92$.

Finally, to complete our analysis, we will also consider Partial Pad\'e Approximants, in which only part of the denominator is given in advance. In particular, we study the PPA sequence $P^L_{1,1}$ \footnote{See sec.~\ref{sec:NpointPade} for notation.} in which the first pole is given by $M_{\rho}^2$ and the other is left free. The best determination of the Taylor coefficients is given by $P^2_{1,1}$, and they yield
\begin{equation}
a_1\, =\, 1.902  \pm 0.029 \,\,\mbox{GeV}^{-2} \, ,  \qquad\qquad
a_2\, =\, 3.28   \pm 0.09  \,\,\mbox{GeV}^{-4} \, ,
\end{equation}
with  the free pole  of the PPA given by $M_{free}^2=(1.6   \pm 0.4 $~GeV$)^2$ and a $\chi^2/\mathrm{dof}=119/92$.

\section{Combined Results}\label{sec:results}

\begin{table}
\centering
\begin{tabular}{|c||c|c|}
\hline
Pad\'{e} & $a_1$ $(\times GeV^2)$ & $a_2$ $(\times GeV^4)$  \\
\hline
$P^4_1$ &     $1.92\pm0.03$ & $3.49\pm0.26$ \\
\hline
$P^2_2$ &     $1.924\pm0.029$ & $3.50\pm0.14$  \\
\hline \hline
$T^5_1$ &    $1.90\pm0.03$ & $3.28\pm0.09$  \\
\hline
$T^3_2$ &    $1.902\pm0.024$ & $3.29\pm0.07$  \\
\hline
$\tilde{T}^3_2$ &    $1.899\pm0.023$ & $3.27\pm0.06$  \\
\hline
$T^3_3$ &    $1.904\pm0.023$ & $3.29\pm0.09$  \\
\hline\hline
$P^2_{1,1}$ &    $1.902\pm0.029$ & $3.28\pm0.09$  \\
\hline
final & $1.91\pm0.01$ &  $3.30\pm0.03$  \\
\hline
\end{tabular}
\caption{Summary of results found for the parameters $a_1$ and $a_2$ defined in Eq.~(\ref{expmodel}) for all the PAs we have worked. See the text for details on the notation.}\label{Tab:VFFpades}
\end{table}

\quad The combination of all the previous rational approximants results
in an average given by
\begin{equation}
a_1\, =\, 1.907\pm 0.010_{\mathrm{stat}} \pm 0.03_{\mathrm{syst}}\,\,\mbox{GeV}^{-2} \  ,  \
a_2\, =\,  3.30 \pm  0.03_{\mathrm{stat}} \pm 0.33_{\mathrm{syst}}\,\,\mbox{GeV}^{-4} \, .
\end{equation}
The first error comes from combining the results of the different fits by means of a weighted average. On
top of that, we have added what we believe to be a conservative estimate of the theoretical (i.e.
systematic) error based on the analysis of the VFF model in Sec.~\ref{sec:modelVFF}. We expect the latter to
give an estimate for the systematic uncertainty due to the approximation of the physical form factor with
rational functions. For comparison with previous analysis, we also provide in Table \ref{table2VMD} the value
of the quadratic radius, which is given by $\langle r^2 \rangle \, =\, 6 \, a_1$ , that would mean $L_9=(6.80\pm0.04)\times 10^{-3}$, \cite{MasjuanPerisSCVFF}. Since the value that we obtain for the quadratic radius is slightly smaller than Amendolia'86 result (\cite{Amendolia}), our result for $L_9$ is also smaller.

We also provide in Table \ref{table2VMD} other determinations of $a_1$ and $a_2$ (this second one when available) found in the literature. Refs.\cite{Colangelo1,Colangelo2,ColangeloB}, named CGL, are based on the use if the Roy equations and Omn\'{e}s dispersion relations. Refs.\cite{Yndurain1,Yndurain2}, named TY, rely on conformal transformations for the joint analysis of both time-like and space-like data. Ref.\cite{op6-VFF}, called BCT, computed the ${\cal O}(p^6)$ Vector Form Factor in Chiral Perturbation Theory and then they fit to experimental data to extract the LECs. Finally, Ref.\cite{Portoles}, named PP, used a model independent parametrization for the Vector Form Factor constrained by unitarity and analyticity. We also include the PDG \cite{PDG} value based on a weighted average of some available experimental data such as \cite{Amendolia}, \cite{Bebek1}, \cite{Dally}.

Recently, two lattice collaboration were able to compute on the lattice the Pion Vector Form Factor. In Ref.\cite{latticeFF}, the RBC-UKQCD collaboration study that Form Factor in a gauge configurations with Domain Wall Fermions and the Iwasaki gauge action. In Ref.\cite{JLQCDlattice}, the JLQCD-TWQCD collaboration used two-flavor lattice QCD with exact chiral symmetry and employing overlap quark action. We would like to remark that this last Ref.\cite{JLQCDlattice} found that the best fit to their lattice data could we obtain using a VMD-like hypothesis:

\begin{equation}\label{PL1lattice}
P(Q^2) \,\, \,= \,\,\,a_1 (-Q^2) \,+\,a_2 (-Q^2)^{2}\,+\,a_3 (-Q^2)^{3}+ \,  \frac{1}{1+\frac{Q^2}{M_{\rho}^2}}\, ,
\end{equation}

where $M_{\rho}^2$ is the value of the $\rho$-meson on the lattice. It is really interesting to notice how close this expression Eq.(\ref{PL1lattice}) is to our Eq.(\ref{PL1}) for $L=4$, which is, actually, our best $P^L_1$ approximant to the experimental data.

\begin{table}
\centering
\begin{tabular}{|c|c|c|c|c|c|c|c|c|}
  \hline
   & $\langle r^2\rangle$   (fm$^2$)  &  $a_2$  (GeV$^{-4}$)     \\ \hline \hline
   PDG~\cite{PDG} & $0.452\pm0.016$ & ...\\
   \hline
  This work (also \cite{MasjuanPerisSCVFF}) &   $0.445\pm 0.002_{\mathrm{stat}}\pm 0.007_{\mathrm{syst}}$ &  $3.30\pm 0.03_{\mathrm{stat}}\pm 0.33_{\mathrm{syst}}$    \\
  CGL~\cite{Colangelo1,Colangelo2,ColangeloB}& $ 0.435\pm 0.005$  & ...   \\
  TY~\cite{Yndurain1,Yndurain2}  & $ 0.432\pm 0.001 $  & $ 3.84\pm 0.02$ \\
  BCT~\cite{op6-VFF} & $0.437\pm 0.016$  & $3.85\pm 0.60$ \\
  PP~\cite{Portoles} & $0.430\pm 0.012$  & $3.79\pm 0.04$ \\
  RBC-UKQCD~\cite{latticeFF} & $0.418\pm 0.031$  & ... \\
  JLQCD-TWQCD~\cite{JLQCDlattice} & $0.409\pm 0.044$  & $3.22\pm0.42$ \\
  \hline
\end{tabular}
\caption{Our results for the quadratic radius $\langle r^2\rangle$ and second derivative $a_2$ are
compared to other determinations~\cite{Colangelo1,Colangelo2,ColangeloB,Yndurain1,Yndurain2,op6-VFF,Portoles,latticeFF,JLQCDlattice}. Our first error is statistical. The second one is systematic, based on the analysis of the VFF model in section \ref{sec:modelVFF}. We also include the PDG \cite{PDG} value based on a weighted average of some available experimental data such as \cite{Amendolia}, \cite{Bebek1}, \cite{Dally}.}
\label{table2VMD}
\end{table}

\chapter{Conclusions}\label{conc}

\def\baselinestretch{1.66}
\quad In this work we have exploited several uses of Pad\'{e} Approximants as approximations to QCD. In particular in the chapter \ref{capitol2} we saw that the resonance saturation in the limit of large-$N_c$ may be reinterpreted within the mathematical Theory of Pad\'{e} Approximants to meromorphic functions. We saw that one may expect convergence of a sequence of Pad\'{e} Approximants to QCD Green's functions in the large-$N_c$ limit in any compact region of the complex $Q^2$ plane except in a region of zero-measure. As the order of the PA grows, the convergence properties guarantees that any given artificial pole of the approximation goes to infinity or is almost canceled by a nearby zero. With the help of a $\langle VV-AA\rangle$ model we have reviewed the main results of this theory where we have seen that while in the Minkowsky region the rational approximant creates the expected artificial poles, in the Euclidean region the description is quite accurate. And this happens hierarchically: although the first poles and residues in a PA may be used to describe the physical mass and decay constants reasonably well, the same is not true for the last ones. In general it will be unreliable to extract properties of individual mesons from an approximation to large-$N_c$ QCD with only a finite number of states.


Studying form factors one can find particular manifestation of that problem. Since a form factor, like a decay constant, is obtained as the residue of a Green's function at the corresponding pole(s), this also means that one may not extract a meson form factor from a rational approximant to a 3-point Green's function, in agreement with Ref.\cite{Lipartia}.  This observation may explain why the analysis of Ref. \cite{theworksCirigliano}, which is based on an extraction of matrix elements such as $\langle \pi|S|P\rangle$ and $\langle \pi|P|S\rangle$ from the 3-point function $\langle SPP\rangle$, finds values for the $K_{\ell 3}$ form factor which are different from those obtained in other analyses \cite{LRoos} or lattice results \cite{latticeSachrajda}.

In spite of all the above problems related to the Minkowski region, our model shows how Pad\'{e}
Approximants may nevertheless be a useful tool in other regions of momentum space. We think that
this is also true in the real case of QCD in the large-$N_c$ limit. In this case one may use the
first few terms of the chiral and operator product expansions of a given Green's function to
construct a Pad\'{e} Approximant which should yield a reasonable description of this function in those
regions of momentum space which are free of poles. In this construction, Pad\'{e} Approximants
containing complex poles, if they appear, should not be dismissed. We exemplify this procedure with the estimate of the LECs at ${\cal O}(p^6)$ of the correlator $\Pi_{LR}(Q^2)$.


However, if not all the residues and masses in a rational approximant are physical, this poses a
challenge to any attempt to use a Lagrangian with a finite number of resonances such as, for
example, the ones in Ref. \cite{swissDRV,swissEGPRressonance}, for describing Green's functions in the large-$N_c$
limit of QCD. Even if these Lagrangians are interpreted in terms of PTAs, with the poles fixed at
the physical value of the meson masses, we have seen how the residues then get very large
corrections with respect to their physical counterparts.

Besides, we also addressed on the different determinations of the mass of first axial state $m_A$ using resonance saturation in the large-$N_c$ limit of QCD. We show how these different determinations may be understood as the position of the pole of a Pad\'{e} Approximant and then, nothing to do with the physical lowest axial state appearing in the Particle Data Book \cite{PDG}. We saw that PAs and PTs leads basically to the same results when the position of their poles may be completely different, even complex for the case of the lowest PA (depending of the value of the combination $\zeta$). In passing we also suggested a way to disentangle the discussion about the sign of the condensate of ${\cal O}(p^{-8})$ and its relation with the low-energy constant $L_{10}$ appearing in the Chiral Lagrangian.


We divided the chapter 5 in two parts to address the issue about Stieltjes functions and Pad\'{e} Approximants. Firstly we tested the reliability of the unitarization of low-energy amplitudes in the context of tha Linear Sigma Model. We saw that the predictions of the sigma mass and width in the LSM with the IAM method are too much out from the original ones. We exploited the possibility of IAM being a Pad\'{e} Approximant sequence which lie out of the application of the theorem of convergence and then leading to inappropriate results. On the contrary, the sequence $P^N_N$ displays a quick convergence behavior with reliable results for the sigma pole and width. For all this, we suggested the use of the $P^N_N$ sequence rather than $P^1_N$. Unfortunately, the study on broad resonances requires to go beyond the tree-level approximation. Thus, in the real world, the $IJ=00$ channel  needs the inclusion of loops, precluding by now the extension to PAs beyond $\cO(p^4)$ in the chiral expansion, i.e. $P^1_1$.  There is also a clear limitation on our experimental knowledge of the low-energy couplings, which barely goes beyond  $\cO(p^4)$. However, our proposal should be still suitable for the analysis of theories with narrow resonances and a relatively good knowledge of the experimental low-energy amplitudes.

In a second part of the chapter we discussed the extraction of parameters from the vacuum polarization function of a heavy quark through Pad\'e Approximants. We presented a clean way to handle the possible ambiguities of an approximation to the vacuum polarization function of a heavy quark at ${\cal O}(\alpha_s^2)$. Since the vacuum polarization function, when considered in terms of the external momentum $q^2$, is a function of the Stieltjes type, the theorem of convergence guarantees, for example, that the poles of your approximation will lie on the real axis mimicking the branch cut present on the original function. Also, with the amount of information obtained on the Taylor expansion around $q^2=0$ for that function we estimate the value for the threshold parameter $K^{(2)}$ which can not be yet computed with Feynmann diagrams.


Finally, in chapter \ref{capitol3} we have explored the reliability of fitting the pion vector form factor with rational approximations. Because these approximants are capable of describing the region of large momentum, they may be better suited than polynomials for a description of the space-like data. As our results in Table \ref{table2VMD} show, the errors achieved with these approximants are competitive with previous analysis existing in the literature. As a straightforward method, can be easily applied for analyzing other form factors in the spacelike region and also to analyze lattice data which usually covers large energy region.


% ...
\appendix

\renewcommand{\theequation}{A.\arabic{equation}} \setcounter{equation}{0}

\chapter{Dispersion relations}\label{secdr}

\quad A dispersion relation is as has been shown by K\"{a}ll\'{e}n and Lehmann \cite{Kallen2}, \cite{Lehmann:1954xi} quite a long time ago a relation that two-point functions have to obey. That dispersion relation follows from the analyticity properties of $\Pi(q^2)$ as a complex function of $q^2$, the only energy-momentum invariant which appears in a two-point function. This tool will be interesting in our calculation of the $\langle VV-AA\rangle$ correlation function in chapter \ref{capitol2}. In full generality $\Pi(q^2)$ is an analytic function in the complex $q^2$-plane but for a cut in the real axis $0\leq q^2 \leq \infty$. It then follows that (\cite{deRafaelSumRules} for a demonstration):

\be
\label{dr}
\Pi(q^2)=\int_0^{\infty}dt\frac{1}{t-q^2-i \epsilon \pi}\frac{1}{\pi}{\mbox{\rm Im}}\Pi(t) + a + bq^2+\cdots
\ee

where the degree of the arbitrary polynomial in the r.h.s. depends on the convergence properties of ${\mbox{\rm Im}}\Pi(t)$ when $t\rightarrow\infty$. The interest of this representation is that $\frac{1}{\pi}{\mbox{\rm Im}}\Pi(t)$ in the integrand, which is usually called \textit{spectral function} $\rho(t)$, is a physical cross-section. For example, with

\be
J^{\mu}(x)=\frac{2}{3}\bar{u}(x)\gamma^{\mu}u(x)-\frac{1}{3}\bar{d}(x)\gamma^{\mu}d(x)-\frac{1}{3}\bar{s}(x)\gamma^{\mu}s(x)
\ee

the electromagnetic hadronic current of light quarks, the relation to the total $e^+e^-$ annihilation cross-section into hadrons is

\be
\sigma(q^2)_{e^+e^-\rightarrow hadrons}=\frac{4 \pi^2 \alpha}{q^2}e^2\frac{1}{\pi}{{\mbox{\rm Im}} \Pi_{EM}}(q^2)
\ee

and

\be
-3\theta(q) q^2 \frac{1}{\pi} \Im \Pi_{EM}(q^2)=\sum_{\Gamma}\langle 0|J_{EM}^{\mu} (0)|\Gamma \rangle \langle \Gamma | J_{\mu,EM} (0)^{\dag} |0\rangle (2\pi)^3 \delta^{(4)}(q-p_{\Gamma})\, ,
\ee

where the sum is extended to all possible physical states, \textit{on-shell states}, with an integration over their corresponding phase space understood.

The physical meaning of the coefficients of the arbitrary polynomial in the r.h.s. of the Eq.~(\ref{dr}) depends on the choice of the local operator $J(x)$ in the two-point function. In some cases the coefficients in question are fixed by low-energy theorems.

If $\Pi(0)$ is known, we can trade the constant $a$ in Eq.~(\ref{dr}):

\be
\label{dr1}
\Pi(q^2)=\Pi(0) + \int_0^{\infty}\frac{dt}{t}\frac{q^2}{t-q^2-i \epsilon \pi}\frac{1}{\pi}{\mbox{\rm Im}}\Pi(t) + bq^2+\cdots
\ee

In other cases the constants are absorbed by renormalization constants. In general, it is always possible to get rid of the polynomial terms by taking an appropriate number of derivatives with respect to $q^2$, which is a substraction process. Eq. (\ref{dr1}), due to the derivative, is called one-substraction relation.

\renewcommand{\theequation}{B.\arabic{equation}}
\setcounter{equation}{0}
\chapter{Weinberg Sum Rules (WSR)}\label{WSR}

In the chiral limit, the first Weinberg sum rule~\cite{We67}

\be
\int_{0}^{\infty} dt \,{\mbox{\rm Im}}\Pi_{LR}(t)=0\,,
\ee

implies that

\be\label{eq:1wsr}
\sum_{V}f_{V}^2 M_{V}^2-\sum_{A}f_{A}^2 M_{A}^2=f_{\pi}^2\,;
\ee

and the second Weinberg sum rule~\cite{We67}

\be
\int_{0}^{\infty} dt\,t \,{\mbox{\rm Im}}\Pi_{LR}(t)=0\,,
\ee

furthermore implies that

\be\label{eq:2wsr}
\sum_{V}f_{V}^2 M_{V}^4-\sum_{A}f_{A}^2 M_{A}^4=0\,.
\ee

Assuming enough convergence in the ressonance region, in QCD~\footnote{For a discussion of the Weinberg sum rules in QCD in the
presence of explicit chiral symmetry breaking see ref.~\cite{FNR79}},
Eq.~(\ref{eq:1wsr}) follows from the fact that there is no local order
parameter of dimension $d=2$ and Eq.~(\ref{eq:2wsr}) from the absence of a
local order parameter of dimension $d=4$. The first possibly non trivial
contribution comes from local order parameters of dimension $d=6$, as shown
in Eq.~(\ref{eq:OPE}).


\addcontentsline{toc}{chapter}{Bibliography}
\bibliographystyle{unsrt}
\bibliography{./bibliografia}

\newpage
\listoffigures

\newpage
\listoftables
\end{document}